# Navigating phase diagram complexity to guide robotic inorganic materials synthesis

Jiadong Chen,[1+] Samuel R. Cross,[2*+] Lincoln J. Miara[2], Jeong-Ju Cho[2], Yan Wang[2], Wenhao Sun[1*]

[1] Department of Materials Science and Engineering; University of Michigan, Ann Arbor, MI, USA
[2] Advanced Materials Lab, Samsung Advanced Institute of Technology–America, Samsung Semiconductor Inc., Cambridge, MA 02138, USA.

[*]Correspondence to: sam.cross@samsung.com (S.R.C.), whsun@umich.edu (W.S.)
[+]Equal Contribution

**Abstract**

Efficient synthesis recipes are needed both to streamline the manufacturing of complex materials and to accelerate the realization of theoretically predicted materials. Oftentimes the solid-state synthesis of multicomponent oxides is impeded by undesired byproduct phases, which can kinetically trap reactions in an incomplete non-equilibrium state. We present a thermodynamic strategy to navigate high-dimensional phase diagrams in search of precursors that circumvent low-energy competing byproducts, while maximizing the reaction energy to drive fast phase transformation kinetics. Using a robotic inorganic materials synthesis laboratory, we perform a large-scale experimental validation of our precursor selection principles. For a set of 35 target quaternary oxides with chemistries representative of intercalation battery cathodes and solid-state electrolytes, we perform 224 reactions spanning 27 elements with 28 unique precursors. Our predicted precursors frequently yield target materials with higher phase purity than when starting from traditional precursors. Robotic laboratories offer an exciting new platform for data-driven experimental science, from which we can develop new inorganic materials synthesis insights for both robot and human chemists.

## Introduction

There is currently a poor scientific understanding of how to design efficient and effective synthesis recipes to target inorganic materials.[1,2,3] As a result, synthesis often becomes a bottleneck in the scalable manufacturing of functional materials,[4] as well as in the laboratory realization of computationally-predicted materials.[5,6] DFT-calculated thermodynamic stability or metastability can often approximate materials synthesizability,[7,8] but finding an optimal synthesis recipe—including temperatures, times and precursors—still requires extensive trial-and-error experimentation. The recent emergence of robotic laboratories[9,10,11] presents an exciting opportunity for high-throughput experiments and sequential-learning algorithms to autonomously optimize materials synthesis recipes.[12,13,14,15,16,17,18,19] However, there remains a poor fundamental understanding of how changing a synthesis recipe affects the underlying thermodynamics and kinetics of a solid-state reaction. Without this scientific understanding, it is difficult to build physics-informed synthesis planning algorithms to guide robotic laboratories, meaning that parameter optimization via high-throughput experiments can end up being unnecessarily resource-intensive and wasteful.

Multicomponent oxides represent an important and challenging space for targeted synthesis. These high-component materials are key to various device technologies—including battery cathodes ($Li(Co,Mn,Ni)O_2$), oxygen evolution catalysts ($Bi_2Sr_2Ca_{n-1}Cu_nO_{2n+4+x}$), high-temperature superconductors ($HgBa_2Ca_2Cu_3O_8$), solid-oxide fuel cells ($La_3SrCr_2Mn_2O_{12}$), and more.[20] Multicomponent oxides are usually synthesized by combining and firing the constituent binary oxide precursors in a furnace. However, this often yields impurity byproduct phases, which arise from incomplete solid-state reactions. From a phase diagram perspective, precursors start at the corners of a phase diagram and combine together towards a target phase in the interior of the phase diagram. If the phase diagram is complicated, *i.e.* with many competing phases between the precursors and the target, undesired phases may form, consuming thermodynamic driving force and kinetically trapping the reaction in an incomplete non-equilibrium state.

High-component oxides reside in high-dimensional phase diagrams, and can be synthesized from many possible precursor combinations. Here we present a thermodynamic strategy to navigate these multidimensional phase diagrams—identifying precursor compositions that circumvent kinetically-competitive byproducts while maximizing the thermodynamic driving force for fast reaction kinetics. We test our principles of precursor selection using a robotic inorganic materials synthesis laboratory, which automates many tedious aspects of the inorganic materials synthesis workflow, such as powder precursor preparation, ball milling, oven firing, and X-ray characterization of reaction products. Using our robotic platform, a single human experimentalist can conduct powder inorganic materials synthesis in both a high-throughput and reproducible manner. Using a diverse target set of 35 quaternary Li-, Na- and K- based oxides, phosphates and borates, which are relevant chemistries for intercalation battery cathodes[21,22] and solid-state electrolytes,[23] we show that precursors identified by our thermodynamic strategy frequently outperform traditional precursors in synthesizing high-purity multicomponent oxides. Our work demonstrates the utility of robotic laboratories not only for advanced materials manufacturing, but also as a platform for large-scale hypothesis validation over a broad and diverse chemical space.

## Principles of Precursor Selection

Recently, we showed that solid-state reactions between three or more precursors initiate at the interfaces between only two precursors at a time.[24] The first pair of precursors to react will usually form an intermediate byproduct, which can consume much of the total reaction energy and leave insufficient driving force to complete a reaction.[25] **Figure 1** illustrates this multi-step reaction progression for an example target compound $LiBaBO_3$, whose simple oxide precursors are $B_2O_3$, BaO, and $Li_2CO_3$. Because $Li_2CO_3$ decomposes to $Li_2O$ upon heating, we can examine the reaction network[26] geometrically upon a pseudo-ternary $Li_2O$-$B_2O_3$-BaO convex hull. Although the overall reaction energy for $Li_2O + BaO + B_2O_3 \rightarrow LiBaBO_3$ is large at $\Delta E$ = -336 meV/atom, there are many low-energy ternary phases along the binary slices $Li_2O$-$B_2O_3$ (**Figure 1b**, blue) and BaO-$B_2O_3$ (**Figure 1b**, green). In the initial pairwise reactions between $Li_2O + BaO + B_2O_3$, we anticipate that stable ternary Li-B-O and Ba-B-O oxides–such as $Li_3BO_3$, $Ba_3(BO_3)_2$ or others—will form rapidly due to large thermodynamic driving forces of $\Delta E \sim -300$ meV/atom. Should these low-energy intermediates form, the ensuing reaction energies to the target become miniscule, *e.g.* $Li_3BO_3 + Ba_3(BO_3)_2 \rightarrow LiBaBO_3$ has only $\Delta E = -22$ meV/atom, (**Figure 1e,** orange),

Instead of allowing the reactions to proceed between the three precursors all at once, we suggest to first synthesize $LiBO_2$, which offers a high-energy starting point for the reaction. **Figure 1g** (purple) shows that $LiBaBO_3$ can be formed directly in the pairwise reaction $LiBO_2 + BaO \rightarrow LiBaBO_3$ with a substantial reaction energy of $\Delta E$ = -192 meV/atom. Moreover, along this reaction isopleth there is a low likelihood of forming impurity phases, as the competing kink of $Li_6B_4O_9 + Ba_2Li(BO_2)_5$ has relatively small formation energy ($\Delta E$ = -55 meV/atom) compared to $LiBaBO_3$. Finally, the inverse hull energy of $LiBaBO_3$, which we define as the energy below the neighboring stable phases on the convex hull, is substantial at $\Delta E_{inv}$ = -153 meV/atom, suggesting that the selectivity of the target $LiBaBO_3$ phase should be much greater than any potential impurity byproducts along the $LiBO_2$-BaO slice.

**Figure 1i** juxtaposes the energy progression between these two precursor pathways. Although both pathways share the same total reaction energy, synthesizing $LiBaBO_3$ from three precursors is likely to first produce low-energy ternary oxide intermediates (**Figure 1a**), leaving little reaction energy to complete the reaction kinetics to the target phase. By first synthesizing a high-energy intermediate ($LiBO_2$), we retain a large fraction of overall reaction energy for the last step of the reaction, promoting the rapid and efficient synthesis of the target phase. We confirm this hypothesis experimentally (**Figure 1j**), where we find that solid-state synthesis of $LiBaBO_3$ from the traditional precursors $Li_2CO_3$, $B_2O_3$ and BaO do not result in any XRD signal of the target phase, whereas $LiBO_2$ + BaO produces $LiBaBO_3$ with high phase purity (**Methods**).

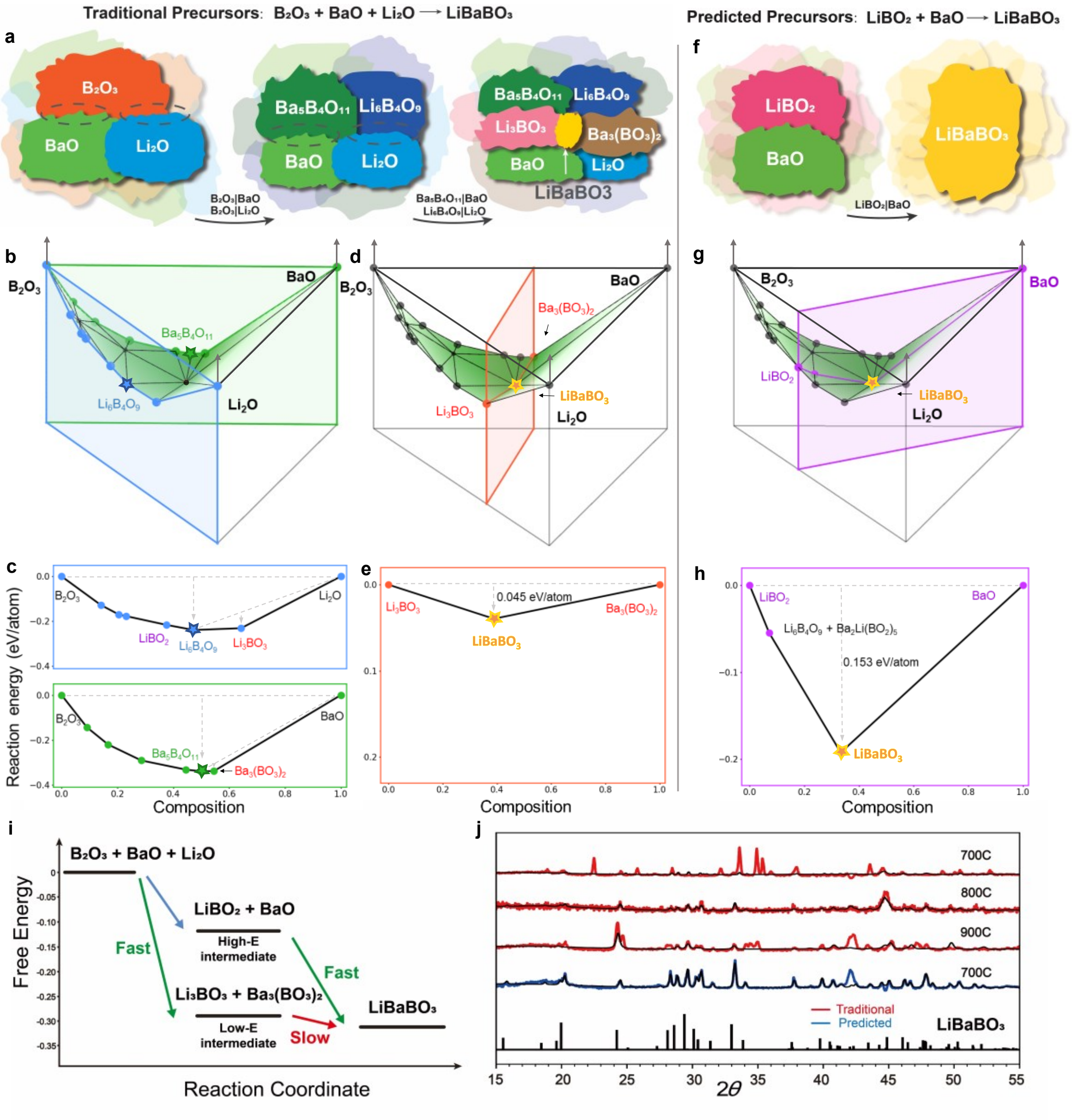

**Figure 1**. Comparison between the traditional reaction ($Li_2O$, $B_2O_3$, and BaO) process and our designed reaction ($LiBO_2$ and BaO) process for $LiBaBO_3$. **a–e)** are for the traditional reaction. **f–h)** are for the predicted reaction. **a**,**f)** Schematic of pairwise reactions process, showing the phase evolution from precursors to the target. **b**,**d**,**g)** are pseudo-ternary $Li_2O$-$B_2O_3$-BaO convex hulls, where reaction convex hulls between precursor pairs are illustrated by colored slices. **c**,**e**,**h)** 2-dimensional slices of the binary reaction convex hulls. Grey arrows show the reaction energy of the corresponding reaction. **i)** Free energy change in a reaction progress, where a relatively high-energy intermediate state saves more energy for the final step in forming the target. **j)** XRD of the solid-state synthesis of $LiBaBO_3$, where red and blue curves are raw XRD data for traditional and predicted precursors, respectively, and the black curve is the fit produced by the Rietveld refinement.

From this instructive $LiBaBO_3$ example, we propose five principles to select effective precursors from a multicomponent convex hull: **1**) Reactions should initiate between only 2 precursors if possible, minimizing the chances of simultaneous pairwise reactions between 3 or more precursors. **2**) Precursors should be relatively high-energy, as these unstable precursors maximize the thermodynamic driving force and thereby the reaction kinetics to the target phase. **3**) The target material should be the deepest point in the reaction convex hull, promoting its kinetic selectivity. **4**) The composition slice formed between the two precursors should intersect as few other competing phases as possible, minimizing the opportunity to form undesired reaction byproducts, and **5**) If byproduct phases are unavoidable, the target phase should have a relatively large inverse hull energy—in other words, the target phase should be substantially lower in energy than its neighboring stable phases in composition space.

On **Figure 2**, we interpret these precursor design principles for an example $LiZnPO_4$ target in the pseudo-ternary $Li_2O$-$P_2O_5$-ZnO phase diagram. If we first synthesize $Zn_2P_2O_7$ to combine with $Li_2O$ (**Figure 2a,b** blue), the deepest point in the reaction convex hull is not $LiZnPO_4$ but rather is ZnO + $Li_3PO_4$, suggesting a kinetic propensity to form these undesired byproducts. If we start from $Zn_3(PO_4)_2$ + $Li_3PO_4$ (**Figure 2c,d**, orange), $LiZnPO_4$ is located at the deepest point along the convex hull; however $Li_3PO_4$ is a low-energy starting precursor, meaning there is a small driving force ($\Delta E$ = -40 meV/atom) left to form $LiZnPO_4$, likely leading to slow reaction kinetics. We suggest that $LiPO_3$ + ZnO (**Figure 2e,f**, purple) are the ideal precursors for $LiZnPO_4$. $LiPO_3$ has a relatively high energy along the $Li_2O$-$P_2O_5$ binary hull, resulting in a large driving force to the target phase of $\Delta E$ = -106 meV/atom. Additionally, there are no competing phases along the $LiPO_3$ + ZnO slice, minimizing the possibility of impurity byproduct phases.

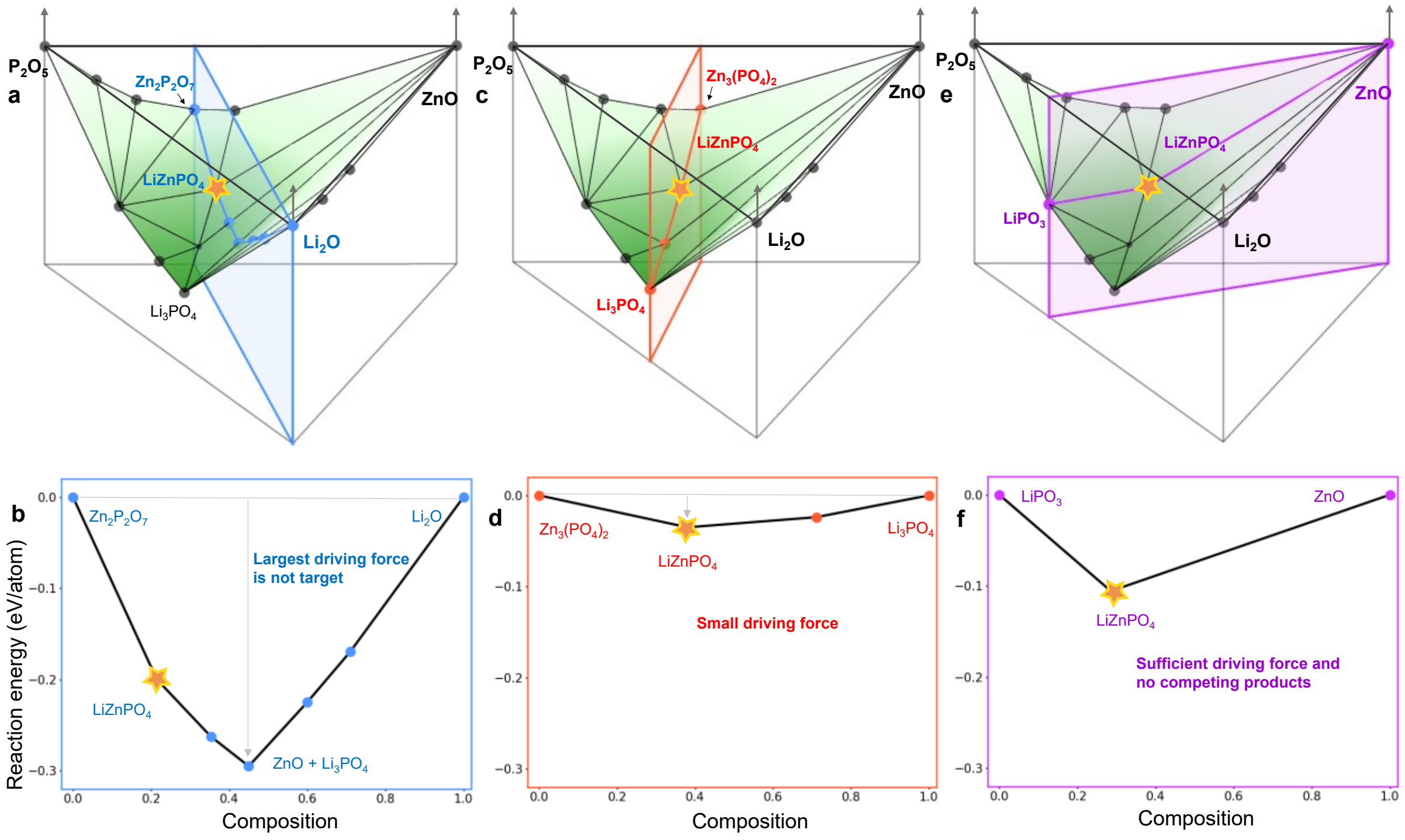

**Figure 2.** Comparison of three pairwise reactions for the synthesis of $LiZnPO_4$ on the pseudo-ternary $Li_2O$-$P_2O_5$-ZnO convex hull. **a**,**c**,**e** The blue, red, and purple slice planes correspond to $Zn_2P_2O_7$ + $Li_2O$, $Zn_3(PO_4)_2$ + $Li_3PO_4$, and $LiPO_3$ + ZnO binary reaction convex hulls, respectively. **b**,**d**,**e** are the corresponding 2-dimensional slices.

### Validation with a robotic ceramic synthesis laboratory

To test our precursor selection hypotheses, we design a large-scale experimental validation effort based in the quaternary Li-, Na-, and K-based oxides, phosphates and borates, which are representative chemistries for intercalation battery materials.[21,23,23] We survey the Materials Project[27] for all known quaternary compounds in this space, then we use our selection principles to predict optimal precursors from the DFT-calculated convex hulls. We also determine the traditional precursors for these reactions, which we previously text-mined from the solid-state synthesis literature.[28] A full list of 3104 reactions in this space are provided in the **Supplementary Data**, and our algorithm to construct these reactions is detailed in **SI.1**. To efficiently maximize the coverage of our experimental validation, we Pareto-optimized our reaction list to select the fewest number of precursors that maximize the number of candidate reactions—resulting in 28 unique precursors for 35 target materials that span 27 elements.

We then compare the phase purity of target materials synthesized from our predicted precursors versus from traditional precursors. We perform this large-scale validation effort using a robotic inorganic materials synthesis laboratory named ASTRAL (Automated Synthesis Testing and Research Augmentation Lab), located at the Samsung Advanced Institute of Technology in Cambridge, Massachusetts. As shown in **Figure 3**, ASTRAL uses a robotic arm to automate sample handling throughout a full ceramic synthesis workflow—from powder precursor preparation to ball milling, to oven firing, to X-ray characterization of reaction products. One tray of 24 samples can pass through the workflow every 72 hours.

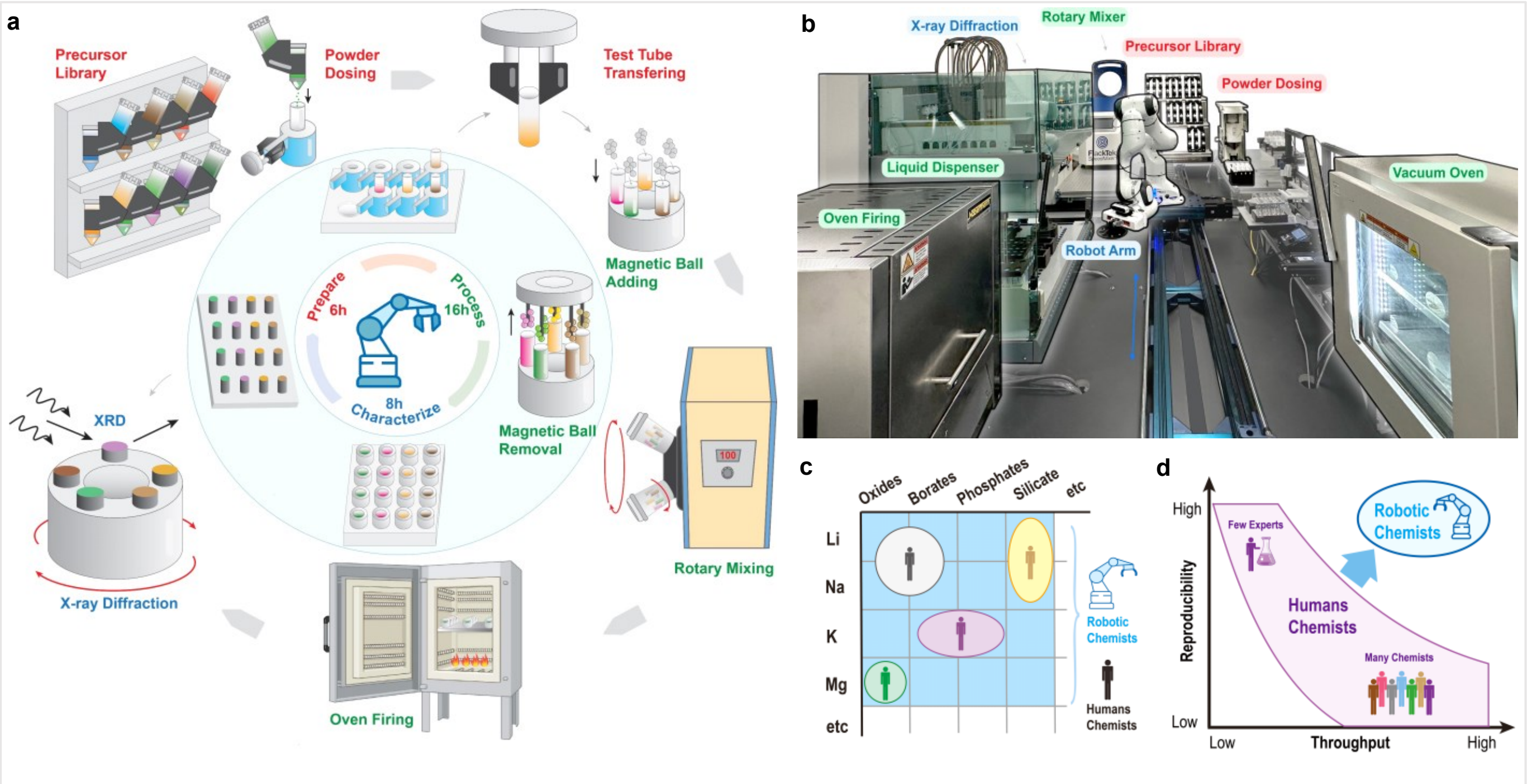

**Figure 3**, **Automated Synthesis Testing and Research Augmentation (ASTRAL) Lab at Samsung's Advanced Materials Lab in Cambridge, Massachusetts. a)** A robot-enabled inorganic materials synthesis workflow—from powder precursor preparation to ball milling, to oven firing, to X-ray characterization of reaction products; **b)** picture of the ASTRAL Lab **c)** Robotic chemists enable a paradigm of large-scale exploration of synthesis hypotheses over a broad chemical space, which normally would have to be undertaken by multiple experimentalist groups. **d)** Human experimentalists have a trade-off between throughput and reproducibility, whereas robotic chemists can achieve both high reproducibility and throughput simultaneously.

ASTRAL automates inorganic materials synthesis from powder precursors, as opposed to previous robotic laboratories that rely on solution-based precursors,[14,15,16,29] inkjet printing[17] or combinatorial thin-film deposition.[13,18] Although it is easier to dose precursor concentrations using these other methods, the resulting products are typically only produced at milligram scale. Powder synthesis, on the other hand, can yield grams of material, which is needed to create ceramic pellets or electrodes for functional property characterization. Moreover, high-temperature powder synthesis is the primary synthesis method of ceramic oxides, so recipes determined from ASTRAL can be upscaled for industrial manufacturing. We overcame major practical challenges in powder precursor processing, which arise primarily from flowability differences between different powders due to varying particle sizes, hardness, hygroscopicity, and compaction. In **Table S2** we summarize the challenges in working with powder precursors, as well as our solutions to these challenges. The essential task is to identify the best dosing head for each precursor, as detailed in **Table S3** for our precursors.

The yield and purity of the target phase was determined using automated Rietveld refinement. Because we have a pre-specified target material with a known crystal structure, the yield of the target phase can be quantified by the ratio of its integrated XRD counts versus the integrated residual. We did not fully characterize all impurity phases,[30,31] as our scientific investigation here is concerned mainly with the relative phase fraction of the target phase. In **Figure S11**, we benchmarked for over 200 pre-solved Rietveld refinement cases (with fully characterized impurity phases) that our automated X-ray refinement classification accurately determines the target phase fraction within 10% of the manually Rietveld-refined phase fraction. For this reason, we ascribe a 10% error bar on the phase purities of our target phases.

In total, we conducted 224 synthesis reactions over 35 target materials, calcined at temperatures from 600°-1000°C. For a target space this diverse, traditional validation of our precursor selection principles would likely have required an extended experimental effort, comprised of multiple human experimentalists working over many years. Once the robotic laboratory is set up, we can comprehensively survey this broad crystal chemistry space in a single experimental campaign (**Figure 3c**). Moreover, a large-scale human effort will inevitably require trade-offs between throughput and reproducibility. Meanwhile, a robotic laboratory produces single-source experimental data with high reproducibility, meaning we can systematically compare synthesis results while minimizing human variability and error (**Figure 3d**). Altogether, the robotic laboratory offers a new platform for data-driven empirical synthesis science, where hypotheses can be investigated rapidly, reproducibly, and comprehensively over diverse crystal chemistries.

## Results and Discussion

For the 35 materials selected, **Figure 4a** shows the relative yield of the target phase starting from computationally-designed versus traditional precursors. **Figure 4b** shows the reaction temperatures attempted, and **Figure 4c** shows the relative performance of the predicted versus traditional precursors. A full list of targets, precursors and reaction results are listed in **Table S4**. For 32 out of 35 compounds (91%), the predicted precursors successfully produce the target phase. In 15 targets, the predicted precursors achieve at least 20% higher phase purity than the traditional precursors (green), and 6 of these 15 target materials could *only* be synthesized by the predicted precursors (dark green). For 16 reactions the precursors have similar target yields (light green), and only in 4 systems do the traditional precursors perform better than the predicted precursors (red). However, we note that even in these 4 systems, the predicted precursors also produce the target materials with moderate to high purities. In 3 systems, neither set of precursors resulted in a target material, which was due to glass formation for $NaBSiO_4$,[32] needing a more reducing atmosphere for $Li_3V_2(PO_4)_3$,[33] and for $NaBaBO_3$ the published reaction temperature was very precise at 790°C,[34] suggesting that perhaps a rounded number like 800°C may be too high. As discussed further in **SI3.4**, these scenarios represent important considerations in future robotic laboratory design.

**a**

| Compound | Best predicted synthesis | Precursor comparison |
|---|---|---|
| $BaLiBO_3$ | High purity | Successful only for predicted |
| $K_2Zr(PO_4)_2$ | | |
| $Li_3Pr_2(BO_3)_3$ | | |
| $KNiPO_4$ | | |
| $KLi(PO_3)_2$ | Moderate purity | |
| $LiNbWO_6$ | Low purity | |
| $Li_3Sc_2(PO_4)_3$ | High purity | Higher purity for predicted |
| $LiGeBO_4$ | | |
| $K_3Bi_2(PO_4)_3$ | | |
| $LiZnBO_3$ | | |
| $K_3Fe_2(PO_4)_3$ | | |
| $KMgPO_4$ | | |
| $K_3LiP_2O_7$ | | |
| $Na_2Al_2B_2O_7$ | | |
| $K_3Al_2(PO_4)_3$ | | |
| $Li_2CuP_2O_7$ | | Similar purities |
| $LiNbGeO_5$ | | |
| $LiZnPO_4$ | | |

| Compound | Best predicted synthesis | Precursor comparison |
|---|---|---|
| $SrLiBO_3$ | High purity | Similar purities |
| $LiMgPO_4$ | | |
| $KTiPO_5$ | | |
| $LiMnPO_4$ | | |
| $KBaPO_4$ | | |
| $Li_3Fe_2(PO_4)_3$ | | |
| $KNbWO_6$ | | |
| $KTiNbO_5$ | Low purity | |
| $Li_3Y_2(BO_3)_3$ | | |
| $KTa_2PO_8$ | | |
| $BaNaBO_3$ | Not detected | |
| $Li_3V_2(PO_4)_3$ | | |
| $NaSiBO_4$ | | |
| $Li_2TiGeO_5$ | High purity | Higher purity for traditional |
| $Li_2TiSiO_5$ | | |
| $NaSrBO_3$ | Moderate purity | |
| $LiSi_2BO_6$ | | |

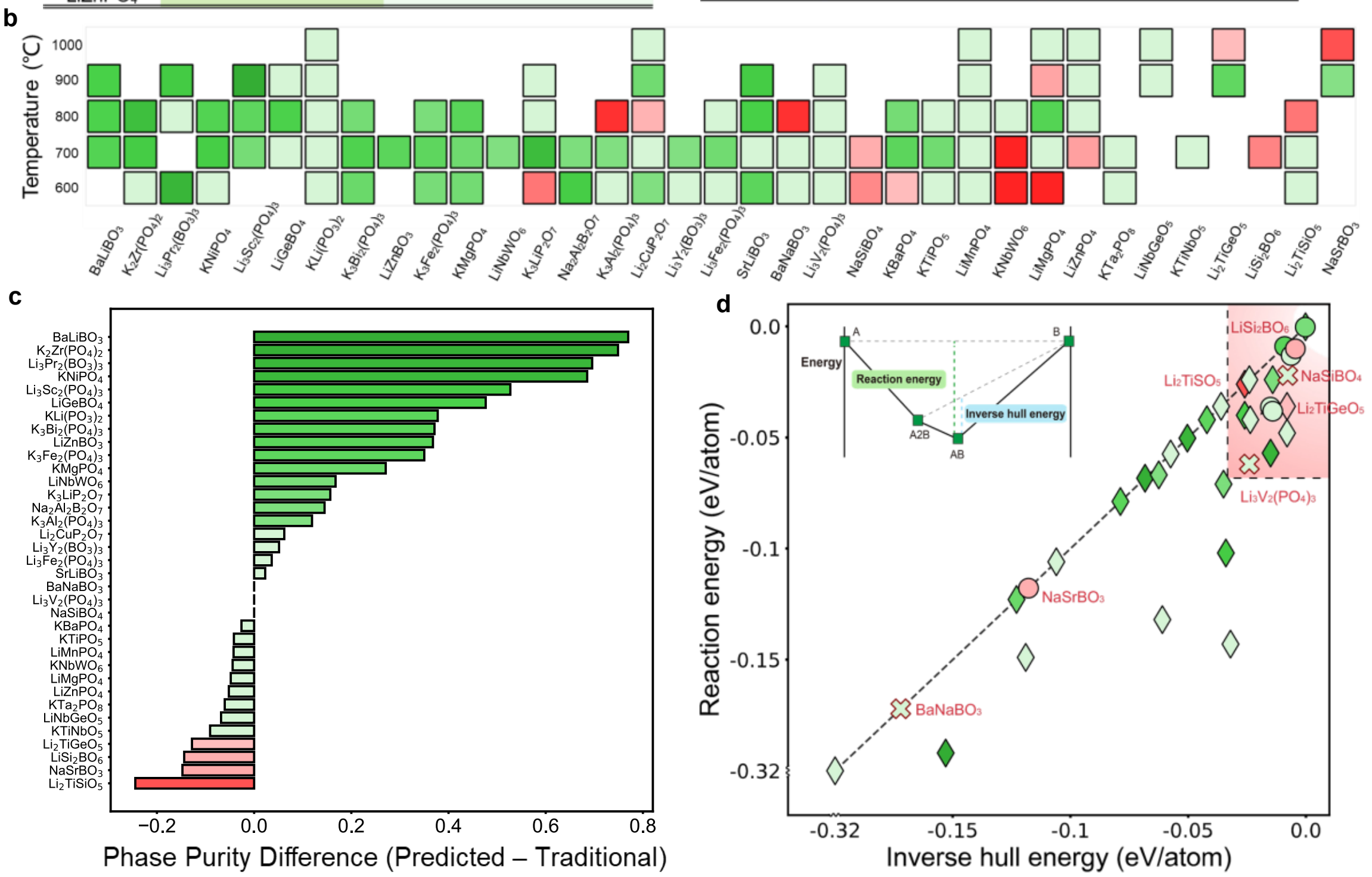

**Figure 4. Robotic synthesis results of target materials from traditional versus predicted precursors. a)** Table of the phase purity of 35 targets obtained from predicted precursors using the highest phase purity from various firing temperatures, compared to traditional precursors. Color of "Precursor comparison" column compares purity from predicted precursors versus traditional, where green means predicted precursors achieve >10% better purity, light green means they have purities within ±10%, and red means traditional precursors achieve >10% better purity. Same color scheme is used in b, c, d. **b)** Heatmap of phase purity of predicted precursors at different calcination temperatures. **c)** shows the target phase purity from predicted precursors versus traditional precursors. Phase purity methods in **SI2.2.3**. **d)** Reaction energies and inverse hull energies for all targets. Marker shape corresponds to best phase purity of predicted precursors, where diamonds are high purity, circles are moderate and low purity, and crosses with red outline means both predicted precursors and traditional precursors failed. The dashed line represents when inverse hull energy equals reaction energy. *Inset*: Convex hull illustrating the reaction energy and the inverse hull energy.

**Figure 4c** shows that our predicted precursors tend to synthesize target materials with higher purity than traditional simple oxide precursors. Many of our predicted ternary oxide precursors are unusual—such as $LiPO_3$, $LiBO_2$, $LiNbO_3$ and more in **Table S4**—as they do not appear as precursors from our previously text-mined database of 19,488 solid-state synthesis recipes.[35] Machine-learning algorithms for synthesis prediction trained on literature datasets would therefore be unlikely to predict our precursors here. This highlights the limitations of machine-learning algorithms in predicting new opportunities in synthesis parameter space, outside the constraints of our anthropogenic biases in chemical reaction data.[36]

Our results show that the success of a reaction was not correlated to the crystal structure or chemistry of the target material—rather, it was primarily determined by the geometry of the underlying convex hull, as well as the magnitude of the thermodynamic driving force. The success of our precursor selection principles is somewhat surprising, considering we evaluate precursor selection using only the DFT-calculated convex hull—which does not account for temperature-dependent effects such as vibrational entropy or oxide decomposition; neglects kinetic considerations such as diffusion rates and nucleation barriers,[37] and has known errors in DFT-calculated formation energies.[38]

Here we rationalize with order-of-magnitude energy arguments why, despite many simplifying assumptions, the DFT-calculated thermodynamic convex hull retains predictive power in identifying effective precursors. First, entropic contributions can generally be neglected because the $\Delta G$ of an oxide synthesis reaction is usually dominated by the $\Delta H$ contribution, rather than the $T\Delta S$ contribution. **Figure S13** compiles a list of 100 experimental ternary oxide reaction energies, and shows that at 1000K the magnitude of $|\Delta G|$ for reactions are ~200 meV/atom, whereas the $|T\Delta S|$ contribution is only ~15 meV/atom. In 60% of the reactions, $|T\Delta S|/|\Delta G| < 10\%$, except in cases where $|\Delta G| < 100$ meV/atom, in which case $T\Delta S$ can be comparable in magnitude to $\Delta H$. The dominance of $\Delta H$ over $T\Delta S$ in oxide synthesis reactions is due to the irreversible exothermic nature of reactions of the form $A + B \rightarrow AB$; as opposed to first-order phase transitions like melting or polymorphic transformations, where $\Delta H \sim T\Delta S$.

Second, ternary convex hulls are often skewed such that certain hull directions are much deeper than others, such as the $Li_2O$-$B_2O_3$ and $BaO$-$B_2O_3$ directions illustrated on the $Li_2O$-$BaO$-$B_2O_3$ convex hull in **Figure 1** (more examples in **S.I.3**). On a high-dimensional phase diagram, there are many combinations of precursor pairs that can slice through a target phase. Even an approximate convex hull, with systematic DFT formation energy errors of 25 meV/atom,[38] can largely capture the relative depths of the convex hull in various compositional directions, as well as the complexity of the hull arising from competing phases. Importantly, DFT is well-poised to capture the very stable phases, which are low-energy thermodynamic sinks to be avoided when designing the reaction isopleths between pairs of precursors.

Finally, although we do not explicitly calculate kinetics here, the magnitude of the thermodynamic driving force is a good proxy for phase transformation kinetics, as $\Delta G_{reaction}$ appears in the denominator of the classical nucleation barrier, as supersaturation in the JMAK theory of crystal growth, and as $d\mu/dx$ in Fick's first law of diffusion.[39] Because we aim to evaluate the *relative* reaction kinetics of different precursors, rather than absolute kinetics, we can usually compare thermodynamic driving forces between different precursor sets without explicitly calculating diffusion barriers[40] or surface energies for nucleation and growth analyses.[41,42]

However, there are limits to this assumption. **Figure 4d** shows the reaction energy and inverse hull energy for all 35 reactions using predicted precursors, among which 2 of the unsuccessful syntheses are marked with a cross, and 4 red markers indicate conditions where the traditional precursors outperformed the predicted precursors. In cases where our predicted precursors were less successful, the reaction energy landscapes were shallow with $\Delta E_{reaction} > -70$ meV/atom, and inverse hull energies of $\Delta E_{IH} > -50$ meV/atom. Because these driving forces are on the order of $k_B T$ at solid-state synthesis temperatures (~1000K), unanticipated kinetic processes may become rate-limiting and disqualify our thermodynamic driving force arguments. These counterexamples provide valuable 'failed synthesis' results[43] to quantify bounds where our precursor selection principles offer less certainty of success.

**Outlook**

Synthesis science is poorly understood, but new theories can be developed by examining falsifiable predictions through empirical validation. In this work, we hypothesized several principles to identify superior precursors for high-purity synthesis of multicomponent oxides. We argued that in high-dimensional phase diagrams with skewed energy landscapes, there is an opportunity to find precursors that are both high in energy and have compositions that circumvent low-energy undesired kinetic byproducts. Using a robotic synthesis laboratory, we validated this hypothesis over 35 target materials with diverse crystal chemistries, producing in this one study as many experimental results as a typical review paper might survey. This work heralds a new paradigm of data-driven experimental synthesis science, where the high throughput and reproducibility of robotic laboratories enable a more comprehensive interrogation of synthesis science hypotheses. This exciting robotic platform can be directed to investigate further fundamental questions, such as the role of temperatures and reaction times in ceramic oxide synthesis. As we use these robotic laboratories to verify human-designed hypotheses, we will deepen our fundamental understanding of the interplay between thermodynamics and kinetics during materials formation. Simultaneously, this scientific understanding will drive the development of physically-informed AI synthesis planning frameworks to enable truly autonomous materials processing and manufacturing.

## METHODS

### DFT convex hulls for precursor identification:

Material phases and formation energies are obtained from the Materials Project[44] using its REST API,[45] retrieved from the December 2020 version of the database. Convex hulls are constructed from the phase diagram package in Pymatgen,[46] reaction convex hulls are calculated from the interfacial reactions package.[47] Software for producing interactive reaction compound convex hulls can be found on Github at the following link: https://github.com/dd-debug/synthesis_planning_algorithm

### Robotic Laboratory:

ASTRAL transports samples between stations using two robots, a 7-axis Panda robotic arm (Franka Emika), and a linear rail (Vention.io). By using the rail system to extend the range of the Panda arm, the system can perform precise laboratory manipulations over a 1.7m × 4m area. Surrounding the central rail system are stations that perform specialized tasks for inorganic materials synthesis, such as dispensing solid powder precursor chemicals and liquid dispersants, a mechanical ball-mill, furnace to calcine and react precursors, and X-ray diffraction to characterize synthesis outcomes. Precursor powders are dispensed sequentially using a Quantos powder dispenser (Mettler Toledo), with sample vials and powder dosing heads exchanged using the robotic arm. Following powder dispensing, 1mL of ethanol is dispensed into each vial using a Freedom EVO 150 liquid handling robot (Tecan Life Sciences), followed by rotary ball milling for 15h at 100rpm to produce a uniform fine mixture of precursor powders. Alumina crucibles (Advalue Technology) are used to hold the mixed precursors. After ball-milling, samples are heated to 80°C for 2h under vacuum to remove ethanol, then transferred to a furnace for calcination in air atmosphere for 8 hours at temperatures from 600°-1000°C. Powders are then characterized via powder XRD (Rigaku Miniflex 600). Further details on the robotic infrastructure are provided in Supplemental Information 2.

### Automated XRD refinement:

Rietveld refinement of data is performed in the BGMN kernel.[48] The target structure is used as the sole input phase for the BMGN kernel, and the Rietveld refinement will split the XRD signal into the target phase, background, and residual. The background XRD pattern is determined from empty sample holders. The fraction of the target phase is estimated by dividing the integrated intensity of the target phase by the combined intensity of the target phase and residual, $I_{target}/(I_{target} + I_{residual})$. Values greater than 0.5 are considered high purity, between 0.2 and 0.5 moderate, and less than 0.2 considered low purity.

## Acknowledgements

This work was supported by the U.S. Department of Energy (DOE), Office of Science, Basic Energy Sciences (BES), under Award #DE-SC0021130. We thank Jihye Morgan for contribution to the development of Samsung ASTRAL. WS thanks S.Y. Chan for important discussions and support. The authors have no conflicts of interest to declare.

[25] Bianchini, Matteo, et al. "The interplay between thermodynamics and kinetics in the solid-state synthesis of layered oxides." *Nature materials* 19.10 (2020): 1088-1095.

[26] McDermott, Matthew J., Shyam S. Dwaraknath, and Kristin A. Persson. "A graph-based network for predicting chemical reaction pathways in solid-state materials synthesis." *Nature communications* 12.1 (2021): 3097.

[27] Jain, Anubhav, et al. "Commentary: The Materials Project: A materials genome approach to accelerating materials innovation." *APL materials* 1.1 (2013): 011002.

[28] He, Tanjin, et al. "Similarity of precursors in solid-state synthesis as text-mined from scientific literature." *Chemistry of Materials* 32.18 (2020): 7861-7873.

[29] Yoshikawa, Naruki, et al. "Digital pipette: Open hardware for liquid transfer in self-driving laboratories." (2023).

[30] Szymanski, Nathan J., et al. "Probabilistic deep learning approach to automate the interpretation of multi-phase diffraction spectra." *Chemistry of Materials* 33.11 (2021): 4204-4215.

[31] Maffettone, Phillip M., et al. "Crystallography companion agent for high-throughput materials discovery." *Nature Computational Science* 1.4 (2021): 290-297.

[32] Feltz, A., and P. Büchner. "Structure and ionic conduction in solids: I. Na+-ion conducting glasses in the systems NaBSiO4-Na2SiO3, NaBSiO4-Na4SiO4 and NaBSiO4-Na3PO4." *Journal of non-crystalline solids* 92.2-3 (1987): 397-406.

[33] Chen, Shanhua, et al. "Kinetic Studies on the Synthesis of Monoclinic Li3V2 (PO4) 3 via Solid-State Reaction." *The Journal of Physical Chemistry A* 118.21 (2014): 3711-3716.

[34] Zhong, Jiyou, et al. "Understanding the blue-emitting orthoborate phosphor NaBaBO 3: Ce 3+ through experiment and computation." Journal of Materials Chemistry C 7.3 (2019): 654-662.

[35] Kononova, Olga, et al. "Text-mined dataset of inorganic materials synthesis recipes." *Scientific data* 6.1 (2019): 203.

[36] Jia, Xiwen, et al. "Anthropogenic biases in chemical reaction data hinder exploratory inorganic synthesis." *Nature* 573.7773 (2019): 251-255.

[37] Aykol, Muratahan, Joseph H. Montoya, and Jens Hummelshøj. "Rational solid-state synthesis routes for inorganic materials." *Journal of the American Chemical Society* 143.24 (2021): 9244-9259.

[38] Hautier, Geoffroy, et al. "Accuracy of density functional theory in predicting formation energies of ternary oxides from binary oxides and its implication on phase stability." *Physical Review B* 85.15 (2012): 155208.

[39] Balluffi, Robert W., Samuel M. Allen, and W. Craig Carter. *Kinetics of materials*. John Wiley & Sons, 2005.

[40] Cosby, Monty R., et al. "Thermodynamic and Kinetic Barriers Limiting Solid-State Reactions Resolved through In Situ Synchrotron Studies of Lithium Halide Salts." *Chemistry of Materials* (2023).

[41] Sun, Wenhao, et al. "Nucleation of metastable aragonite CaCO3 in seawater." *Proceedings of the National Academy of Sciences* 112.11 (2015): 3199-3204.

[42] Sun, Wenhao, et al. "Non-equilibrium crystallization pathways of manganese oxides in aqueous solution." *Nature communications* 10.1 (2019): 573.

[43] Raccuglia, Paul, et al. "Machine-learning-assisted materials discovery using failed experiments." *Nature* 533.7601 (2016): 73-76.

[44] Jain, Anubhav, et al. "Commentary: The Materials Project: A materials genome approach to accelerating materials innovation." *APL materials* 1.1 (2013): 011002.

[45] Ong, Shyue Ping, et al. "The Materials Application Programming Interface (API): A simple, flexible and efficient API for materials data based on REpresentational State Transfer (REST) principles." *Computational Materials Science* 97 (2015): 209-215.

[46] Ong, Shyue Ping, et al. "Python Materials Genomics (pymatgen): A robust, open-source python library for materials analysis." *Computational Materials Science* 68 (2013): 314-319.

[47] Richards, William D., et al. "Interface stability in solid-state batteries." *Chemistry of Materials* 28.1 (2016): 266-273.

[48] Doebelin, Nicola, and Reinhard Kleeberg. "Profex: a graphical user interface for the Rietveld refinement program BGMN." *Journal of applied crystallography* 48.5 (2015): 1573-1580.

**Navigating phase diagram complexity to guide robotic inorganic materials synthesis**

Jiadong Chen,[1+] Samuel R. Cross,[2*+] Lincoln J. Miara[2], Jeong-Ju Cho[2], Yan Wang[2], Wenhao Sun[1*]

[1] Department of Materials Science and Engineering; University of Michigan, Ann Arbor, MI, USA
[2] Advanced Materials Lab, Samsung Advanced Institute of Technology–America, Samsung Semiconductor Inc., Cambridge, MA 02138, USA.

*Corresponding Authors: sam.cross@samsung.com, whsun@umich.edu
[+]Equal Contribution

# Supplementary Information

**Contents**

# SI1. Synthesis planning algorithm

The code used to predict precursors for more efficient synthesis is open-sourced at [https://github.com/dd-debug/synthesis_planning_algorithm](https://github.com/dd-debug/synthesis_planning_algorithm). The code is built in python, and leverages the Materials Project Application Programming Interface (API) and the pymatgen code base, specifically, the `pymatgen.analysis.phase_diagram` and `pymatgen.analysis.interface_reactions` modules.

Compositions and energies of various materials systems were retrieved from the Materials Project using the REST API in December 2020.

To determine the list of 3104 reactions in the supplementary data, along with the precursors predicted using our design principles, we first collect all quaternary oxides with Li-, Na-, and K- cations, including quaternary oxides that have complex phosphate $(PO_4)^{3-}$ and borate $(BO_3)^{3-}$ anions.

For a given *A*-*B*-*C*-O quaternary oxide convex hull, for each quaternary oxide, we enumerate all pairwise precursor combinations that can form these candidate target phases. In this study, we only considered candidate targets that fall on an isopleth between a pair of precursors. It is not generally the case that two precursors will be available for each target oxide. We exclude reactions that consider elemental $O_2$ as a precursor. In the convex hull, each pairwise reaction corresponds to the slice plane between the pairwise precursors, which intersects the target. This approach determines all compositionally feasible pairwise reactions for the formation of all candidate quaternary oxide targets.

The list is the further sieved by identifying reactions where the target material is the deepest point in the reaction convex hull (as calculated from the interface_reactions module). We also evaluate the *inverse hull energy* of each phase, defined as the energetic extent by which the target phase is below its neighboring stable phases in the convex hull. The Inverse Hull Energy is illustrated in Figure **S1** for the target $Li_3Sc_2(PO4)_3$ phase from the precursors $LiPO_3$ + $Sc_2O_3$. Of the two possible reactions that could form $Li_3Sc_2(PO4)_3$, which are $3LiPO_3 + Sc_2O_3 \rightarrow Li_3Sc(PO_4)_3$ and $2ScPO_4 + Li_3PO_4 \rightarrow Li_3Sc_2(PO_4)_3$, we hypothesize that $3LiPO_3 + Sc_2O_3$ will be the best precursors, due to its large inverse hull energy.

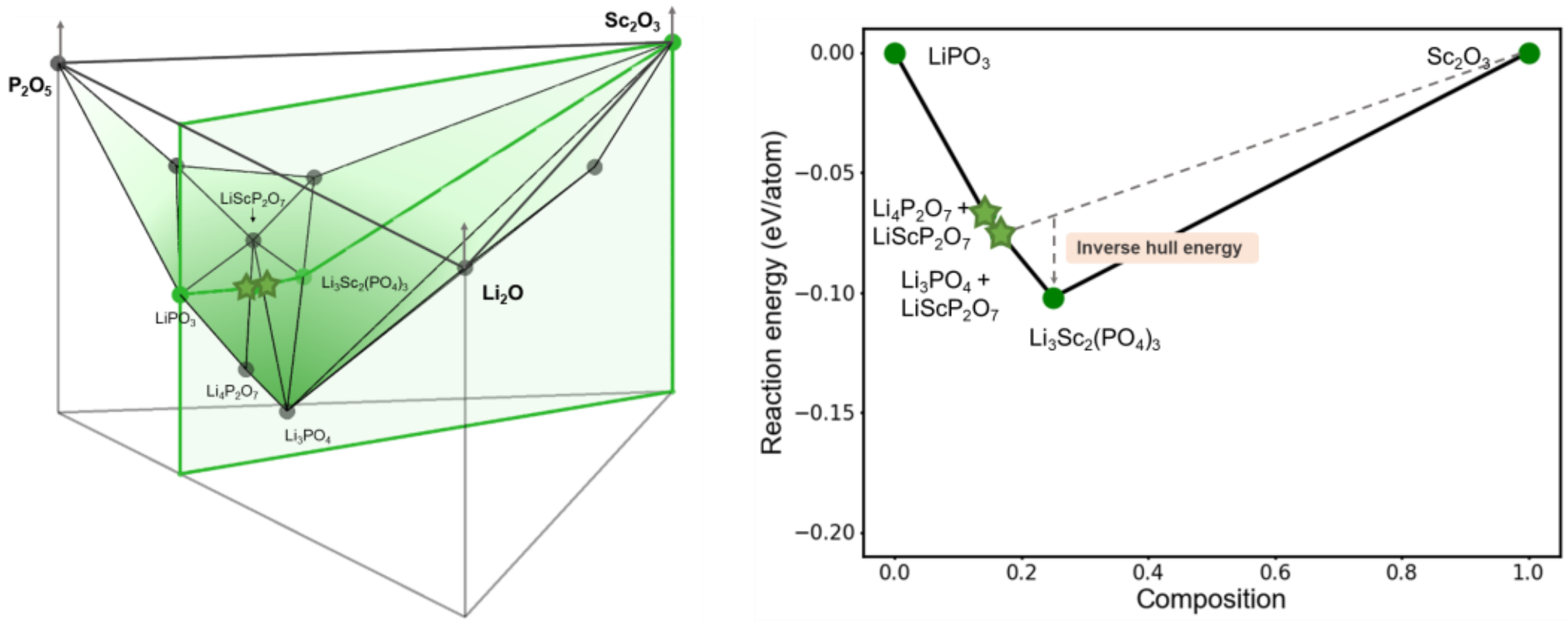

**Figure S1.** Reaction compound convex hull of $Li_3Sc_2(PO_4)_3$. **Left.)** the convex hull of $P_2O_5$, $Sc_2O_3$, and $Li_2O$, where two kinks (green stars) represent the decomposition reactions that might happen at given compositions. The equilibrium phase is a 2-phase coexistence. The green slice plane corresponds to **Right.)** $LiPO_3|Sc_2O_3$ convex hull.

The inverse hull energy is computed using the reaction convex hull from interface_reactions, where we identify the kinks in the convex hull that compete with the target compound. Because this is a 1-dimensional compositional intersection with a 3-dimensional quaternary phase diagram, the intersection can include critical compositions that correspond to single phases, or tie lines between 2 phases.

If the intersected tie line is the deepest point in the reaction convex hull, we anticipate the reaction will form the terminal phases of the tie line, such as green stars will decompose to $Li_4P_2O_7$ + $LiScP_2O_7$ and $Li_3PO_4$ + $LiScP_2O_7$ in **Figure S1**.

In executing this algorithm over the Li-, Na- and K- containing quaternary oxides, borates and phosphates, we identified 3104 reactions. We then determined the minimum set of precursors that would maximize the number of potential candidate reactions, whilst also considering the available precursors available on hand at Samsung. This process led to the target materials and precursor selections presented in this work.

## SI2. Robotic laboratory setup and procedures

### SI2.1. ASTRAL system overview

The ASTRAL platform developed at the Samsung Advanced Materials Lab (AML) is a robotic system designed to perform high-throughput automated synthesis of inorganic materials, in order to accelerate the research and development of new materials of technological interest. To the best of our knowledge, ASTRAL is the first robotic system that automates inorganic ceramic synthesis from powder precursors. To develop ASTRAL, we overcame major practical challenges in powder precursor processing, and the challenges and solutions in powder ceramic synthesis for automated laboratory are shown in **Table S2**.

The ASTRAL system is centered around a flexible collaborative robot arm mounted on a linear rail, which is able to perform dexterous manipulation tasks and transport samples throughout the system. Surrounding the central rail system are several stations that perform specialized tasks needed for the synthesis process, such as dispensing solid powder precursor chemicals, dispensing liquid chemicals, heat treatment to calcine and react precursors, and X-ray diffraction to characterize synthesis outcomes. The layout of the ASTRAL platform and robotic coverage area are illustrated in 3D-model of the system shown in **Figure S4**.

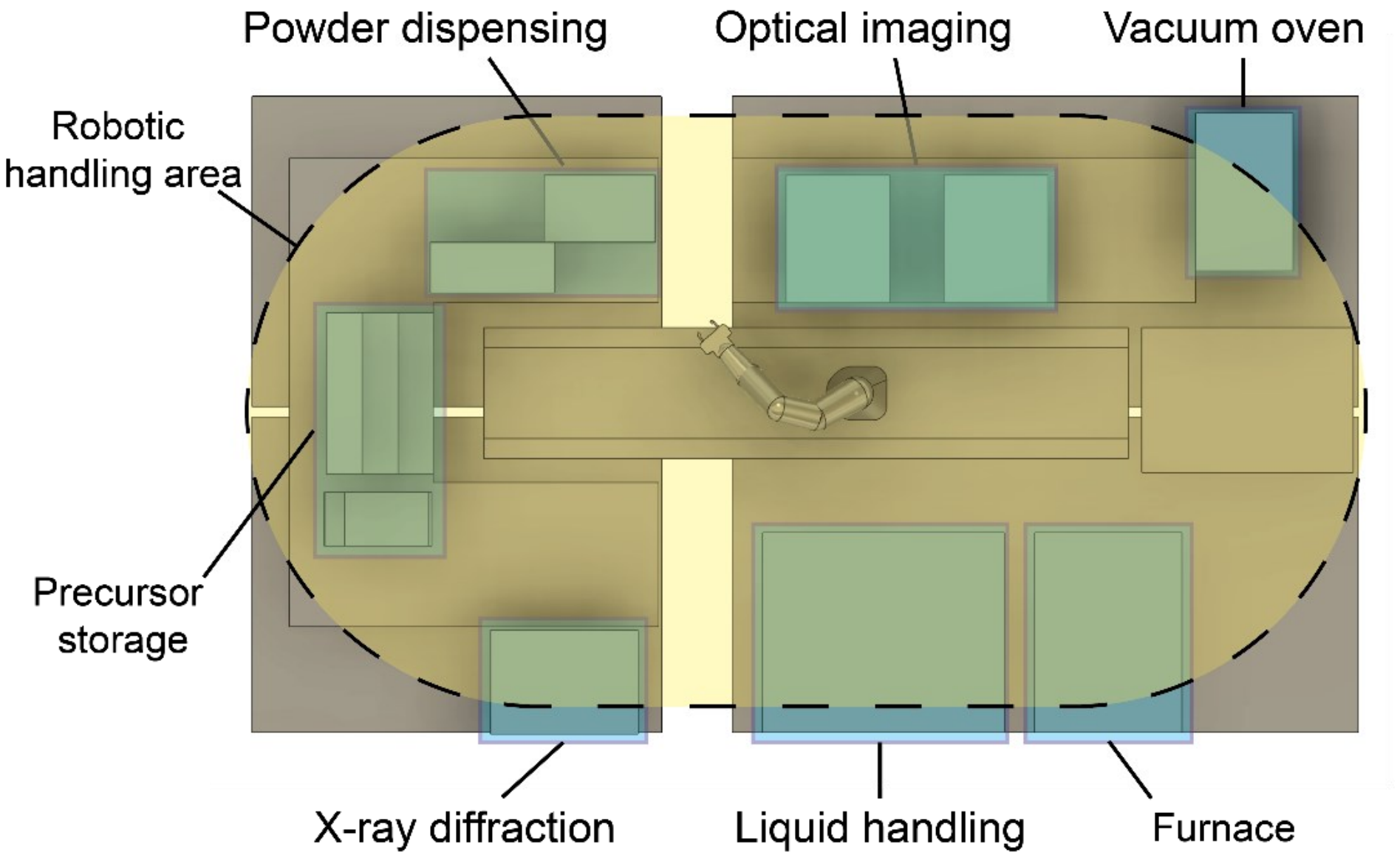

**Figure S4:** 3D-model of ASTRAL automated synthesis platform. Stations for storage, characterization, and synthesis operations, marked with blue rectangles, are arranged around the perimeter of the platform. The Panda robotic arm and rail in the center of the platform transports samples between stations throughout the robotic handling area marked in yellow.

#### SI2.1.1. Mobile robotic arm

Transport of samples between stations is accomplished by two robots, a 7-axis Panda robotic arm supplied by Franka Emika, and a linear rail supplied by Vention.io. The Panda arm is a highly flexible collaborative robot with a reach of 855mm and a payload of 3kg, with positional repeatability of 0.5mm allowing high reliability for manipulating small objects. The Panda arm is mounted on the linear rail system, which uses the Vention.io MachineMotion controller and a rack-and-pinion actuator to transport the arm over a linear distance of 2320mm, with positional repeatability of 0.1mm. By utilizing the rail system to extend the range of the Panda arm, the system is able to accomplish highly precise generalized manipulation tasks over the approximately 1710mm x 4030mm area shown in **Figure S4**.

In addition to performing repeatable precise movements, the Panda arm is equipped with force and position sensors that allow it to detect collisions and allow it to apply controlled gentle force to objects that are being manipulated. This allows the robot to handle samples and interface with a wide variety of equipment in a manner similar to a human researcher. For example, the Panda arm can be used to press buttons, turn handles, and open doors without risk of damage, allowing the system to easily and safely interface with equipment that is designed for human use.

### SI2.1.2. Sample handling

Disposable 5mL glass test tubes (Corning) are used to hold precursors during dispensing, mixing, and vacuum drying. Alumina crucibles (Advalue Technology) are used to hold the mixed precursors during high temperature heat treatment. As the ASTRAL system is designed to accommodate handling of individual as well as trays of 8 or 24 sample holders, the platform uses a wide range of custom holders and adapters to enable reliable robotic handling, shown in **Figure S5**.

Sample holders intended for room temperature use are manufactured out of Acrylonitrile styrene acrylate (ASA), and utilize embedded magnets to reversibly and accurately locate parts during handling. In addition, aluminum plates are used to hold sets of 24 glass tubes during vacuum drying, and alumina plates are used to hold sets of 24 crucibles during high temperature heat treatment. All holder plates include a grip pad matching the geometry of the Panda arm grip surface to maximize the reliability of robotic handling.

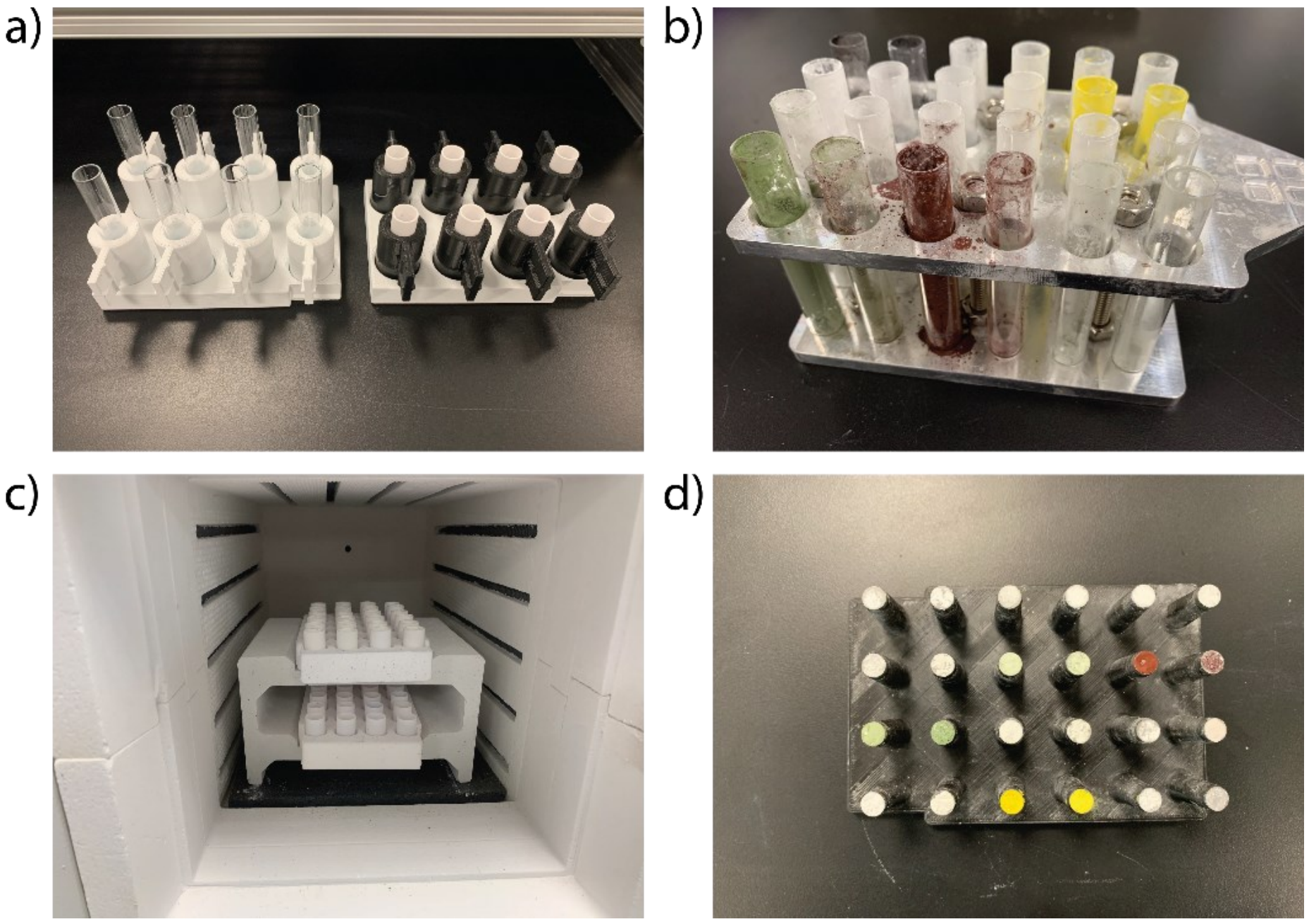

**Figure S5:** Custom holders used by ASTRAL platform for sample handling. All plates and individual holders have customized grip surfaces optimized for handling by the Panda robotic arm. (a) ASA plates holding 8 ASA sample holders, which each hold an individual glass test tube or alumina crucible. Both the plates and sample holders have embedded magnets to securely locate parts during robotic handling. (b) Aluminum plate holding 24 test tubes suitable for solvent mixing, wet chemistry, and heat treatment up to 250C. (c) Cast alumina plates holding 24 alumina crucibles, suitable for heat treatment up to 1200C. (d) ASA plate holding 24 stainless steel stubs with mounted powder samples for X-ray diffraction. Embedded magnets are used to hold the stubs in place during handling operations.

### SI2.1.3. Powder dispensing

The ASTRAL platform uses a Quantos solid dispensing unit, supplied by Mettler Toledo, to dispense the precursor powders used for synthesis experiments. The Quantos dispenser uses gravimetric dispensing to dose powders into sample containers, with fully automated operation enabled by an RS-232 interface. To accomplish the many-to-many dispensing required for synthesis experiments, the ASTRAL platform uses the Panda arm to sequentially load sample holders and precursor dosing heads into the Quantos dispenser to complete each dispense operation.

The ASTRAL platform includes storage for up to 63 separate powder precursor chemicals in the storage rack shown in **Figure S6a**. The precursor powders are stored in dosing heads designed to interface with the Quantos solid dispensing unit, supplied by Mettler Toledo. To facilitate handling by the Panda robot arm, the dosing heads are equipped with custom designed grip pads, which clamp securely onto the exterior and provide a reliable grip surface that the Panda uses to insert and retrieve dosing heads from the dispenser.

Powders are stored in dosing heads supplied by Mettler Toledo designed to interface with the Quantos powder dispenser. In order to facilitate handling using the Panda robot arm, the dosing heads are equipped with custom designed grip pads as indicated in **Figure S6b**, which clamp securely onto the exterior and provide an optimal shape for controlled handling by the gripper used by the Panda arm.

Powder dispensing presents several practical challenges, particularly due to the variability of physical characteristics such as particle sizes and flowability. We found that no single model of dosing head could successfully dispense all of the powders, but by selecting between three different models we are able to reliably dose all of the precursors. The full list of precursors used in this study, with associated dosing head and manufacturer information, is included in **Table S3**.

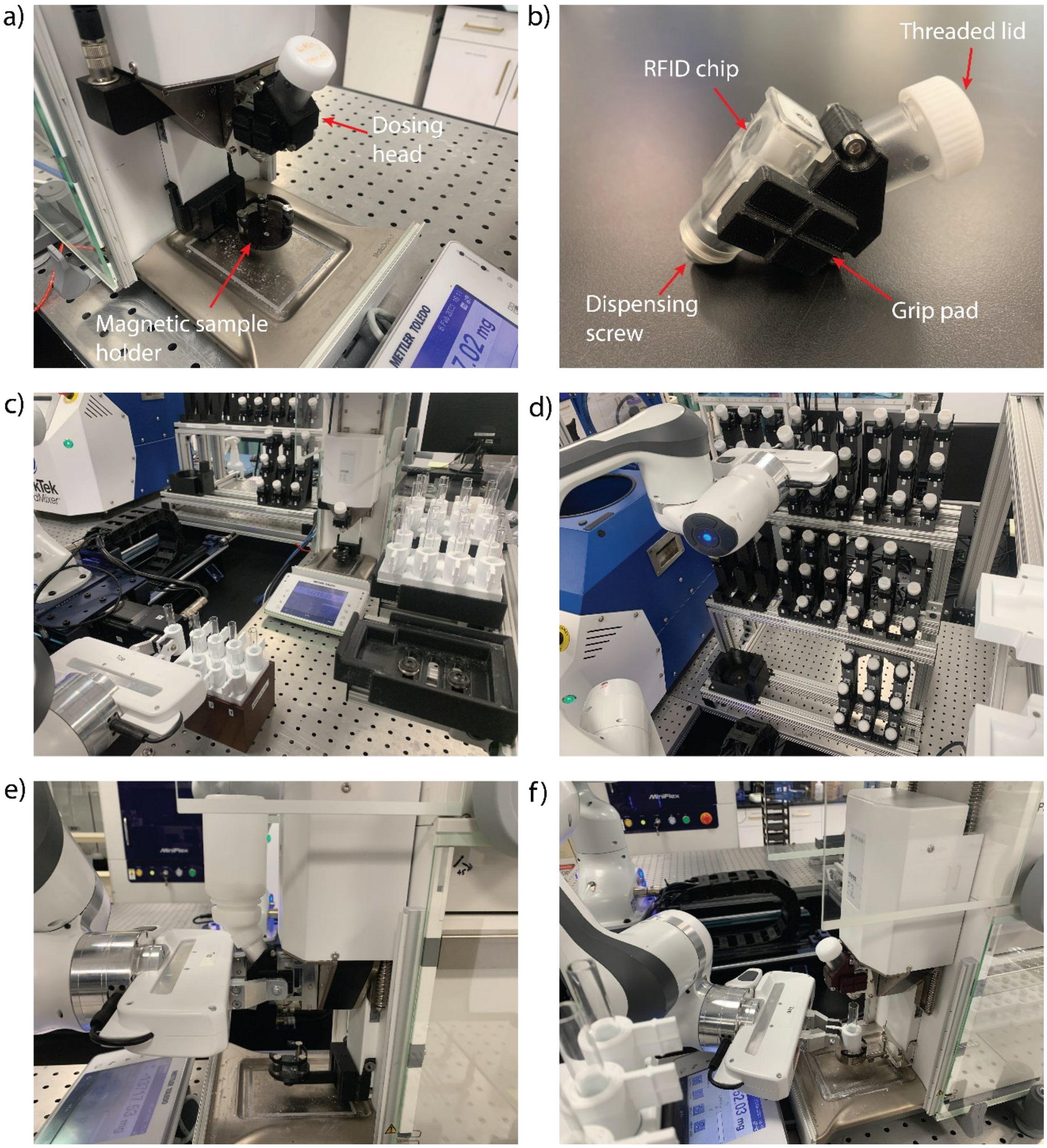

**Figure S6:** Components and workflow for the ASTRAL platform automated powder dispensing system. (a) Quantos powder dispenser, used for gravimetric dispensing of precursor powders. (b) Dosing head used by Quantos to store and dispense powders, with attached grip pad for Panda arm. (c) Powder dispensing station in use by ASTRAL to dose precursors into 24 glass sample holders. (d) Powder inventory with 63 storage slots for precursor dosing heads. (e) Panda arm loading dosing head into Quantos to prepare for precursor dispensing. (f) Panda arm loading sample holder into Quantos to receive dispensed powder.

### SI2.1.4. Liquid handling

Dispensing of liquid solvents and precursor solutions is accomplished by a Freedom EVO 100 liquid handling robot, supplied by Tecan Life Sciences. The liquid handler uses a set of 8 reusable pipettes with 1mL solution capacity, mounted on a 3-axis gantry system to allow movement throughout the deck of the robot. The liquid handler employs the pipettes to aspirate and dispense liquids with 1µL accuracy from an installed storage rack containing 64 prepared solutions, allowing any arbitrary mixture of liquids to be added to samples. All containers are sealed with septum caps to allow pipetting operations while minimizing evaporation of stored liquids.

The Freedom EVO liquid handler is also equipped with an additional Pick and Place (PnP) arm, which is used for fast and accurate movement of small objects between containers on the deck of the liquid handler. In the ASTRAL platform, the PnP arm is used for robotic manipulation of objects that are too small to be accomplished by the Panda robot arm, such as individual test tubes, crucibles, and XRD sample holders.

While the ASTRAL platform is capable of using the liquid handler for a broad range of wet chemistry, in particular using Pechini method for inorganic sol-gel synthesis, in the present study the only liquid dispensed was 1.5mL of ethanol added to each sample to act as a milling solvent. The PnP arm was additionally used for all handling operations transferring individual crucibles, test tubes, and XRD sample stubs between holder plates.

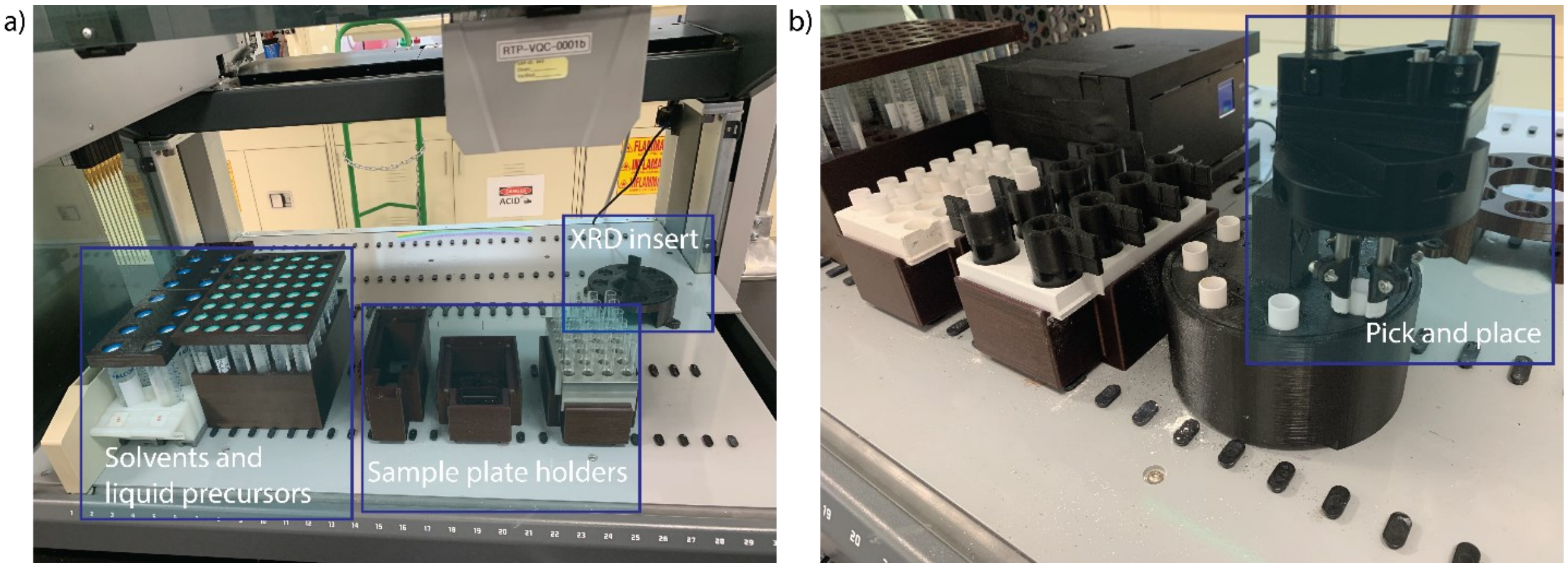

**Figure S7:** Freedom EVO 100 liquid handling robot installed in the ASTRAL platform. (a) Configuration of the deck of the liquid handler. The storage racks on the left hold 16 50mL Falcon tubes and 48 15mL Falcon tubes with solvents and liquid precursor chemicals for use in synthesis experiments. The holders on the right side of the deck are accessible for the Panda arm to place any of the holder plates for liquid dispensing or pick and place operations. (b) The liquid handler pick and place arm transferring crucibles between holder plates.

### SI2.1.5. Ball milling

Powder mixing in the ASTRAL platform is done by an SFM-2 rotary ball mill supplied by MTI Corporation, using 3mm stainless steel mixing balls to break up and mix powders contained inside sealed test tubes. At the time of writing, the powder mixing by the ASTRAL platform is not fully automated, but is instead accomplished using high-throughput handling methods that allow a human researcher to efficiently and reproducibly process batches of samples. The high-throughput ball milling system consists of (1) a specialized jig used to dispense a consistent number of mixing balls into 24 test tubes at a time, (2) customized inserts allowing groups of 8 samples to be mixed simultaneously in a single chamber of the mill, (3) a magnetic extraction tool to allow simultaneous removal of mixing balls from 8 samples at a time, and (4) a customized funnel allowing simultaneous transfer of 24 samples from glass test tubes to alumina crucibles in preparation for heat treatment.

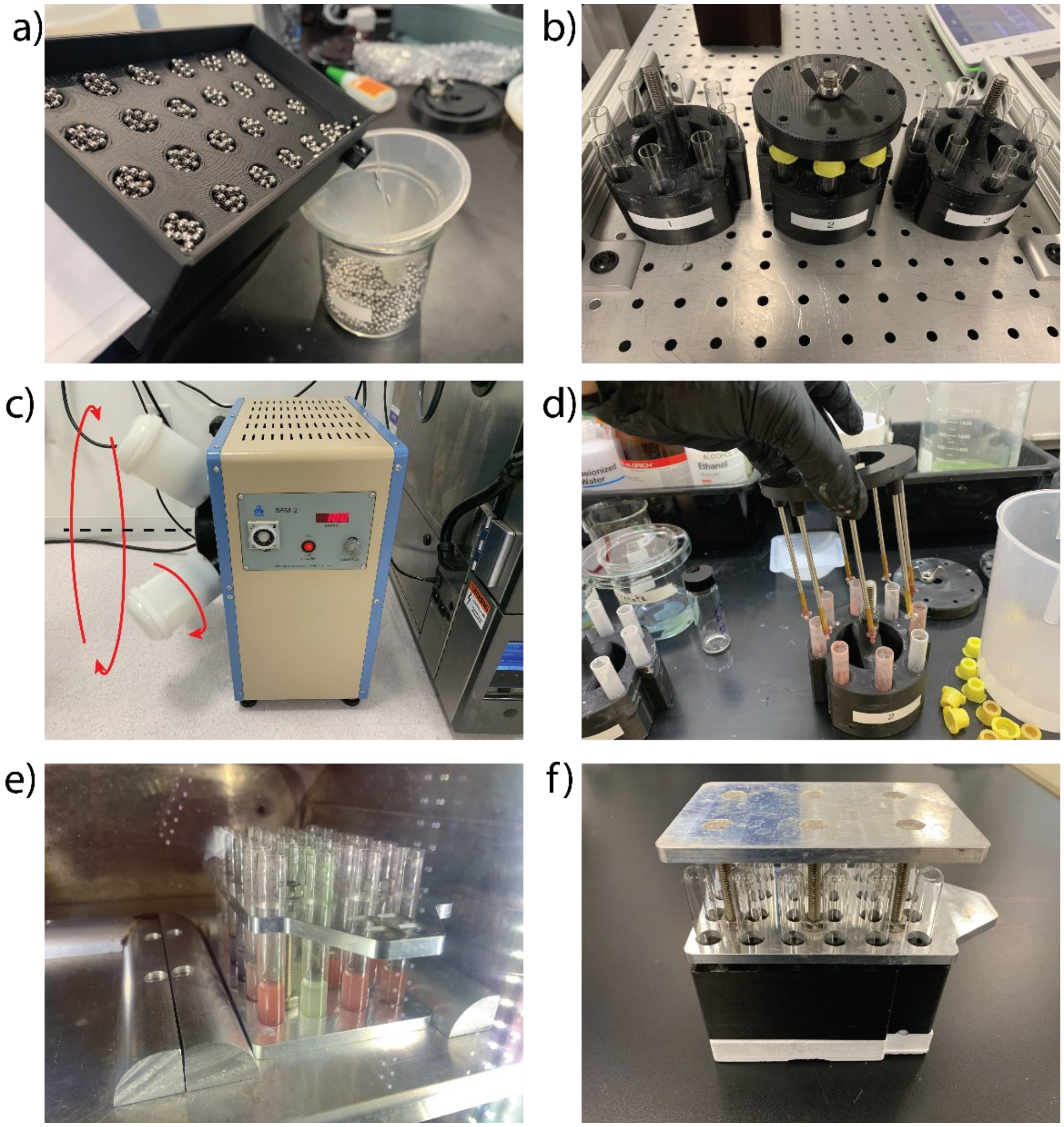

**Figure S8:** High-throughput ball milling for ASTRAL synthesis experiments. (a) Fixture for controlled dispensing of 3mm stainless steel mixing balls into plate of 24 glass tubes. (b) Custom inserts to hold 8 test tubes in a single chamber of the rotary mixer. (c) Mixing precursors in rotary ball mill.. (d) Magnetic extraction of stainless steel mixing balls. (e) Vacuum drying for solvent removal. (f) Powder transfer from glass tubes to alumina crucibles via customized funnel plate.

### SI2.1.6. Solvent removal

Sample drying and solvent removal is accomplished using a vacuum oven (Across International AT09e), with an attached diaphragm vacuum pump (Welch DryFast), with trays of samples loaded and unloaded using the Panda arm. The temperature controls for the vacuum oven are controlled automatically by the ASTRAL controller using Modbus communications over an RS485-USB interface. The gas intake for the vacuum chamber can be diverted between vacuum and intake of ambient air using an automated solenoid diverting valve actuated by a digital i/o module attached to the Vention MachineMotion controller. To allow gentle but fast solvent drying, throughout the drying process the intake valve is cycled between 120s of vacuum and 10s of ambient air refill. The periodic refill cycles allow for efficient displacement and exhaust of accumulated solvent vapors, and decreases the chance of powders being expelled from test tubes due to overly aggressive solvent vaporization.

### SI2.1.7. High temperature heat treatment

High temperature calcination and heat treatment of samples is accomplished using a Nabertherm p480 box furnace, with a controllable temperature range of up to 1200C. Automated control of the heating profile by the Astral controller is done using a Modbus/TCP communication interface. Batches of samples are held in 12mm x 29mm alumina crucibles, loaded into alumina holder plates that can be handled by the Panda arm to transport samples in and out of the furnace, as illustrated in **Figure S9**.

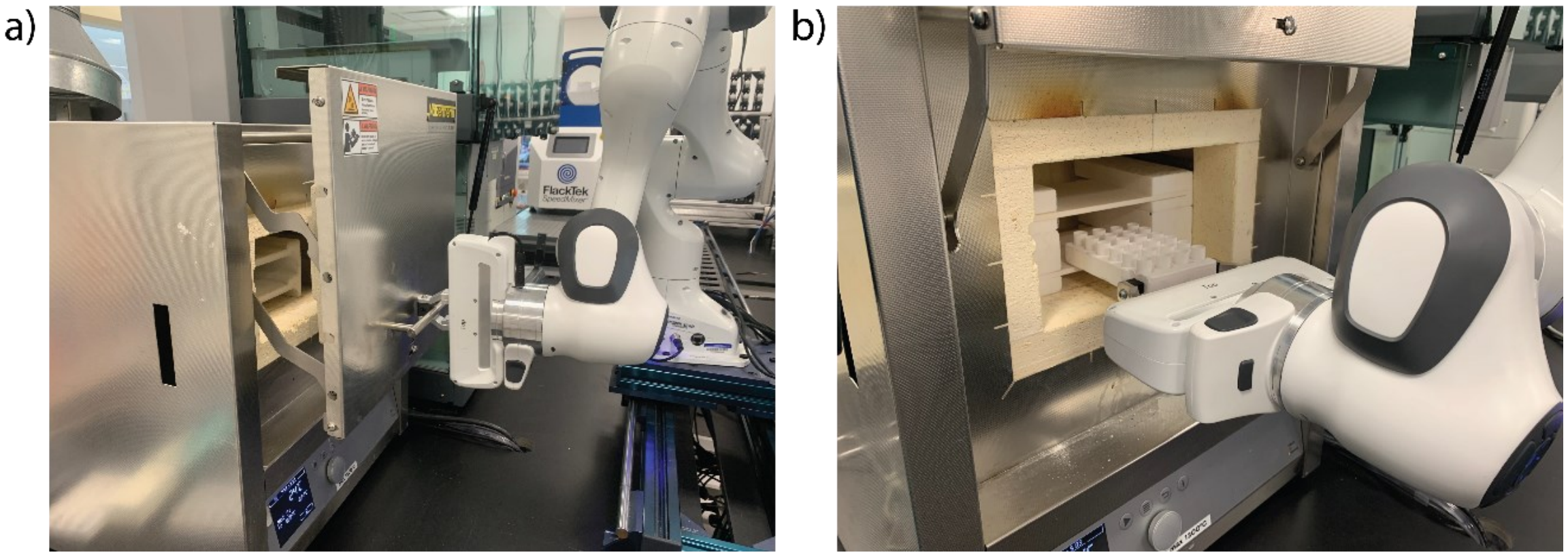

**Figure S9:** Heat treatment and calcination at up to 1200C using Nabertherm p480 box furnace. (a) Panda arm opening furnace door. (b) Panda arm loading tray of 24 samples into furnace for heat treatment.

### SI2.1.8. X-ray diffraction

The ASTRAL platform uses automated X-ray diffraction in order to characterize the outcome of synthesis experiments. As with the ball milling process, preparation for the X-ray diffraction requires some manual sample preparation by a human researcher, utilizing a high-throughput processing setup designed to produce high efficiency and consistent results. The high-throughput setup consists of (1) a plate containing 24 embedded stainless steel dowel pins, and (2) a matching plate with 24 posts, each containing an embedded magnet securing a matching stainless-steel stub, with a thin layer of vacuum grease applied to the top surface. In order to prepare samples for X-ray diffraction, the dowel plate is first pressed repeatedly into the crucible plate to break up the fired powders into small loose particles. Once the powders are broken up, the sample holder plate is inverted and pressed into the plate of powders, causing a thin layer of powder for each sample to adhere to the vacuum grease in a flat layer appropriate for X-ray diffraction. Sample holders and key process steps for the XRD sample preparation are illustrated in **Figure S10 (a-c)**.

Following the transfer of powders to the XRD holder, the remainder of the X-ray diffraction measurements are accomplished using fully automated robotic handling. To do this, each plate of 24 mounted samples is transported by the Panda arm to the deck of the liquid handler. The PnP arm is then used to transfer the next group of 8 samples into a custom insert designed to fit into the 8-sample changer used by the Rigaku Miniflex XRD. The insert is then loaded into the XRD, using the Panda robot arm to release the interlock, open the door, and transport the samples. The required measurements are then automatically set up and executed, using the pyautogui python package to interface with the GUI of the Miniflex Guidance control software through simulated mouse and keyboard actions.

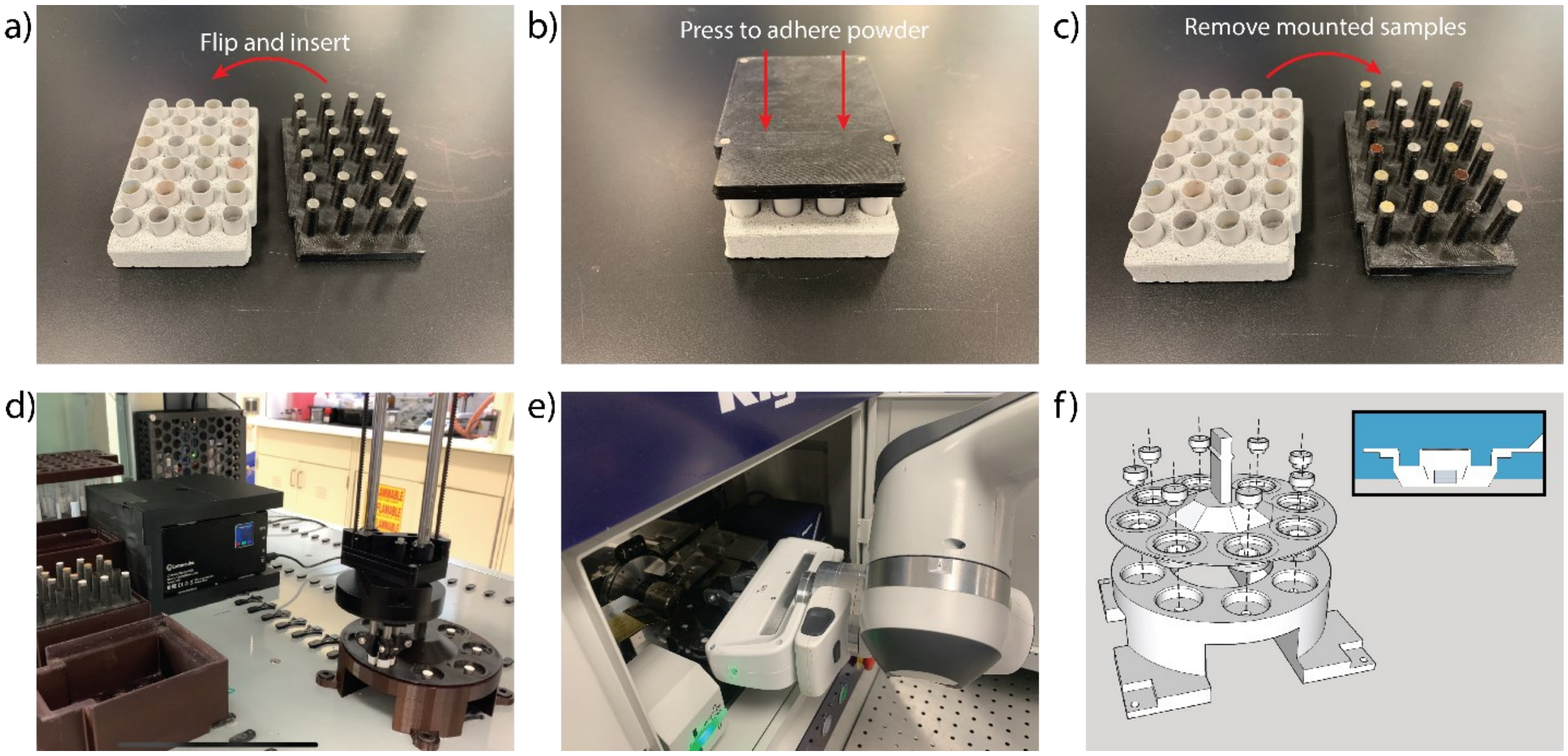

**Figure S10:** Workflow for characterization of synthesis outcomes by the ASTRAL platform via X-ray diffraction. (a-c) High-throughput preparation of powder samples for XRD. A thin layer of vacuum grease is applied to the stainless steel XRD stubs prior to pressing to allow a thin layer of calcined powder to adhere to the top surface. XRD stubs are secured to ASA holder plate during pressing operation by embedded magnets. (d) PnP arm transferring XRD stubs between holder plate and insert. (e) Panda arm transferring insert into XRD for measurement. (f) Design detail of XRD insert, showing magnetic holding system that is used to locate powder samples in XRD measurement plane.

## SI2.2. ASTRAL platform software implementation

### SI2.2.1. ASTRAL python controller

The ASTRAL platform manages the scheduling and control of robotic movements using central control software written in-house using python, with a graphical interface implemented through the PyQt5 package. The ASTRAL python controller uses a modular, multithreaded programming approach to permit simultaneous operation of any number of attached robots, with each robot controlled by a separate thread. The thread controlling each robot is responsible for receiving and executing commands and monitoring the status of the equipment.

For operation of multiple experiments in parallel, the ASTRAL controller uses a reservation system to allocate resources and avoid conflicts. For each experiment step, the controller determines which robots and inventory resources are required for execution, and will wait until all resources are available to begin. Once the process step has started, all required resources are reserved for the exclusive use of that experiment for the duration of the step, then released back to the pool of available resources.

The scheduling approach used by the ASTRAL platform allows parallel operation of multiple experiments, which is particularly critical for inorganic materials as synthesis methods typically require multiple days to complete. For example, during typical operation, the system will be dispensing powders for one plate of samples, while running heat treatment on a second and X-ray diffraction on a third, allowing the system to maintain maximum possible throughput without requiring human attention to avoid conflicts.

### SI2.2.2. Experiment planning

ASTRAL experiments are specified using a JSON format, which describes the experimental procedure as a series of general, human-readable steps, such as "Dispense powders", "Mix". For execution of each experiment, the JSON used as input by the ASTRAL controller to generate a list of all of the required robot tasks required to complete the specified experiment. The intention of the JSON format is to use a flexible, platform-independent specification that mimics the manner in which procedures are typically described in scientific literature. This enables a high level of flexibility and portability of the experiment plans, and acts analogously to a thorough and consistently formatted digital lab notebook.

Each experiment JSON contains specifications for running a synthesis experiment on a batch of up to 24 samples, which will be processed in parallel for mixing, heat treatment, and XRD. For each sample an independent target composition, yield, and list of precursors are provided, allowing preparation of 24 different mixtures. The appropriate quantities of each precursor to dispense for each sample are calculated using the pymatgen reaction_calculator module to generate a balanced reaction from the specified precursors. For syntheses that use a heat treatment of 600C or more, it is assumed that volatile compounds such as $CO_2$, $H_2O$, and $O_2$ can be freely removed, and so the reaction calculator is adjusted to permit loss of these compounds when generating a balanced reaction from the precursor list.

### SI2.2.3. Automated XRD analysis

X-ray diffraction is the primary characterization method used by the ASTRAL platform to determine the outcome of synthesis experiments. Standard methods used for XRD analysis require two steps: (1) identification of phases present in the sample, and (2) pattern fitting through methods such as Rietveld refinement to quantify the lattice parameters and weight percent of the phases. The traditional method for phase identification is to compare collected XRD patterns to a database of reference structures such as the Inorganic Crystal Structure Database (ICSD), most often using a search-match algorithm to compare peak positions and determine likely matches. While this approach is very effective at detecting matches to known structures, it requires a human to review candidate structures to exclude false positives and select true matches. More recently, several research groups have presented machine learning algorithms that can be trained on a set of reference structures to identify phases in experimental XRD [Manuscript References 30, 31]. These machine learning approaches offers great potential for improving automated phase identification, but requires additional steps to construct an appropriate training data set consistent with the characteristics of the experimental setup and chemical spaces. The training of these machine-learning methods is also reported in Ref 30 to take up to 20 hours for a system, also requiring GPU-accelerated machines.

Given the 35 systems that we are investigating here, we were not able to use these machine-learning methods to fully quantify all impurity phases detected in XRD for all samples processed on the ASTRAL platform. However, some quantification of synthesis outcomes is necessary to efficiently analyze trends over large data sets. Therefore, we have adopted a semi-quantitative approach to evaluating synthesis outcomes, based on Rietveld refinement of the XRD using only the crystal structure of the target material.

Rietveld refinement of data is accomplished using the BGMN kernel, with python scripts used for the automated generation of the necessary input files, execution of the Rietveld refinement via the command line, and extraction of the fitting data from the output files. The target structure is used as the sole input phase for the BMGN kernel, and as such, in an ideal case, the Rietveld refinement will split the XRD signal into components associated with the target phase, background, and residual. The fraction of the target phase can then be estimated by dividing the integrated intensity of the target phase by the combined intensity of the target phase and residual, $I_{target}/(I_{target} + I_{residual})$. In this work we considered values greater than 0.5 to be high purity, between 0.2 and 0.5 moderate purity, and less than 0.2 considered low purity. To minimize the excess residual, for each sample the algorithm supplies a background XRD pattern taken on an empty sample holder, to increase the effectiveness of the BGMN background fitting. As the background differs slightly for different sample holders, this procedure is repeated for each of 16 XRD patterns for empty sample holders, and the lowest residual is used as the final value for the calculation.

The primary limitations of this method are: (1) it neglects the different scattering factors of the target and impurity phases, (2) it can underestimate phase fraction due to any components of the residual that are not associated with impurity phases, and (3) it can overestimate phase fraction due to incorrect fitting of peaks for the target phase to impurity peaks. Due to the possibility of false positives due to (3), a value of $0.2 \times 10^6$ counts is used as a detection threshold, so the target phase is considered not detected for any samples where the target phase intensity is lower than this value.

Despite these potential limitations, we validated that our procedure produces adequate results on a wide range of data, and is suitably accurate for detecting successful or failed synthesis outcomes in the great majority of cases. To perform this validation, we used a set of 255 previously-obtained experimental XRD patterns collected using the ASTRAL platform for which all impurity phases were identified. We then compared our approach of calculating $I_{target}/(I_{target} + I_{residual})$, versus the fully Rietveld refined XRD phase fractions.

**Figure S11** shows a comparison of the XRD quantification results using both our semi-quantitative method described above and full quantitative Rietveld refinement. The color of the dots correspond to the weighted R-factor $R_{wp}/R_{exp}$ returned by the BGMN kernel as a goodness-of-fit metric, with higher values indicating greater discrepancy between the theoretical and experimental curves. The semi-quantitative method produced an estimated phase fraction that was on average 22.9% lower than the full quantitative refinement, but otherwise the two measures produced good agreement with a root mean squared difference of 4.6%. Therefore, we consider that the phase purity estimates produced by the semi-quantitative method are likely to be conservative, but generally effective for discriminating synthesis outcomes within 10% accuracy.

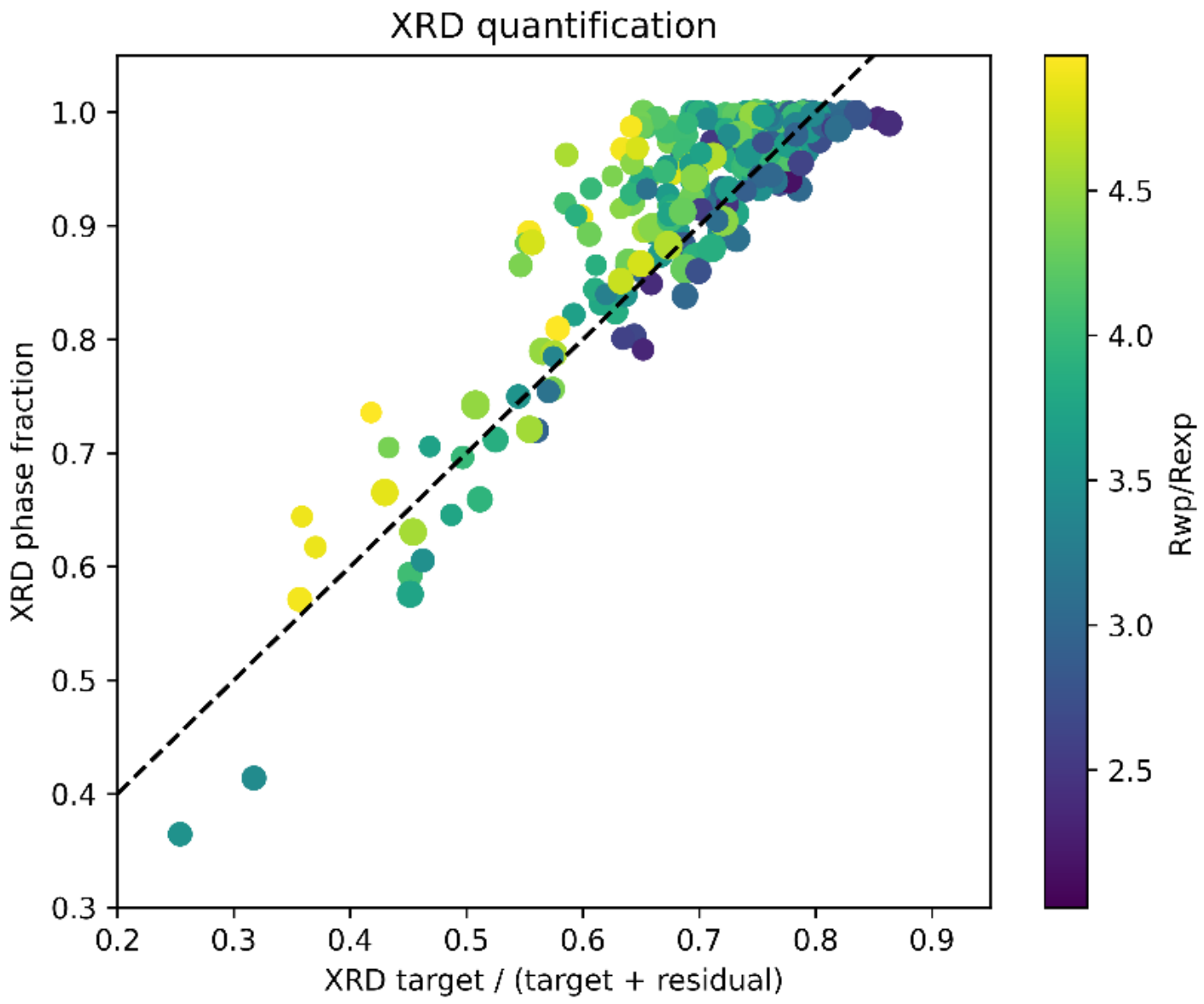

**Figure S11:** Comparison between semi-quantitative XRD analysis (x-axis) and full quantitative Rietveld refinement (y-axis) on a test data set of 255 samples synthesized on the ASTRAL platform. The size of the points is determined by the integrated XRD signal, while the color of the dots are determined by the quality-of-fit metric ($R_{wp}/R_{exp}$) output by BGMN for the full Rietveld refinement.

## SI2.3. Challenges and solutions of automated powder ceramic synthesis.

**Table S2:** Problems of powder ceramic synthesis for automated laboratory and our solutions.

| Challenges | Solutions |
|---|---|
| Powders are difficult to handle for automated dispensing due to varying size and physical properties | The ASTRAL platform using a Quantos powder dispenser supplied by Mettler Toledo, which uses gravimetric dispensing to dose precursor powders with high accuracy. To accommodate broad variety of powder types needed for synthesis experiments, each precursor powder is assigned one of three models of dosing heads for reliable dispensing. |
| Hygroscopic precursors | For handling hygroscopic precursors, we use the Quantos dosing heads for short-term storage, tightly sealed to minimize moisture infiltration. Hygroscopic powders are replaced on a schedule to maintain the quality of the dispensed precursors. |
| Powders are much more difficult to mix than liquid precursors | Successful synthesis requires that precursors are mixing intimately and homogeneously before heat treatment to produce a uniform and consistent product.<br><br>The ASTRAL platform accomplishes mixing of powders using a high-throughput ball milling system, consisting of the following components:<br>• High-throughput dispensing of mixing balls<br>• Automated addition of milling solvent<br>• High-throughput milling holders<br>• Magnetic mixing ball extraction<br>• High-throughput powder transfer to crucibles using funnel plate |
| Powders react and/or fuse with crucibles during high temperature heat treatment | During high-temperature calcination, many precursors or reaction products may become molten, and react with the alumina crucible, resulting in contamination with aluminum and/or fusing of the sample to the crucible walls.<br><br>We apply a boron nitride coating to the alumina crucibles for materials that are susceptible to this behavior. The boron nitride coating is highly non-reactive and resists wetting by most molten oxides, minimizing reactivity and fusing between the samples and crucibles. |
| Difficulties in preparing and mounting powders for XRD characterization | It is challenging to automate preparation of powders for characterization, due to varying physical properties and lack of a solvent to assist with dispersal.<br><br>To address this, the ASTRAL platform performs characterization using a high-throughput system for XRD measurement, consisting of:<br>• Custom magnetic sample stubs for XRD measurements<br>• High-throughput mounting of powders onto XRD stubs by full plate<br>• Fully automated robotic XRD loading and measurement execution |

**Table S3:** List of precursor chemicals and associated Quantos dosing heads stored in the ASTRAL inventory

| Formula | CAS number | Dosing head | Distributor | Purity |
|---|---|---|---|---|
| $Al_2O_3$ | 1344-28-1 | QH012-LNLX | Sigma Aldrich | 99.99% |
| $B_2O_3$ | 1303-86-2 | QH012-LNMW | Sigma Aldrich | 99% |
| $Bi_2O_3$ | 1304-76-3 | QH012-LNLT | Alfa Aesar | 99.9% |
| BaO | 1304-28-5 | QH012-LNMW | Alfa Aesar | 99.5% |
| CaO | 1305-78-8 | QH012-LNMW | Alfa Aesar | 99% |
| CuO | 1317-38-0 | QH012-LNMW | Acros Organics | 99% |
| $Fe_2O_3$ | 1309-37-1 | QH012-LNLT | Alfa Aesar | 99.9% |
| $GeO_2$ | 1310-53-8 | QH012-LNMW | Alfa Aesar | 99.999% |
| $In_2O_3$ | 1312-43-2 | QH012-LNMW | Thermo Fisher | 99.99% |
| $K_2CO_3$ | 584-08-7 | QH012-LNLT | Alfa Aesar | 99% |
| $K_3PO_4$ | 7778-53-2 | QH012-LNMW | Alfa Aesar | 99% |
| $KNbO_3$ | 12030-85-2 | QH012-LNMW | Strem Chemicals | 99.999% |
| $KPO_3$ | 7790-53-6 | QH012-LNLT | Strem Chemicals | 98% |
| $Li_2CO_3$ | 554-13-2 | QH012-LNMW | Alfa Aesar | 99% |
| $Li_2TiO_3$ | 12031-82-2 | QH012-LNMW | Sigma Aldrich | 99% |
| $LiBO_2$ | 13453-69-5 | QH012-LNMW | Alfa Aesar | 99% |
| $LiNbO_3$ | 12031-63-9 | QH012-LNMW | Alfa Aesar | 99.99% |
| $LiPO_3$ | 13762-75-9 | QH012-LNLT | American Elements | 99% |
| MgO | 1309-48-4 | QH012-LNMW | Sigma Aldrich | 99% |
| MnO | 1344-43-0 | QH012-LNMW | Alfa Aesar | 99.99% |
| $Na_2CO_3$ | 497-19-8 | QH012-LNMW | Alfa Aesar | 98% |
| $NaBO_2$ | 10555-76-7 | QH012-LNMW | Alfa Aesar | 98% |
| $NaPO_3$ | 68915-31-1 | QH012-LNMW | Acros Organics | 99% |
| $Nb_2O_5$ | 1313-96-8 | QH012-LNMW | Sigma Aldrich | 99.99% |
| $NH_4H_2PO_4$ | 7722-76-1 | QH012-LNMW | Alfa Aesar | 99% |
| NiO | 1313-99-1 | QH012-LNMW | Alfa Aesar | 99% |
| $Pr_6O_{11}$ | 12037-29-5 | QH012-LNMW | Sigma Aldrich | 99.9% |
| $Sc_2O_3$ | 12060-08-1 | QH012-LNMW | Sigma Aldrich | 99.9.% |
| $SiO_2$ | 60676-86-0 | QH012-LNMW | Sigma Aldrich | 99.5% |
| SrO | 1314-11-0 | QH012-LNMW | Sigma Aldrich | 99.9% |
| $Ta_2O_5$ | 1314-61-0 | QH012-LNLT | Sigma Aldrich | 99% |
| $TiO_2$ | 1317-80-2 | QH012-LNLT | Alfa Aesar | 99.5% |
| $V_2O_3$ | 1314-34-7 | QH012-LNMW | Alfa Aesar | 95% |
| $WO_3$ | 1314-35-8 | QH012-LNLT | Sigma Aldrich | 99.9% |
| $Y_2O_3$ | 1314-36-9 | QH012-LNMW | Sigma Aldrich | 99.99% |
| ZnO | 1314-13-2 | QH012-LNLT | Acros Organics | 99.5% |

# SI3. Synthesis recipes and XRD results

We summarize the synthesis recipes and results for both traditional and predicted reactions in 35 systems we synthesized in **Table S4**. The reaction energy and inverse hull energy of each predicted reaction is shown in **Figure S12**. The details of the synthesis for each system are also provided. For each system, we prepare four sections of data, 1) traditional precursors reaction compound convex hull; 2) predicted precursors reaction compound convex hull; 3) XRD results of traditional and predicted synthesis; 4) Table of phase fraction of traditional and predicted synthesis.

## SI3.1. Summary of synthesis results

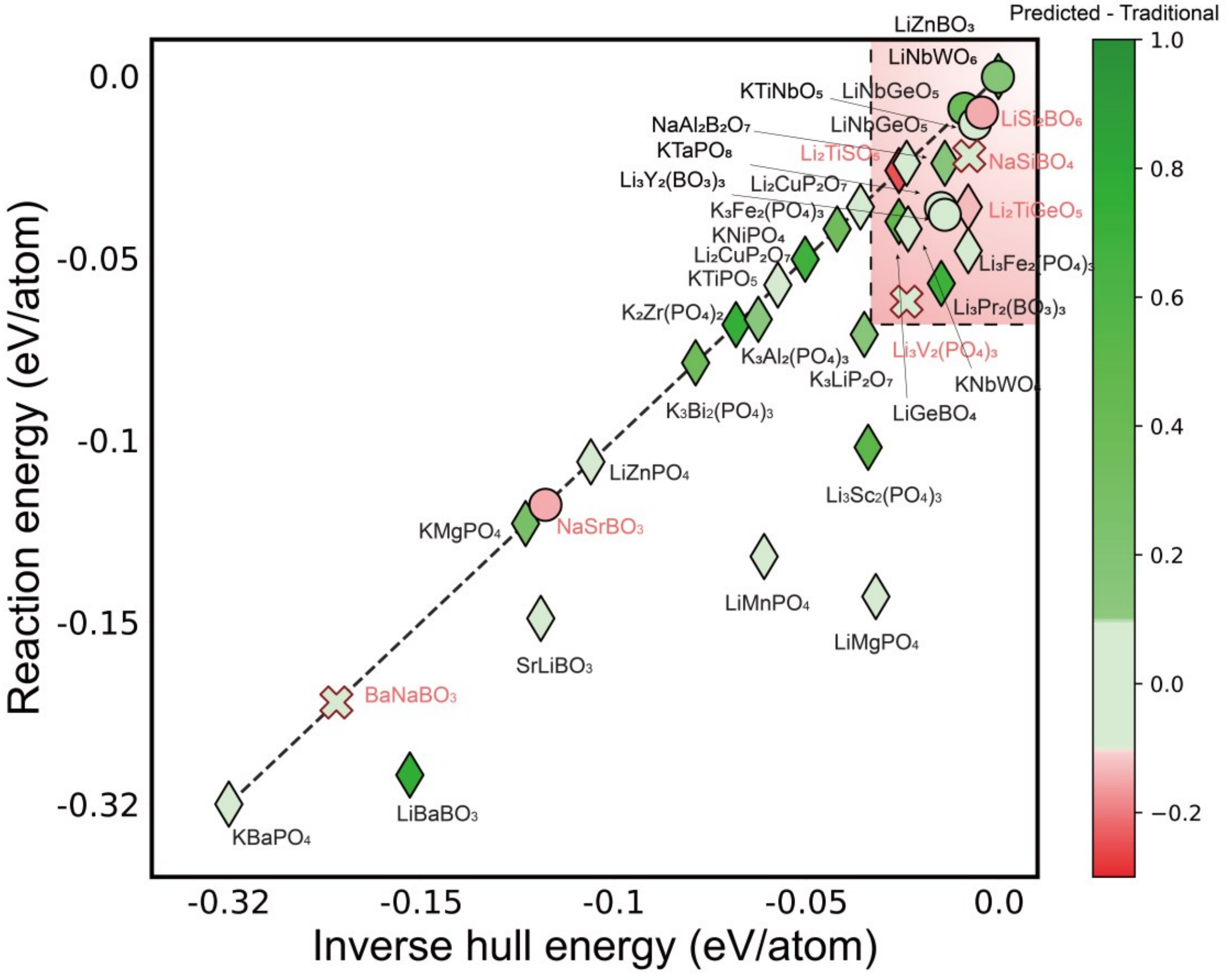

**Figure S12:** Labeled plot of reaction energy and inverse hull energy for all targets. Marker shape corresponds to best phase purity of predicted precursors, where diamonds are high purity, circles are moderate and low purity, and crosses with red outline means both predicted precursors and traditional precursors failed. The dashed line represents when inverse hull energy equals reaction energy.

**Table S4.** Traditional and predicted precursors for different targets. The colors in the first four columns represent shows how much better the predicted precursors over traditional, where green means predicted precursors perform better, light green means they perform similarly, and red means traditional precursors perform slightly better. The color in the "Best predicted Synthesis" column represents what is the best phase purity the predicted precursors can get, where green means high phase purity, light blue means moderate purity, yellow means low purity, and gray means both traditional and predicted precursors failed with no XRD signal. The "Best Temperature" column shows the reaction temperature to get the best synthesis results using predicted precursors. The last two columns show the inverse hull energies and reaction energies for predicted precursors.

| | **Target** | **Traditional Precursors** | **Predicted Precursors** | **Best Predicted Synthesis** | **Best Temperature (C)** | **For predicted precursors (eV/atom)** | |
|---|---|---|---|---|---|---|---|
| | | | | | | **Inverse hull energy** | **Reaction Energy** |
| 1 | $BaLiBO_3$ | $Li_2CO_3$, $B_2O_3$, BaO | BaO, $LiBO_2$ | | 800 | -0.153 | -0.192 |
| 2 | $K_2Zr(PO_4)_2$ | $K_2CO_3$, $NH_4H_2PO_4$, $ZrO_2$ | $KPO_3$, $ZrO_2$ | | 800 | -0.068 | -0.068 |
| 3 | $Li_3Pr_2(BO_3)_3$ | $Li_2CO_3$, $B_2O_3$, $Pr_6O_{11}$ | $LiBO_2$, $Pr_6O_{11}$ | High purity | 600 | -0.015 | -0.057 |
| 4 | $KNiPO_4$ | $K_2CO_3$, $NH_4H_2PO_4$, NiO | $KPO_3$, NiO | | 800 | -0.050 | -0.050 |
| 5 | $Li_3Sc_2(PO_4)_3$ | $Sc_2O_3$, $Li_2CO_3$, $NH_4H_2PO_4$ | $Sc_2O_3$, $LiPO_3$ | | 900 | -0.034 | -0.102 |
| 6 | $LiGeBO_4$ | $Li_2CO_3$, $B_2O_3$, $GeO_2$ | $LiBO_2$, $GeO_2$ | | 800 | -0.026 | -0.040 |
| 7 | $KLi(PO_3)_2$ | $Li_2CO_3$, $K_2CO_3$, $NH_4H_2PO_4$ | $LiPO_3$, $KPO_3$ | Moderate | 800 | -0.009 | -0.009 |
| 8 | $LiNbWO_6$ | $Li_2CO_3$, $Nb_2O_5$, $WO_3$ | $LiNbO_3$, $WO_3$ | Low purity | 700 | 0.000 | 0.000 |
| 9 | $LiZnBO_3$ | $Li_2CO_3$, ZnO, $B_2O_3$ | $LiBO_2$, ZnO | | 700 | 0.000 | 0.000 |
| 10 | $K_3Fe_2(PO_4)_3$ | $K_2CO_3$, $NH_4H_2PO_4$, $Fe_2O_3$ | $KPO_3$, $Fe_2O_3$ | | 700 | -0.042 | -0.042 |
| 11 | $KMgPO_4$ | $K_2CO_3$, $NH_4H_2PO_4$, MgO | MgO, $KPO_3$ | | 800 | -0.123 | -0.123 |
| 12 | $K_3Bi_2(PO_4)_3$ | $K_2CO_3$, $NH_4H_2PO_4$, $Bi_2O_3$ | $Bi_2O_3$, $KPO_3$ | | 700 | -0.079 | -0.079 |
| 13 | $K_3LiP_2O_7$ | $Li_2CO_3$, $NH_4H_2PO_4$, $K_2CO_3$ | $LiPO_3$, $K_3PO_4$ | | 700 | -0.035 | -0.071 |
| 14 | $Na_2Al_2B_2O_7$ | $Na_2CO_3$, $Al_2O_3$, $B_2O_3$ | $Al_2O_3$, $NaBO_2$ | | 700 | -0.014 | -0.024 |
| 15 | $K_3Al_2(PO_4)_3$ | $K_2CO_3$, $NH_4H_2PO_4$, $Al_2O_3$ | $KPO_3$, $Al_2O_3$ | | 700 | -0.063 | -0.067 |
| 16 | $Li_2CuP_2O_7$ | $Li_2CO_3$, $NH_4H_2PO_4$, CuO | $LiPO_3$, CuO | | 700 | -0.036 | -0.036 |
| 17 | $LiNbGeO_5$ | $GeO_2$, $Li_2CO_3$, $Nb_2O_5$ | $GeO_2$, $LiNbO_3$ | High purity | 1000 | -0.024 | -0.024 |
| 18 | $Li_3Fe_2(PO_4)_3$ | $Li_2CO_3$, $NH_4H_2PO_4$, $Fe_2O_3$ | $LiPO_3$, $Fe_2O_3$ | | 700 | -0.008 | -0.048 |
| 19 | $SrLiBO_3$ | $Li_2CO_3$, $B_2O_3$, SrO | $LiBO_2$, SrO | | 600 | -0.119 | -0.149 |
| 20 | $KNbWO_6$ | $K_2CO_3$, $Nb_2O_5$, $WO_3$ | $WO_3$, $KNbO_3$ | | 800 | -0.024 | -0.042 |
| 21 | $LiMgPO_4$ | $Li_2CO_3$, $NH_4H_2PO_4$, MgO | $LiPO_3$, MgO | | 800 | -0.032 | -0.143 |
| 22 | $LiZnPO_4$ | $Li_2CO_3$, $NH_4H_2PO_4$, ZnO | $LiPO_3$, ZnO | | 800 | -0.106 | -0.106 |
| 23 | $KBaPO_4$ | $K_2CO_3$, $NH_4H_2PO_4$, BaO | $KPO_3$, BaO | | 700 | -0.316 | -0.316 |
| 24 | $KTiPO_5$ | $K_2CO_3$, $NH_4H_2PO_4$, $TiO_2$ | $TiO_2$, $KPO_3$ | | 800 | -0.057 | -0.057 |
| 25 | $LiMnPO_4$ | $Li_2CO_3$, $NH_4H_2PO_4$, MnO | $LiPO_3$, MnO | | 700 | -0.061 | -0.132 |
| 26 | $KTa_2PO_8$ | $K_2CO_3$, $NH_4H_2PO_4$, $Ta_2O_5$ | $KPO_3$, $Ta_2O_5$ | | 700 | -0.015 | -0.036 |
| 27 | $Li_3Y_2(BO_3)_3$ | $Li_2CO_3$, $Y_2O_3$, $B_2O_3$ | $LiBO_2$, $Y_2O_3$ | Low purity | 700 | -0.014 | -0.038 |
| 28 | $KTiNbO_5$ | $K_2CO_3$, $TiO_2$, $Nb_2O_5$ | $TiO_2$, $KNbO_3$ | | 700 | -0.006 | -0.013 |
| 29 | $BaNaBO_3$ | $Na_2CO_3$, BaO, $B_2O_3$ | BaO, $NaBO_2$ | | 600 | -0.172 | -0.172 |
| 30 | $Li_3V_2(PO_4)_3$ | $Li_2CO_3$, $NH_4H_2PO_4$, $V_2O_3$ | $LiPO_3$, $V_2O_3$ | Not detected | 900 | -0.024 | -0.062 |
| 31 | $NaSiBO_4$ | $Na_2CO_3$, $SiO_2$, $B_2O_3$ | $SiO_2$, $NaBO_2$ | | 600 | -0.008 | -0.022 |
| 32 | $Li_2TiGeO_5$ | $GeO_2$, $Li_2CO_3$, $TiO_2$ | $GeO_2$, $Li_2TiO_3$ | High purity | 1000 | -0.008 | -0.036 |
| 33 | $Li_2TiSiO_5$ | $SiO_2$, $TiO_2$, $Li_2CO_3$ | $SiO_2$, $Li_2TiO_3$ | | 1000 | -0.026 | -0.026 |
| 34 | $NaSrBO_3$ | $Na_2CO_3$, SrO, $B_2O_3$ | SrO, $NaBO_2$ | Moderate | 700 | -0.118 | -0.118 |
| 35 | $LiSi_2BO_6$ | $Li_2CO_3$, $SiO_2$, $B_2O_3$ | $LiBO_2$, $SiO_2$ | | 700 | -0.004 | -0.010 |

## SI3.2. Metastable materials synthesis efficacy

In this work, we also considered 4 target materials that are calculated in DFT to be metastable relative to the convex hull, meaning they have an energy above the hull. These metastable materials are listed in **Table S5**. We aimed to investigate if the materials were calculated to be metastable, if they were still synthesizable using predicted precursors. We chose $LiZnBO_3$, which is calculated in DFT to be metastable with respect to our predicted precursors ZnO + $LiBO_2$. We also chose $LiNbWO_6$, $KTiNbO_5$, and $Li_3Y_2(BO_3)_3$, which are metastable with respect to decomposition products that are not our precursors. We hypothesized that by starting with precursors that are in a different 'compositional direction', we might be able to synthesize these metastable phases.

Of these four systems, we obtained a reasonably high target yield for $LiZnBO_3$, whereas the three metastable targets received low yields from both the predicted and traditional precursors. All three metastable materials were synthesized with low sample purity, ostensibly within the noise of the XRD characterization method. This illustrates that our algorithm is better suited to predict precursors for target materials that are convex hull stable, rather than metastable.

**Table S5.** Target materials that are not thermodynamic stable on the convex hull.

| Target | Energy above hull (meV/atom) | Decomposition products | Target phase fraction | |
|---|---|---|---|---|
| | | | From predicted precursors | From traditional precursors |
| $LiZnBO_3$ | 8 | 1/3 ZnO + 2/3 $LiBO_2$ | 0.52 | 0.15 |
| $LiNbWO_6$ | 10 | $LiNb_3O_8$ + $Li_2WO_4$ + $WO_3$ | 0.17 | 0.05 |
| $KTiNbO_5$ | 1 | $K_4Nb_6O_{17}$ + $K_2Ti_6O_{13}$ | 0.18 | 0.27 |
| $Li_3Y_2(BO_3)_3$ | 39 | 19/34 $Li_6Y(BO_3)_3$ + 15/34 $YBO_3$ | 0.17 | 0.12 |

## SI3.3. Comparison of energy contribution between $T\Delta S$ and $\Delta H$

In this section, we compared the magnitude of the entropy contribution, $T\Delta S$, to the overall ΔG of a reaction. We used experimental thermochemical data queried through Materials Project API in the 'Experimental Data' field. This experimental thermochemical data originated from NIST JANAF[1], Materials Thermochemistry[2], and the CODATA Key Values for Thermodynamics[3]. We collected entropy ($S$) and formation enthalpy ($H_f$) data at 298K for all convex hull stable binary and ternary oxides among 49 common metal elements. Then, using the selected binary metal oxides as precursors, we generated all possible pairwise combination reactions for the formation of the selected ternary oxides, resulting in exactly 100 reactions total. The energy contributions of $T\Delta S$, $\Delta H_f$, reaction formation energy $\Delta G$, are plotted in **Figure S13a**, **S13b**, and **S13c**, respectively. The ratio of the magnitude of the entropy contribution to total reaction energy magnitude ($|T\Delta S\ /\ \Delta G|$) was also calculated for each individual reaction, shown in **Figure S13d**. Full oxide reactions are presented in **Table S5**.

Altogether, **Figure S13** indicates that for the majority of reactions, the energy contribution of entropy at 1000K is considerably smaller in magnitude than the total reaction energy. By choosing a characteristic synthesis temperature of 1000K, the distribution peak of |TΔS| term is ~15 meV/atom, while that of ΔH term is -185 meV/atom. Specifically, 60% of reactions have $|T\Delta S\ /\ \Delta G|$ values less than 0.1. Among the remaining 40% of reactions where $|T\Delta S\ /\ \Delta G|$ values are greater than or equal to 0.1, approximately half have a relatively low reaction formation energy $\Delta G$ (~100 meV/atom). Therefore, in the context of oxide synthesis reactions, entropic contributions are usually negligible due to the dominant contribution of the enthalpy ΔH to the free energy ΔG.

[1] Chase, Malcolm W., and National Information Standards Organization (US). NIST-JANAF thermochemical tables. Vol. 9. Washington, DC: American Chemical Society, 1998.

[2] Kubaschewski, Oswald, Charles B. Alcock, and P. J. Spencer. "Materials thermochemistry. revised." Pergamon Press Ltd, Headington Hill Hall, Oxford OX 3 0 BW, UK, 1993. 363 (1993).

[3] Cox, J. D., Wagman, D. D., and Medvedev, V. A., CODATA Key Values for Thermodynamics, Hemisphere Publishing Corp., New York, 1989.

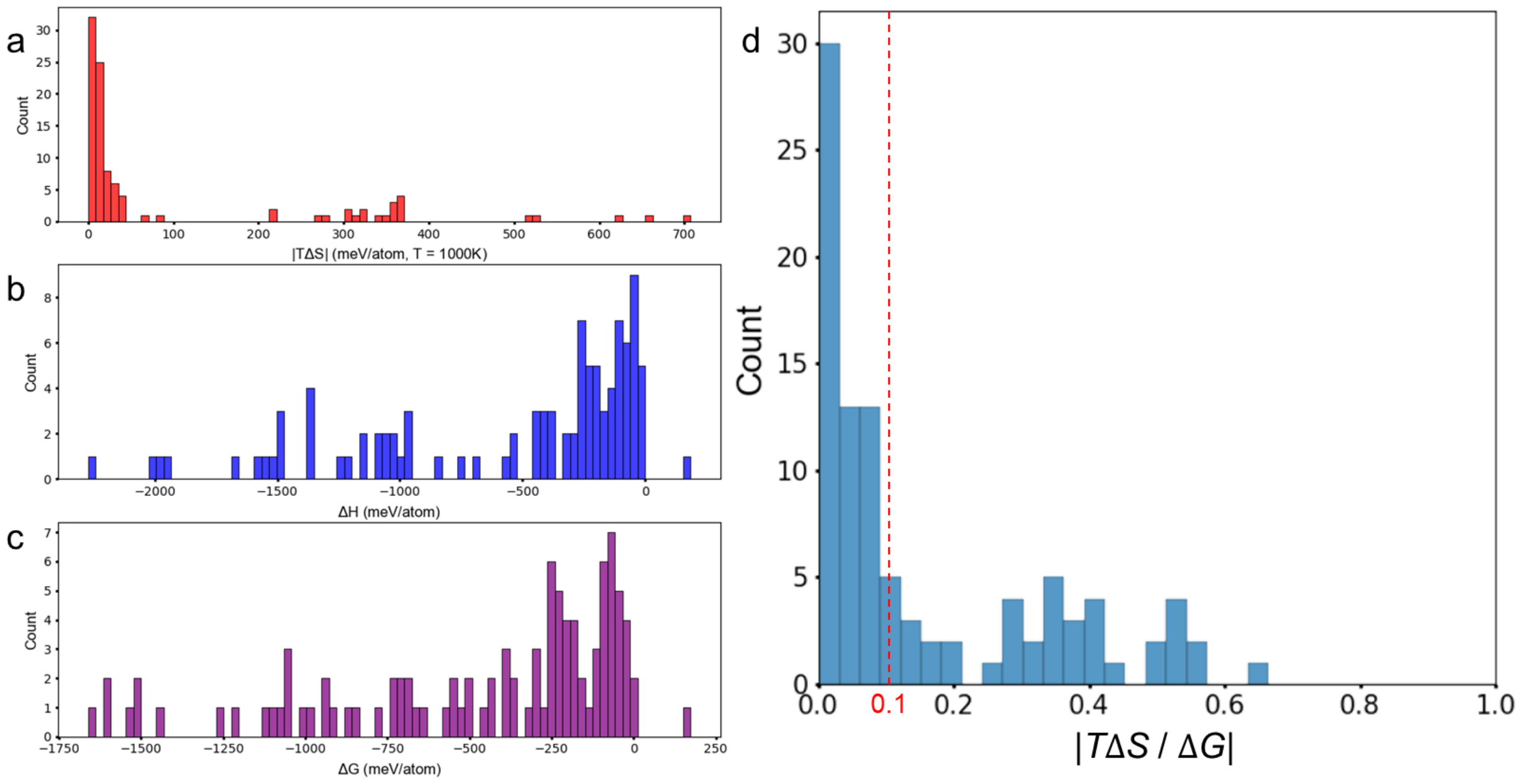

**Figure S13.** Histograms of **a**. $|T\Delta S|$, **b**. $\Delta H$, **c**. $\Delta G$, **d**. $|T\Delta S/\Delta G|$ of 100 reactions which uses binary metal oxides as reactants to synthesize ternary metal oxides in Materials Projects database. The entropy and enthalpy data we use is experimental data in room temperature (298K). The synthesis temperature T we choose is 1000K.

**Table S5.** Thermodynamic data table of 100 reactions which uses binary metal oxides as precursors to synthesize ternary metal oxides.

| | Reactions | ΔS (meV/atom/K) | TΔS (meV/atom, T = 1000K) | ΔH (meV/atom) | ΔG (meV/atom) | \|TΔS/ΔG\| |
|---|---|---|---|---|---|---|
| 1 | 0.5 K2O + 0.5 Fe2O3 -> KFeO2 | -0.00729 | -7.29 | -1668 | -1661 | 0.00439 |
| 2 | SrO + WO3 -> SrWO4 | -0.66 | -660 | -2271 | -1610 | 0.41 |
| 3 | SrO2 + MoO2 -> SrMoO4 | 0.0405 | 40.5 | -1567 | -1607 | 0.0252 |
| 4 | Na2O2 + MoO2 -> Na2MoO4 | 0.0269 | 26.9 | -1499 | -1526 | 0.0176 |
| 5 | BaO2 + MoO2 -> BaMoO4 | 0.0127 | 12.7 | -1509 | -1521 | 0.00834 |
| 6 | 2 SrO + SiO2 -> Sr2SiO4 | -0.523 | -523 | -2024 | -1501 | 0.349 |
| 7 | 2 SrO + TiO2 -> Sr2TiO4 | -0.519 | -519 | -1953 | -1434 | 0.362 |
| 8 | NiO + WO3 -> NiWO4 | -0.707 | -707 | -1978 | -1271 | 0.556 |
| 9 | MnO2 + MoO2 -> MnMoO4 | 0.0629 | 62.9 | -1144 | -1207 | 0.0522 |
| 10 | SrO + TiO2 -> SrTiO3 | -0.355 | -355 | -1483 | -1128 | 0.315 |
| 11 | SrO + SiO2 -> SrSiO3 | -0.367 | -367 | -1475 | -1107 | 0.332 |
| 12 | K2O + WO3 -> K2WO4 | -0.303 | -303 | -1370 | -1067 | 0.284 |
| 13 | 2 NaO2 + 3 TiO -> Na2Ti3O7 | -0.0881 | -88.1 | -1150 | -1062 | 0.0829 |
| 14 | SrO + MoO3 -> SrMoO4 | -0.309 | -309 | -1365 | -1056 | 0.292 |
| 15 | SrO2 + TiO -> SrTiO3 | 0.0305 | 30.5 | -1025 | -1055 | 0.0289 |
| 16 | SrO + ZrO2 -> SrZrO3 | -0.356 | -356 | -1353 | -997.7 | 0.356 |
| 17 | SrO + HfO2 -> SrHfO3 | -0.366 | -366 | -1352 | -986.9 | 0.37 |
| 18 | BaO2 + TiO -> BaTiO3 | -0.0412 | -41.2 | -975.3 | -934.1 | 0.0441 |
| 19 | PbO + WO3 -> PbWO4 | -0.619 | -619 | -1552 | -933.1 | 0.663 |
| 20 | Na2O + WO3 -> Na2WO4 | -0.321 | -321 | -1230 | -908.4 | 0.354 |
| 21 | CaO + WO3 -> CaWO4 | -0.342 | -342 | -1202 | -860.3 | 0.397 |
| 22 | Li2O2 + TiO -> Li2TiO3 | 0.00117 | 1.17 | -853.8 | -855 | 0.00137 |
| 23 | Cs2O + MoO3 -> Cs2MoO4 | -0.218 | -218 | -1003 | -784.3 | 0.278 |
| 24 | Li2O + WO3 -> Li2WO4 | -0.312 | -312 | -1054 | -741.5 | 0.421 |
| 25 | MnO + WO3 -> MnWO4 | -0.355 | -355 | -1084 | -729.7 | 0.486 |
| 26 | MgO + WO3 -> MgWO4 | -0.366 | -366 | -1076 | -709.5 | 0.516 |
| 27 | 0.5 Li2O2 + NbO2 -> LiNbO3 | 0.00555 | 5.55 | -700.4 | -705.9 | 0.00787 |
| 28 | CoO + WO3 -> CoWO4 | -0.367 | -367 | -1056 | -688.8 | 0.533 |
| 29 | SrO + Al2O3 -> SrAl2O4 | -0.279 | -279 | -961.7 | -682.8 | 0.408 |
| 30 | ZnO + WO3 -> ZnWO4 | -0.346 | -346 | -1014 | -668.9 | 0.517 |
| 31 | Cs2O + SiO2 -> Cs2SiO3 | -0.321 | -321 | -963 | -642.3 | 0.499 |
| 32 | K2O + SiO2 -> K2SiO3 | 0.015 | 15 | -549.8 | -564.8 | 0.0265 |
| 33 | 1.5 Na2O + 0.5 V2O5 -> Na3VO4 | -0.017 | -17 | -568.9 | -551.9 | 0.0309 |
| 34 | 3 CaO + WO3 -> Ca3WO6 | -0.213 | -213 | -761.3 | -548.1 | 0.389 |
| 35 | Na2O + MoO3 -> Na2MoO4 | -0.0148 | -14.8 | -549.3 | -534.5 | 0.0277 |
| 36 | 0.5 Na2O + 0.5 Al2O3 -> NaAlO2 | 0.271 | 271 | -237.4 | -508.5 | 0.533 |
| 37 | 0.5 Na2O2 + VO2 -> NaVO3 | 0.04 | 40 | -458.3 | -498.3 | 0.0803 |
| 38 | Na2O2 + Ti3O5 -> Na2Ti3O7 | 0.00839 | 8.39 | -439.9 | -448.3 | 0.0187 |
| 39 | 0.5 Na2O + 0.5 V2O5 -> NaVO3 | 0.00569 | 5.69 | -430 | -435.6 | 0.0131 |
| 40 | Na2O + 2 MoO3 -> | 0.00323 | 3.23 | -426.3 | -429.6 | 0.00752 |
| | Na2Mo2O7 | | | | | |
| 41 | K2O + 2 SiO2 -> K2Si2O5 | 0.0112 | 11.2 | -378.6 | -389.9 | 0.0288 |
| 42 | 2 BaO + SiO2 -> Ba2SiO4 | -0.0144 | -14.4 | -396.3 | -382 | 0.0376 |
| 43 | BaO + MoO3 -> BaMoO4 | -0.00506 | -5.06 | -385.1 | -380 | 0.0133 |
| 44 | 2 Na2O + SiO2 -> Na4SiO4 | -0.0355 | -35.5 | -410.9 | -375.4 | 0.0945 |
| 45 | Na2O + SiO2 -> Na2SiO3 | -0.0366 | -36.6 | -408.9 | -372.3 | 0.0984 |
| 46 | BaO + SiO2 -> BaSiO3 | -0.0225 | -22.5 | -334.8 | -312.4 | 0.072 |
| 47 | BaO + TiO2 -> BaTiO3 | -0.0291 | -29.1 | -327.8 | -298.7 | 0.0975 |
| 48 | CaO + MoO3 -> CaMoO4 | 0.0114 | 11.4 | -286.2 | -297.6 | 0.0385 |
| 49 | 2 BaO + TiO2 -> Ba2TiO4 | 0.00385 | 3.85 | -289.8 | -293.6 | 0.0131 |
| 50 | 3 CaO + V2O5 -> Ca3V2O8 | 0.0238 | 23.8 | -256.8 | -280.6 | 0.0847 |
| 51 | BaO2 + 2 VO2 -> BaV2O6 | 0.00737 | 7.37 | -253.4 | -260.7 | 0.0283 |
| 52 | 2 CaO + V2O5 -> Ca2V2O7 | 0.0128 | 12.8 | -247.2 | -260 | 0.0493 |
| 53 | Na2O + 2 SiO2 -> Na2Si2O5 | -0.0161 | -16.1 | -275.1 | -259 | 0.0623 |
| 54 | BaO + ZrO2 -> BaZrO3 | 0.00466 | 4.66 | -248.8 | -253.5 | 0.0184 |
| 55 | 0.5 Na2O + 0.5 Cr2O3 -> NaCrO2 | -0.0128 | -12.8 | -259.4 | -246.6 | 0.0518 |
| 56 | Li2O + SiO2 -> Li2SiO3 | -0.00233 | -2.33 | -244.3 | -241.9 | 0.00963 |
| 57 | BaO + HfO2 -> BaHfO3 | -0.021 | -21 | -257.5 | -236.5 | 0.0889 |
| 58 | 0.5 Li2O + 0.5 Nb2O5 -> LiNbO3 | -0.00451 | -4.51 | -240.5 | -236 | 0.0191 |
| 59 | Li2O + TiO2 -> Li2TiO3 | 0.00699 | 6.99 | -223.9 | -230.9 | 0.0303 |
| 60 | 2 CaO + SiO2 -> Ca2SiO4 | 0.0116 | 11.6 | -208.9 | -220.4 | 0.0525 |
| 61 | K2O + 4 SiO2 -> K2Si4O9 | -0.00133 | -1.33 | -220.6 | -219.3 | 0.00608 |
| 62 | BaO + V2O5 -> BaV2O6 | -0.0102 | -10.2 | -211.5 | -201.3 | 0.0506 |
| 63 | Na2O + 3 TiO2 -> Na2Ti3O7 | -0.0064 | -6.4 | -207 | -200.6 | 0.0319 |
| 64 | 0.5 Li2O + 0.5 Ta2O5 -> LiTaO3 | -0.00098 | -0.984 | -200.2 | -199.2 | 0.00494 |
| 65 | 0.5 Na2O + 0.5 Fe2O3 -> NaFeO2 | -0.00312 | -3.12 | -198.3 | -195.2 | 0.016 |
| 66 | CaO + SiO2 -> CaSiO3 | 0.012 | 12 | -175.3 | -187.4 | 0.0643 |
| 67 | CaO + TiO2 -> CaTiO3 | 0.0116 | 11.6 | -174.3 | -185.9 | 0.0625 |
| 68 | 0.5 Li2O + 0.5 Fe2O3 -> LiFeO2 | 0.0328 | 32.8 | -147.3 | -180.2 | 0.182 |
| 69 | CaO + V2O5 -> CaV2O6 | 0.012 | 12 | -165.2 | -177.2 | 0.0676 |
| 70 | CaO + GeO2 -> CaGeO3 | 0.0197 | 19.7 | -145.5 | -165.2 | 0.119 |
| 71 | BaO + Al2O3 -> BaAl2O4 | 0.0136 | 13.6 | -148.6 | -162.2 | 0.084 |
| 72 | 0.5 Li2O + 0.5 Al2O3 -> LiAlO2 | 0.00196 | 1.96 | -133.3 | -135.3 | 0.0145 |
| 73 | MgO + MoO3 -> MgMoO4 | 0.0245 | 24.5 | -93.22 | -117.7 | 0.208 |
| 74 | Li2O + ZrO2 -> Li2ZrO3 | 0.00578 | 5.78 | -108.4 | -114.2 | 0.0507 |
| 75 | MnO + MoO3 -> MnMoO4 | -0.00273 | -2.73 | -106.2 | -103.5 | 0.0264 |
| 76 | CaO + Cr2O3 -> CaCr2O4 | 0.00941 | 9.41 | -88.67 | -98.08 | 0.096 |
| 77 | Li2O + HfO2 -> Li2HfO3 | -0.00181 | -1.81 | -99.75 | -97.94 | 0.0185 |
| 78 | 0.5 La2O3 + 0.5 Al2O3 -> LaAlO3 | -0.0246 | -24.6 | -121.4 | -96.82 | 0.254 |
| 79 | 2 MgO + SiO2 -> Mg2SiO4 | -0.00281 | -2.81 | -97.86 | -95.05 | 0.0296 |
| 80 | 2 MgO + V2O5 -> Mg2V2O7 | 0.0149 | 14.9 | -77.25 | -92.13 | 0.162 |
| 81 | ZnO + Cr2O3 -> ZnCr2O4 | -0.0118 | -11.8 | -93.09 | -81.33 | 0.145 |
| 82 | 2 MnO + SiO2 -> Mn2SiO4 | 0.000296 | 0.296 | -78.14 | -78.44 | 0.00377 |
| 83 | CaO + ZrO2 -> CaZrO3 | 0.0106 | 10.6 | -63.42 | -74.01 | 0.143 |
| 84 | MgO + GeO2 -> MgGeO3 | 0.00816 | 8.16 | -56.07 | -64.23 | 0.127 |
| 85 | MgO + V2O5 -> MgV2O6 | 0.00386 | 3.86 | -56.82 | -60.68 | 0.0636 |
| 86 | CaO + HfO2 -> CaHfO3 | 0.00334 | 3.34 | -56.59 | -59.92 | 0.0557 |
| 87 | MgO + Cr2O3 -> MgCr2O4 | -0.00237 | -2.37 | -61.95 | -59.58 | 0.0397 |
| 88 | MgO + TiO2 -> MgTiO3 | -0.00445 | -4.45 | -62.08 | -57.62 | 0.0773 |
| 89 | 2 ZnO + SiO2 -> Zn2SiO4 | 0.00118 | 1.18 | -50.02 | -51.21 | 0.0231 |
| 90 | 2 MgO + TiO2 -> Mg2TiO4 | 0.017 | 17 | -30.37 | -47.39 | 0.359 |
| 91 | CdO + SiO2 -> CdSiO3 | -0.00145 | -1.45 | -46.93 | -45.48 | 0.0319 |
| 92 | MgO + Al2O3 -> MgAl2O4 | -0.00798 | -7.98 | -53.22 | -45.24 | 0.176 |
| 93 | 2 CoO + SiO2 -> Co2SiO4 | 0.0141 | 14.1 | -22.08 | -36.14 | 0.389 |
| 94 | ZrO2 + SiO2 -> ZrSiO4 | -0.0161 | -16.1 | -44.2 | -28.14 | 0.571 |
| 95 | MnO + Al2O3 -> MnAl2O4 | -0.0344 | -34.4 | -58.9 | -24.51 | 1.4 |
| 96 | 2 BeO + SiO2 -> Be2SiO4 | -0.00968 | -9.68 | -31.08 | -21.4 | 0.452 |
| 97 | CaO + 2 Al2O3 -> CaAl4O7 | 0.00443 | 4.43 | -10.89 | -15.32 | 0.289 |
| 98 | BeO + Al2O3 -> BeAl2O4 | -0.0218 | -21.8 | -25.38 | -3.611 | 6.03 |
| 99 | Al2O3 + SiO2 -> Al2SiO5 | -0.0188 | -18.8 | -7.123 | 11.71 | 1.61 |
| 100 | MgO + 2 TiO2 -> MgTi2O5 | 0.0116 | 11.6 | 183.9 | 172.4 | 0.0672 |

## SI3.4. Failed synthesis: Summary/discussion

For a number of compounds, neither set of precursors produced an XRD signal matching the target crystal structure, and as such these synthesis attempts are classified as failures based on the XRD quantification method used in this study.

While it is often not possible to determine the exact reason for an unsuccessful synthesis, there are several common factors that can result in failed synthesis even for a thermodynamically stable target:

1) **Insufficient synthesis temperature**
   - If the calcination temperature is insufficient, some of the precursors may not fully decompose and react, and as a result does not form a uniform product.
   - Likely applies to $KTiNbO_5$ and $Li_3Y_2(BO_3)_2$ in the present study.
2) **Evaporation of precursors**
   - Some precursors have significant vapor pressure and are prone to being lost to evaporation during calcination, resulting in deficiency of the affected components.
   - Well known to occur with Li, P, B precursors.
   - The great majority of studies on Li-oxide synthesis for example add excess Li precursor, most often 10%, to hedge against evaporation. We did not do this here, since it was difficult to apply this uniformly over such a broad chemical space, including Na- and K- based compounds.
   - It is hard to determine *a prioi* which samples evaporation could apply to here – usually this will affect purity more than overall success/failure, but it can be very impactful in cases where (a) formation of the target phase requires high temperature and/or longer times, or (b) there is a small composition window for the target phase.
3) **Excessive oxidation during synthesis**
   - For all ASTRAL synthesis experiments presented in this study, calcination was performed in ambient air, and as such each element will attain the most energetically favorable oxidation state based on reaction with oxygen gas at high temperature.
   - For materials containing transition metals, this can result in incorrect oxidation states during synthesis, preventing formation of the target phase.
   - Likely applies to $Li_3V_2(PO_4)_3$ – in the literature report, a reducing atmosphere (Ar + $H_2$) is used for the final synthesis reaction. Also, V has many available oxidation states (+2, +3, +4, +5), and for the target material we need $V^{3+}$, so it is reasonable to suspect that $V^{5+}$ formation could be the cause of the failure.
4) **Amorphous synthesis products**
   - ASTRAL classifies synthesis outcomes based on powder XRD, and so any amorphous phases present are not detected or used for quantification. This can result in an apparently failed synthesis, even in cases where the sample has formed a homogeneous mixture of the correct composition.
   - For such glass forming compositions, successful crystallization requires controlled cooling, typically with a period of annealing at an appropriate temperature to nucleate and grow crystals.
   - Likely applies to $KLi(PO_3)_2$ and $NaSiBO_4$ in the present study, as each (1) contains a high proportion of glass-forming elements, (2) each formed fully fused samples with a glassy appearance, and (3) despite adequate yield of reaction product, almost no crystalline peaks were observed on XRD.

## LiZnPO4

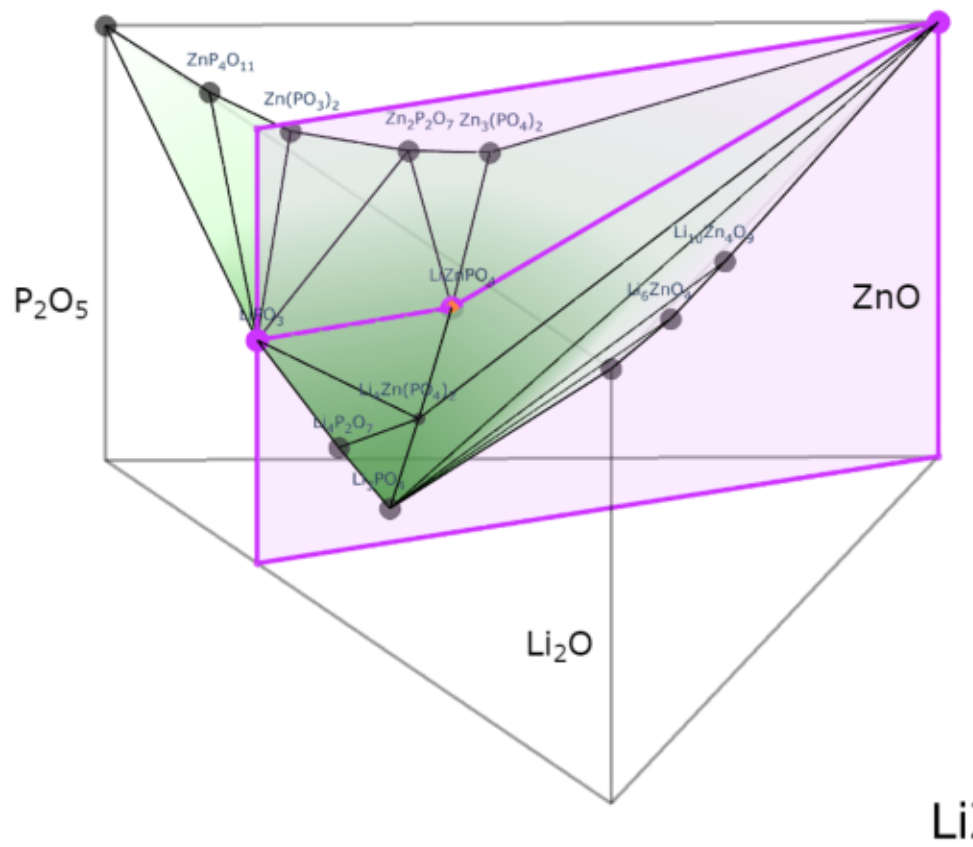

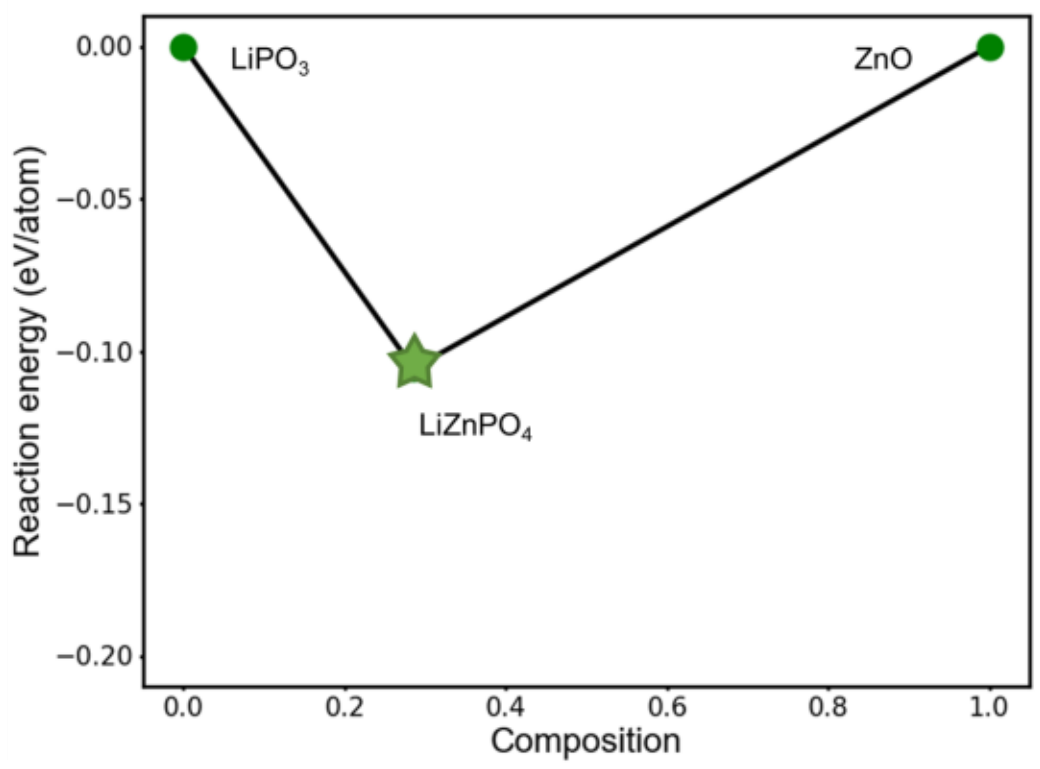

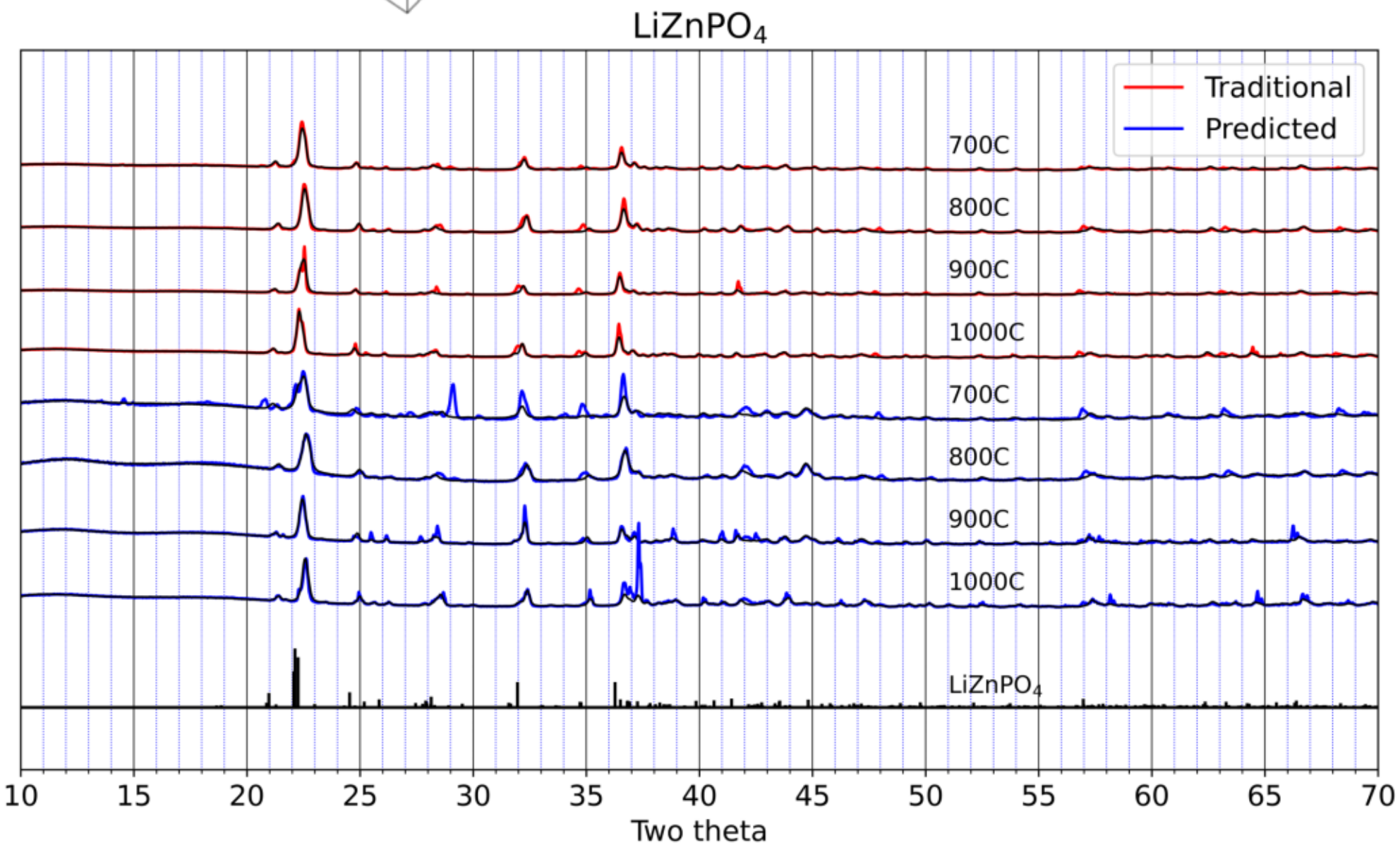

| Compound | Precursor type | Precursors | Temp (°C) | Target intensity (e6) | Residual intensity (e6) | Target phase fraction |
|---|---|---|---|---|---|---|
| $LiZnPO_4$ | Traditional | $Li_2CO_3$, $NH_4H_2PO_4$, ZnO | 700 | 3.78 | 1.35 | 0.74 |
| | | | 800 | 5.01 | 1.62 | 0.76 |
| | | | 900 | 3.53 | 1.70 | 0.67 |
| | | | 1000 | 3.40 | 1.18 | 0.74 |
| | Predicted | $LiPO_3$, ZnO | 700 | 1.83 | 1.34 | 0.58 |
| | | | 800 | 1.54 | 0.65 | 0.70 |
| | | | 900 | 2.12 | 0.99 | 0.68 |
| | | | 1000 | 2.54 | 1.26 | 0.67 |

## LiMnPO4

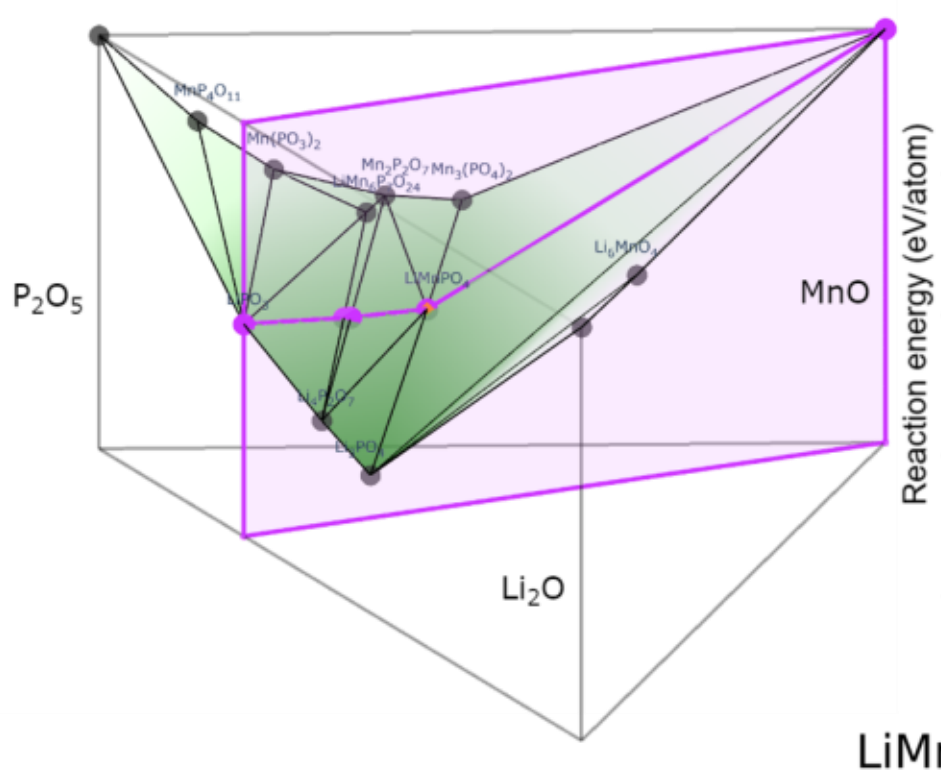

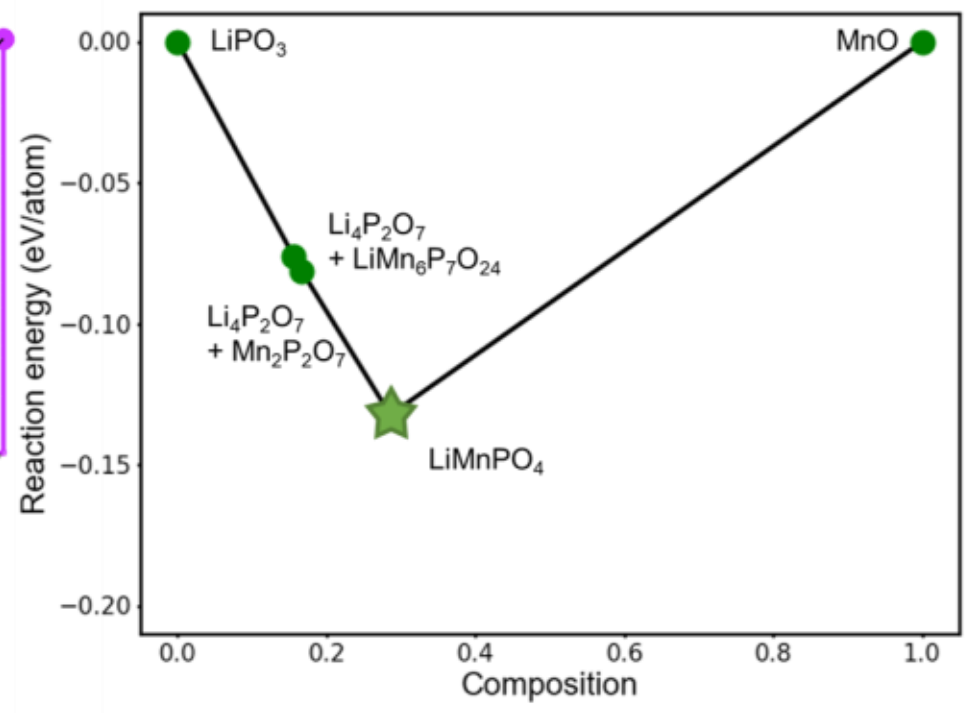

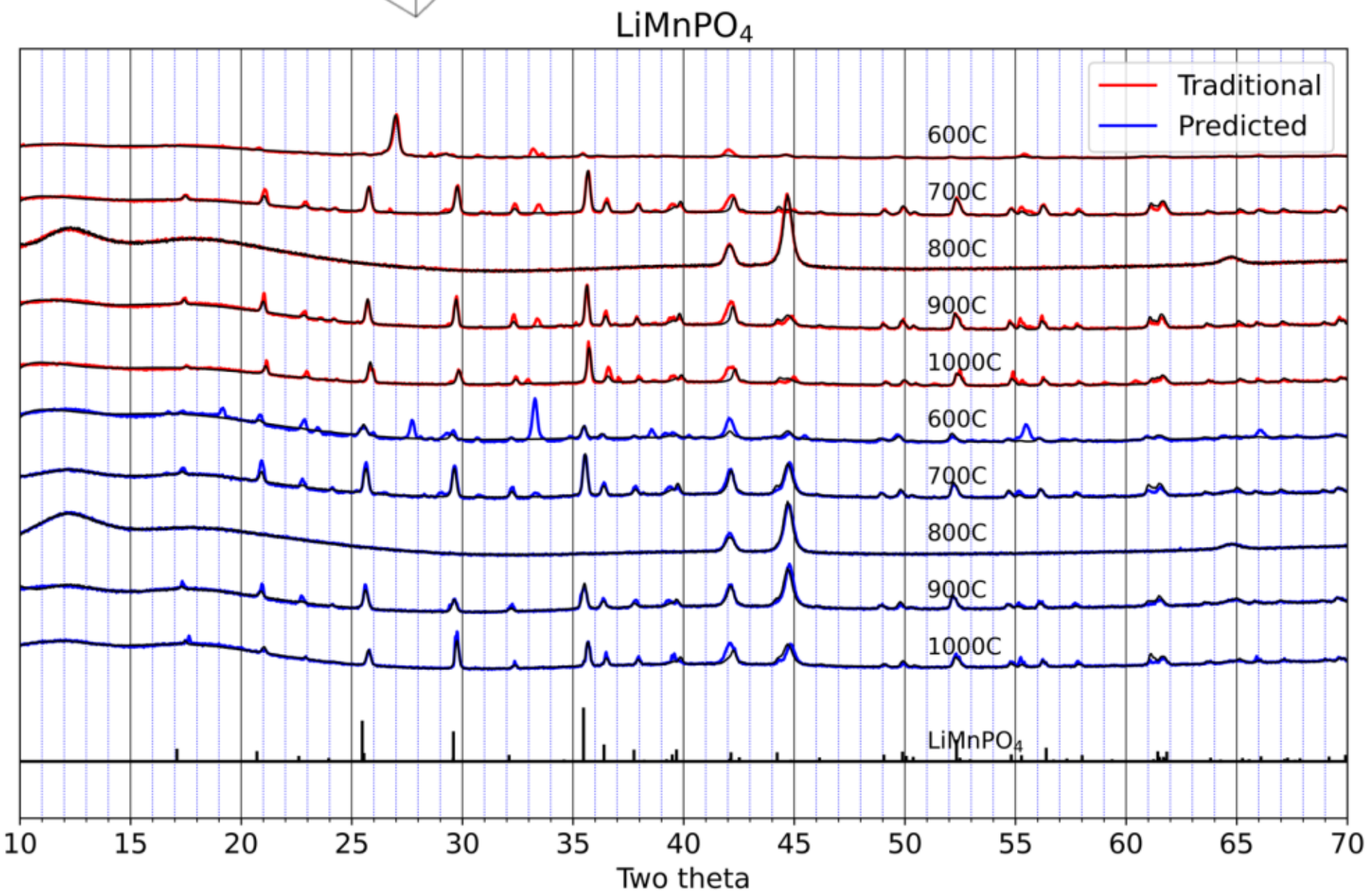

| Compound | Precursor type | Precursors | Temp (°C) | Target intensity (e6) | Residual intensity (e6) | Target phase fraction |
|---|---|---|---|---|---|---|
| LiMnPO4 | Traditional | $Li_2CO_3$, MnO, $NH_4H_2PO_4$ | 600 | 0.28 | 0.78 | 0.26 |
| | | | 700 | 1.00 | 0.59 | 0.63 |
| | | | 800 | 0.00 | 0.29 | 0.01 |
| | | | 900 | 0.60 | 0.47 | 0.56 |
| | | | 1000 | 0.51 | 0.46 | 0.52 |
| | Predicted | MnO, $LiPO_3$ | 600 | 0.32 | 0.90 | 0.27 |
| | | | 700 | 0.66 | 0.46 | 0.59 |
| | | | 800 | 0.00 | 0.26 | 0.02 |
| | | | 900 | 0.42 | 0.36 | 0.54 |
| | | | 1000 | 0.40 | 0.38 | 0.51 |

# $Li_2CuP_2O_7$

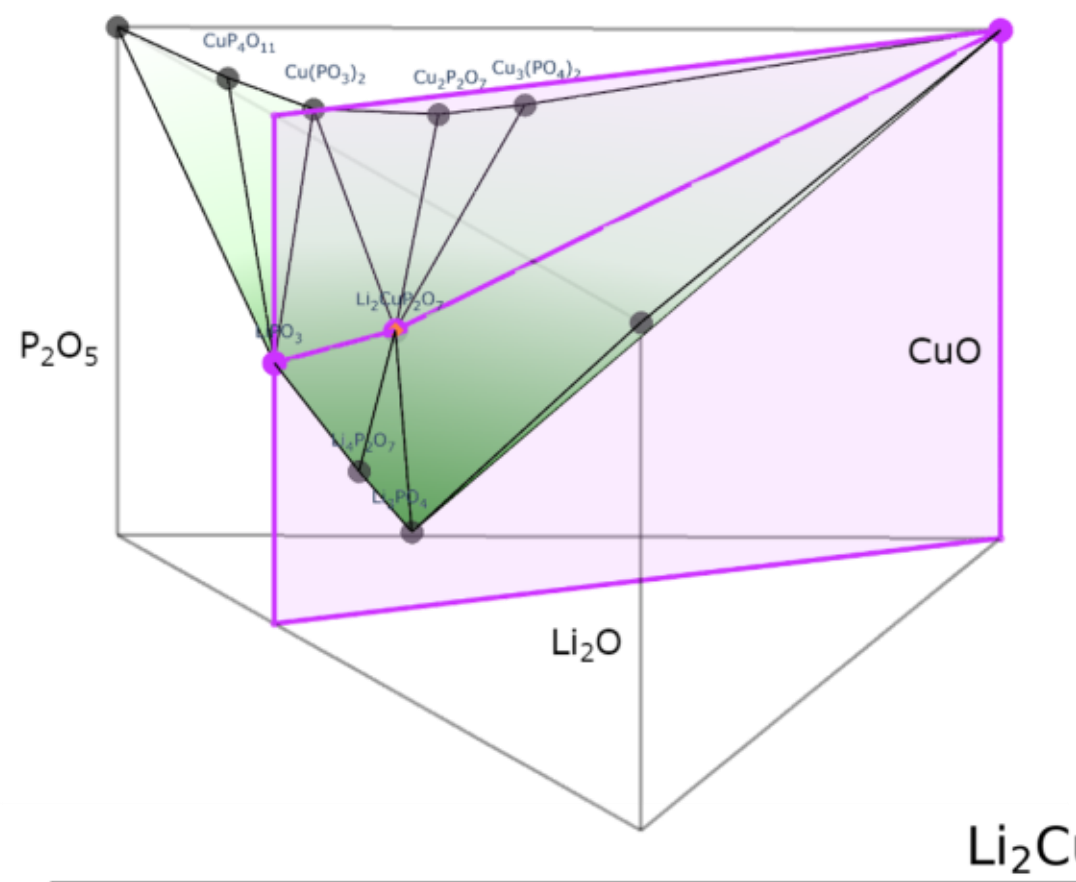

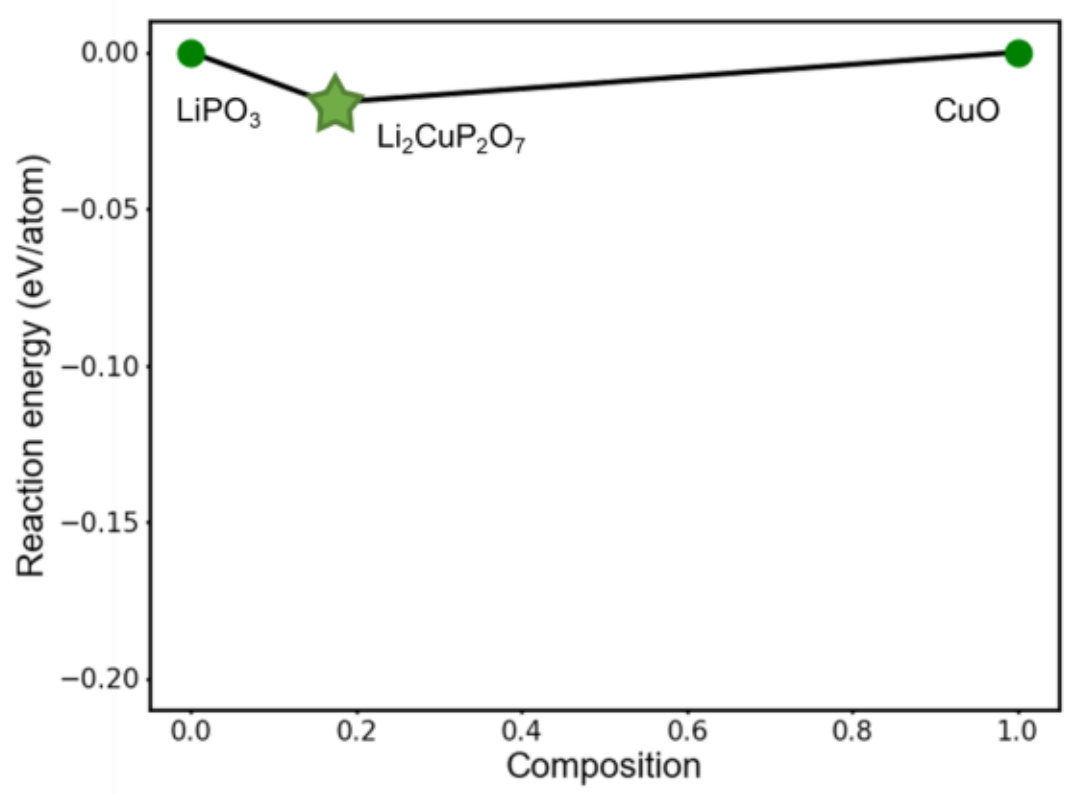

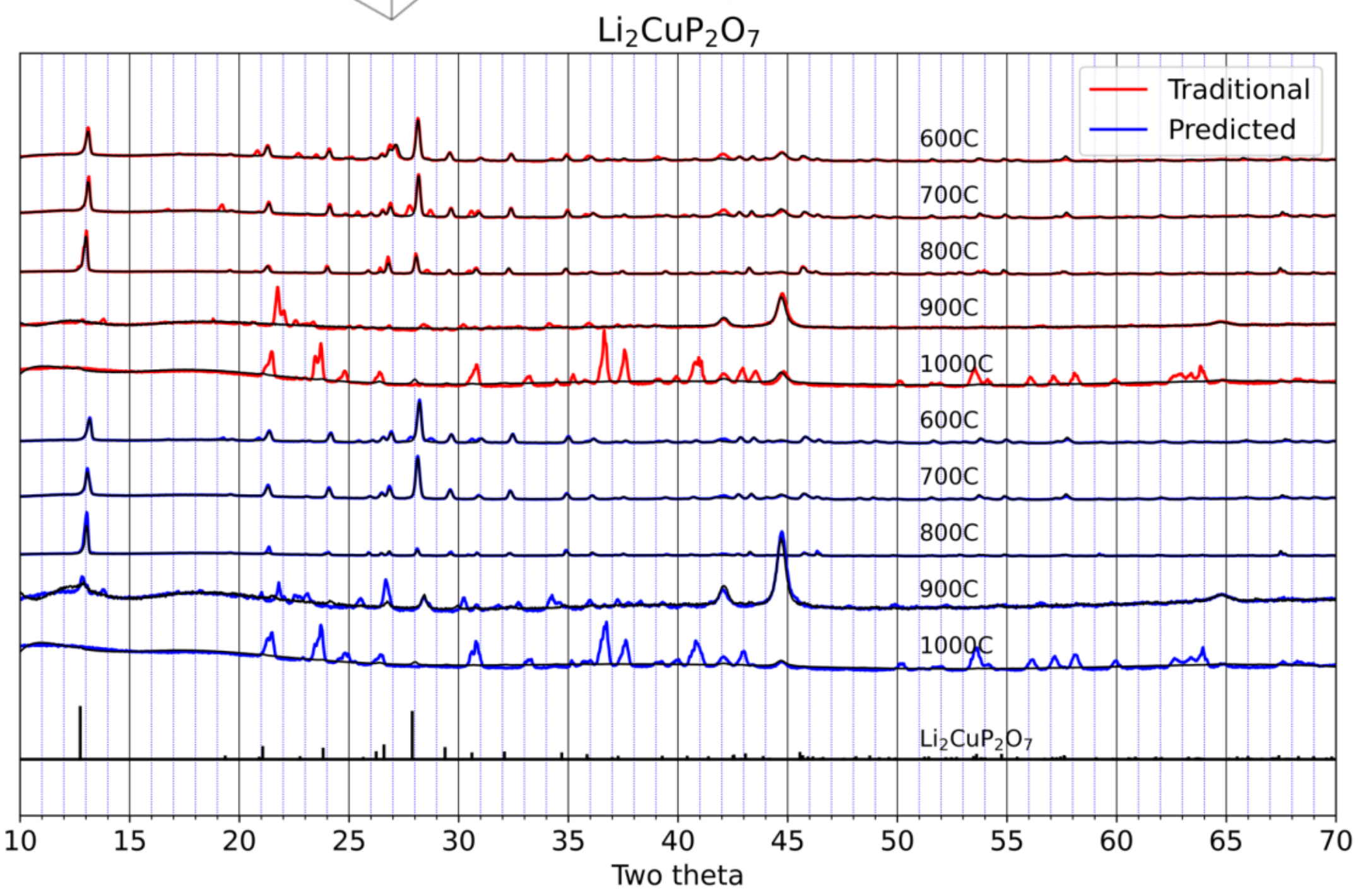

| Compound | Precursor type | Precursors | Temp (°C) | Target intensity (e6) | Residual intensity (e6) | Target phase fraction |
|---|---|---|---|---|---|---|
| $Li_2CuP_2O_7$ | Traditional | CuO, $Li_2CO_3$, $NH_4H_2PO_4$ | 600 | 1.90 | 1.10 | 0.63 |
| | | | 700 | 1.93 | 1.09 | 0.64 |
| | | | 800 | 3.15 | 1.22 | 0.72 |
| | | | 900 | 0.00 | 0.84 | 0.00 |
| | | | 1000 | 0.10 | 1.59 | 0.06 |
| | Predicted | CuO, $LiPO_3$ | 600 | 3.37 | 1.11 | 0.75 |
| | | | 700 | 3.00 | 0.83 | 0.78 |
| | | | 800 | 2.83 | 2.04 | 0.58 |
| | | | 900 | 0.20 | 0.70 | 0.22 |
| | | | 1000 | 0.05 | 1.46 | 0.03 |

## $LiMgPO_4$

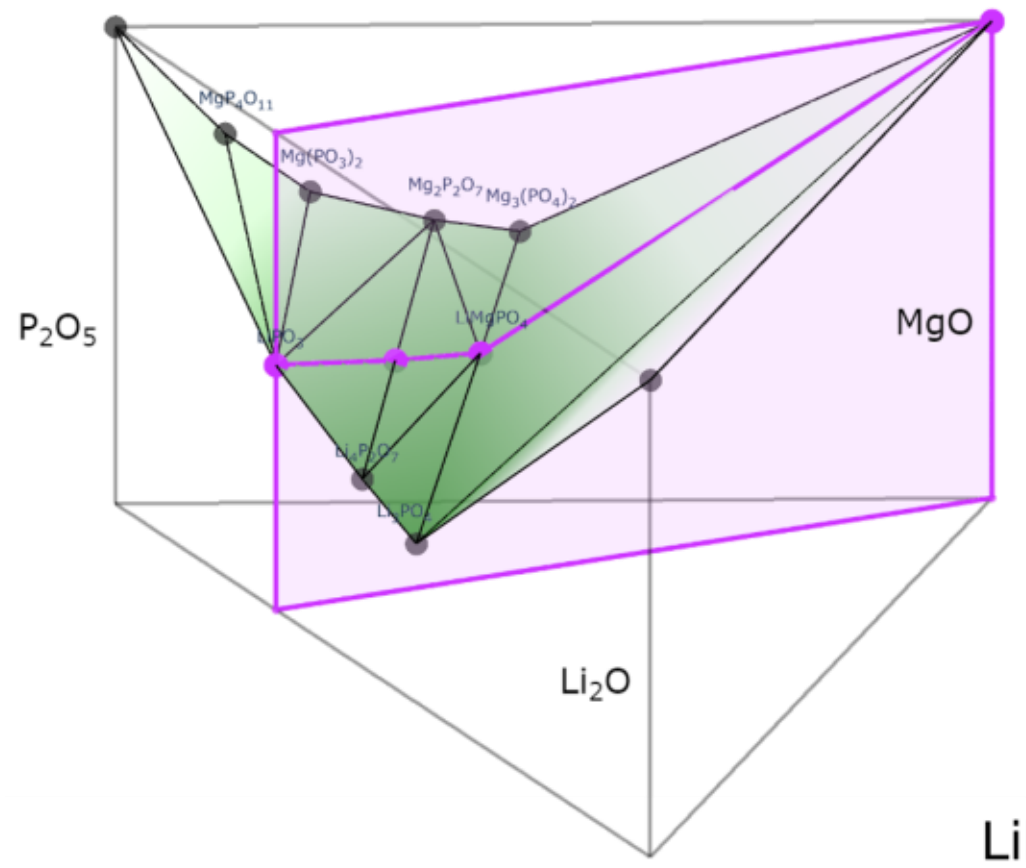

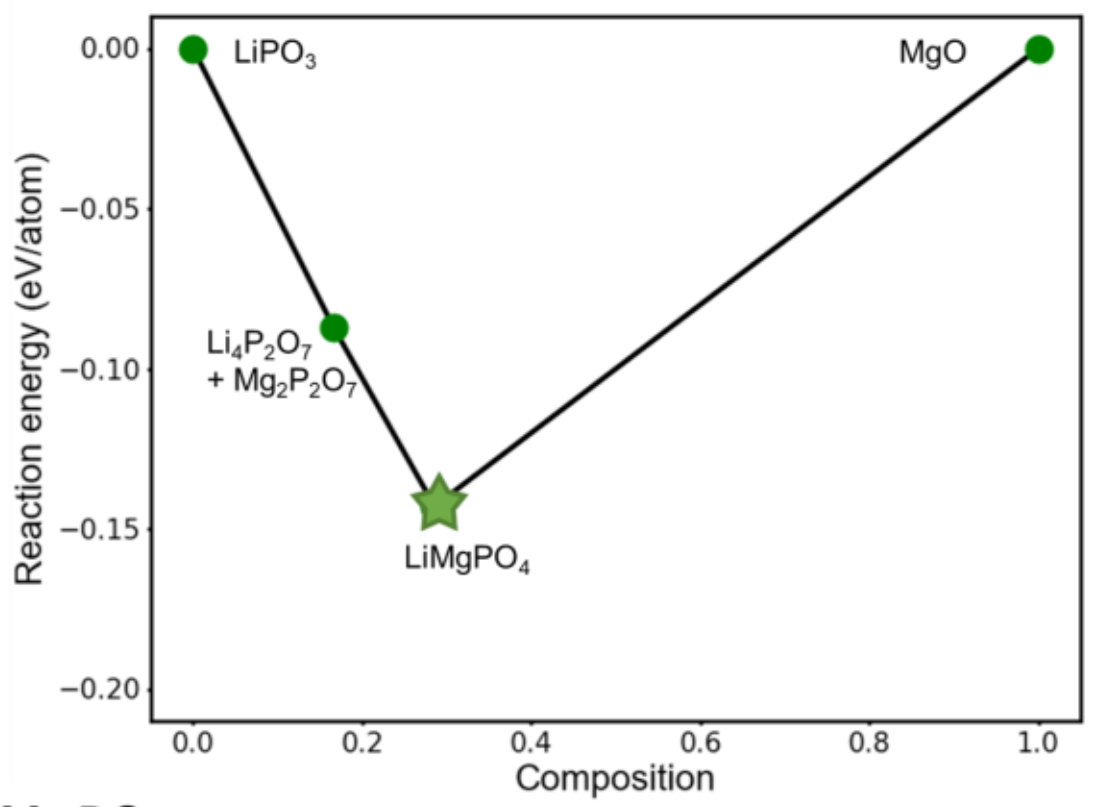

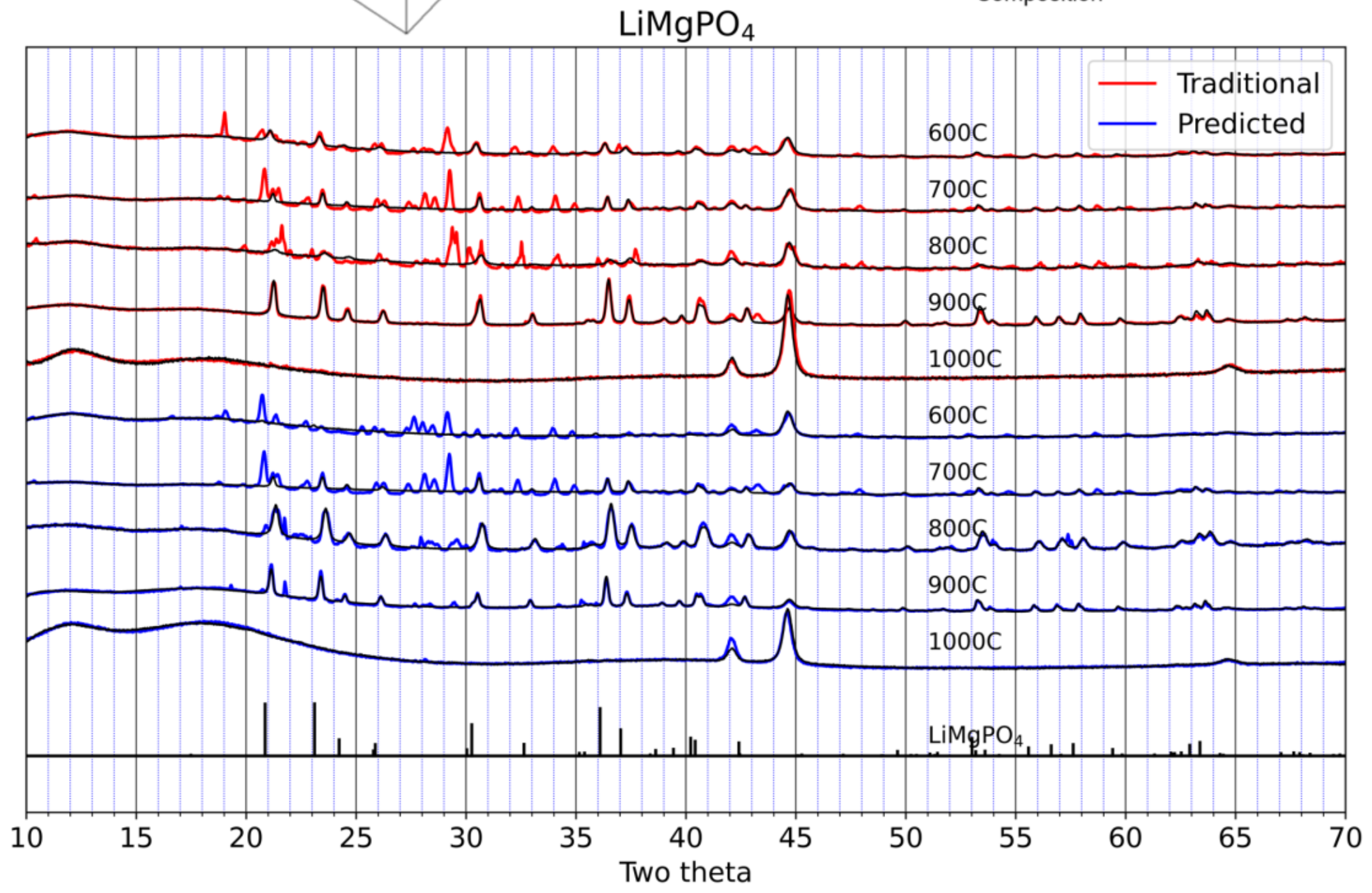

| Compound | Precursor type | Precursors | Temp (°C) | Target intensity (e6) | Residual intensity (e6) | Target phase fraction |
|---|---|---|---|---|---|---|
| $LiMgPO_4$ | Traditional | $Li_2CO_3$, MgO, $NH_4H_2PO_4$ | 600 | 0.65 | 1.04 | 0.39 |
| | | | 700 | 0.13 | 0.95 | 0.12 |
| | | | 800 | 0.32 | 0.81 | 0.29 |
| | | | 900 | 1.61 | 0.53 | 0.75 |
| | | | 1000 | 0.03 | 0.38 | 0.07 |
| | Predicted | MgO, $LiPO_3$ | 600 | 0.06 | 1.13 | 0.05 |
| | | | 700 | 0.14 | 0.77 | 0.15 |
| | | | 800 | 1.33 | 0.56 | 0.70 |
| | | | 900 | 1.37 | 0.92 | 0.60 |
| | | | 1000 | 0.11 | 0.52 | 0.17 |

## $BaLiBO_3$

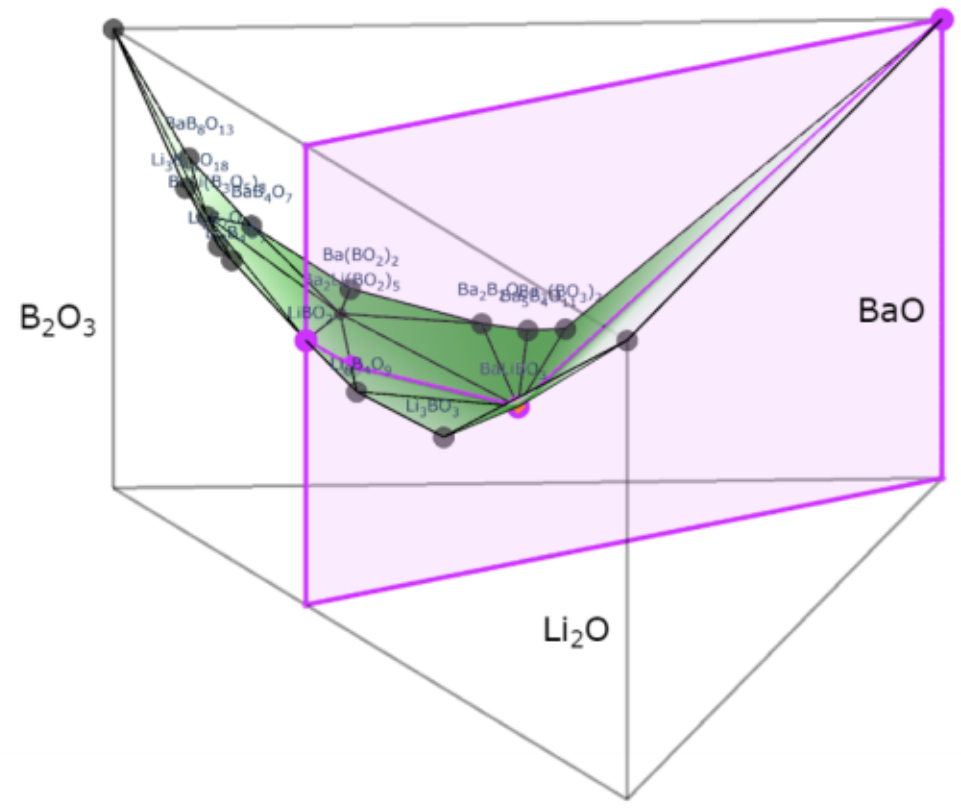

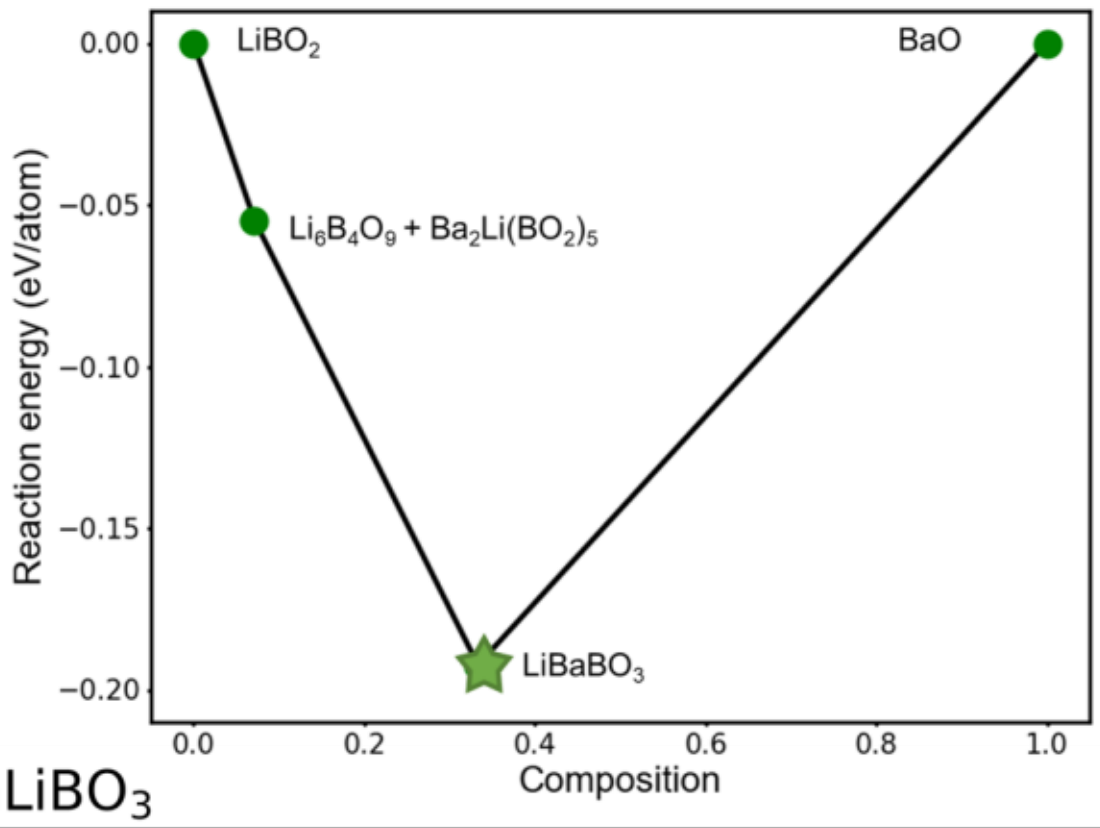

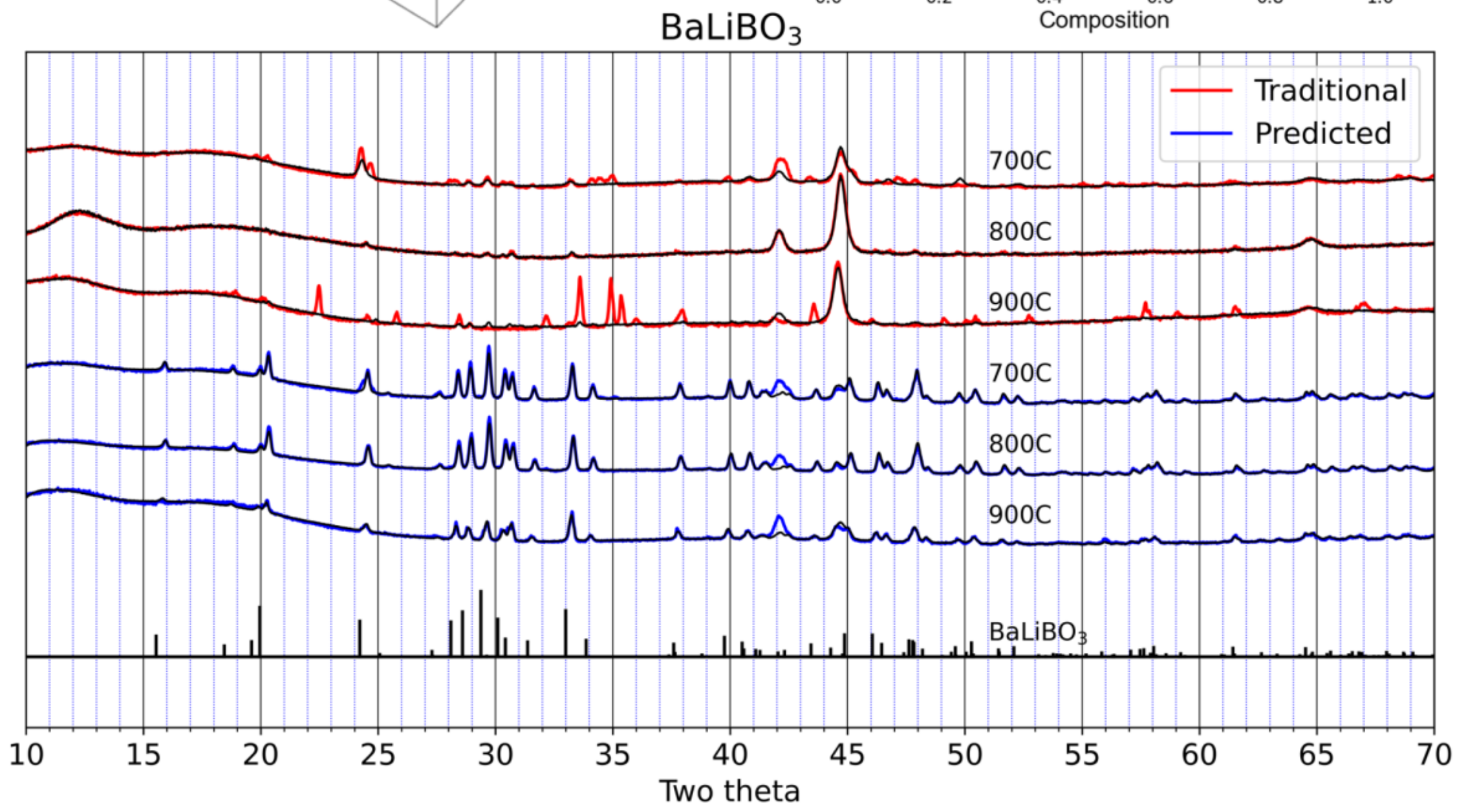

| Compound | Precursor type | Precursors | Temp (°C) | Target intensity (e6) | Residual intensity (e6) | Target phase fraction |
|---|---|---|---|---|---|---|
| $BaLiBO_3$ | Traditional | $Li_2CO_3$, BaO, $B_2O_3$ | 700 | 0.23 | 0.49 | 0.32 |
| | | | 800 | 0.11 | 0.25 | 0.30 |
| | | | 900 | 0.08 | 0.69 | 0.10 |
| | Predicted | BaO, $LiBO_2$ | 700 | 0.98 | 0.35 | 0.74 |
| | | | 800 | 1.14 | 0.34 | 0.77 |
| | | | 900 | 0.54 | 0.36 | 0.60 |

## $SrLiBO_3$

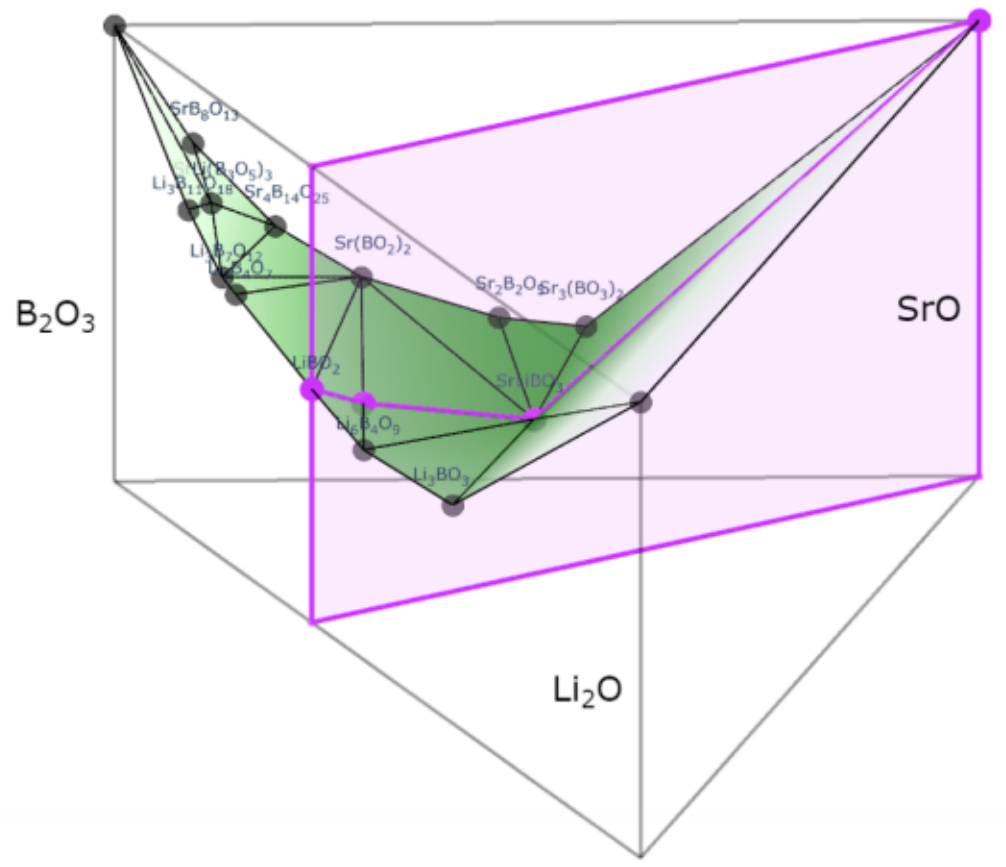

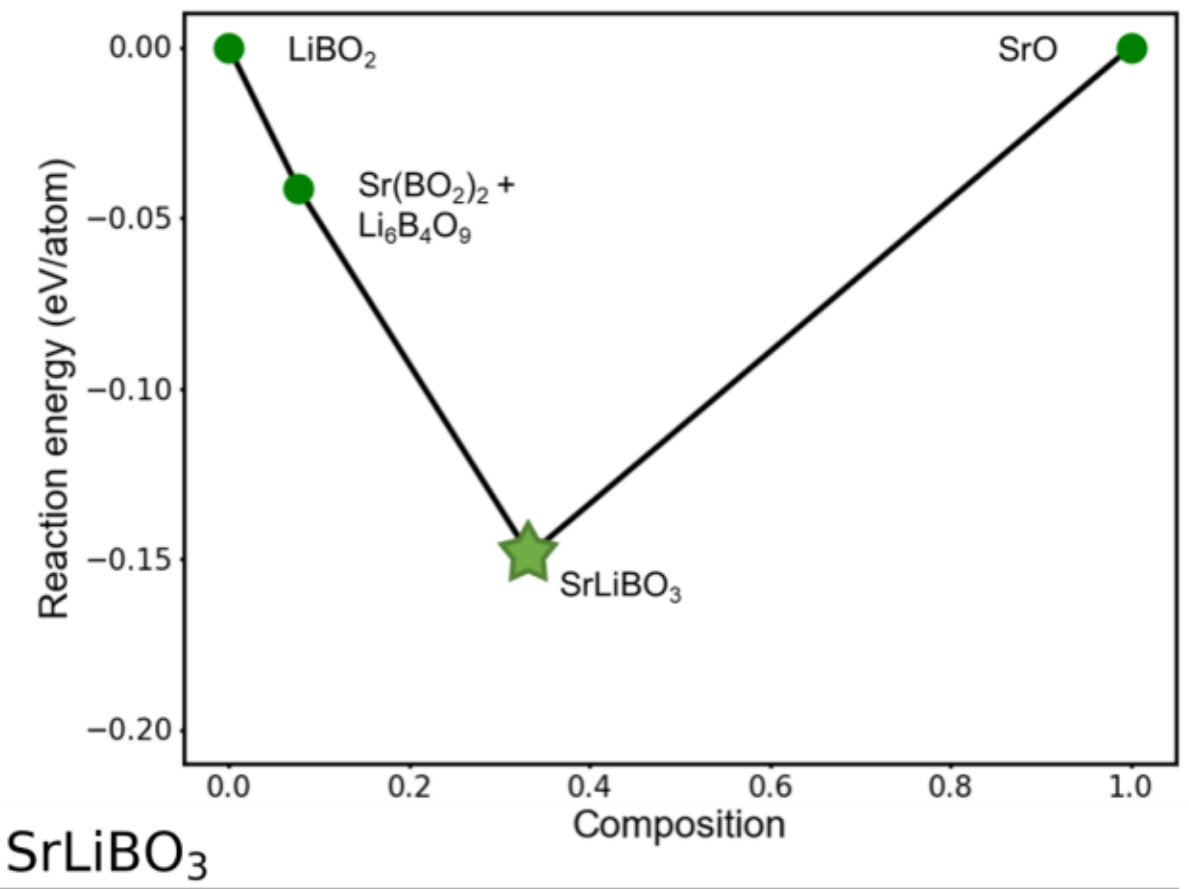

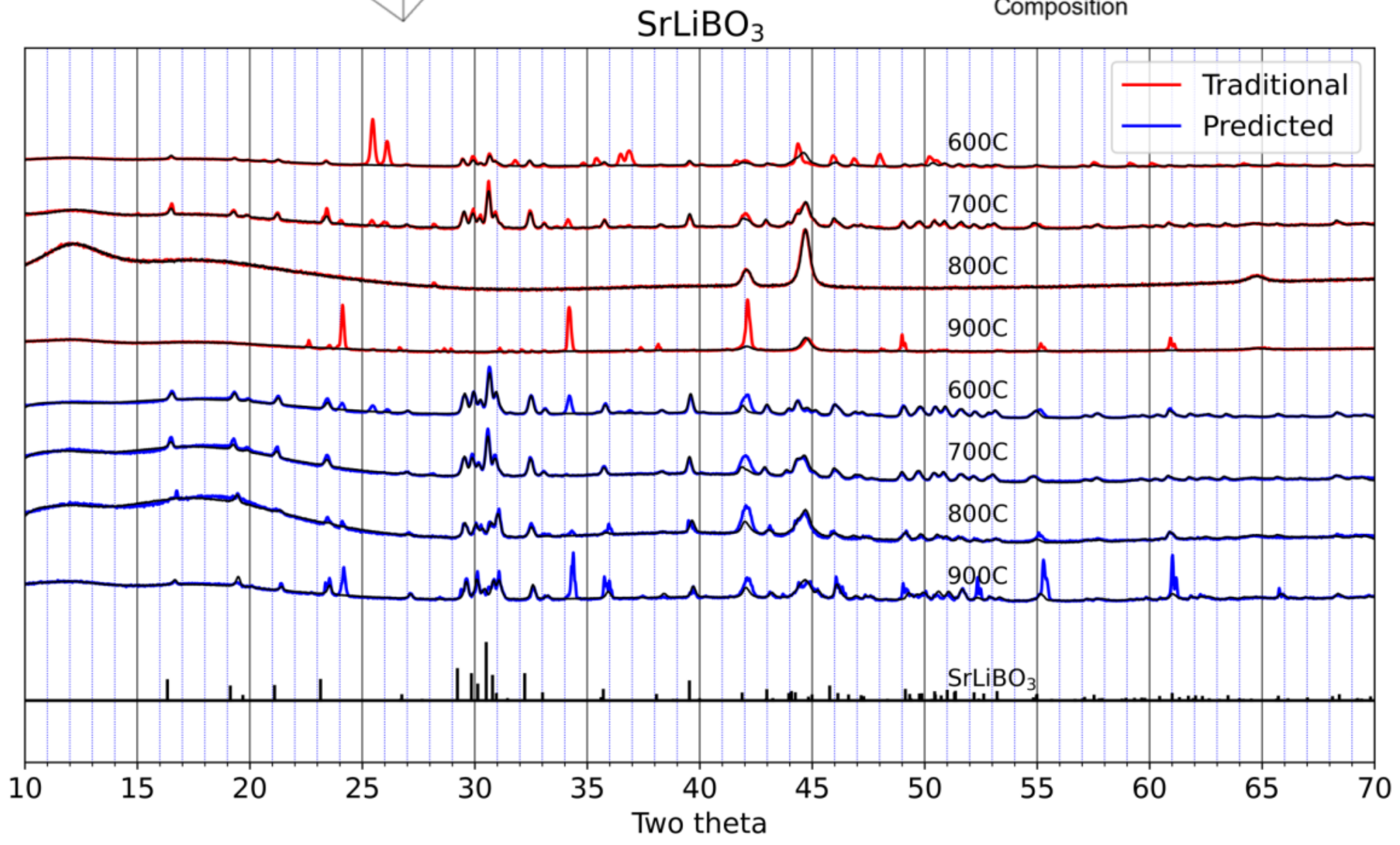

| Compound | Precursor type | Precursors | Temp (°C) | Target intensity (e6) | Residual intensity (e6) | Target phase fraction |
|---|---|---|---|---|---|---|
| $SrLiBO_3$ | Traditional | $Li_2CO_3$, SrO, $B_2O_3$ | 600 | 1.02 | 1.98 | 0.34 |
| | | | 700 | 1.22 | 0.58 | 0.68 |
| | | | 800 | 0.00 | 0.25 | 0.00 |
| | | | 900 | 0.00 | 1.00 | 0.00 |
| | Predicted | SrO, $LiBO_2$ | 600 | 3.28 | 1.39 | 0.70 |
| | | | 700 | 1.47 | 0.77 | 0.66 |
| | | | 800 | 0.71 | 0.70 | 0.50 |
| | | | 900 | 0.92 | 0.81 | 0.53 |

## $Li_3Pr_2(BO_3)_3$

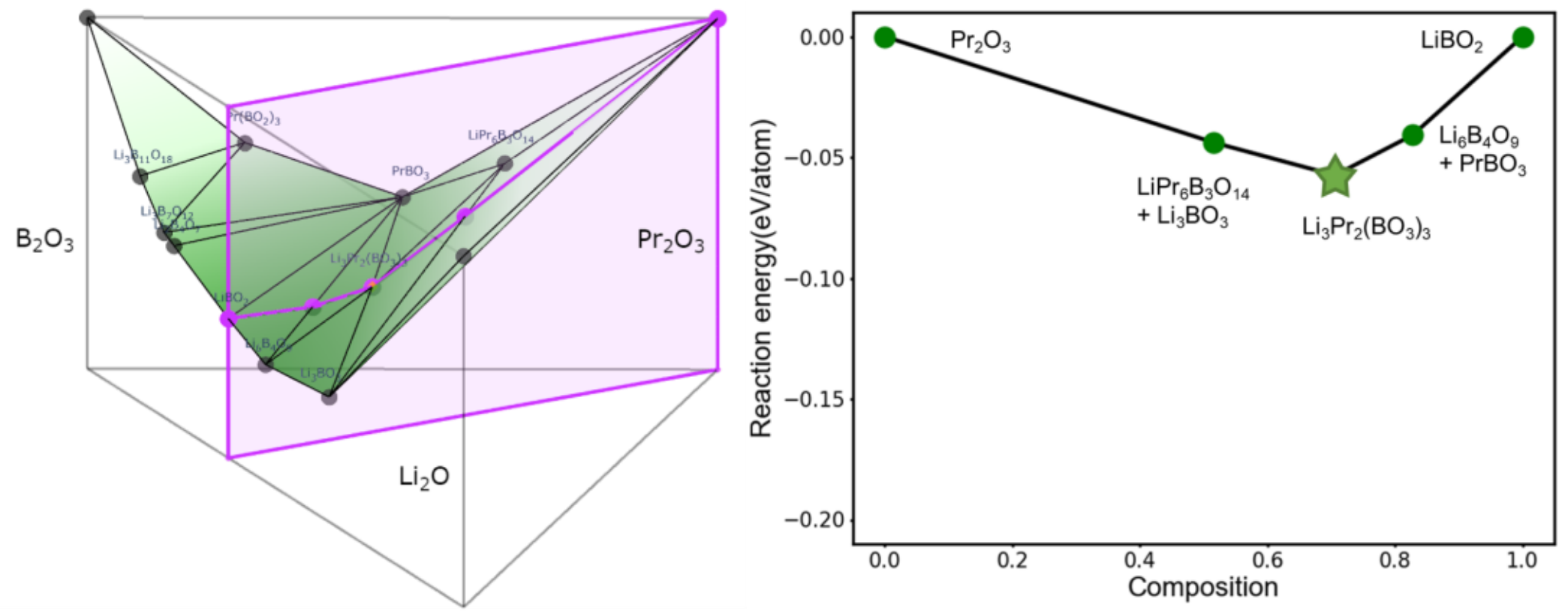

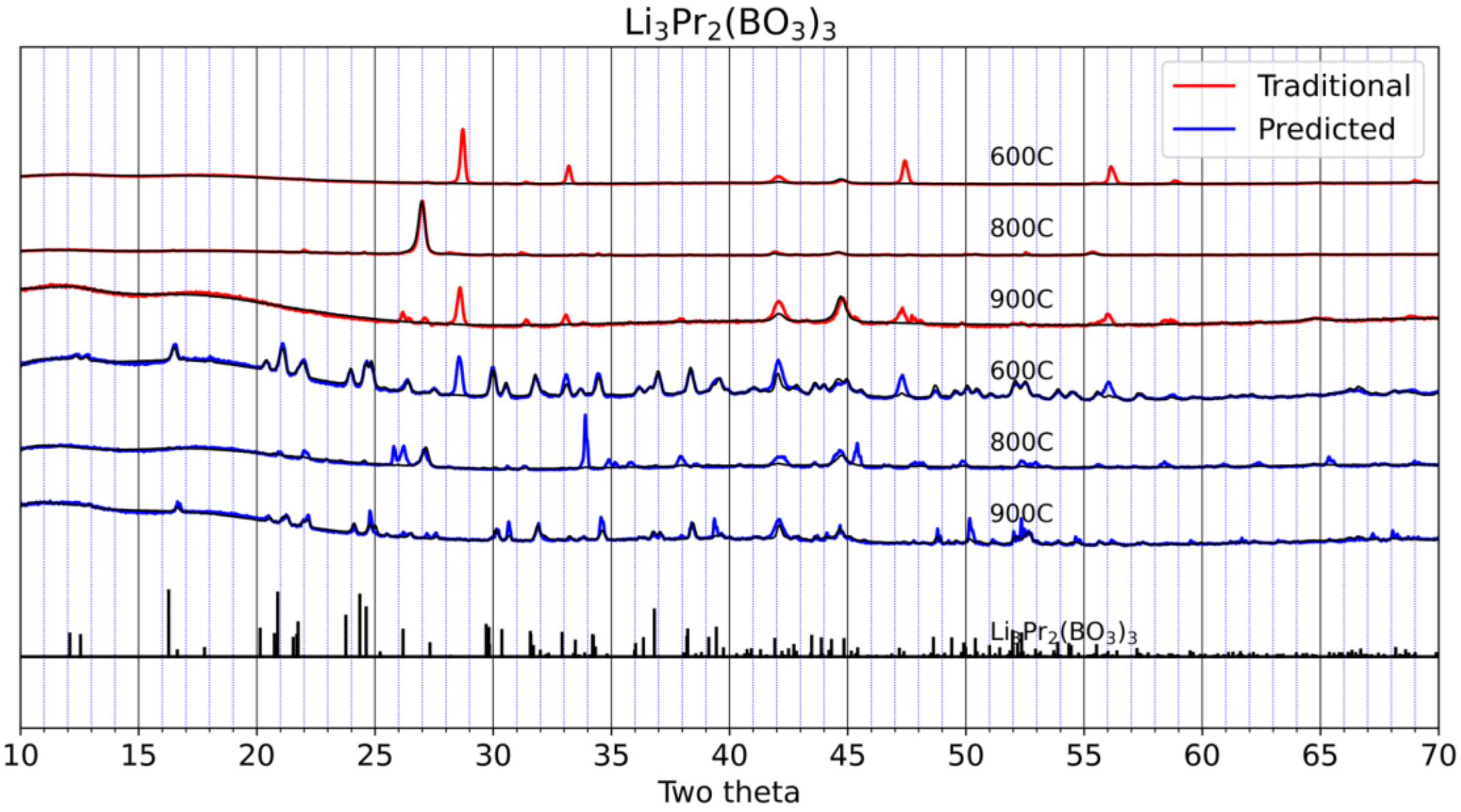

| Compound | Precursor type | Precursors | Temp (°C) | Target intensity (e6) | Residual intensity (e6) | Target phase fraction |
|---|---|---|---|---|---|---|
| $Li_3Pr_2(BO_3)_3$ | Traditional | $Li_2CO_3$, $Pr_6O_{11}$, $B_2O_3$ | 600 | 0.00 | 1.32 | 0.00 |
| | | | 800 | 0.08 | 0.53 | 0.13 |
| | | | 900 | 0.00 | 0.48 | 0.00 |
| | Predicted | $Pr_6O_{11}$, $LiBO_2$ | 600 | 1.40 | 0.61 | 0.70 |
| | | | 800 | 0.06 | 0.55 | 0.10 |
| | | | 900 | 0.52 | 0.43 | 0.55 |

## LiGeBO4

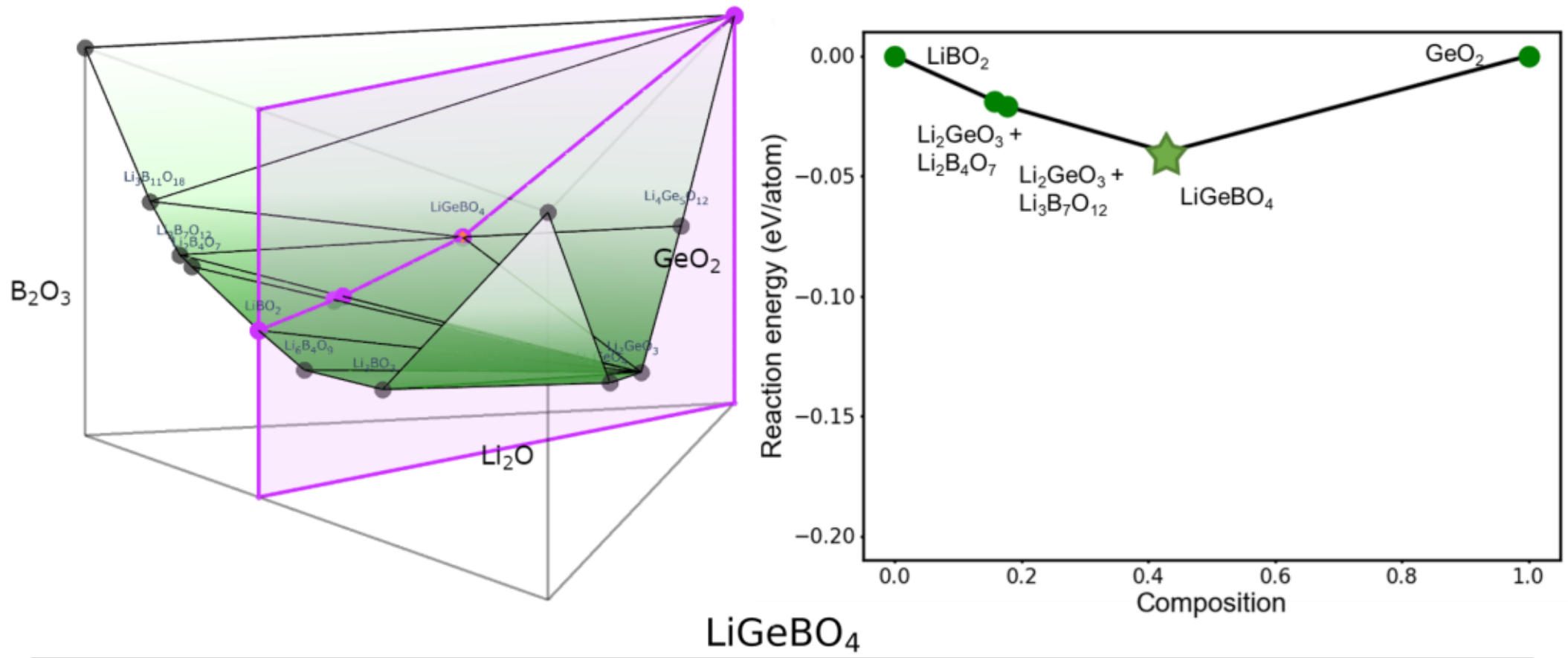

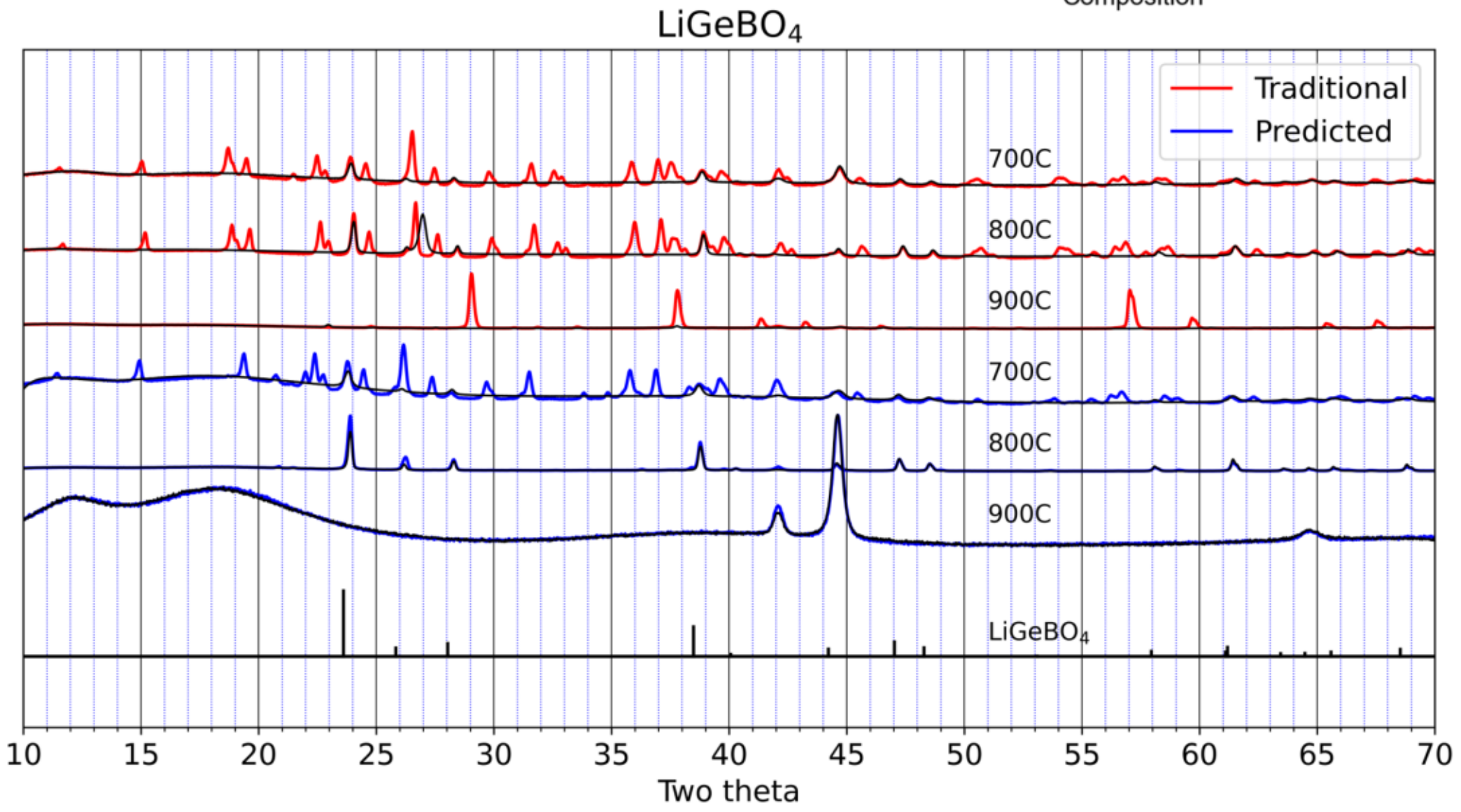

| Compound | Precursor type | Precursors | Temp (°C) | Target intensity (e6) | Residual intensity (e6) | Target phase fraction |
|---|---|---|---|---|---|---|
| LiGeBO4 | Traditional | $Li_2CO_3$, $GeO_2$, $B_2O_3$ | 700 | 0.44 | 2.80 | 0.14 |
| | | | 800 | 0.96 | 5.36 | 0.15 |
| | | | 900 | 0.16 | 3.21 | 0.05 |
| | Predicted | $GeO_2$, $LiBO_2$ | 700 | 0.58 | 3.17 | 0.15 |
| | | | 800 | 3.80 | 2.25 | 0.63 |
| | | | 900 | 0.00 | 0.41 | 0.00 |

## $K_3LiP_2O_7$

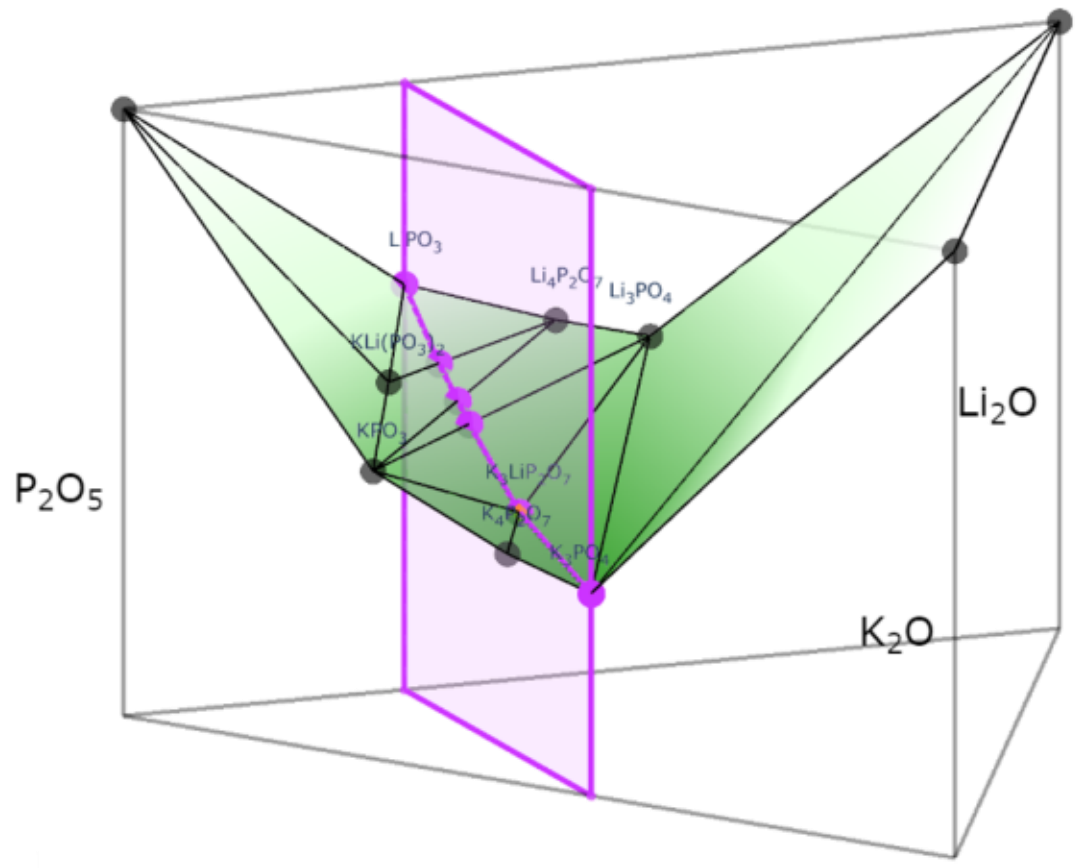

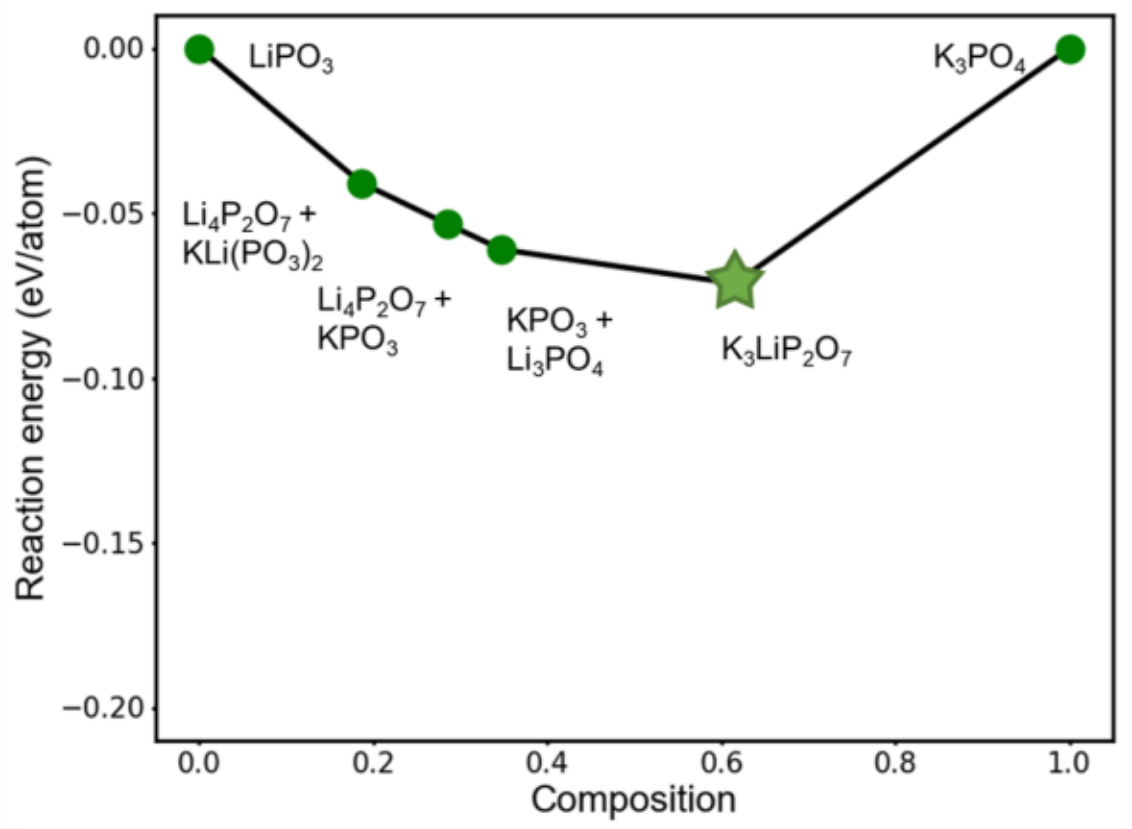

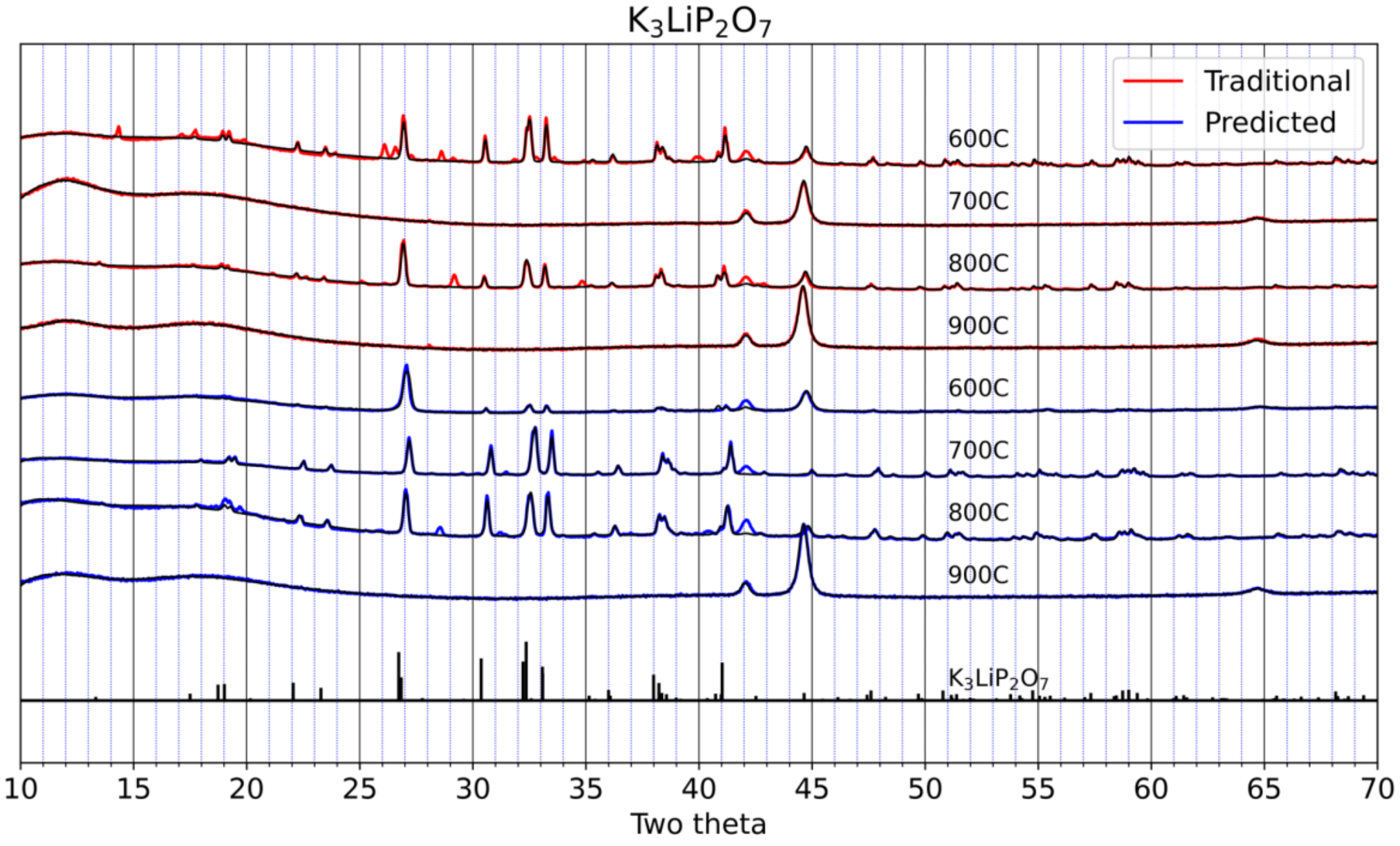

| Compound | Precursor type | Precursors | Temp (°C) | Target intensity (e6) | Residual intensity (e6) | Target phase fraction |
|---|---|---|---|---|---|---|
| $K_3LiP_2O_7$ | Traditional | $K_2CO_3$, $Li_2CO_3$, $NH_4H_2PO_4$ | 600 | 0.96 | 0.71 | 0.58 |
| | | | 700 | 0.00 | 0.94 | 0.00 |
| | | | 800 | 0.90 | 0.56 | 0.62 |
| | | | 900 | 0.00 | 0.19 | 0.02 |
| | Predicted | $K_3PO_4$, $LiPO_3$ | 600 | 0.32 | 0.54 | 0.37 |
| | | | 700 | 1.17 | 0.60 | 0.66 |
| | | | 800 | 1.05 | 0.54 | 0.66 |
| | | | 900 | 0.01 | 0.21 | 0.05 |

# $Li_3Sc_2(PO_4)_3$

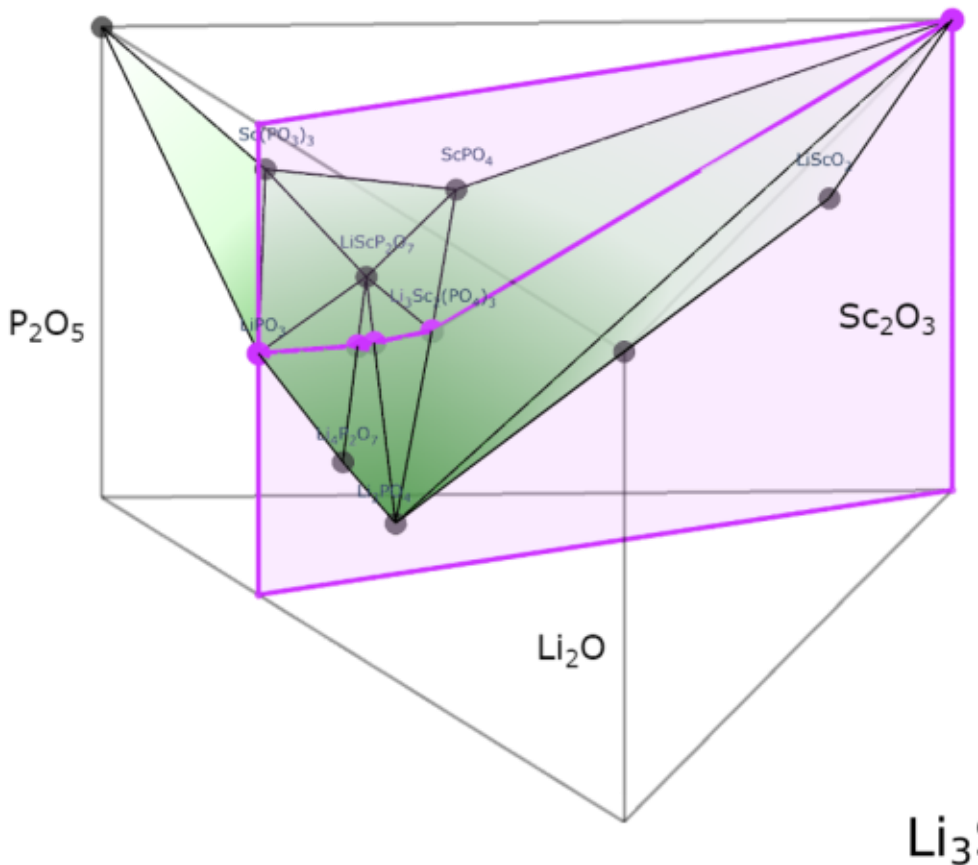

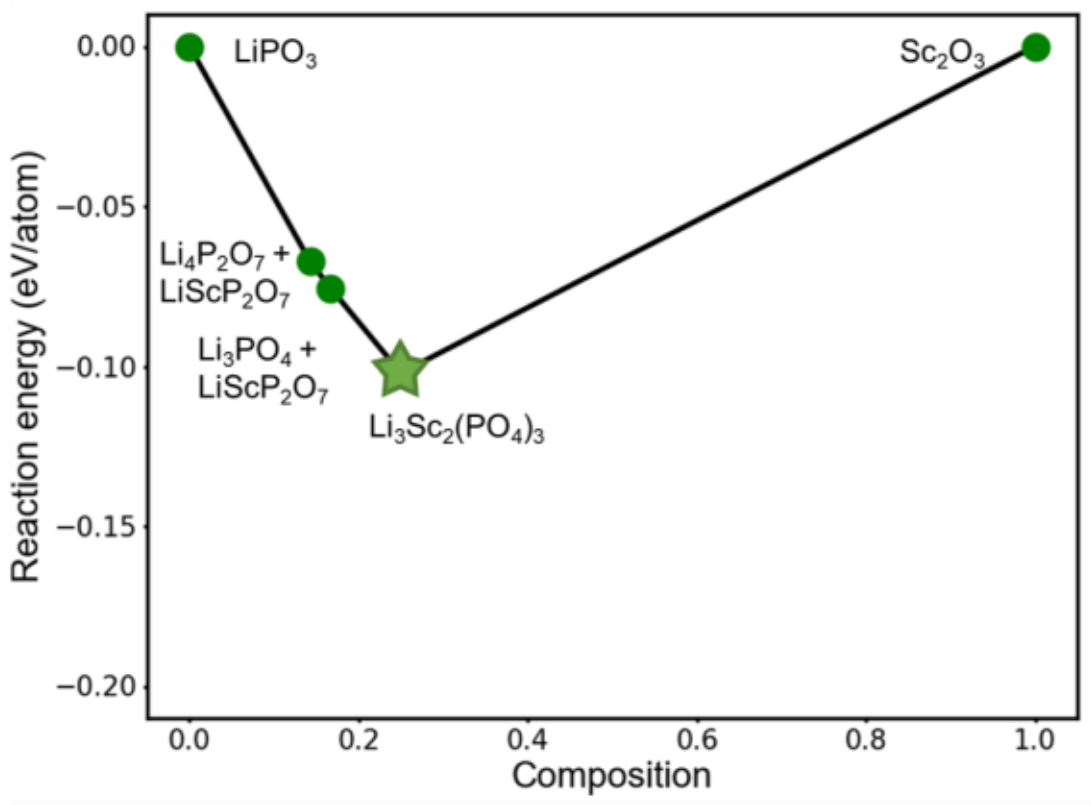

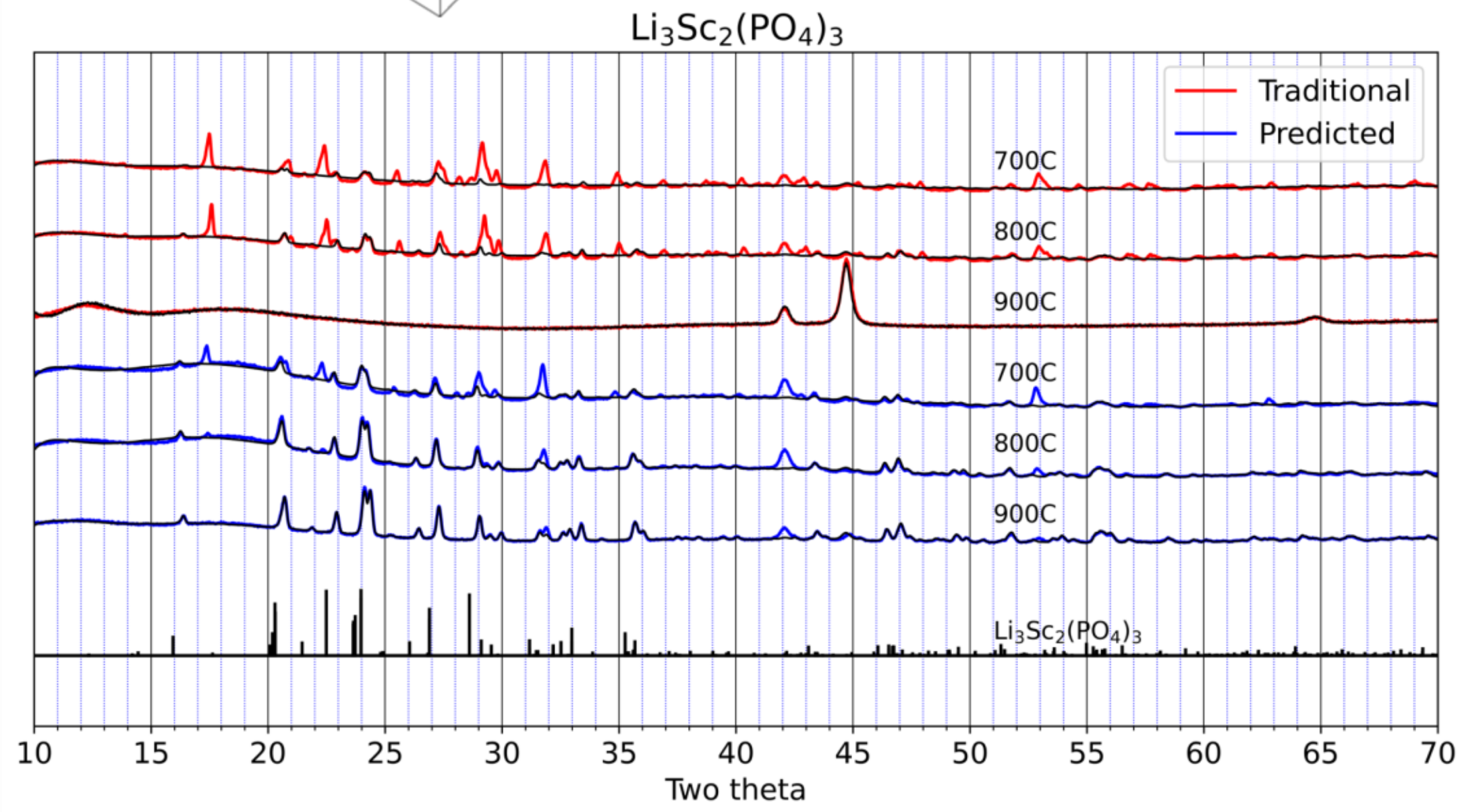

| Compound | Precursor type | Precursors | Temp (°C) | Target intensity (e6) | Residual intensity (e6) | Target phase fraction |
|---|---|---|---|---|---|---|
| $Li_3Sc_2(PO_4)_3$ | Traditional | $Sc_2O_3$, $Li_2CO_3$, $NH_4H_2PO_4$ | 700 | 0.24 | 1.35 | 0.15 |
| | | | 800 | 0.46 | 1.20 | 0.28 |
| | | | 900 | 0.01 | 0.23 | 0.04 |
| | Predicted | $Sc_2O_3$, $LiPO_3$ | 700 | 0.72 | 1.37 | 0.34 |
| | | | 800 | 1.64 | 0.95 | 0.63 |
| | | | 900 | 1.09 | 0.26 | 0.81 |

## $Li_3V_2(PO_4)_3$

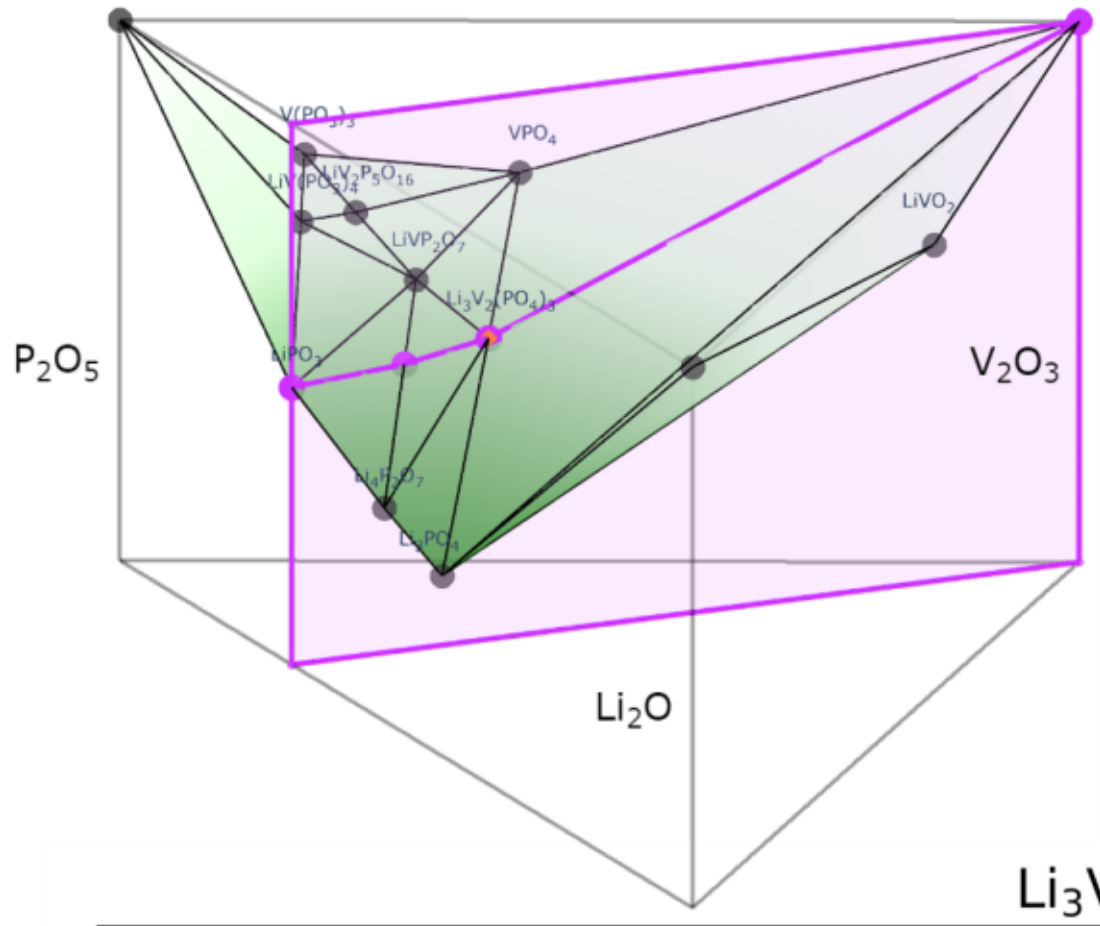

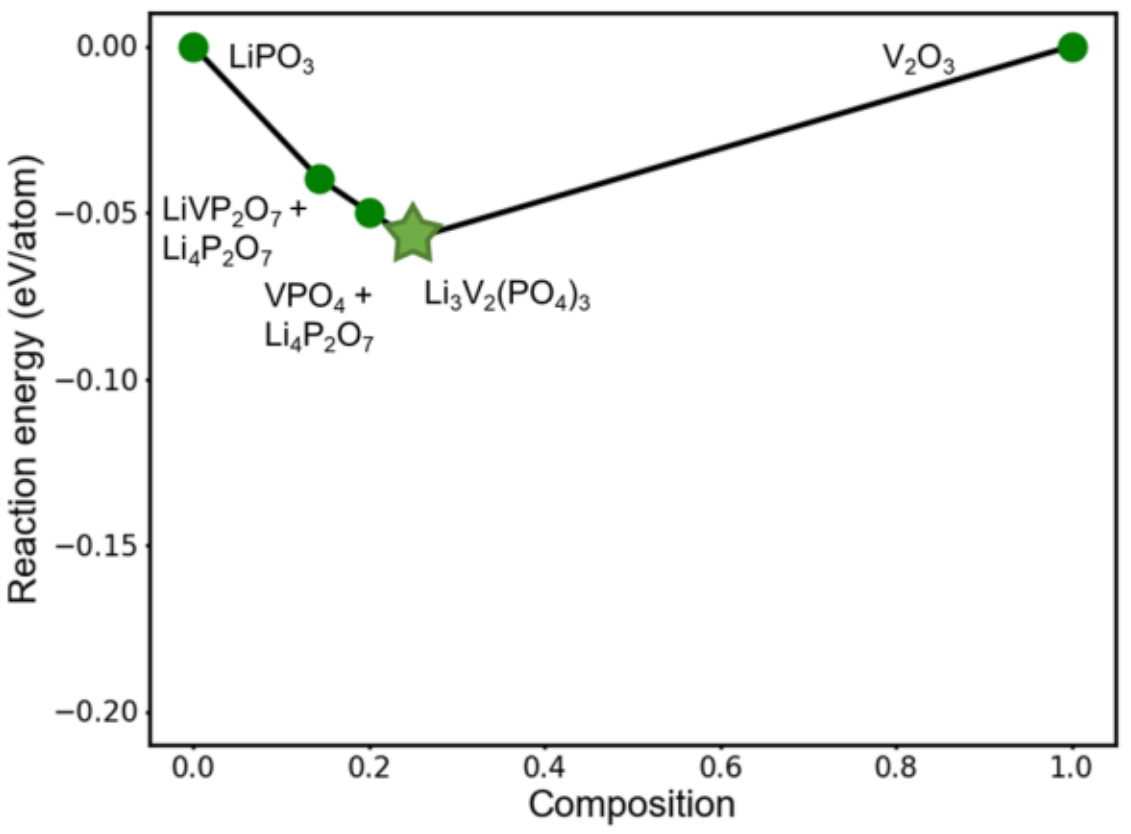

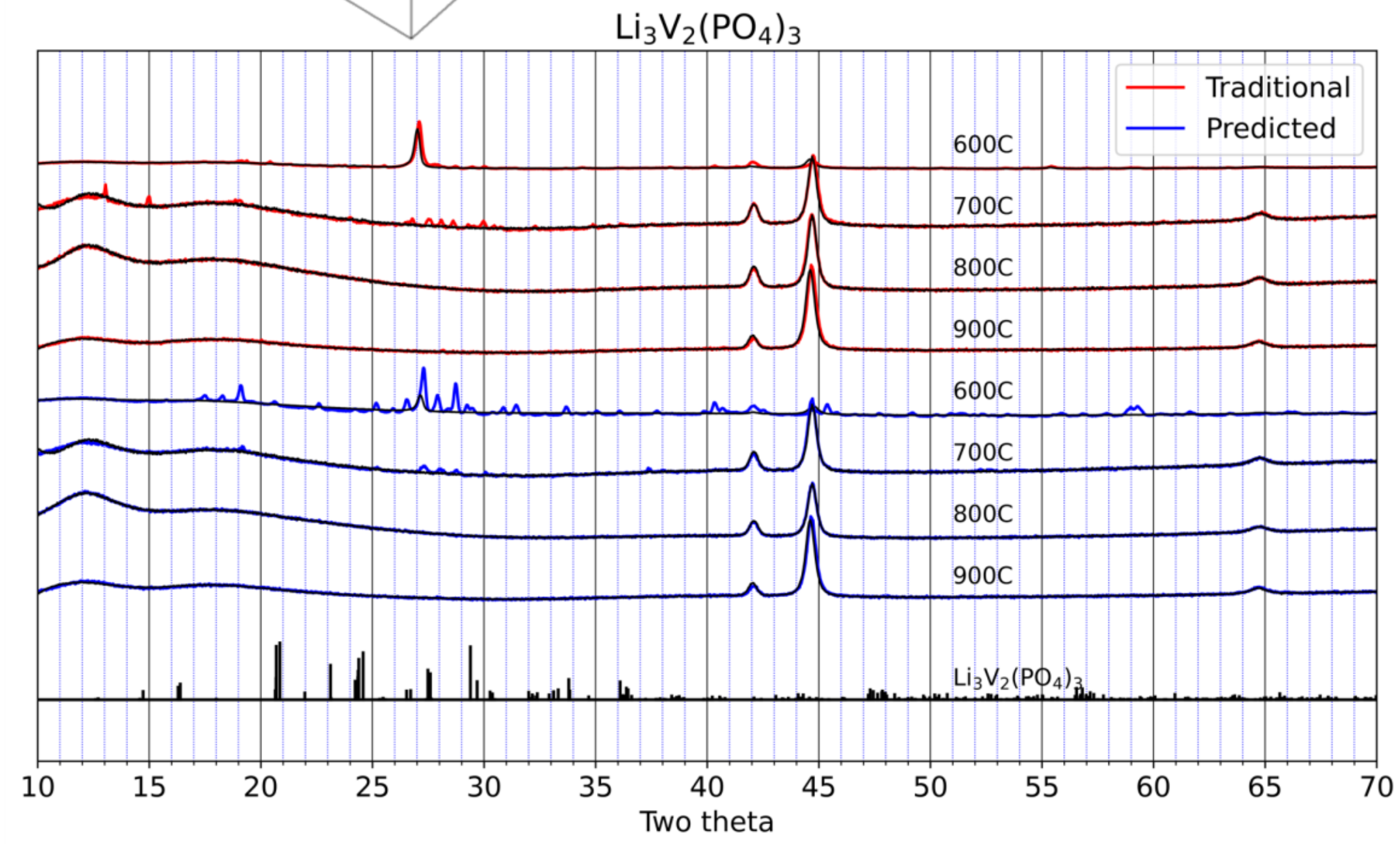

| Compound | Precursor type | Precursors | Temp (°C) | Target intensity (e6) | Residual intensity (e6) | Target phase fraction |
|---|---|---|---|---|---|---|
| $Li_3V_2(PO_4)_3$ | Traditional | $V_2O_3$, $Li_2CO_3$, $NH_4H_2PO_4$ | 600 | 0.00 | 0.84 | 0.00 |
| | | | 700 | 0.00 | 0.67 | 0.00 |
| | | | 800 | 0.00 | 0.36 | 0.00 |
| | | | 900 | 0.01 | 0.24 | 0.06 |
| | Predicted | $V_2O_3$, $LiPO_3$ | 600 | 0.00 | 1.32 | 0.00 |
| | | | 700 | 0.00 | 1.07 | 0.00 |
| | | | 800 | 0.00 | 0.27 | 0.00 |
| | | | 900 | 0.02 | 0.24 | 0.07 |

## $KLi(PO_3)_2$

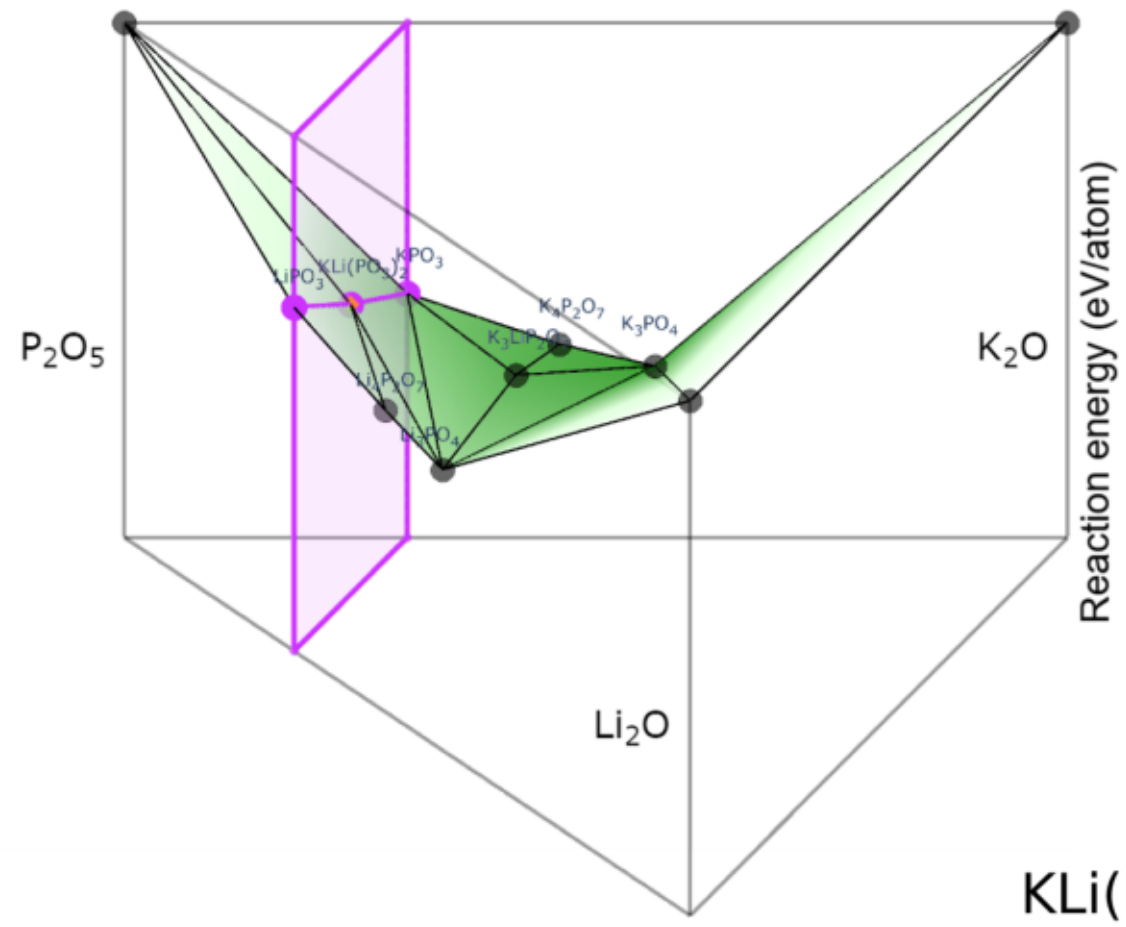

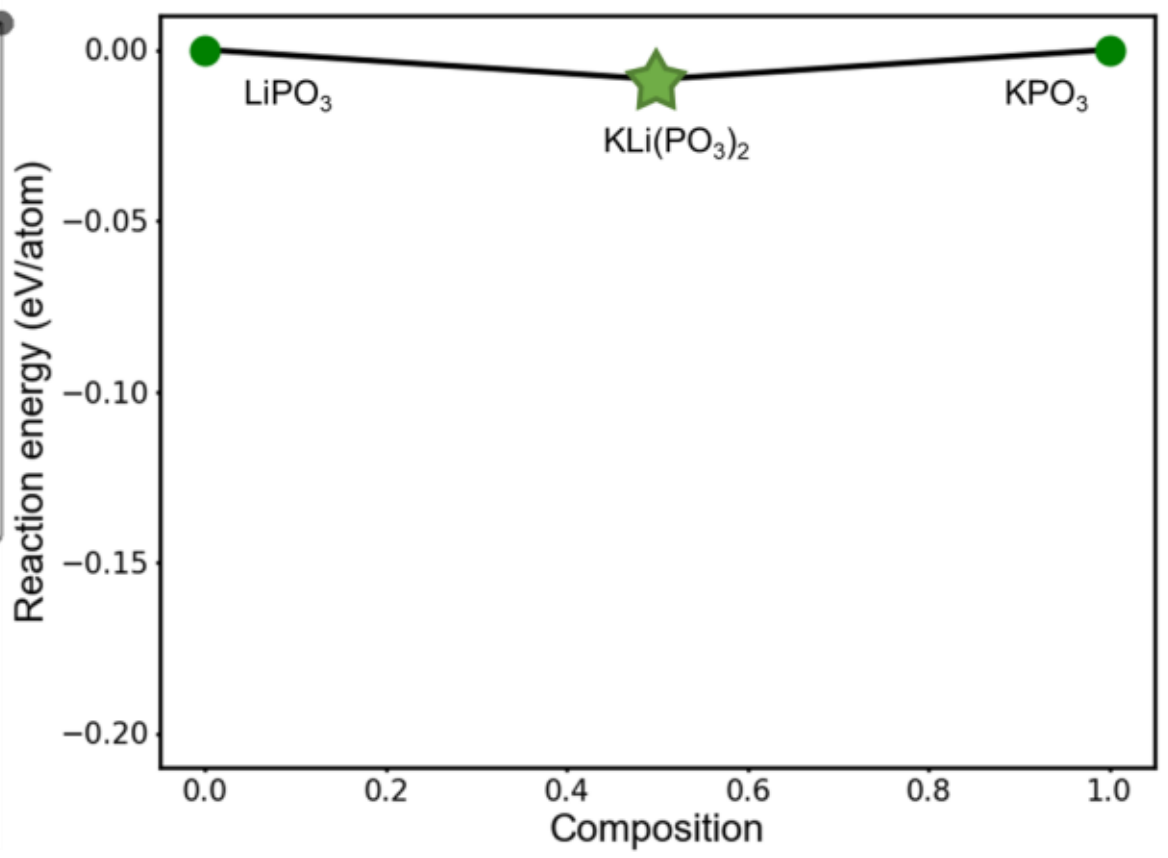

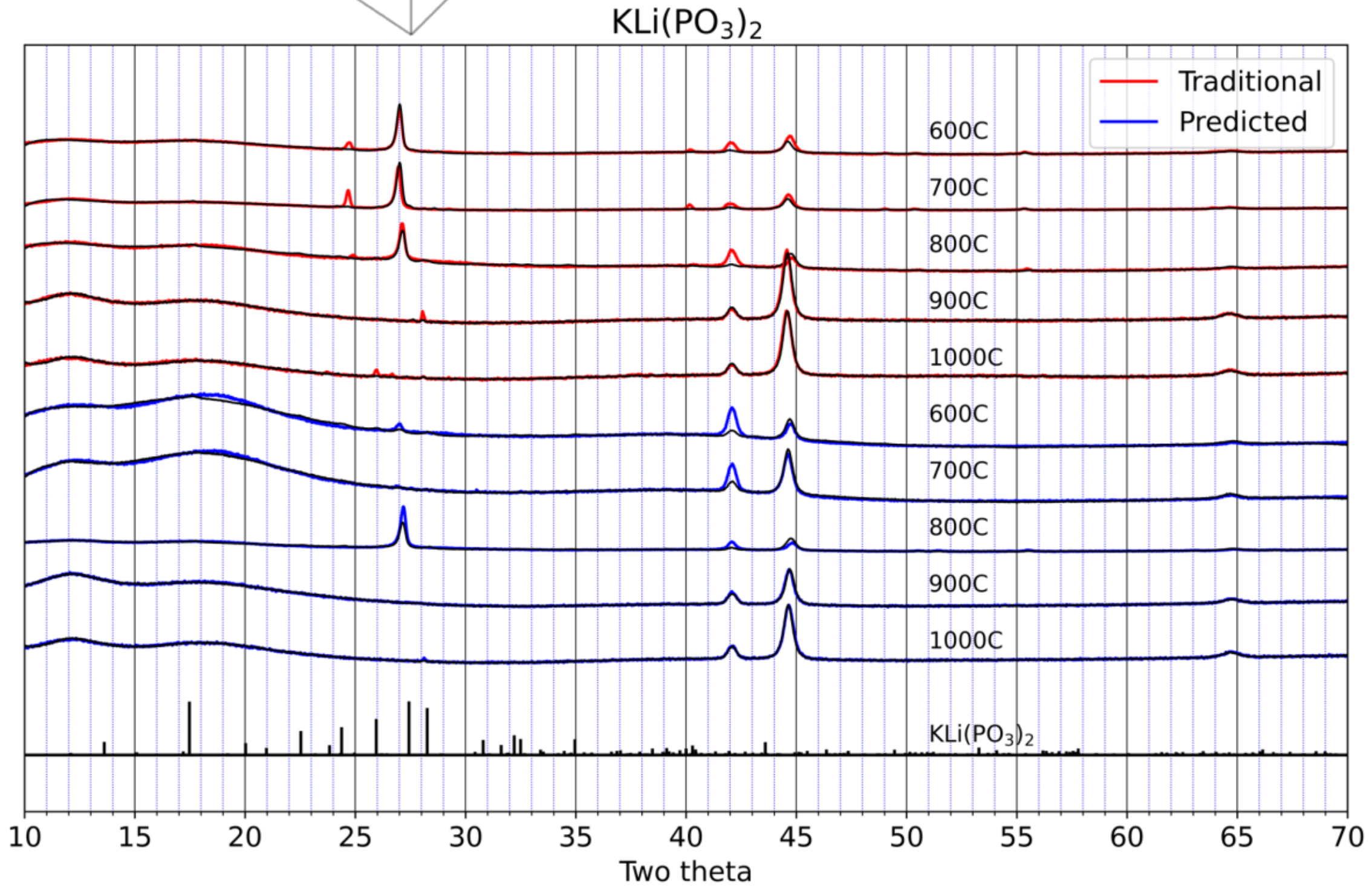

| Compound | Precursor type | Precursors | Temp (°C) | Target intensity (e6) | Residual intensity (e6) | Target phase fraction |
|---|---|---|---|---|---|---|
| $KLi(PO_3)_2$ | Traditional | $K_2CO_3$, $Li_2CO_3$, $NH_4H_2PO_4$ | 600 | 0.25 | 0.51 | 0.33 |
| | | | 700 | 0.15 | 0.81 | 0.16 |
| | | | 800 | 0.22 | 0.41 | 0.35 |
| | | | 900 | 0.03 | 0.24 | 0.12 |
| | | | 1000 | 0.02 | 0.25 | 0.09 |
| | Predicted | $KPO_3$, $LiPO_3$ | 600 | 0.27 | 0.75 | 0.27 |
| | | | 700 | 0.12 | 0.57 | 0.17 |
| | | | 800 | 0.33 | 0.54 | 0.38 |

## $Li_2TiSiO_5$

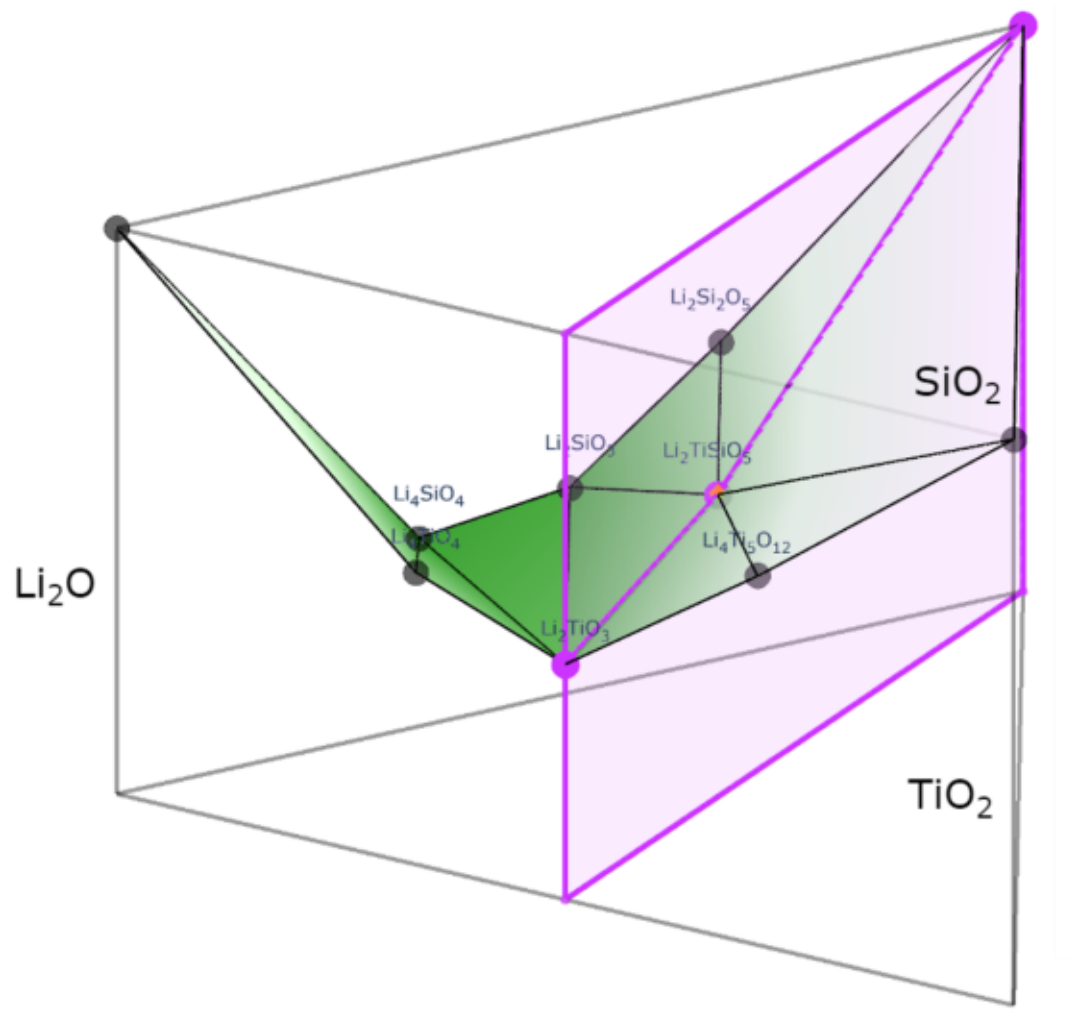

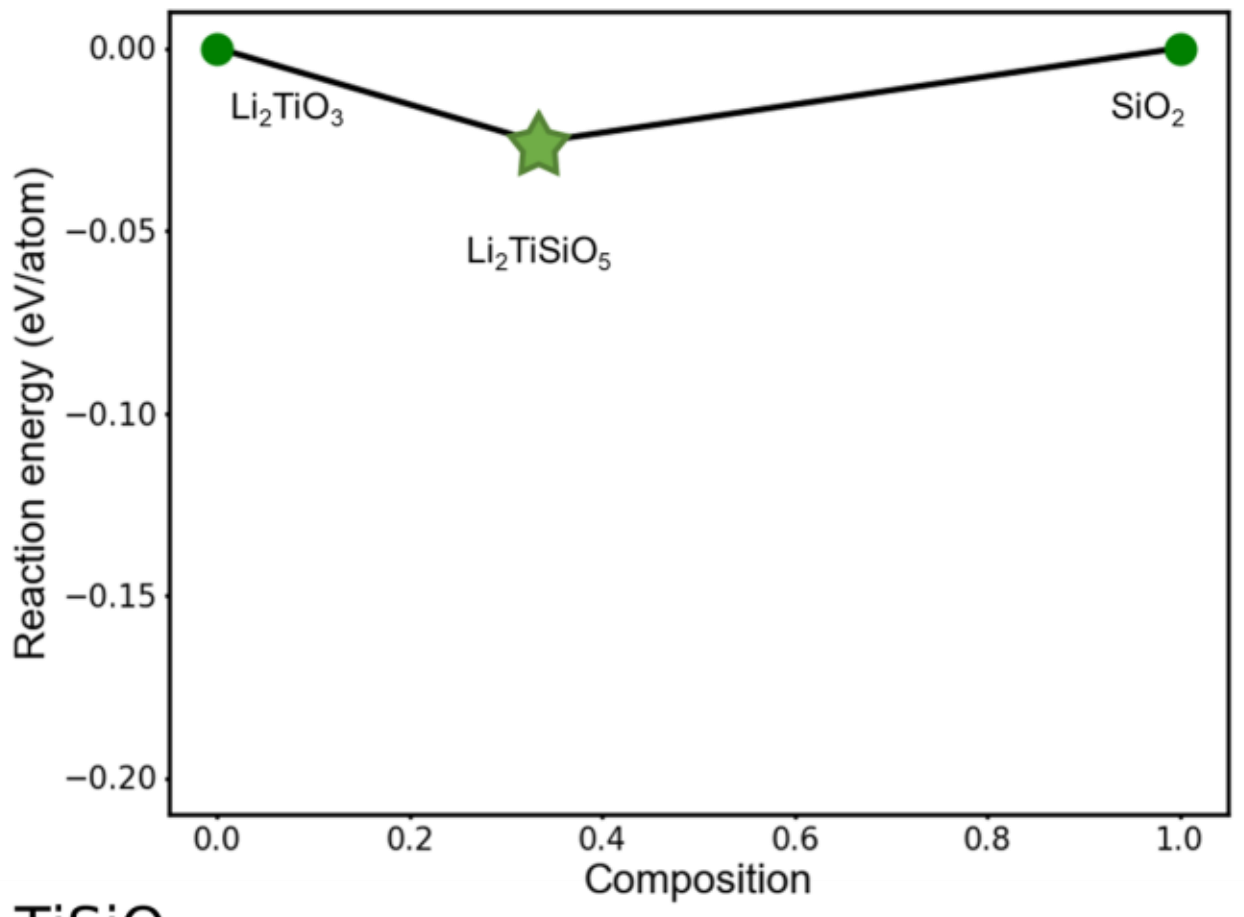

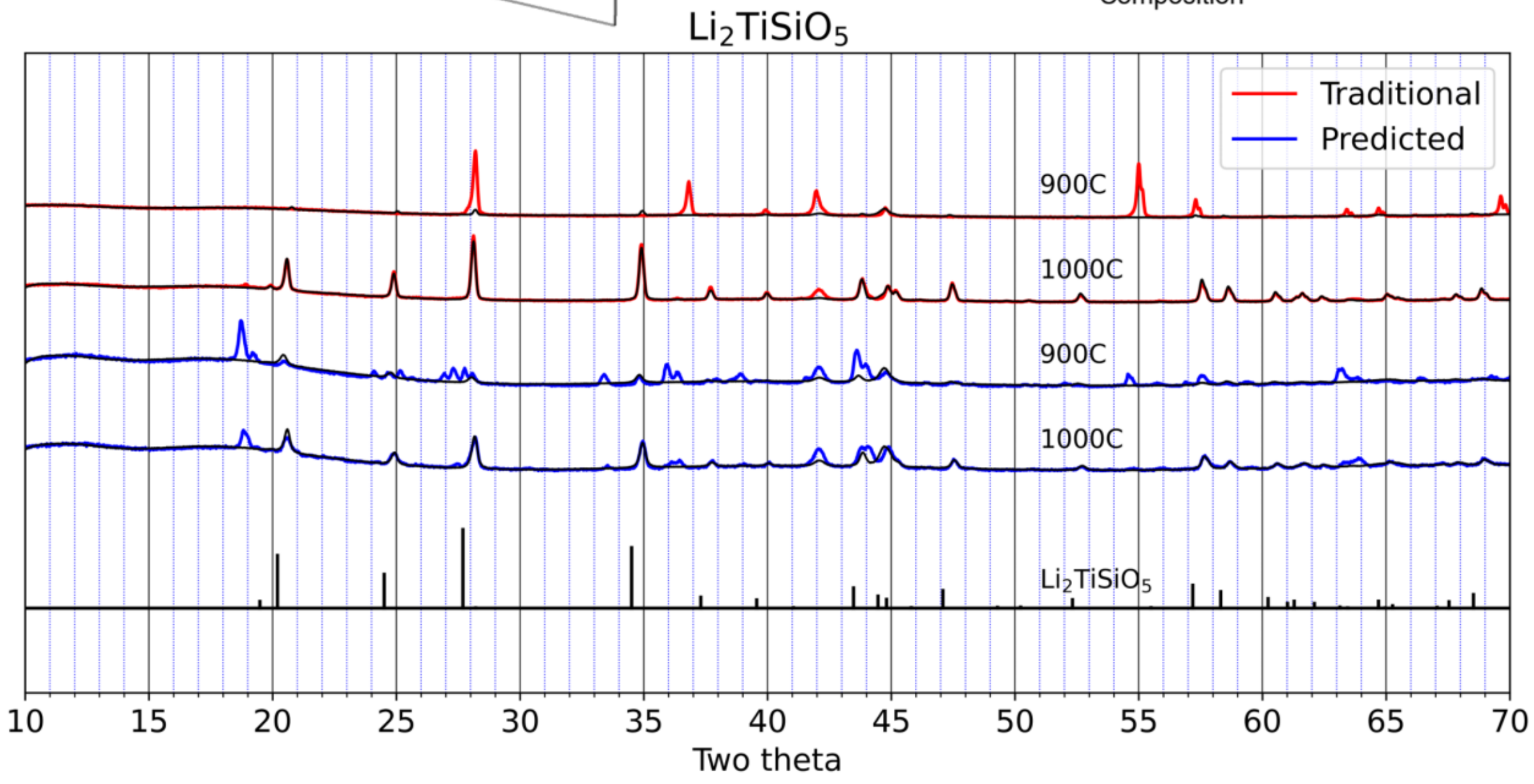

| Compound | Precursor type | Precursors | Temp (°C) | Target intensity (e6) | Residual intensity (e6) | Target phase fraction |
|---|---|---|---|---|---|---|
| $Li_2TiSiO_5$ | Traditional | $SiO_2$, $TiO_2$, Li2CO3 | 900 | 0.06 | 0.68 | 0.08 |
| | | | 1000 | 0.84 | 0.28 | 0.75 |
| | Predicted | $SiO_2$, $Li_2TiO_3$ | 900 | 0.14 | 0.62 | 0.18 |
| | | | 1000 | 0.34 | 0.33 | 0.51 |

## $LiNbGeO_5$

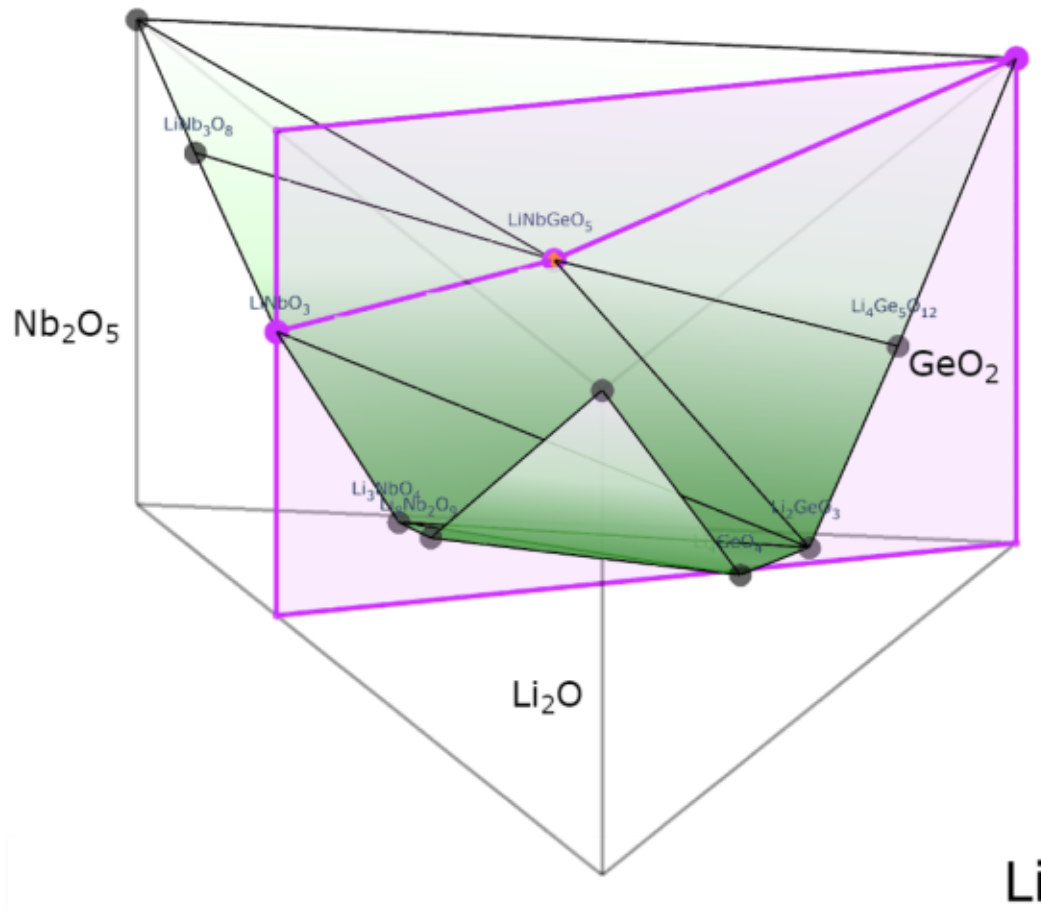

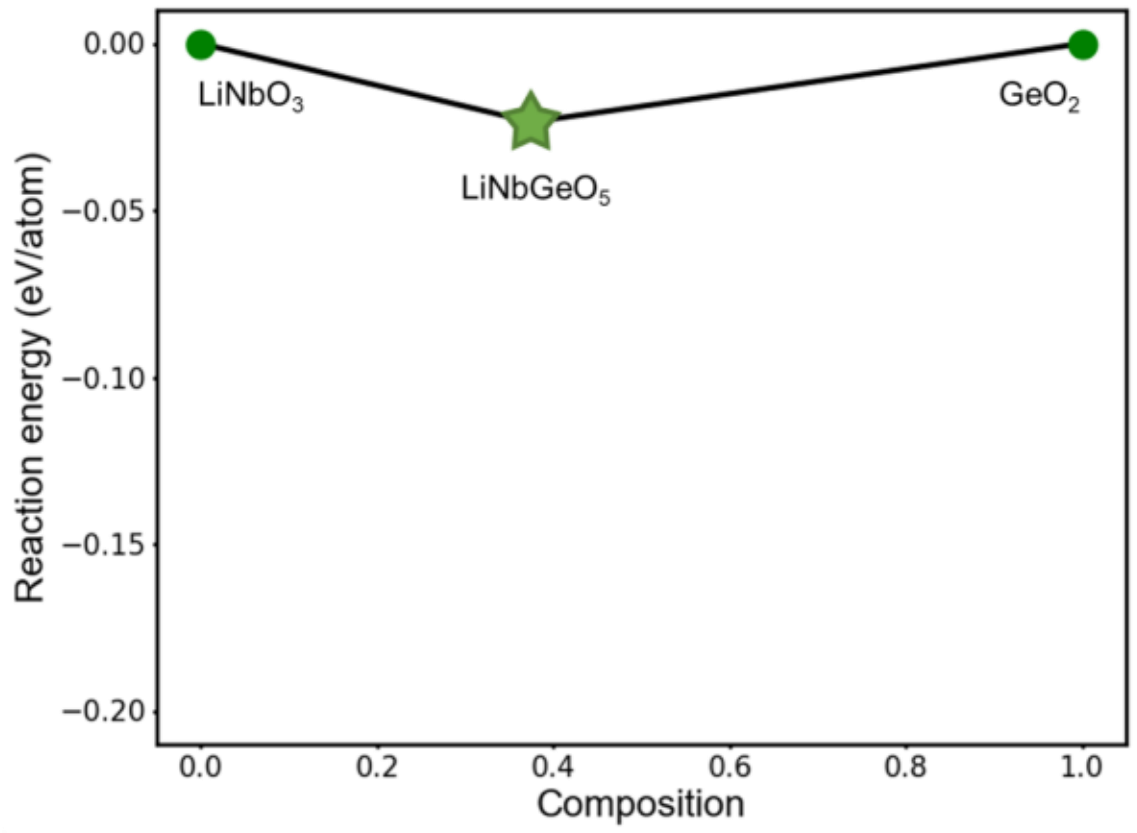

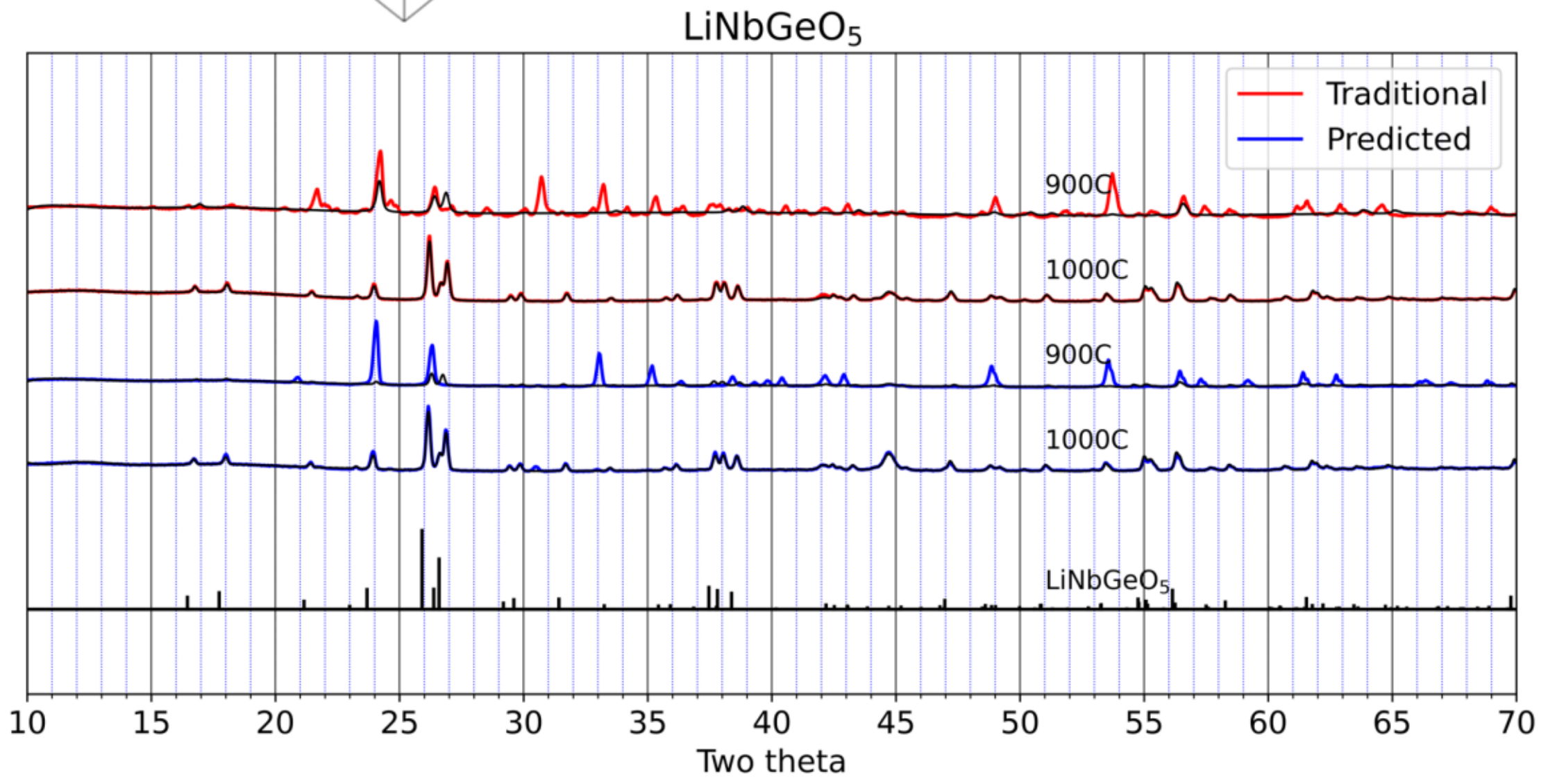

| Compound | Precursor type | Precursors | Temp (°C) | Target intensity (e6) | Residual intensity (e6) | Target phase fraction |
|---|---|---|---|---|---|---|
| $LiNbGeO_5$ | Traditional | $GeO_2$, $Li_2CO_3$, $Nb_2O_5$ | 900 | 0.47 | 1.52 | 0.24 |
| | | | 1000 | 1.33 | 0.24 | 0.85 |
| | Predicted | $GeO_2$, $LiNbO_3$ | 900 | 0.41 | 1.79 | 0.19 |
| | | | 1000 | 1.25 | 0.36 | 0.78 |

# $Li_2TiGeO_5$

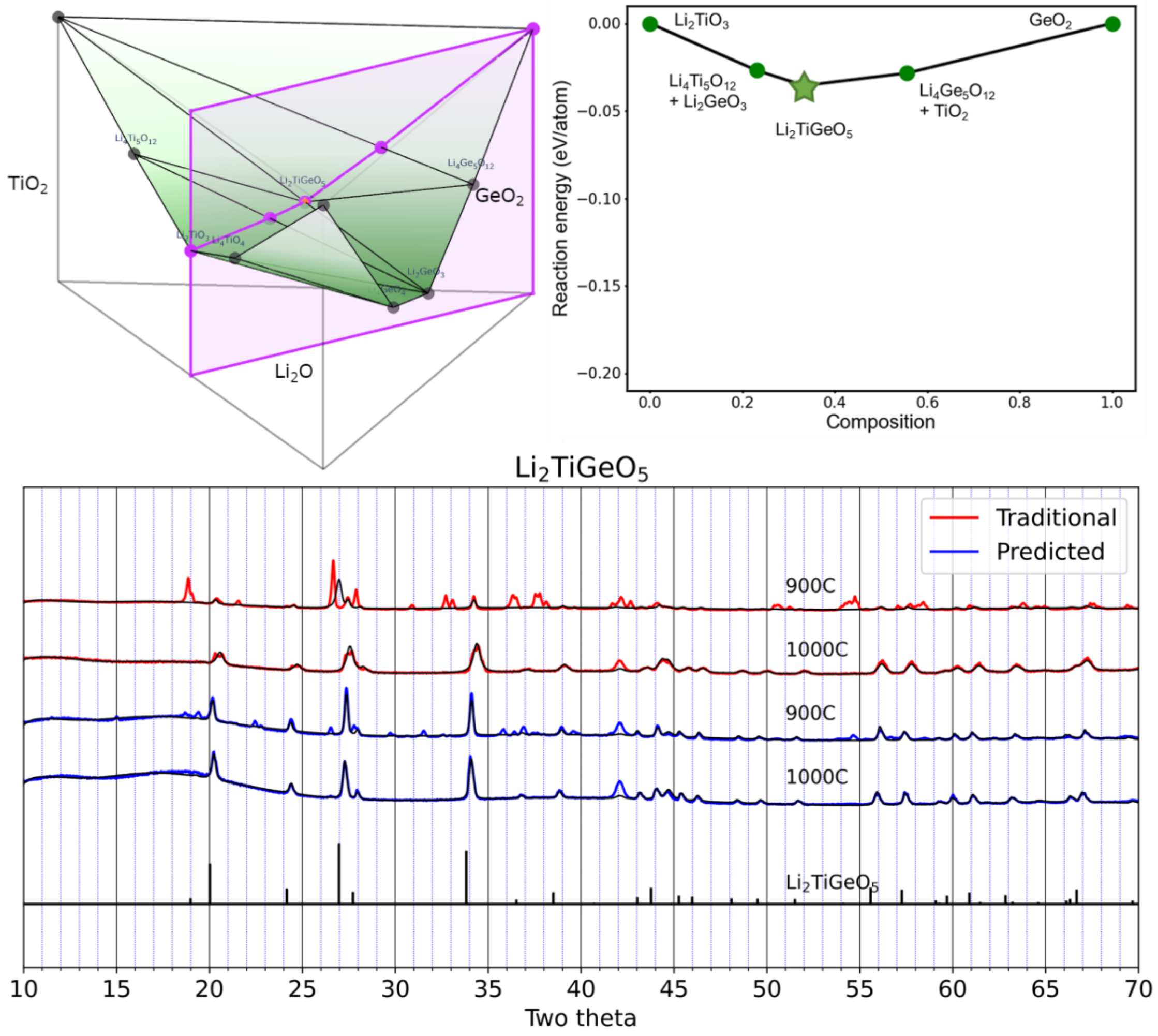

| Compound | Precursor type | Precursors | Temp (°C) | Target intensity (e6) | Residual intensity (e6) | Target phase fraction |
|---|---|---|---|---|---|---|
| $Li_2TiGeO_5$ | Traditional | $GeO_2$, $Li_2CO_3$, $TiO_2$ | 900 | 0.28 | 1.56 | 0.15 |
| | | | 1000 | 0.73 | 0.25 | 0.75 |
| | Predicted | $GeO_2$, $Li_2TiO_3$ | 900 | 0.73 | 0.80 | 0.48 |
| | | | 1000 | 0.76 | 0.47 | 0.62 |

## $Li_3Fe_2(PO_4)_3$

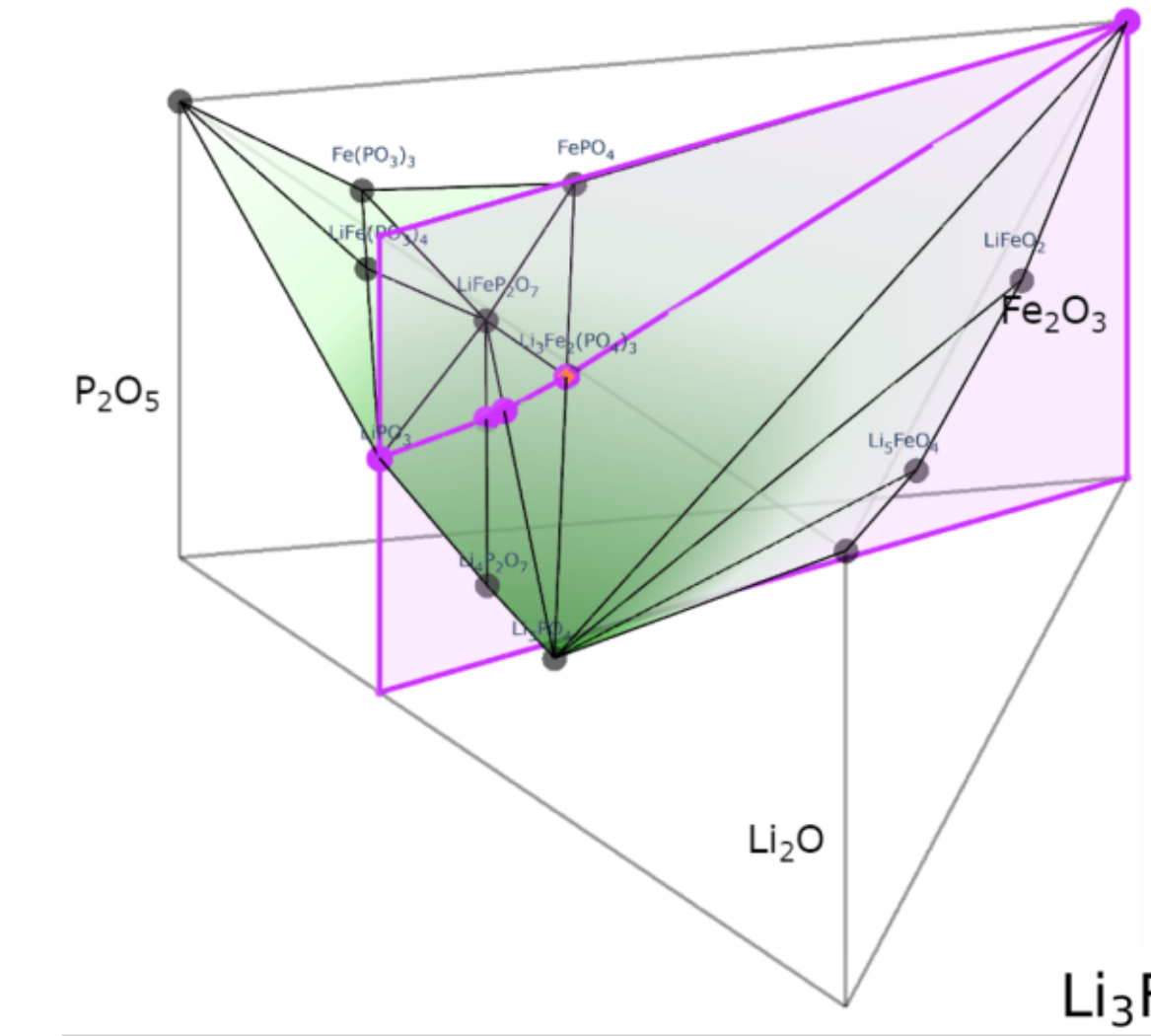

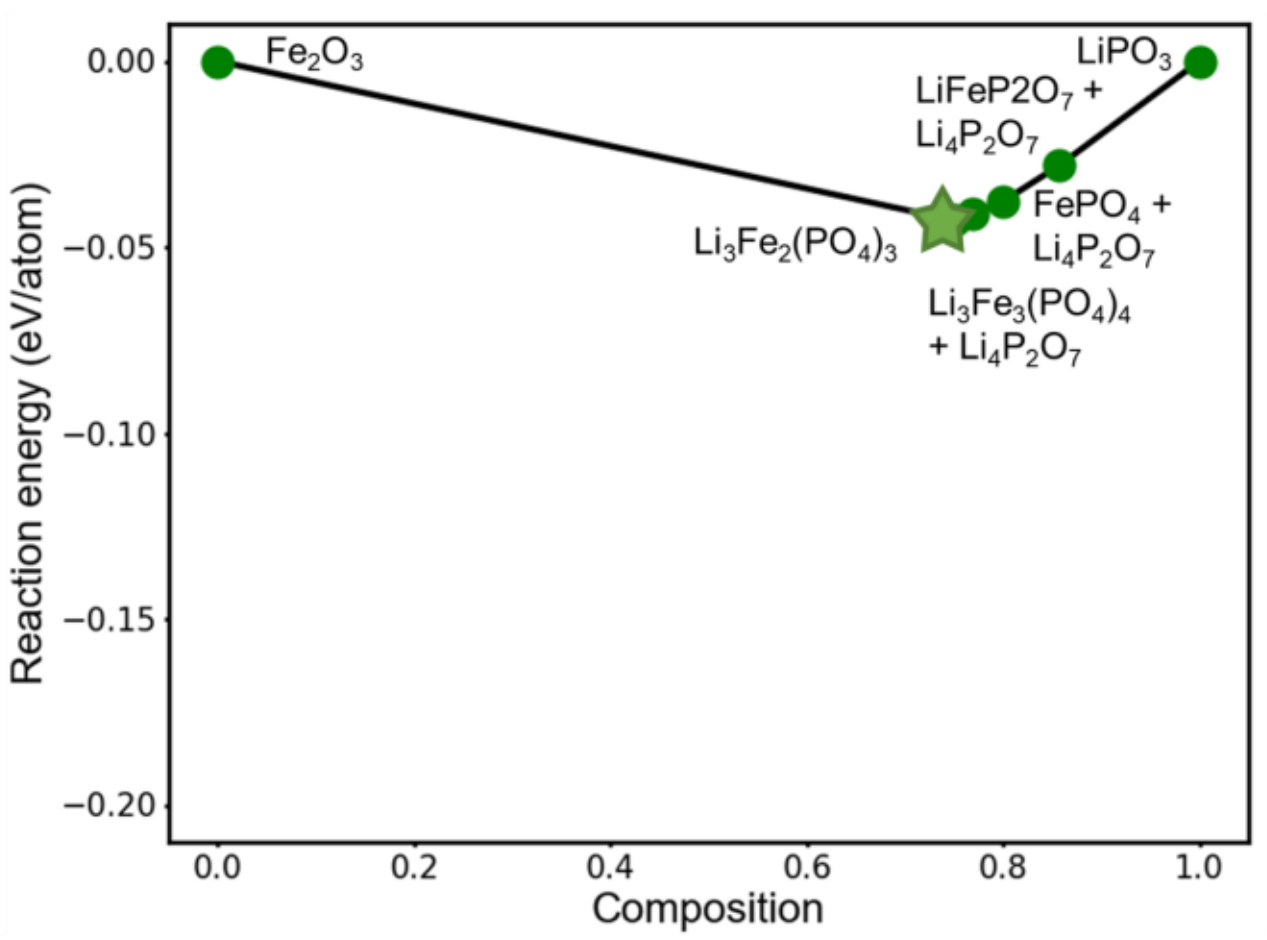

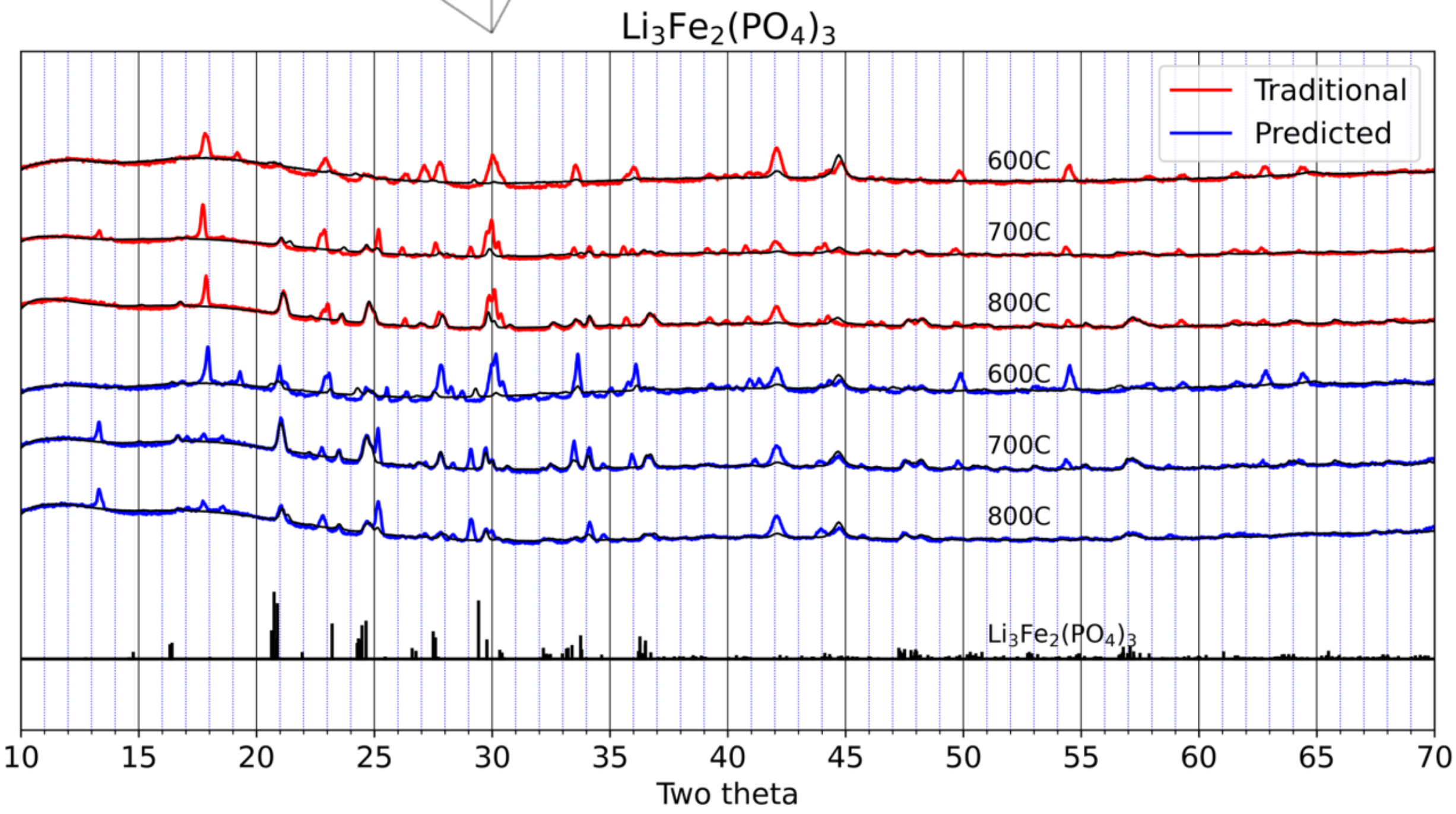

| Compound | Precursor type | Precursors | Temp (°C) | Target intensity (e6) | Residual intensity (e6) | Target phase fraction |
|---|---|---|---|---|---|---|
| $Li_3Fe_2(PO_4)_3$ | Traditional | $Fe_2O_3$, $Li_2CO_3$, $NH_4H_2PO_4$ | 600 | 0.09 | 0.88 | 0.09 |
| | | | 700 | 0.39 | 1.20 | 0.25 |
| | | | 800 | 0.59 | 0.77 | 0.43 |
| | Predicted | $Fe_2O_3$, $LiPO_3$ | 600 | 0.24 | 1.20 | 0.17 |
| | | | 700 | 0.69 | 0.78 | 0.47 |
| | | | 800 | 0.40 | 0.75 | 0.35 |

# $KBaPO_4$

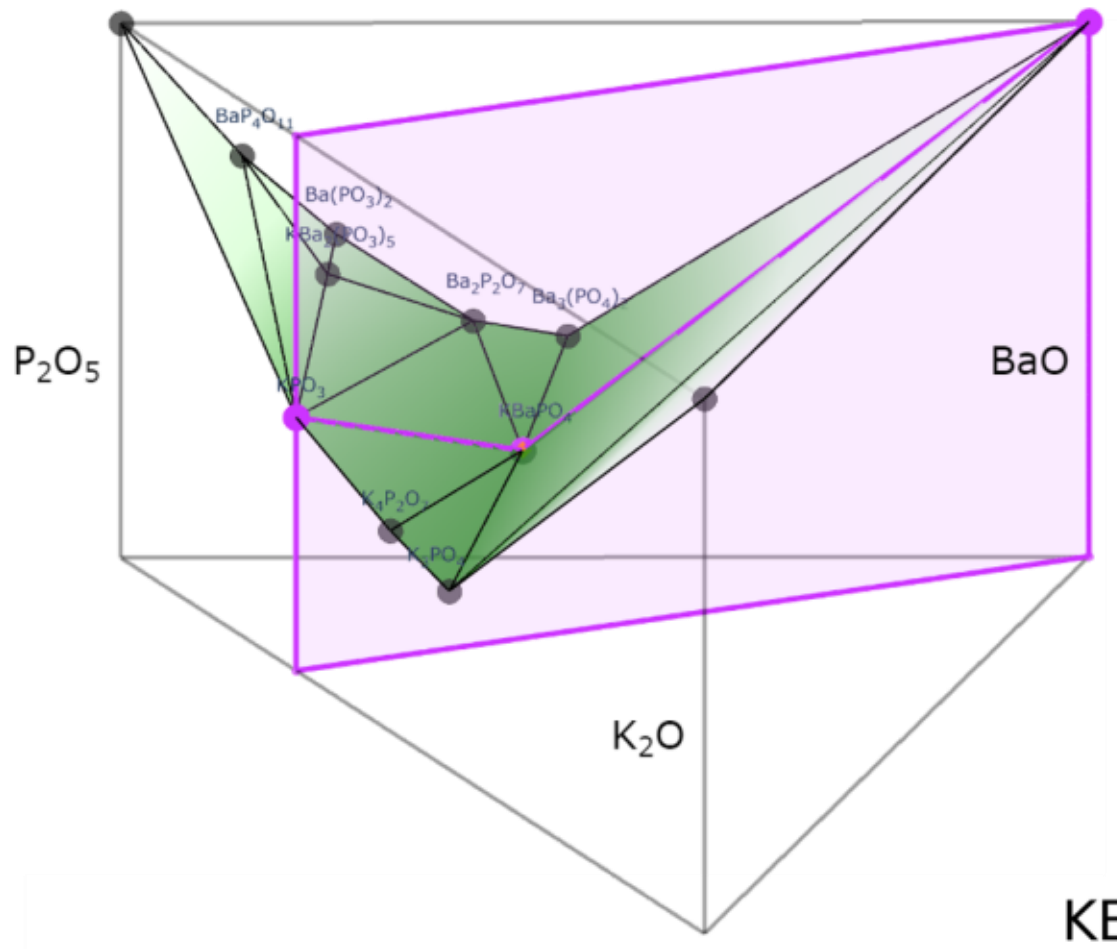

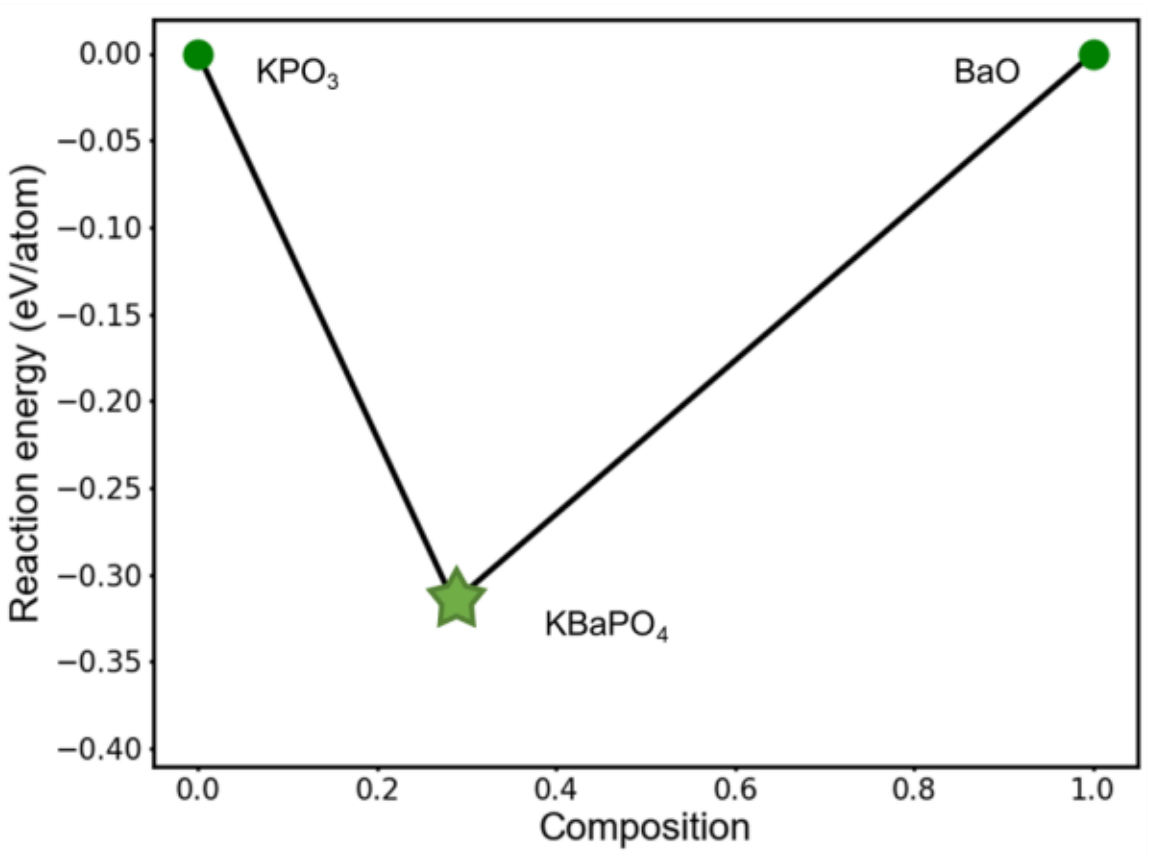

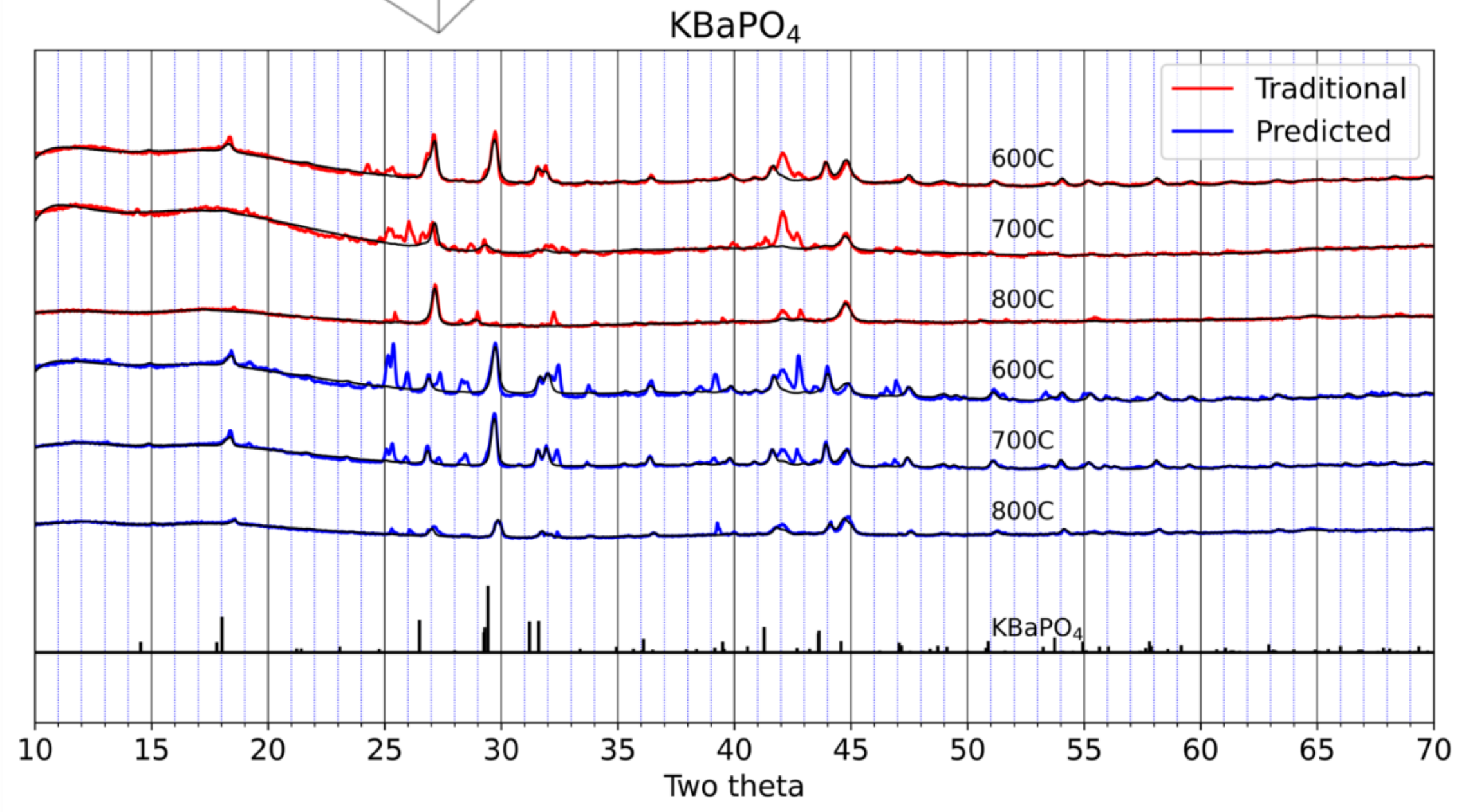

| Compound | Precursor type | Precursors | Temp (°C) | Target intensity (e6) | Residual intensity (e6) | Target phase fraction |
|---|---|---|---|---|---|---|
| $KBaPO_4$ | Traditional | BaO, $K_2CO_3$, $NH_4H_2PO_4$ | 600 | 0.88 | 0.62 | 0.59 |
| | | | 700 | 0.17 | 0.71 | 0.20 |
| | | | 800 | 0.13 | 0.28 | 0.32 |
| | Predicted | BaO, $KPO_3$ | 600 | 0.82 | 0.97 | 0.46 |
| | | | 700 | 1.03 | 0.81 | 0.56 |
| | | | 800 | 0.23 | 0.22 | 0.52 |

## $BaNaBO_3$

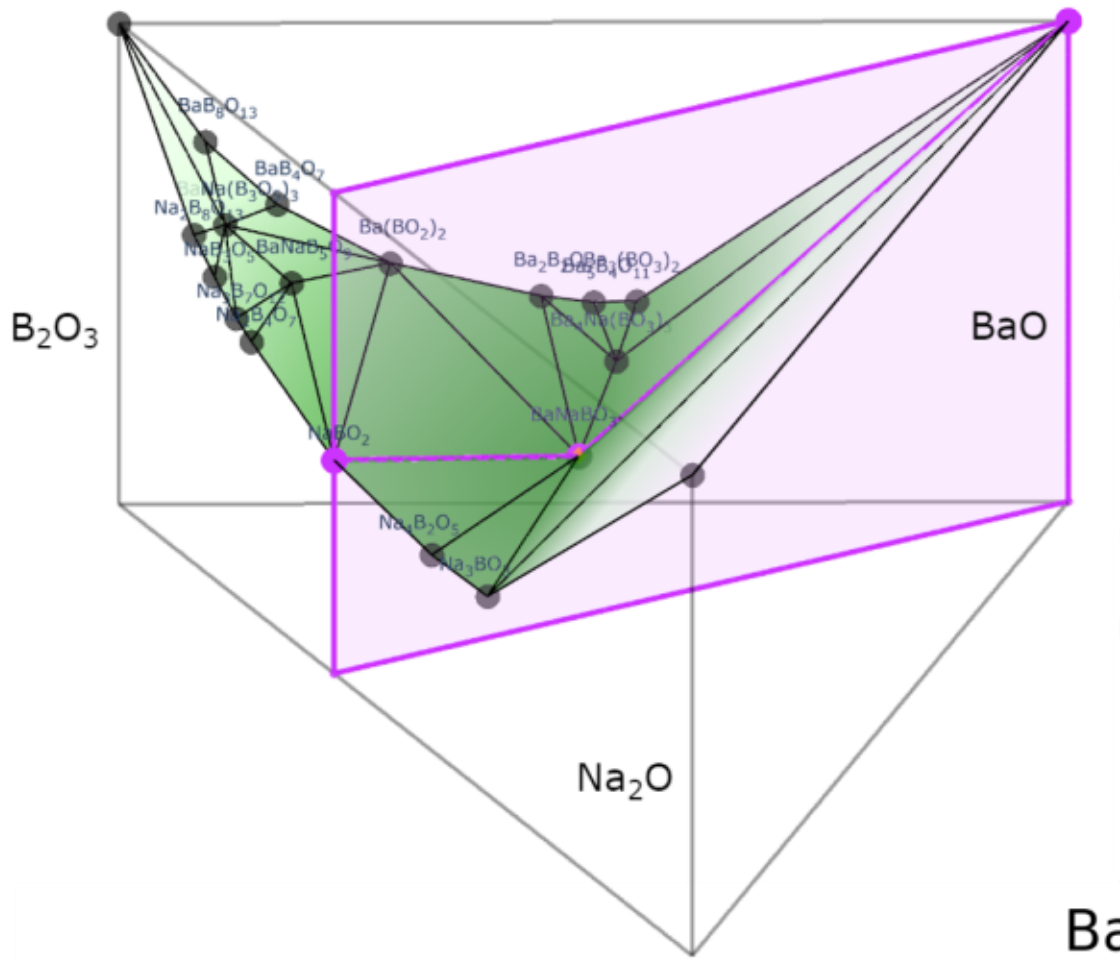

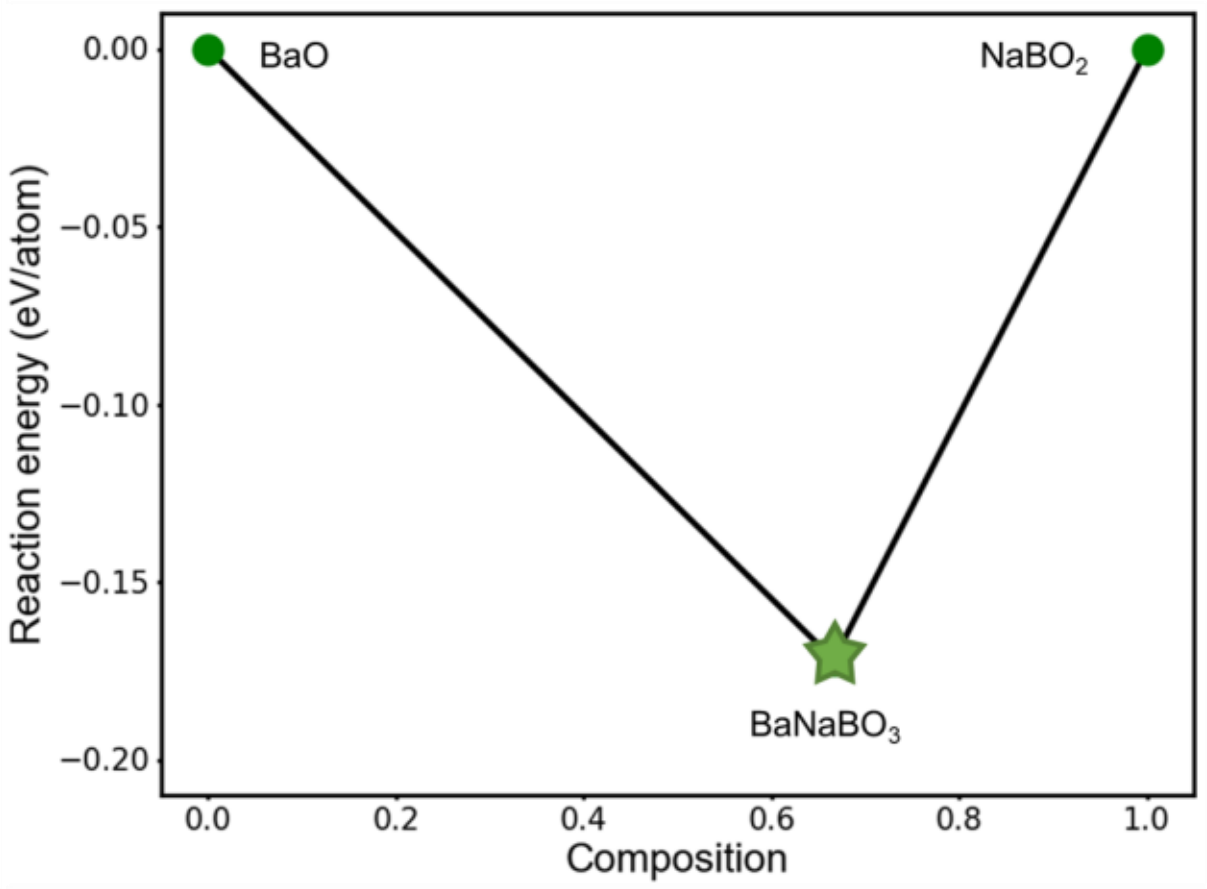

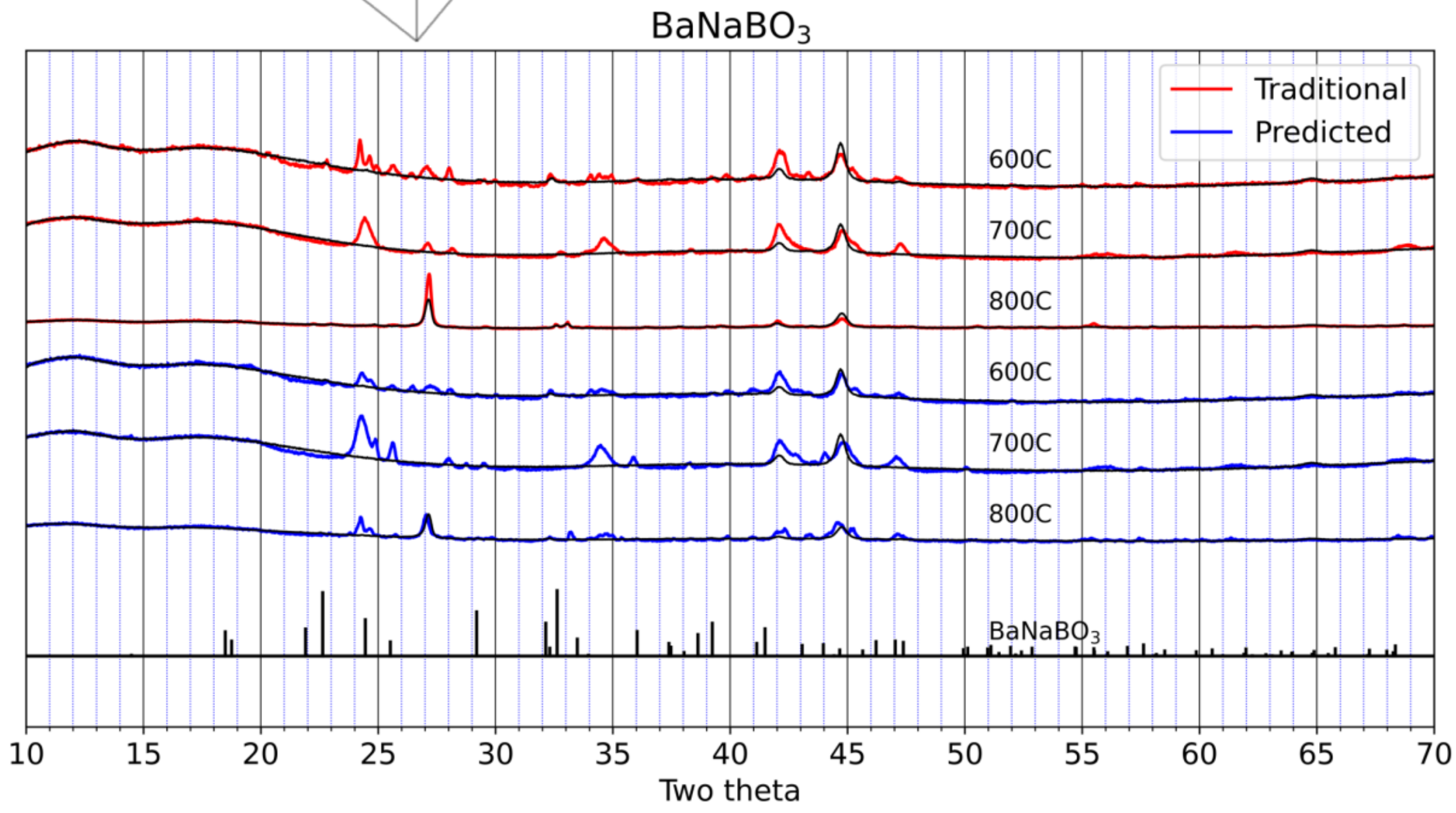

| Compound | Precursor type | Precursors | Temp (°C) | Target intensity (e6) | Residual intensity (e6) | Target phase fraction |
|---|---|---|---|---|---|---|
| $BaNaBO_3$ | Traditional | $B_2O_3$, BaO, $Na_2CO_3$ | 600 | 0.06 | 0.66 | 0.08 |
| | | | 700 | 0.00 | 0.68 | 0.00 |
| | | | 800 | 0.17 | 0.34 | 0.33 |
| | Predicted | BaO, $NaBO_2$ | 600 | 0.03 | 0.55 | 0.05 |
| | | | 700 | 0.00 | 0.95 | 0.00 |
| | | | 800 | 0.02 | 0.37 | 0.05 |

## $KMgPO_4$

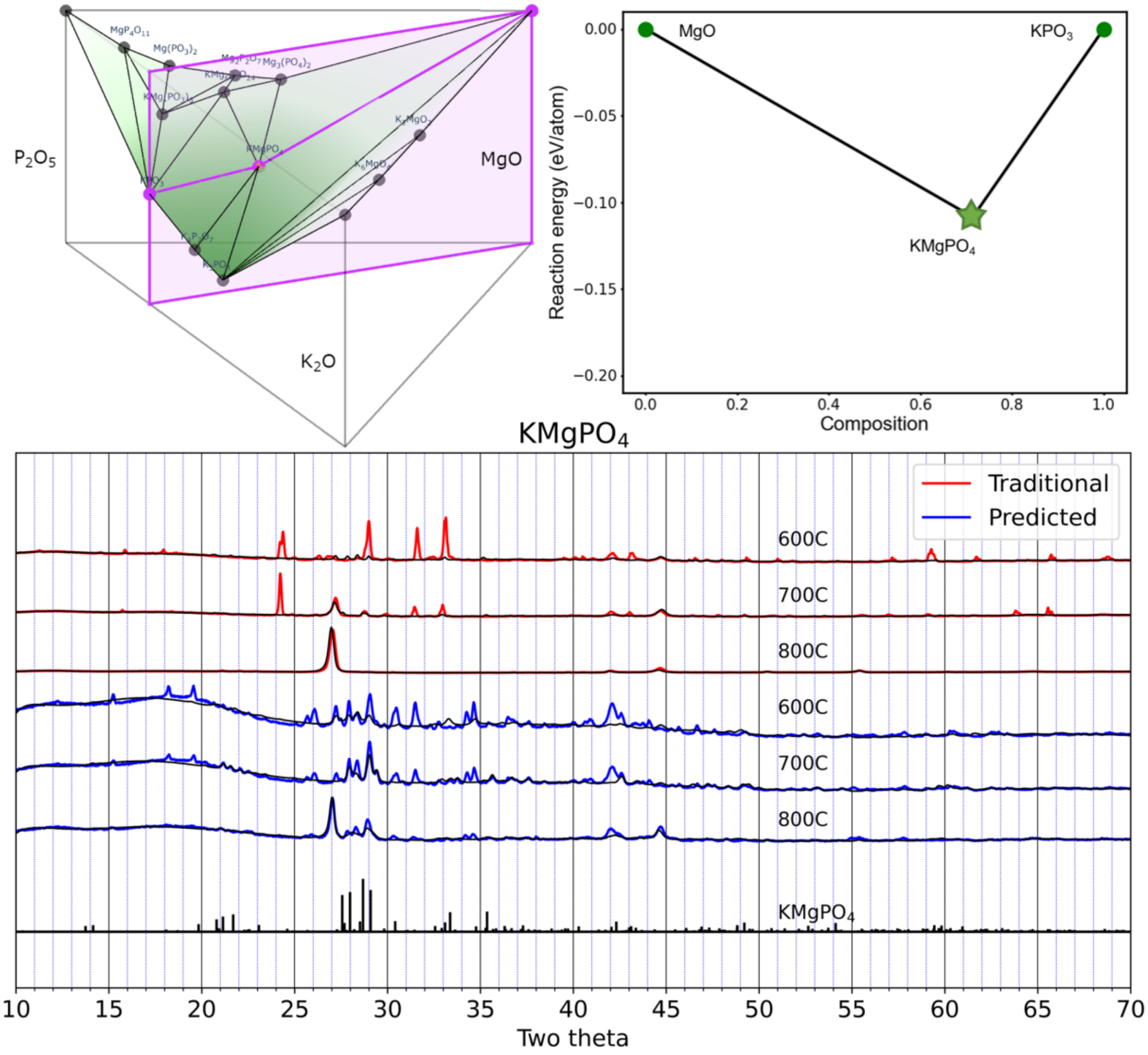

| Compound | Precursor type | Precursors | Temp (°C) | Target intensity (e6) | Residual intensity (e6) | Target phase fraction |
|---|---|---|---|---|---|---|
| $KMgPO_4$ | Traditional | $K_2CO_3$, MgO, $NH_4H_2PO_4$ | 600 | 0.29 | 1.56 | 0.15 |
| | | | 700 | 0.39 | 1.39 | 0.22 |
| | | | 800 | 0.07 | 0.62 | 0.11 |
| | Predicted | $KPO_3$, MgO | 600 | 0.99 | 1.70 | 0.37 |
| | | | 700 | 1.29 | 1.53 | 0.46 |
| | | | 800 | 0.55 | 0.57 | 0.49 |

# $NaSrBO_3$

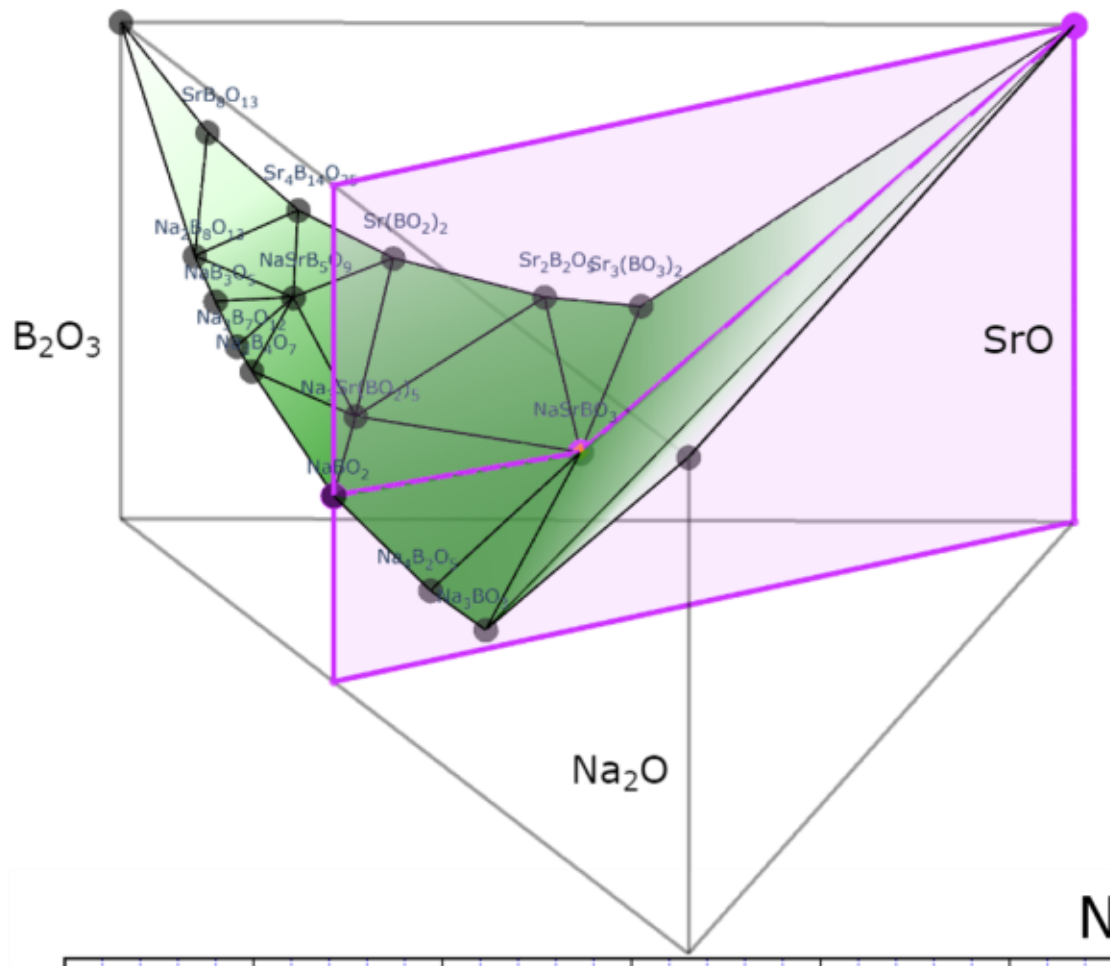

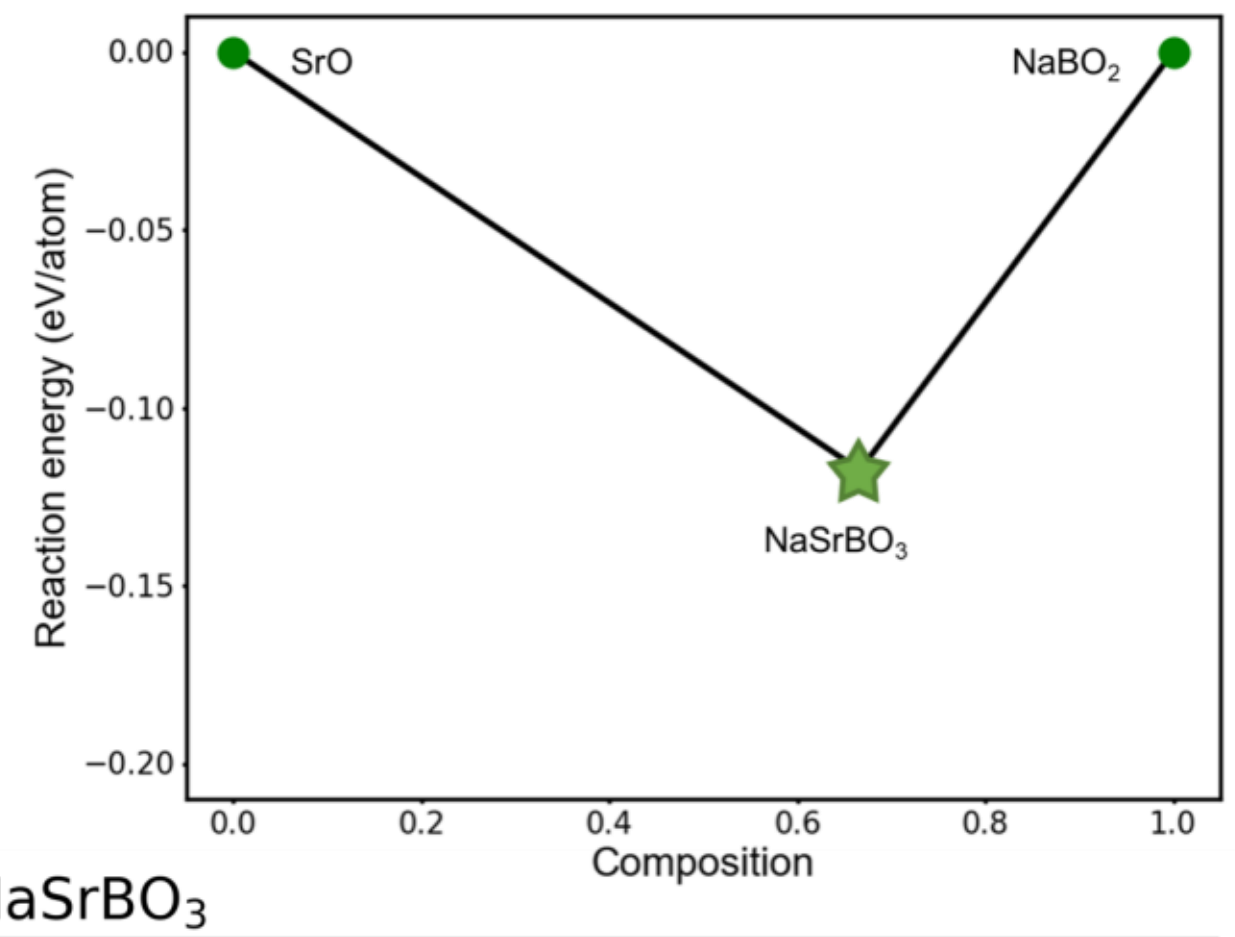

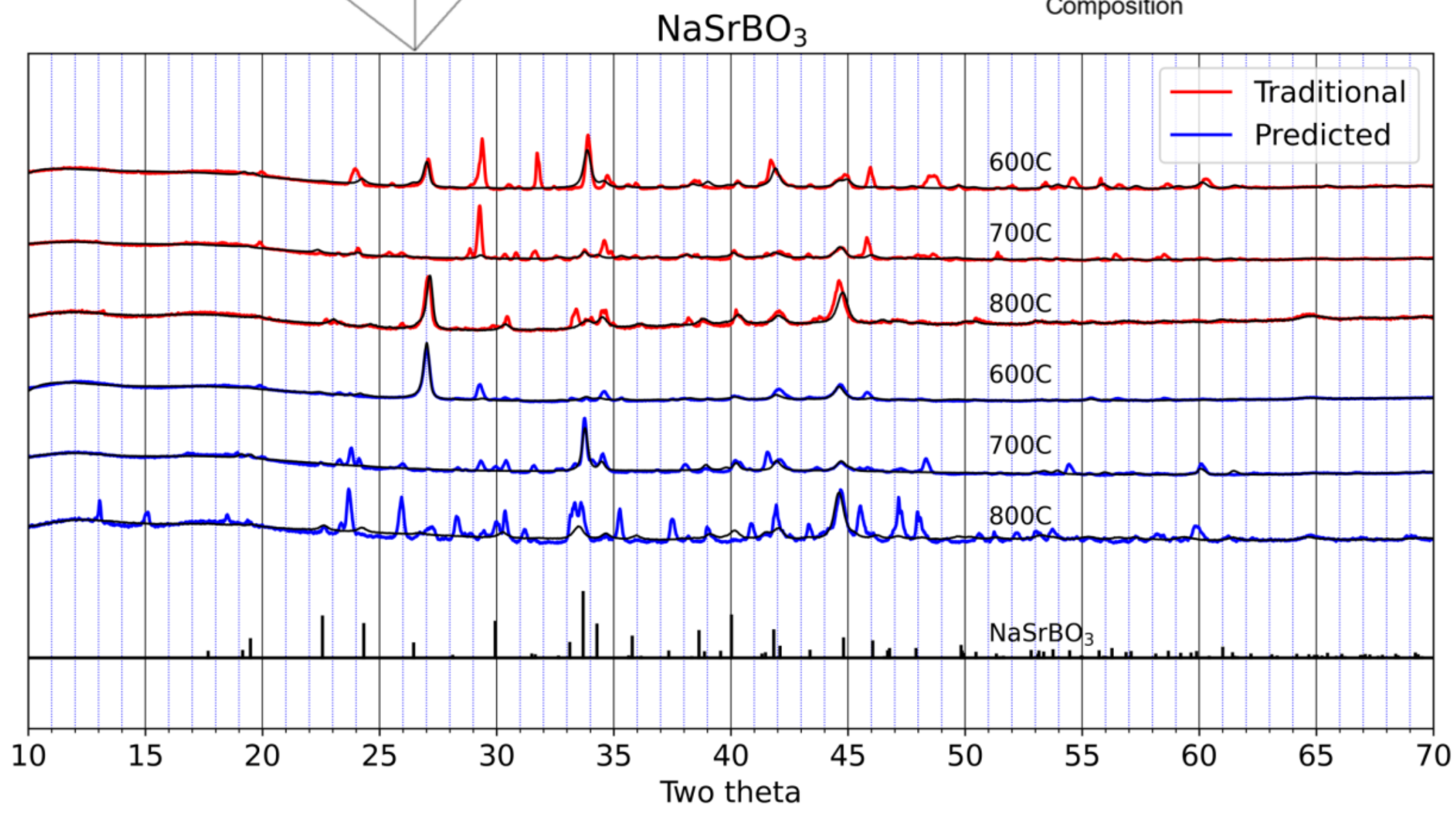

| Compound | Precursor type | Precursors | Temp (°C) | Target intensity (e6) | Residual intensity (e6) | Target phase fraction |
|---|---|---|---|---|---|---|
| $NaSrBO_3$ | Traditional | $B_2O_3$, $Na_2CO_3$, SrO | 600 | 1.13 | 1.58 | 0.42 |
| | | | 700 | 0.57 | 1.30 | 0.30 |
| | | | 800 | 0.57 | 0.57 | 0.50 |
| | Predicted | $NaBO_2$, SrO | 600 | 0.49 | 0.92 | 0.35 |
| | | | 700 | 0.69 | 1.28 | 0.35 |
| | | | 800 | 0.49 | 1.19 | 0.29 |

## $K_3Bi_2(PO_4)_3$

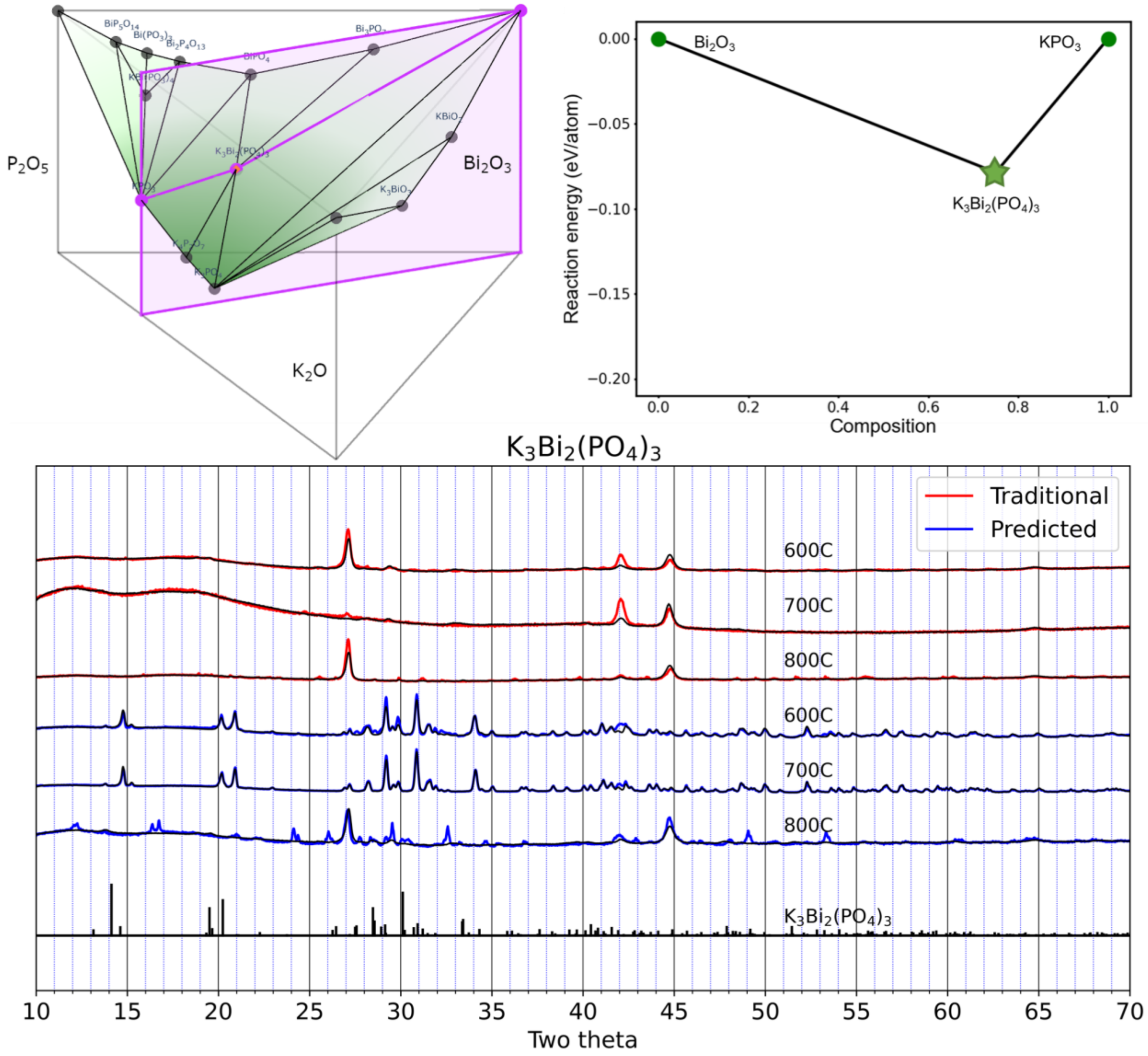

| Compound | Precursor type | Precursors | Temp (°C) | Target intensity (e6) | Residual intensity (e6) | Target phase fraction |
|---|---|---|---|---|---|---|
| $K_3Bi_2(PO_4)_3$ | Traditional | $Bi_2O_3$, $K_2CO_3$, $NH_4H_2PO_4$ | 600 | 0.56 | 0.60 | 0.48 |
| | | | 700 | 0.27 | 0.40 | 0.41 |
| | | | 800 | 0.07 | 0.43 | 0.14 |
| | Predicted | $Bi_2O_3$, $KPO_3$ | 600 | 2.67 | 0.88 | 0.75 |
| | | | 700 | 3.37 | 0.58 | 0.85 |
| | | | 800 | 0.31 | 0.58 | 0.34 |

## $K_2Zr(PO_4)_2$

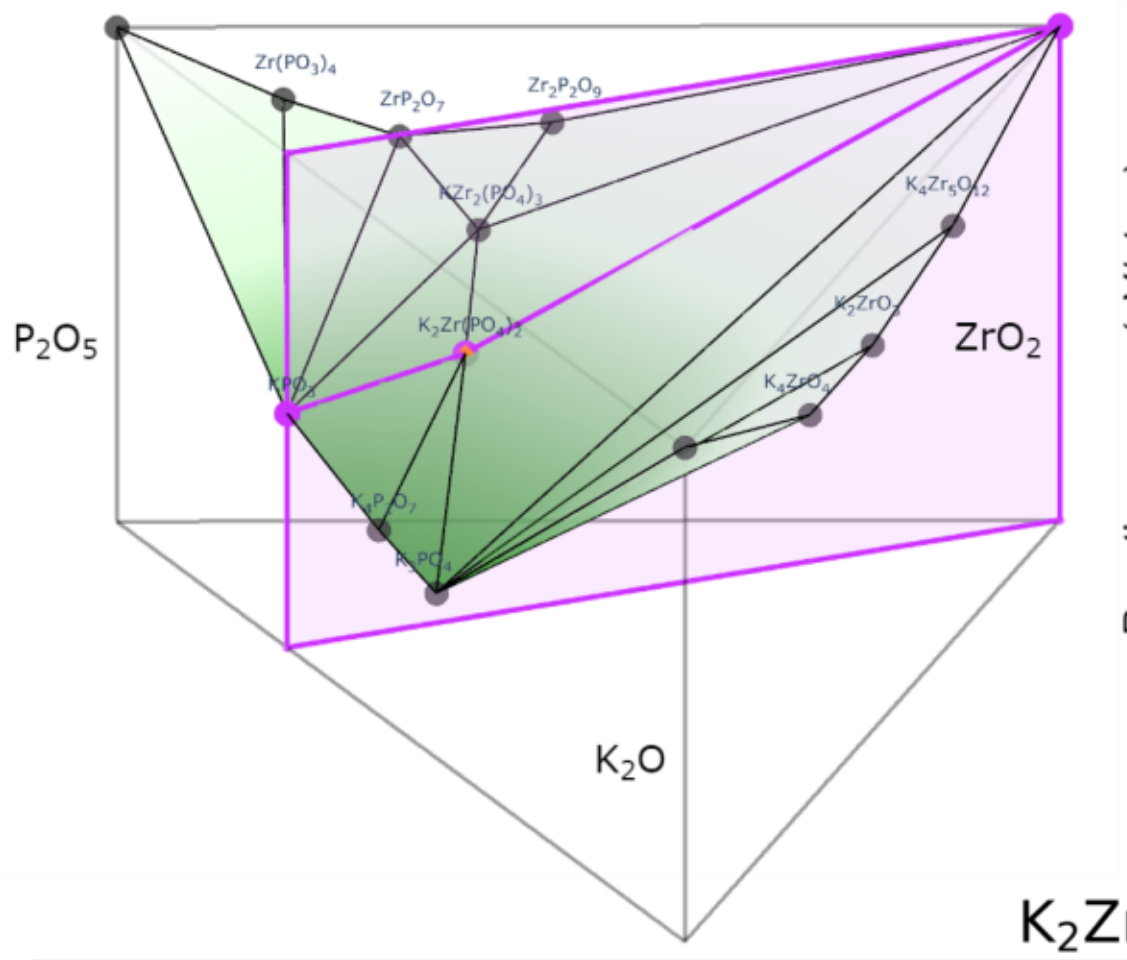

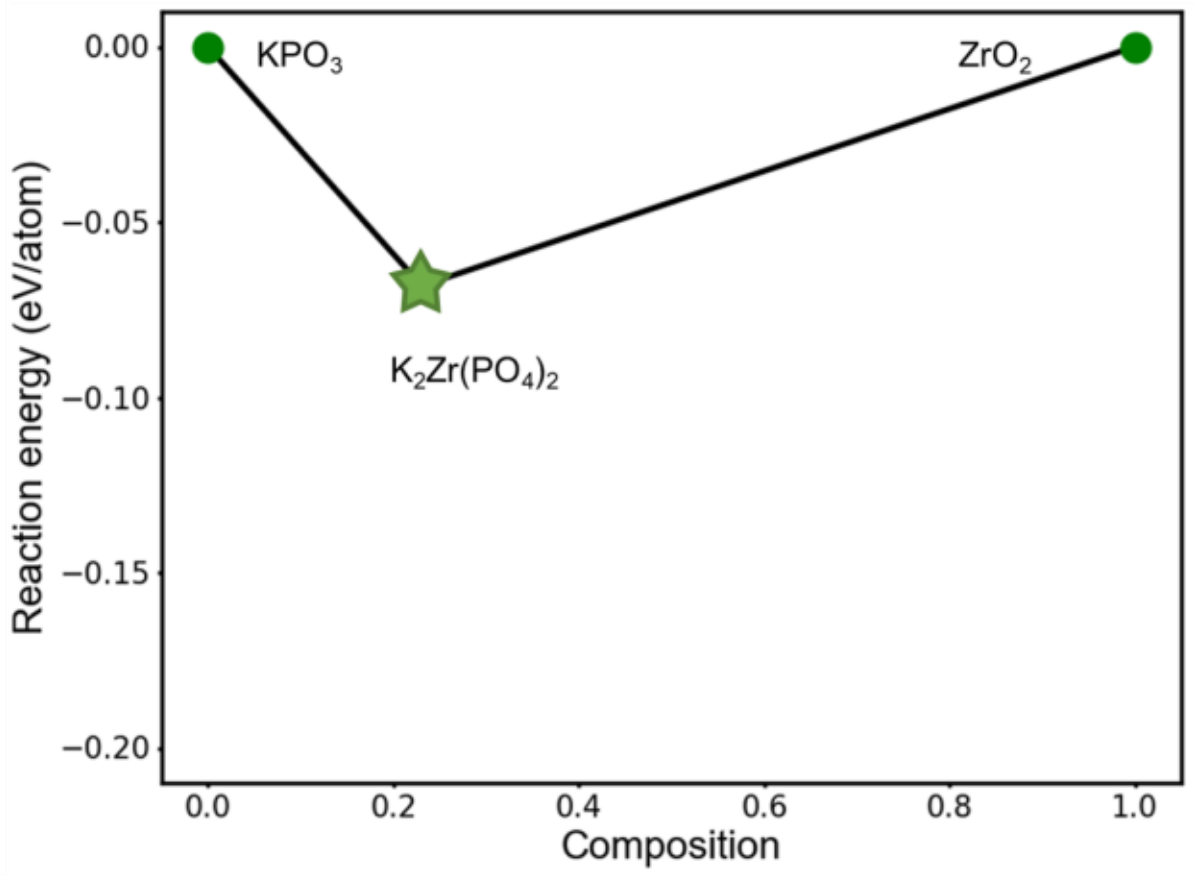

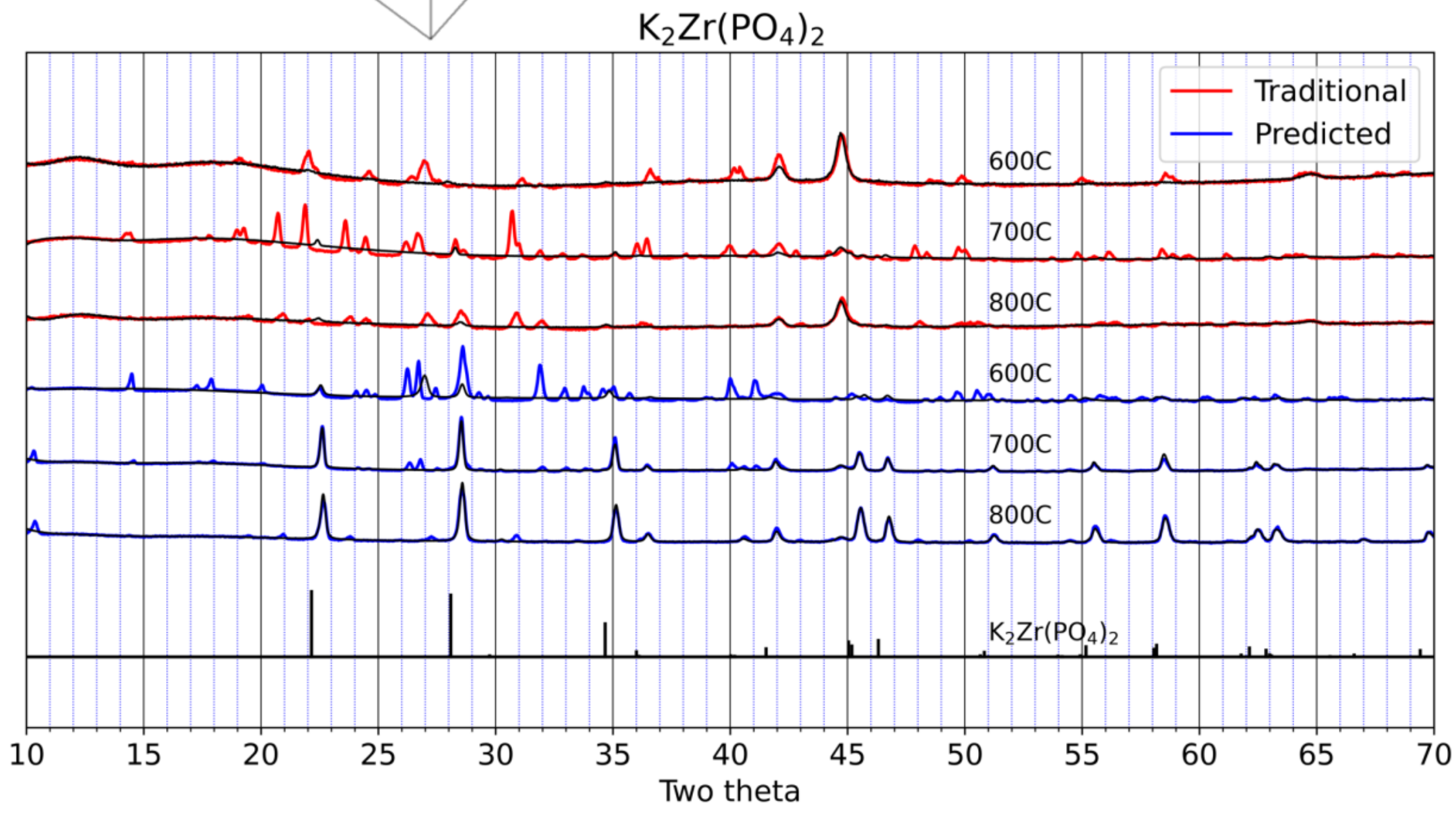

| Compound | Precursor type | Precursors | Temp (°C) | Target intensity (e6) | Residual intensity (e6) | Target phase fraction |
|---|---|---|---|---|---|---|
| $K_2Zr(PO_4)_2$ | Traditional | $K_2CO_3$, $NH_4H_2PO_4$, $ZrO_2$ | 600 | 0.04 | 0.62 | 0.06 |
| | | | 700 | 0.09 | 1.67 | 0.05 |
| | | | 800 | 0.06 | 0.44 | 0.12 |
| | Predicted | $KPO_3$, $ZrO_2$ | 600 | 0.44 | 2.42 | 0.15 |
| | | | 700 | 1.31 | 0.92 | 0.59 |
| | | | 800 | 0.93 | 0.31 | 0.75 |

## $K_3Al_2(PO_4)_3$

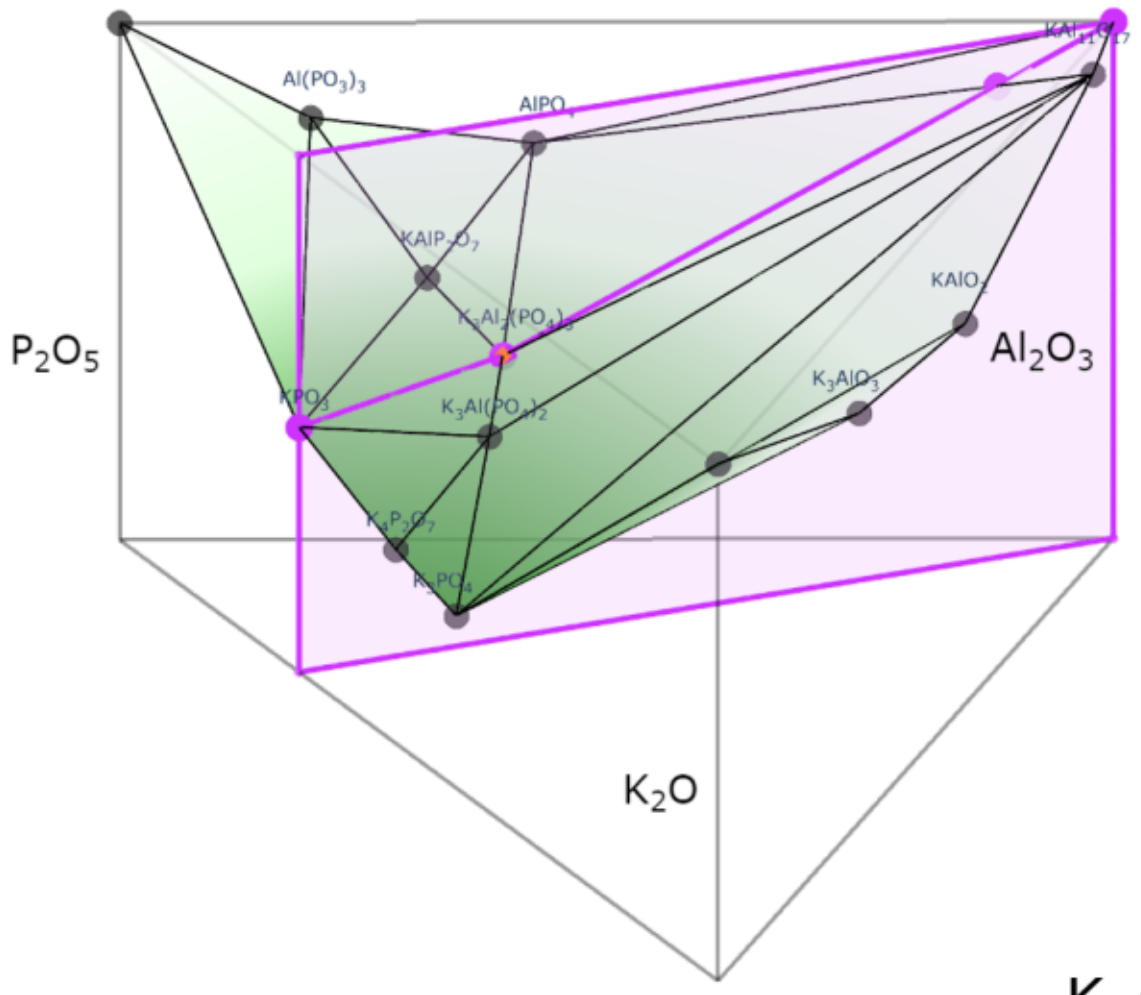

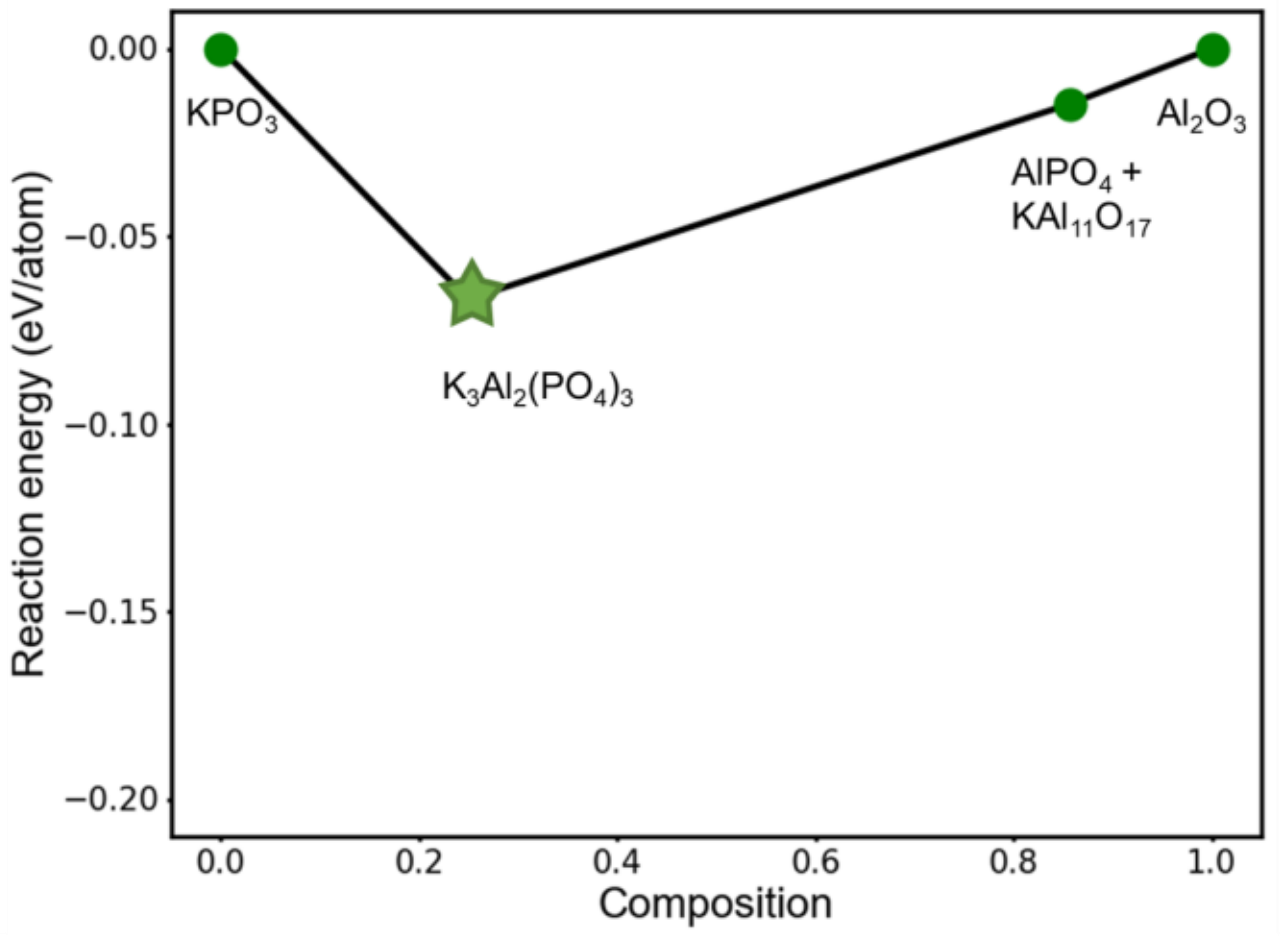

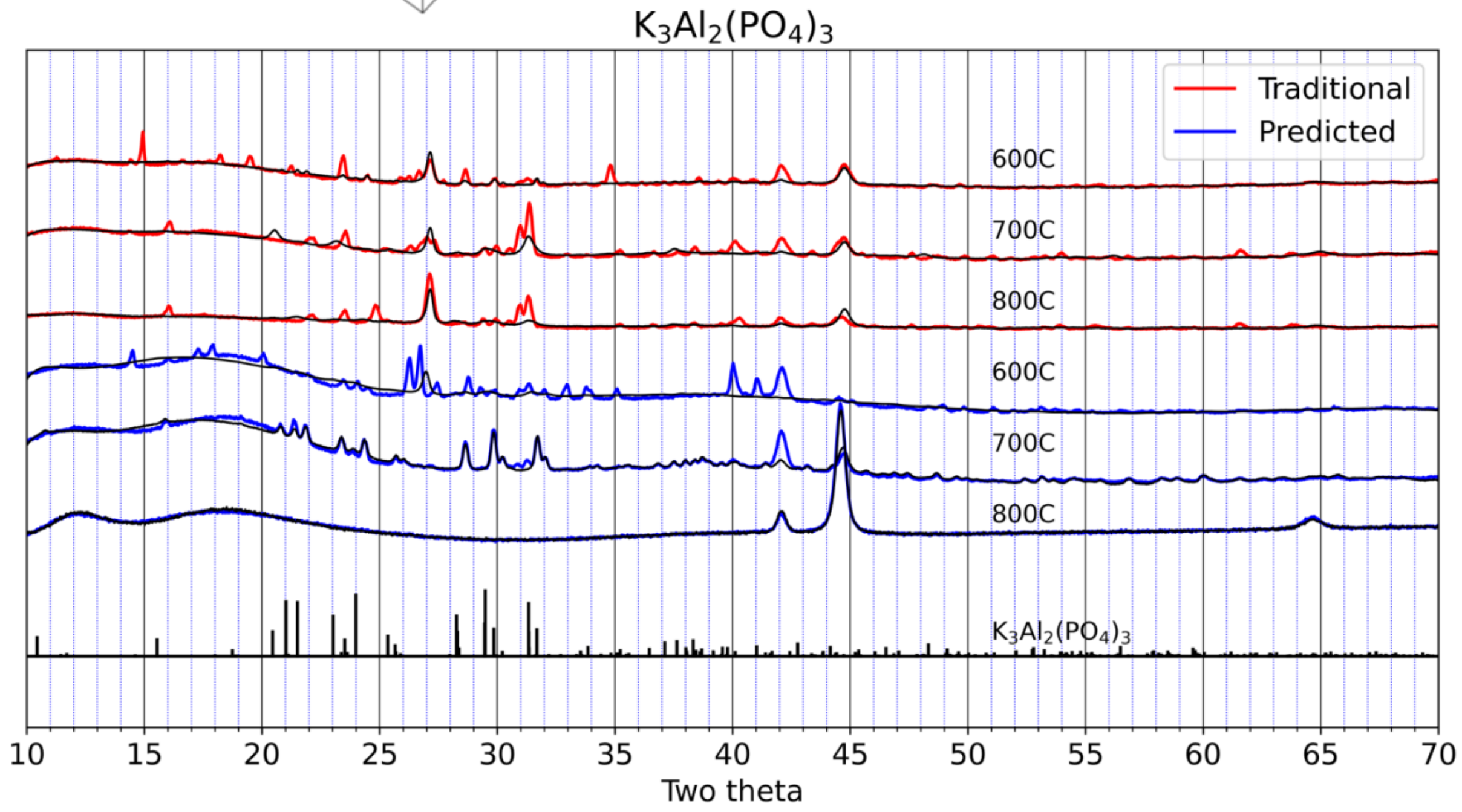

| Compound | Precursor type | Precursors | Temp (°C) | Target intensity (e6) | Residual intensity (e6) | Target phase fraction |
|---|---|---|---|---|---|---|
| $K_3Al_2(PO_4)_3$ | Traditional | $Al_2O_3$, $K_2CO_3$, $NH_4H_2PO_4$ | 600 | 0.17 | 0.78 | 0.18 |
| | | | 700 | 0.83 | 0.94 | 0.47 |
| | | | 800 | 0.40 | 0.63 | 0.39 |
| | Predicted | $Al_2O_3$, $KPO_3$ | 600 | 0.52 | 1.44 | 0.27 |
| | | | 700 | 1.14 | 0.80 | 0.59 |
| | | | 800 | 0.04 | 0.31 | 0.11 |

## $KTiPO_5$

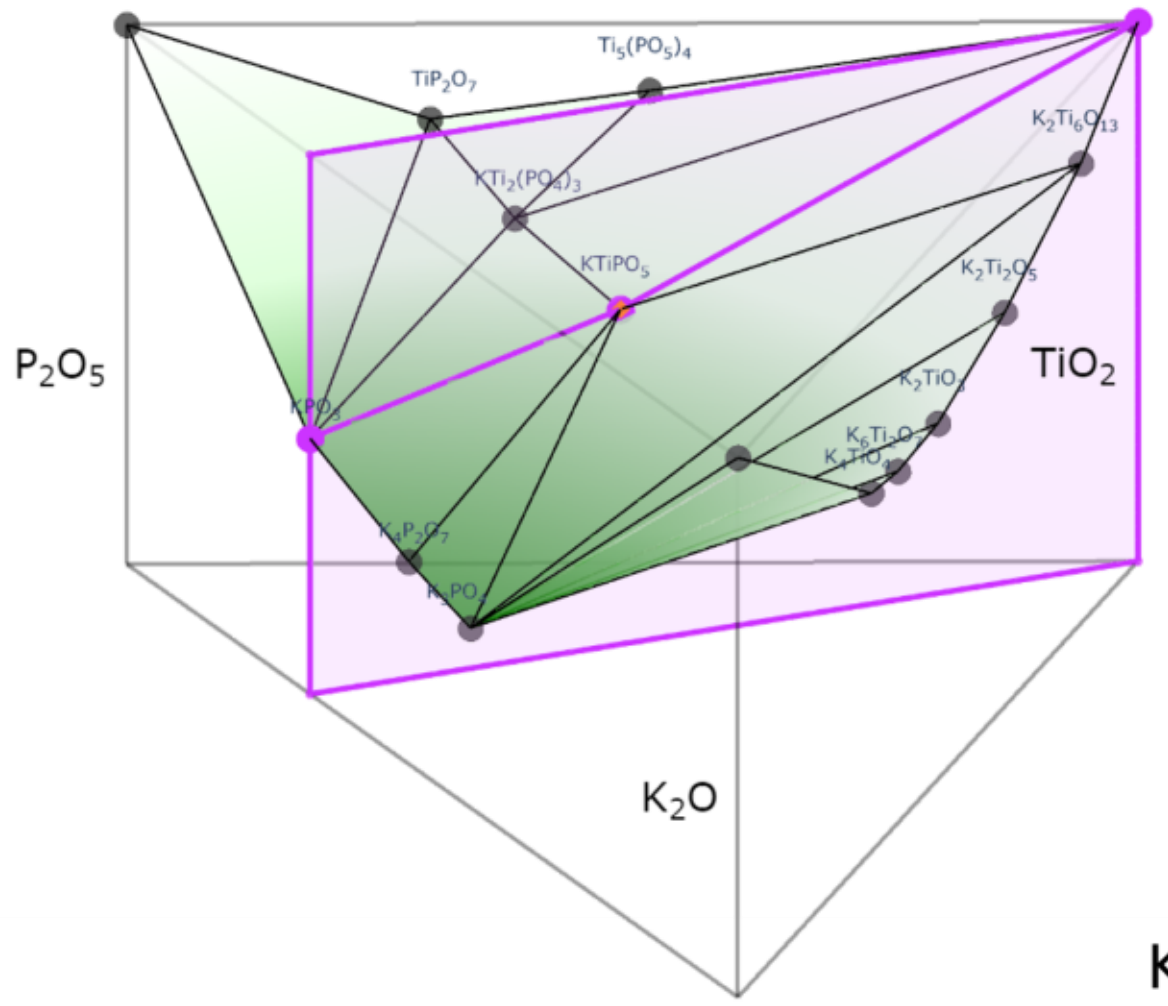

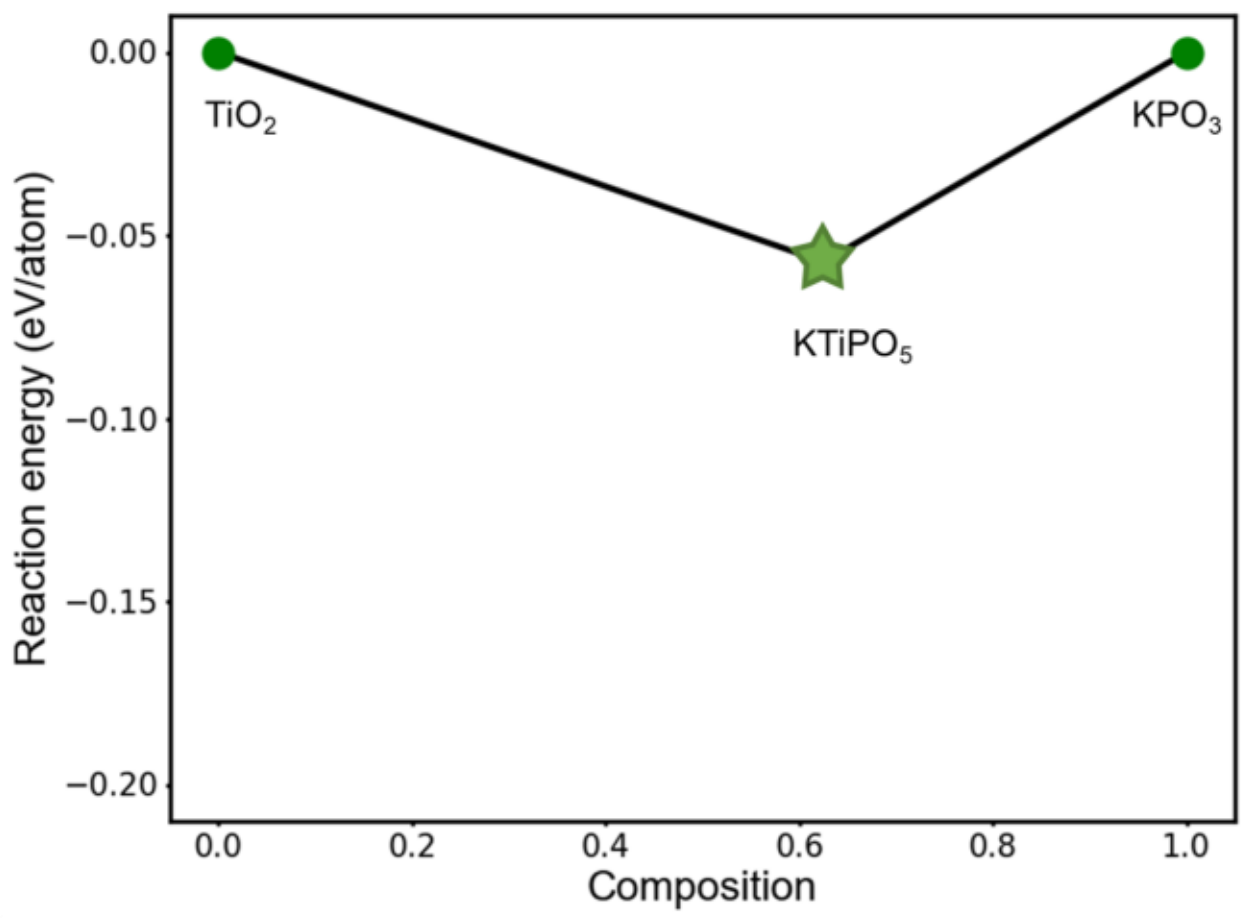

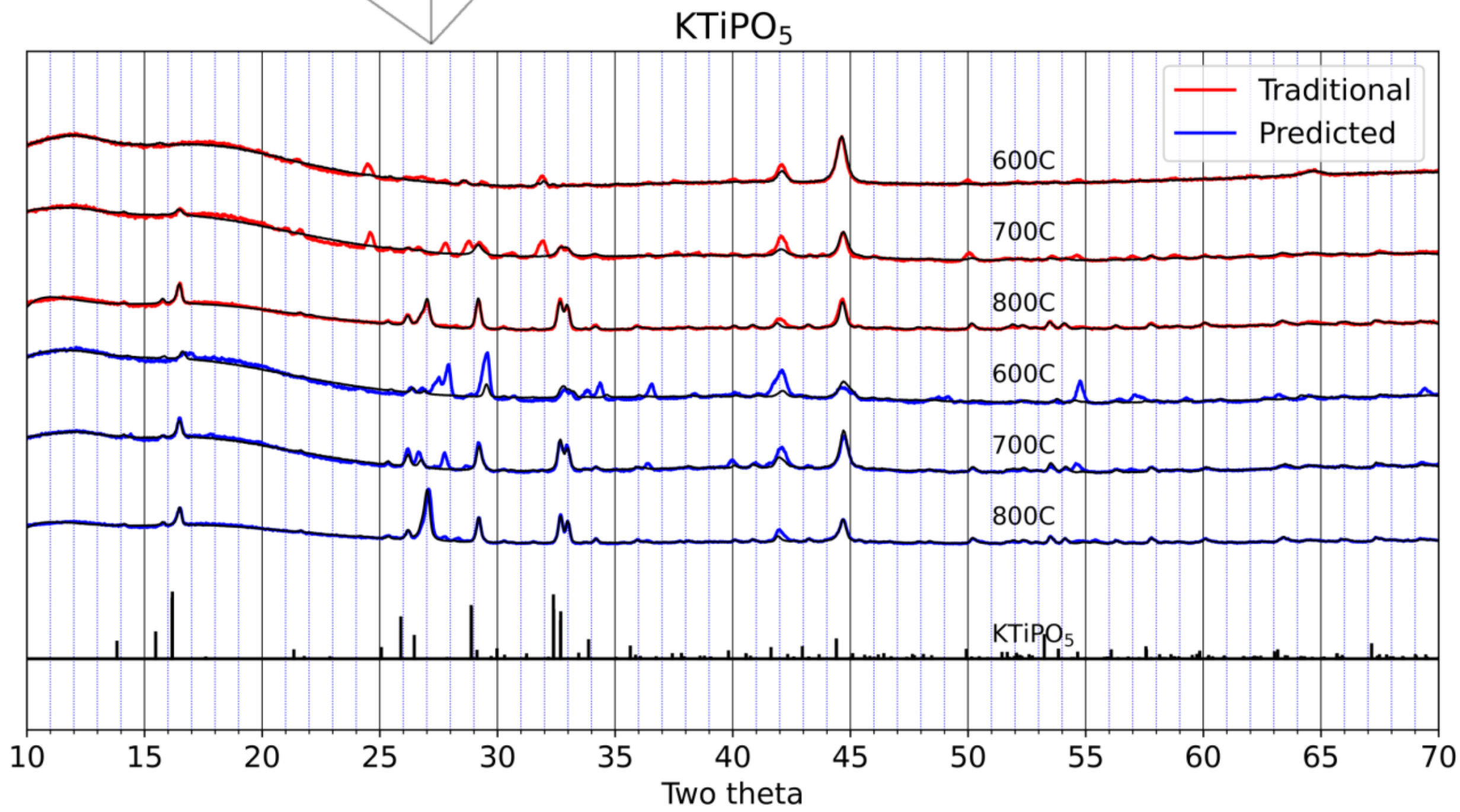

| Compound | Precursor type | Precursors | Temp (°C) | Target intensity (e6) | Residual intensity (e6) | Target phase fraction |
|---|---|---|---|---|---|---|
| $KTiPO_5$ | Traditional | $K_2CO_3$, $NH_4H_2PO_4$, $TiO_2$ | 600 | 0.09 | 0.42 | 0.17 |
| | | | 700 | 0.28 | 0.58 | 0.33 |
| | | | 800 | 0.52 | 0.27 | 0.66 |
| | Predicted | $KPO_3$, $TiO_2$ | 600 | 0.21 | 0.81 | 0.21 |
| | | | 700 | 0.68 | 0.56 | 0.55 |
| | | | 800 | 0.52 | 0.31 | 0.62 |

## KNiPO4

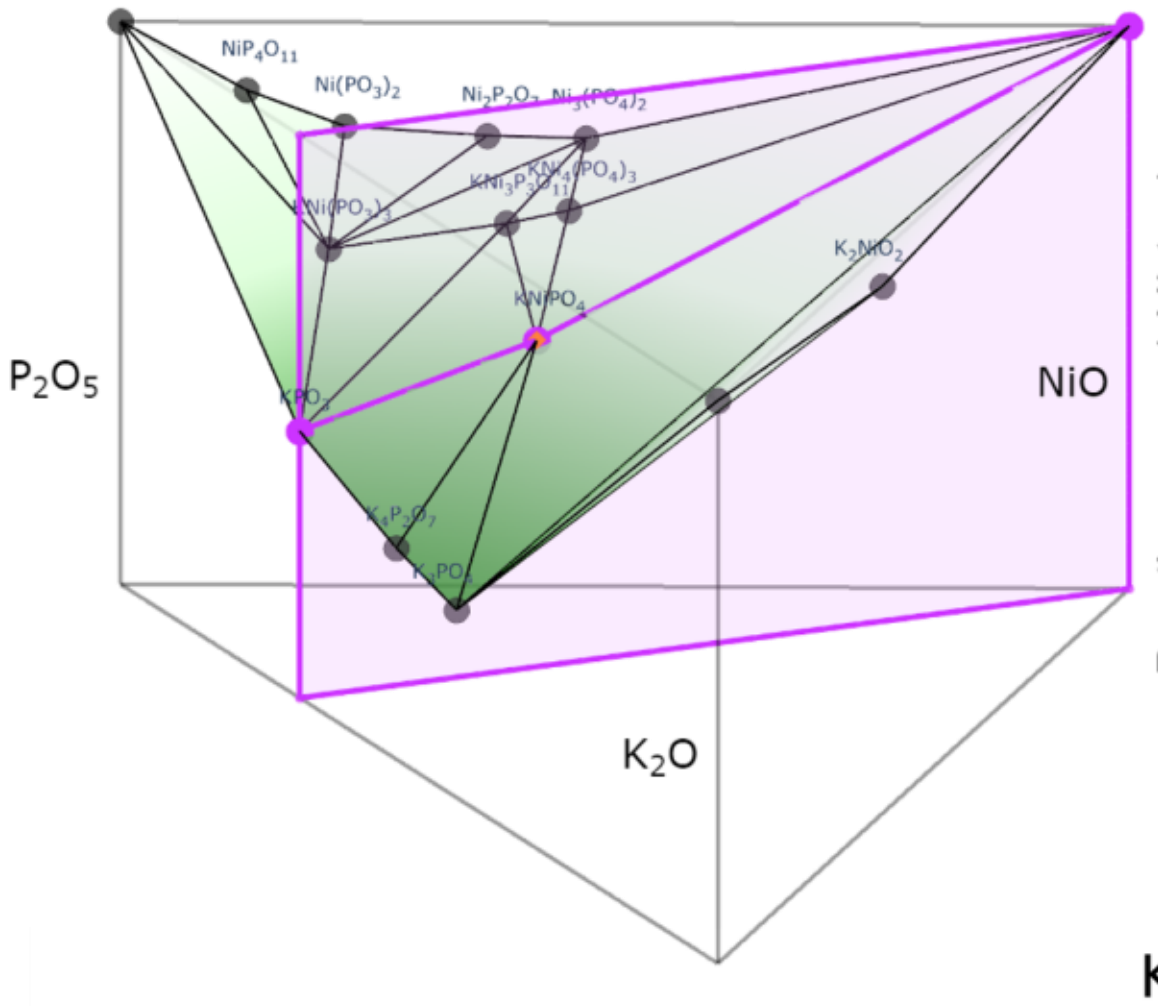

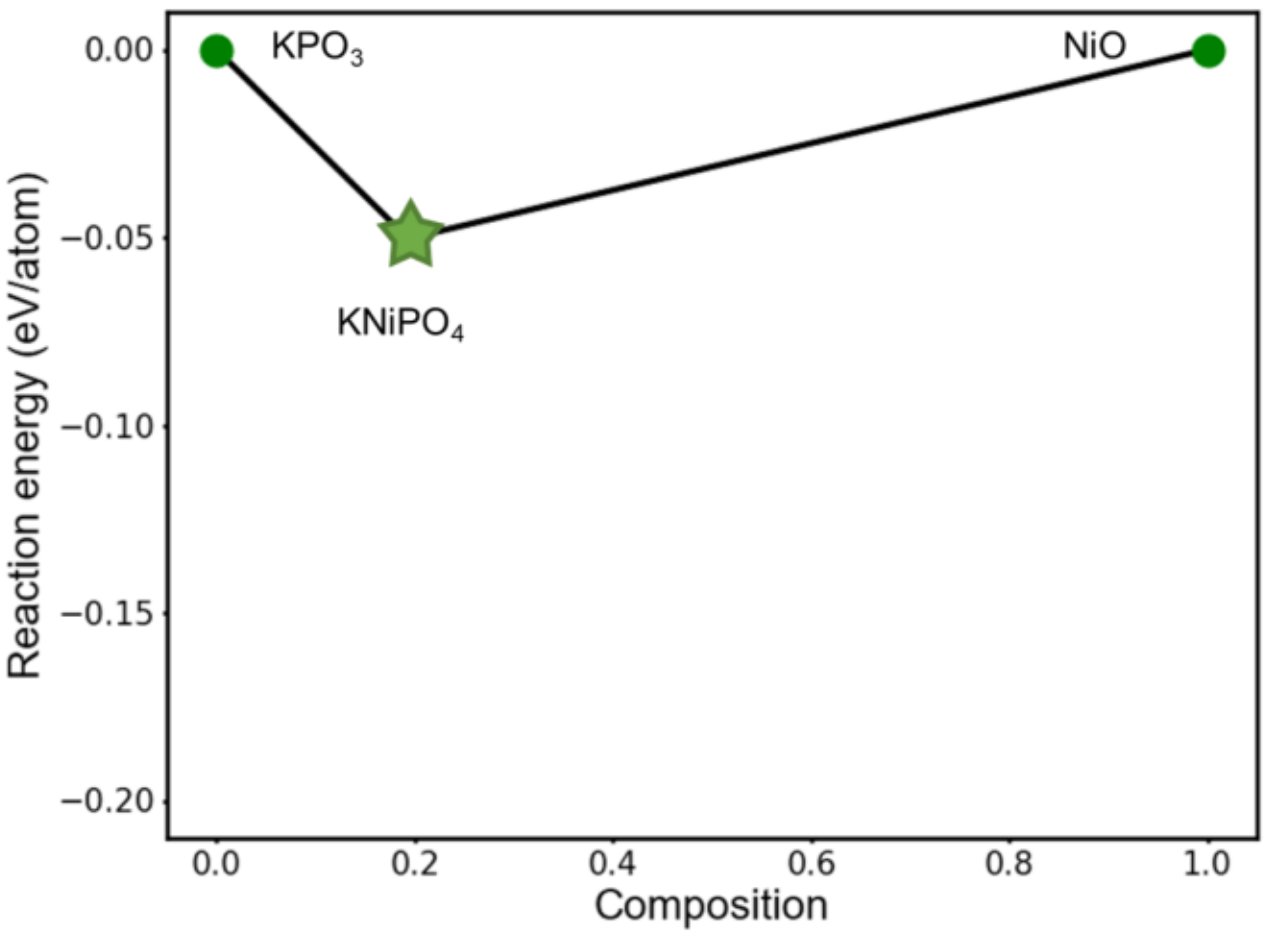

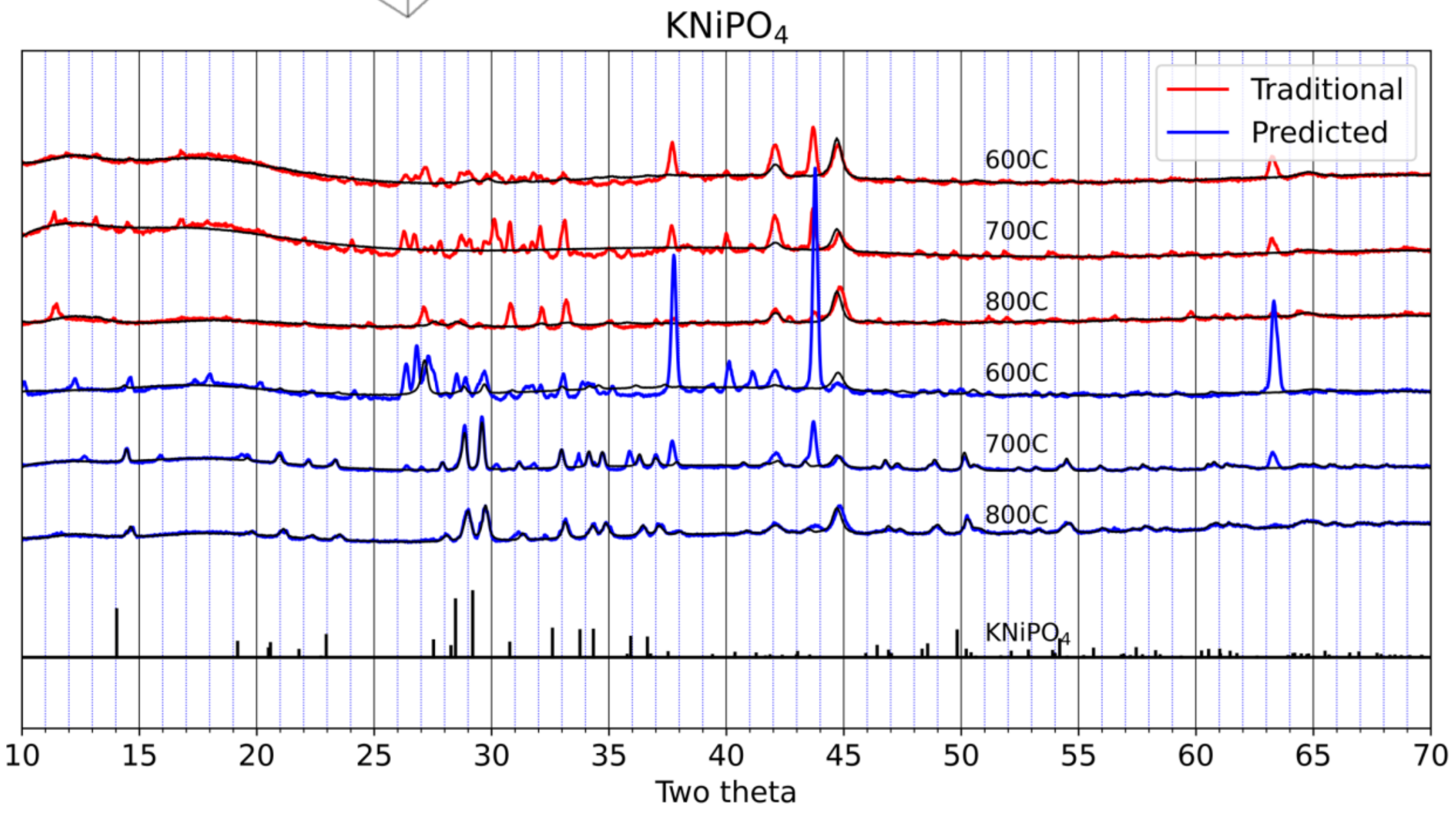

| Compound | Precursor type | Precursors | Temp (°C) | Target intensity (e6) | Residual intensity (e6) | Target phase fraction |
|---|---|---|---|---|---|---|
| KNiPO4 | Traditional | $K_2CO_3$, $NH_4H_2PO_4$, NiO | 600 | 0.15 | 0.87 | 0.15 |
| | | | 700 | 0.00 | 1.07 | 0.00 |
| | | | 800 | 0.21 | 0.62 | 0.25 |
| | Predicted | $KPO_3$, NiO | 600 | 0.23 | 2.26 | 0.09 |
| | | | 700 | 1.13 | 1.04 | 0.52 |
| | | | 800 | 0.75 | 0.34 | 0.69 |

## $K_3Fe_2(PO_4)_3$

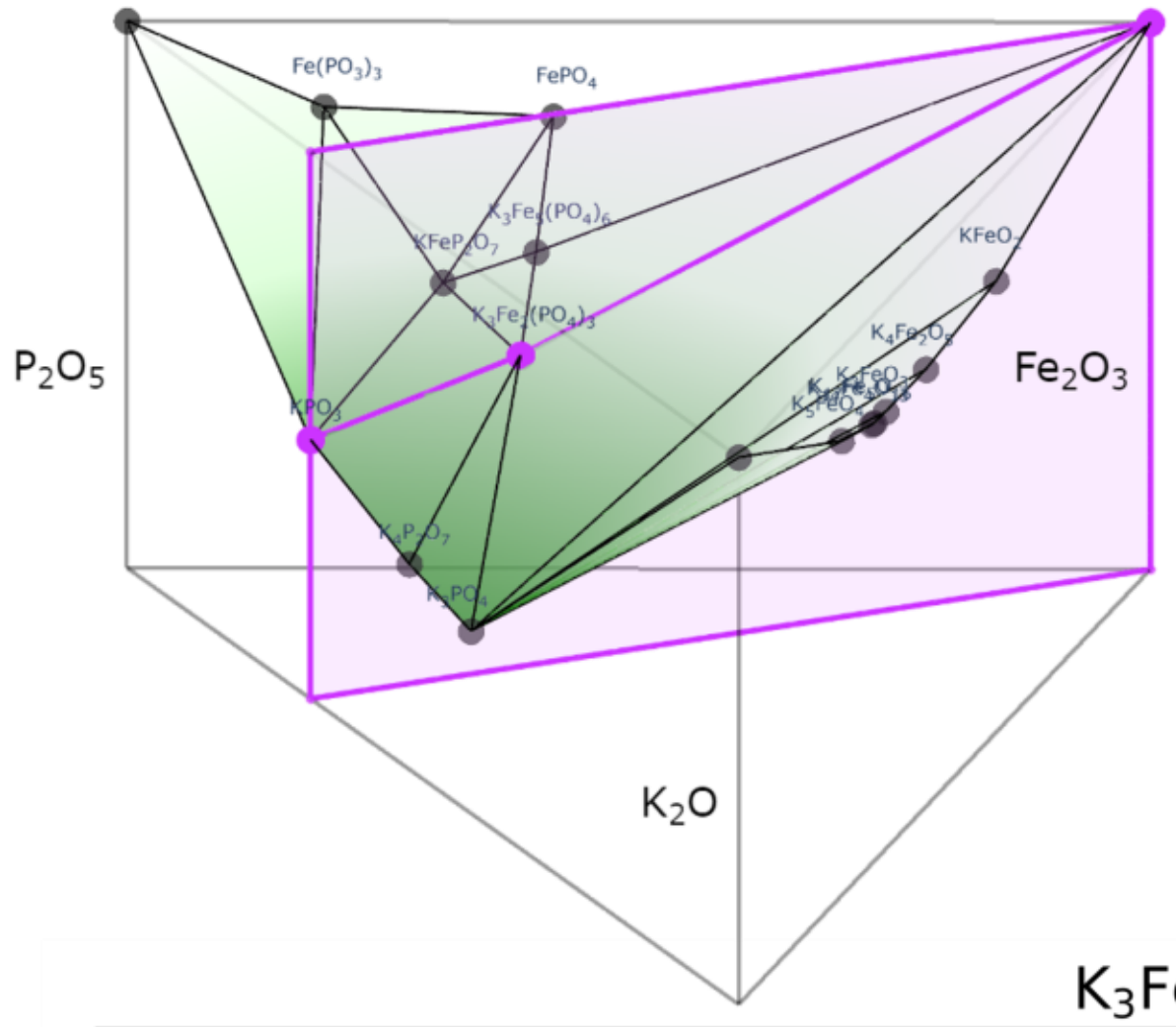

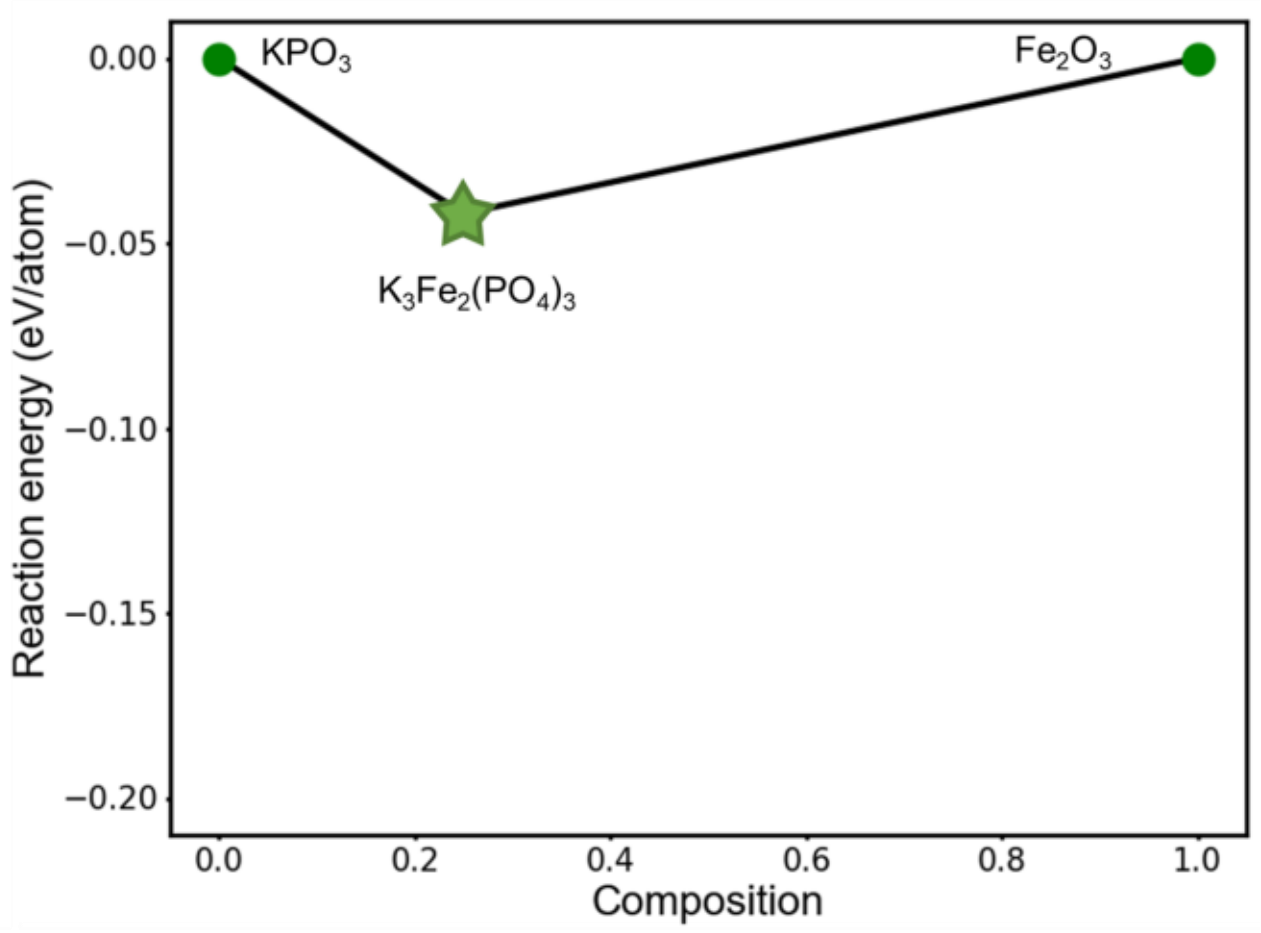

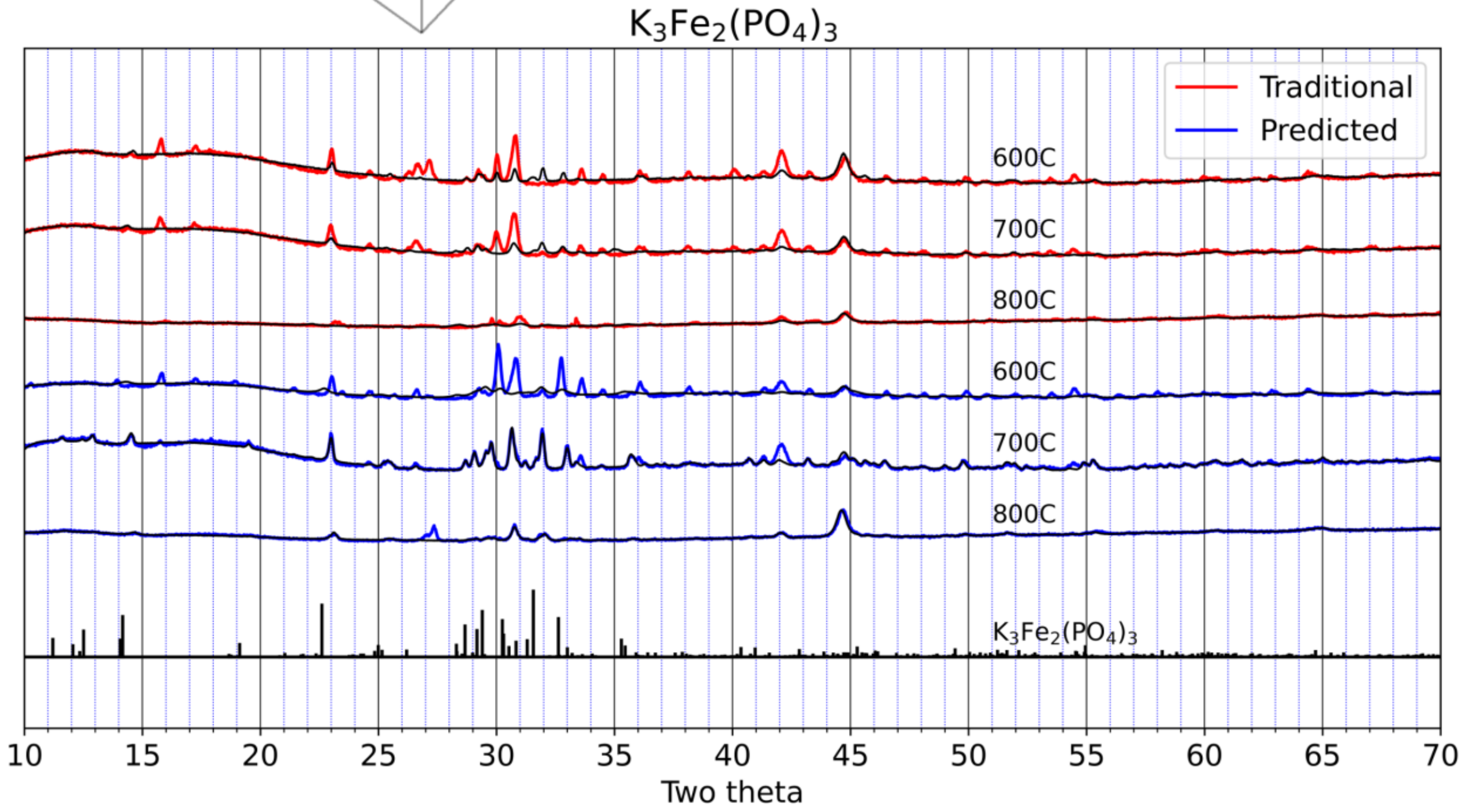

| Compound | Precursor type | Precursors | Temp (°C) | Target intensity (e6) | Residual intensity (e6) | Target phase fraction |
|---|---|---|---|---|---|---|
| $K_3Fe_2(PO_4)_3$ | Traditional | $Fe_2O_3$, $K_2CO_3$, $NH_4H_2PO_4$ | 600 | 0.24 | 0.76 | 0.24 |
| | | | 700 | 0.31 | 0.65 | 0.33 |
| | | | 800 | 0.14 | 0.22 | 0.39 |
| | Predicted | $Fe_2O_3$, $KPO_3$ | 600 | 0.77 | 1.09 | 0.41 |
| | | | 700 | 0.87 | 0.42 | 0.68 |
| | | | 800 | 0.22 | 0.19 | 0.55 |

# $KNbWO_6$

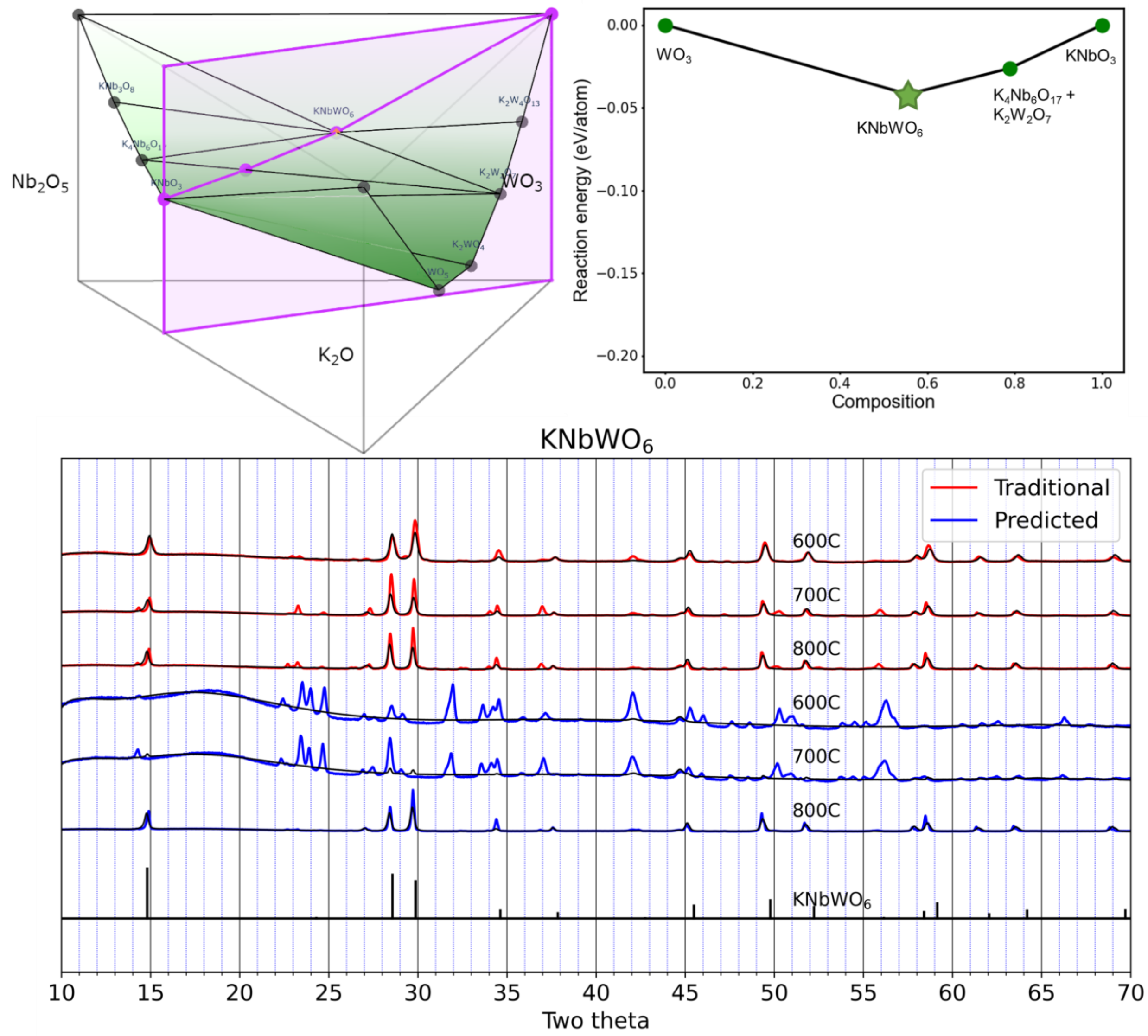

| Compound | Precursor type | Precursors | Temp (°C) | Target intensity (e6) | Residual intensity (e6) | Target phase fraction |
|---|---|---|---|---|---|---|
| $KNbWO_6$ | Traditional | $K_2CO_3$, WO3 $Nb_2O_5$ | 600 | 1.96 | 1.02 | 0.66 |
| | | | 700 | 2.65 | 2.49 | 0.52 |
| | | | 800 | 1.37 | 1.18 | 0.54 |
| | Predicted | $KNbO_3$, $WO_3$ | 600 | 0.00 | 2.18 | 0.00 |
| | | | 700 | 0.09 | 2.43 | 0.04 |
| | | | 800 | 2.97 | 1.86 | 0.61 |

## $KTa_2PO_8$

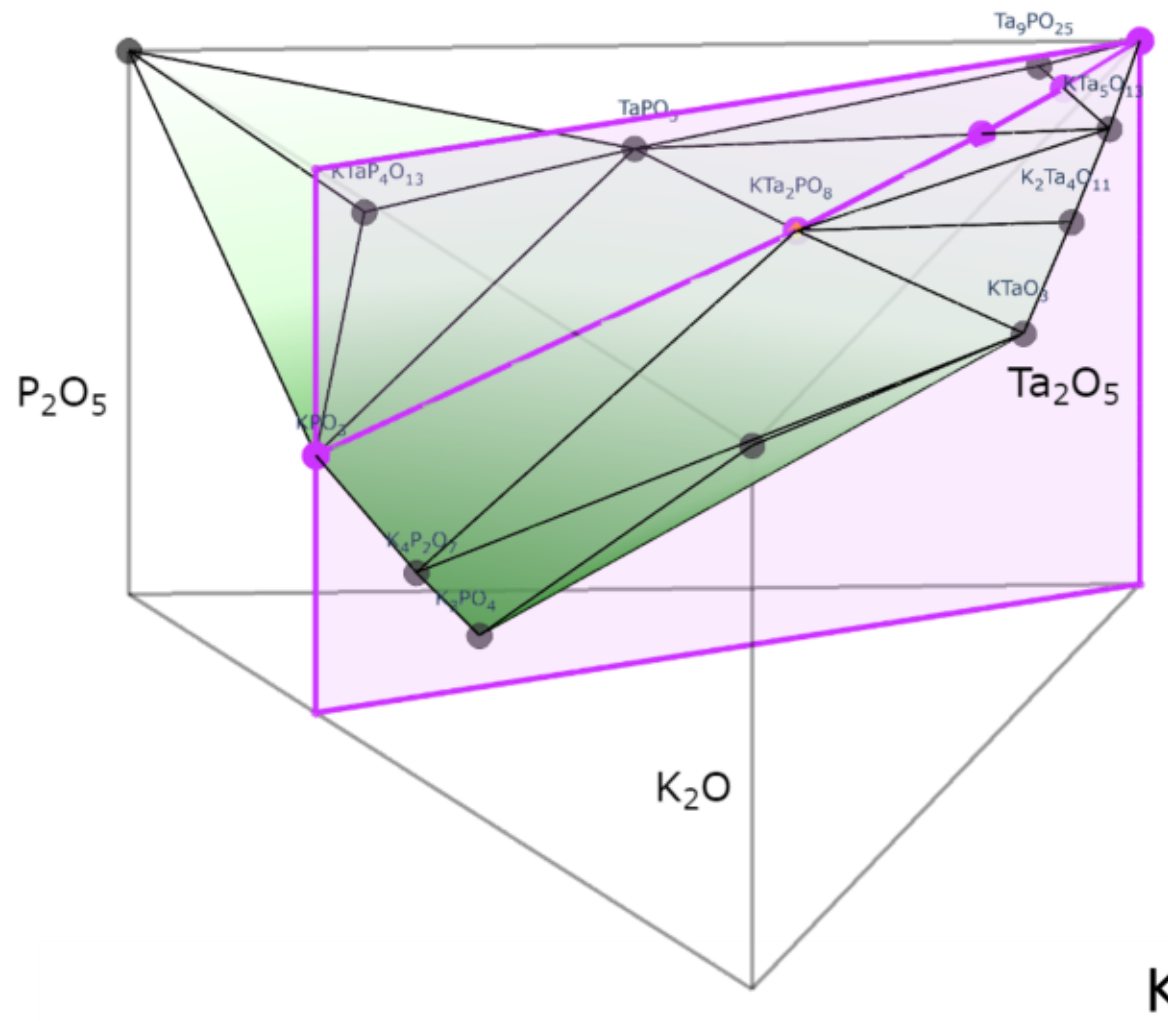

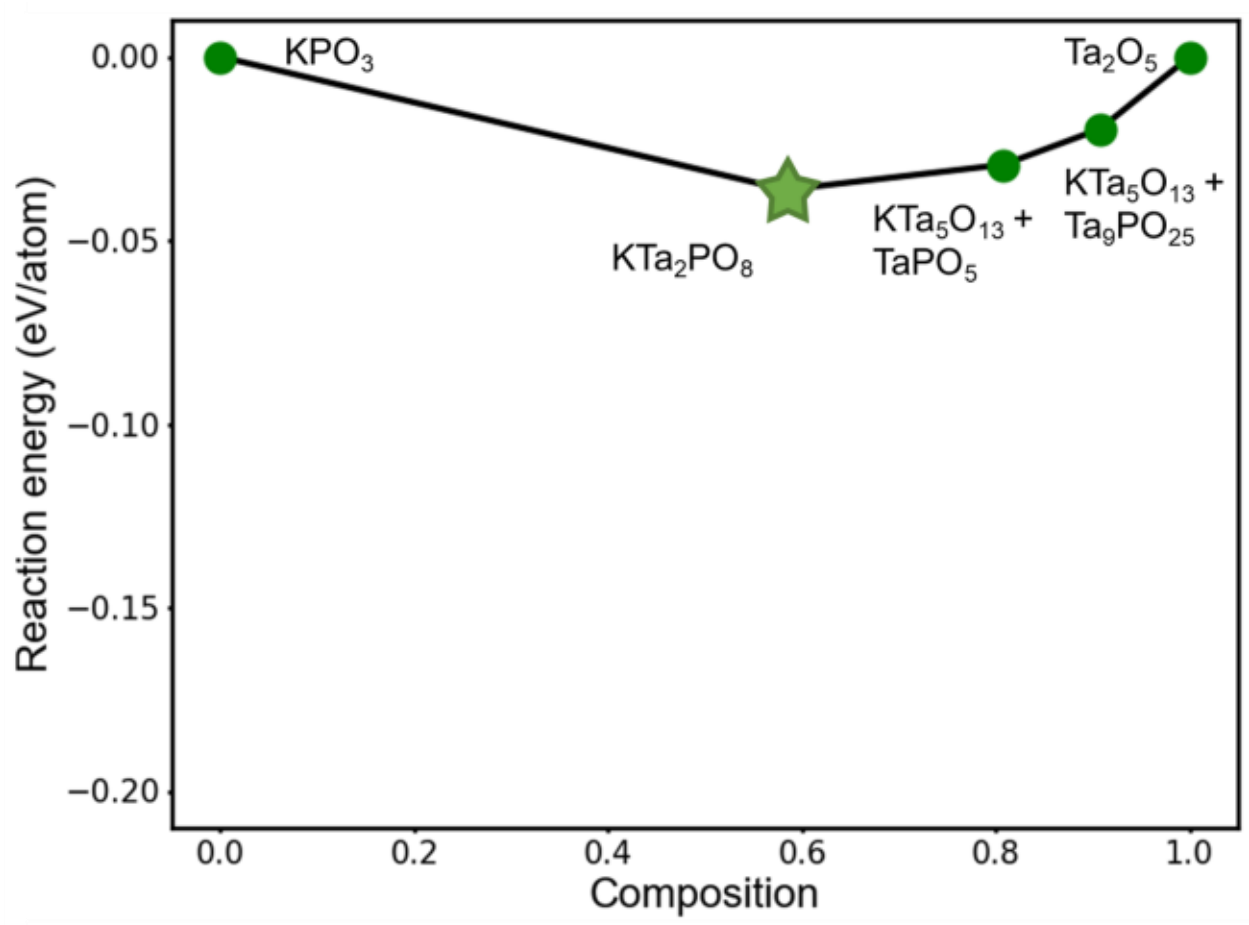

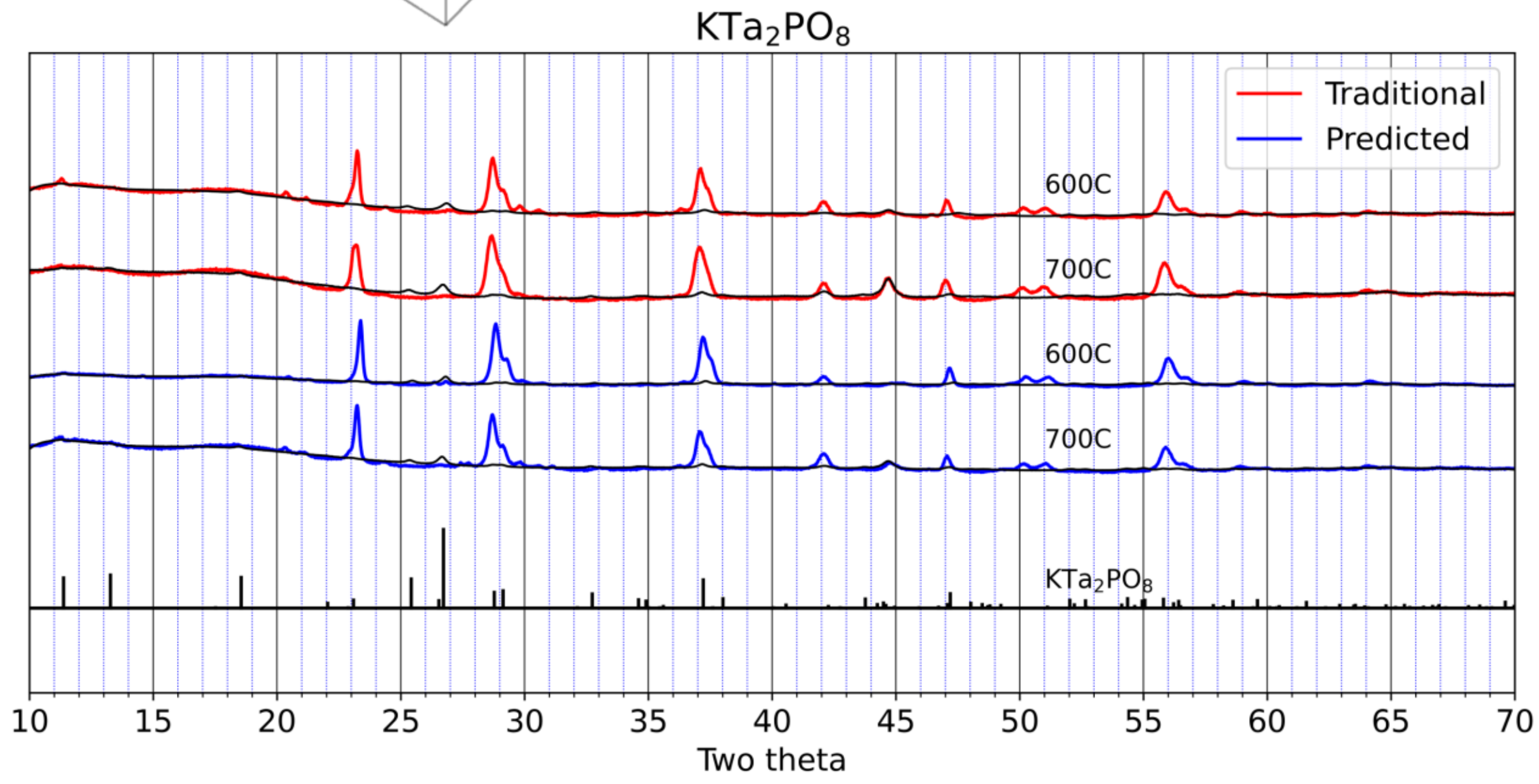

| Compound | Precursor type | Precursors | Temp (°C) | Target intensity (e6) | Residual intensity (e6) | Target phase fraction |
|---|---|---|---|---|---|---|
| $KTa_2PO_8$ | Traditional | $K_2CO_3$, $Ta_2O_5$, $NH_4H_2PO_4$ | 600 | 0.36 | 1.50 | 0.19 |
| | | | 700 | 0.46 | 1.77 | 0.21 |
| | Predicted | $KPO_3$, $Ta_2O_5$ | 600 | 0.46 | 2.71 | 0.14 |
| | | | 700 | 0.30 | 1.52 | 0.16 |

## $Li_3Y_2(BO_3)_3$

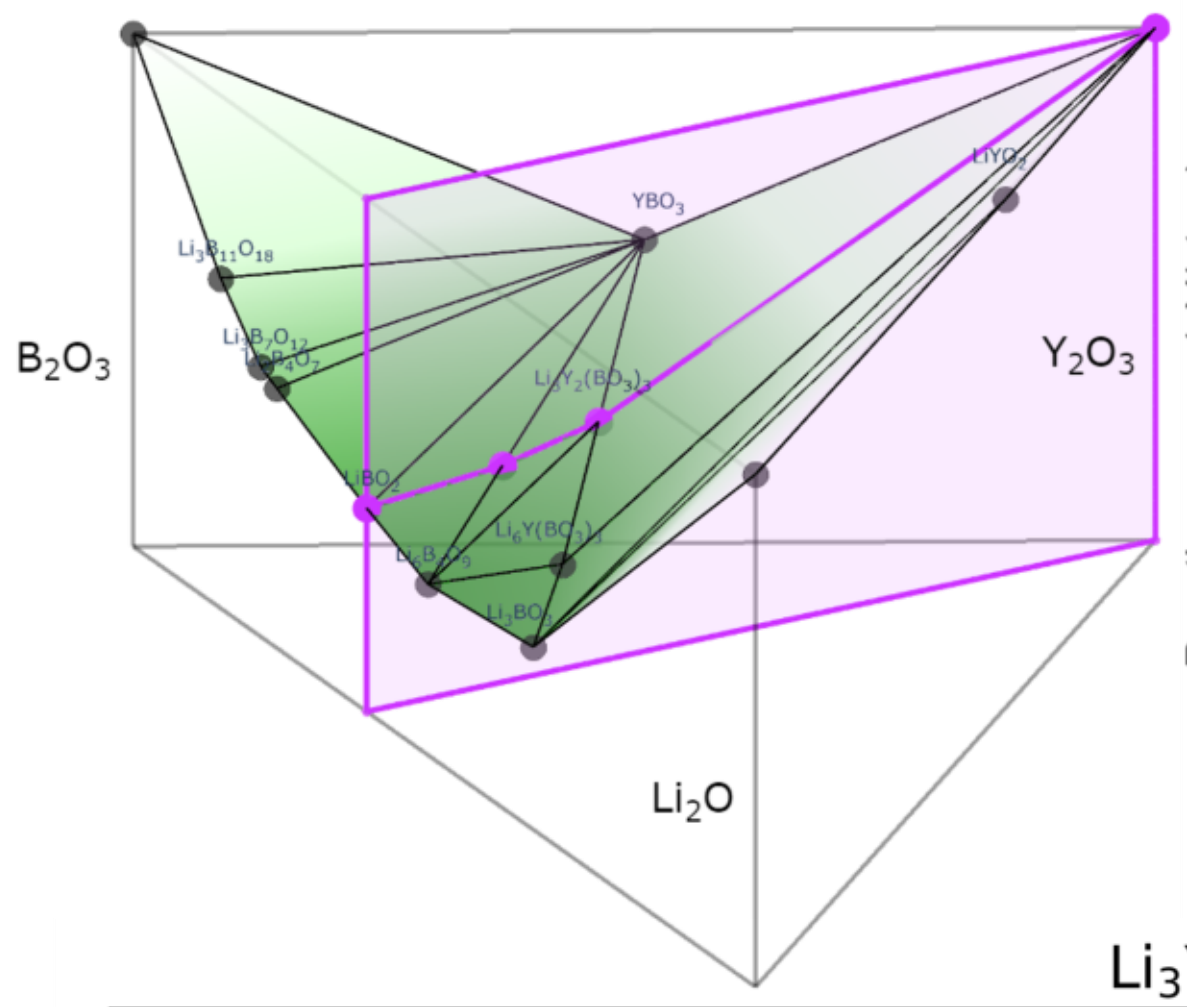

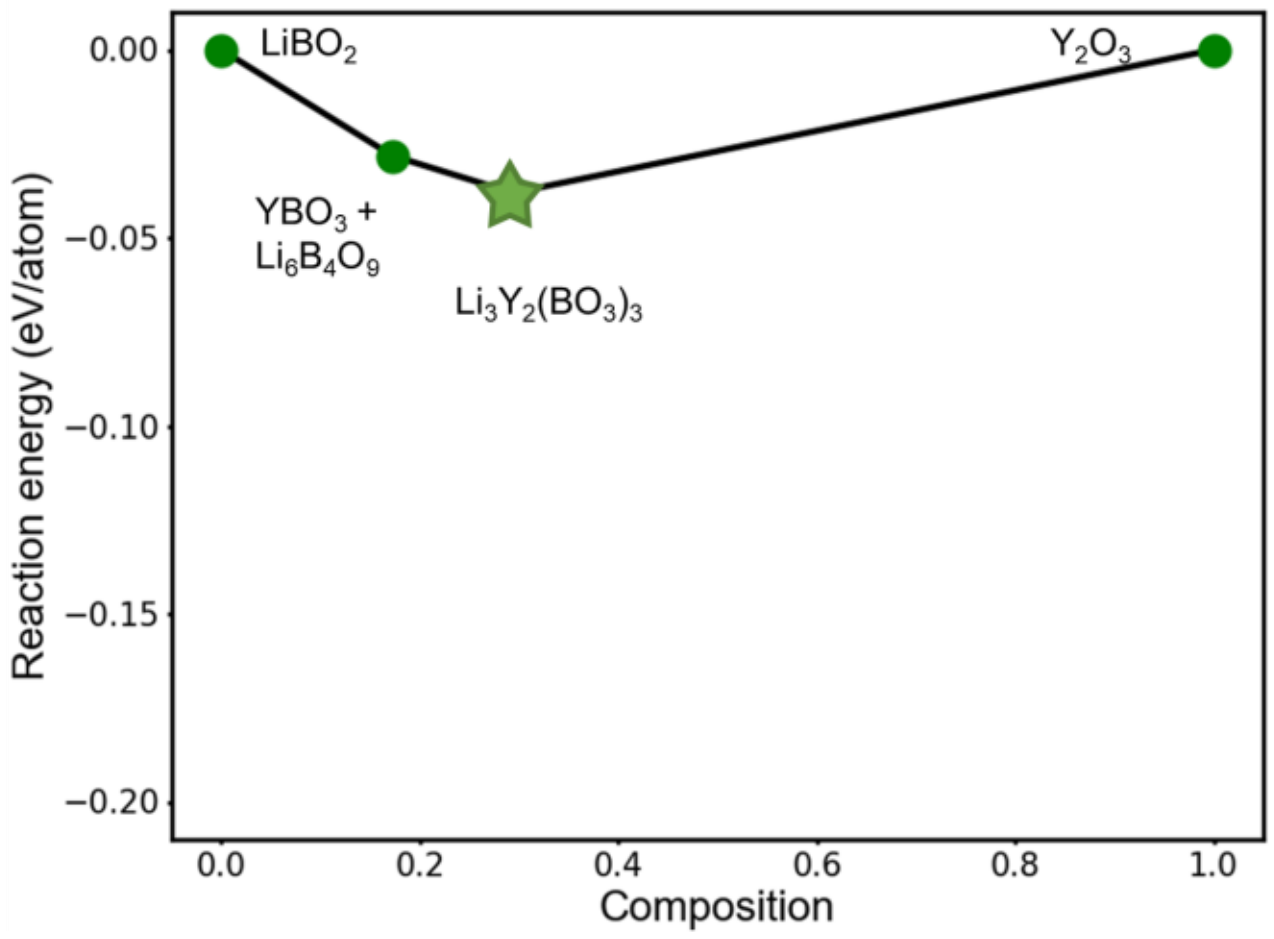

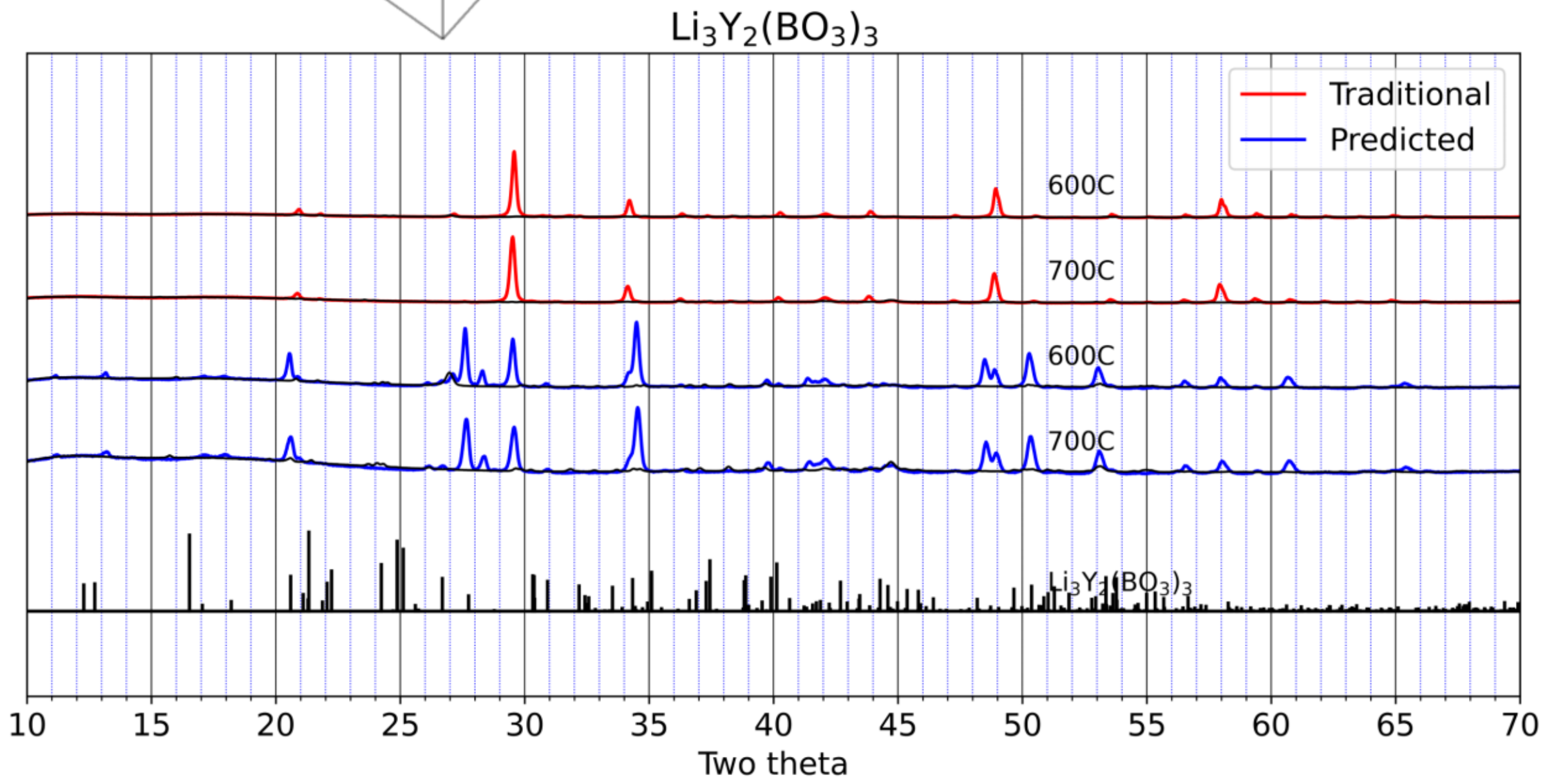

| Compound | Precursor type | Precursors | Temp (°C) | Target intensity (e6) | Residual intensity (e6) | Target phase fraction |
|---|---|---|---|---|---|---|
| $Li_3Y_2(BO_3)_3$ | Traditional | $B_2O_3$, $Li_2CO_3$, $Y_2O_3$ | 600 | 0.56 | 4.00 | 0.12 |
| | | | 700 | 0.20 | 2.88 | 0.07 |
| | Predicted | $LiBO_2$, $Y_2O_3$ | 600 | 0.40 | 3.56 | 0.10 |
| | | | 700 | 0.52 | 2.45 | 0.17 |

## $Na_2Al_2B_2O_7$

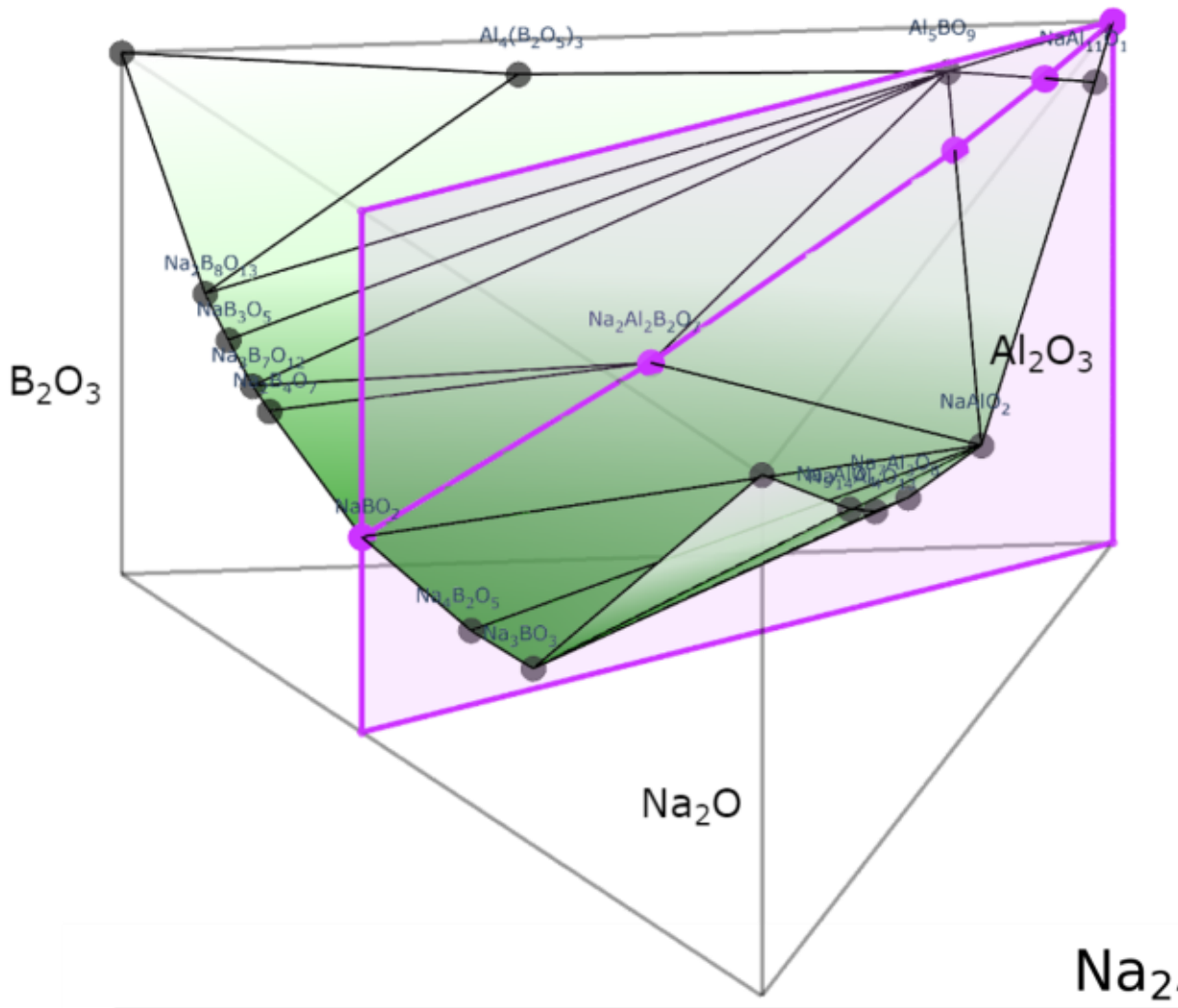

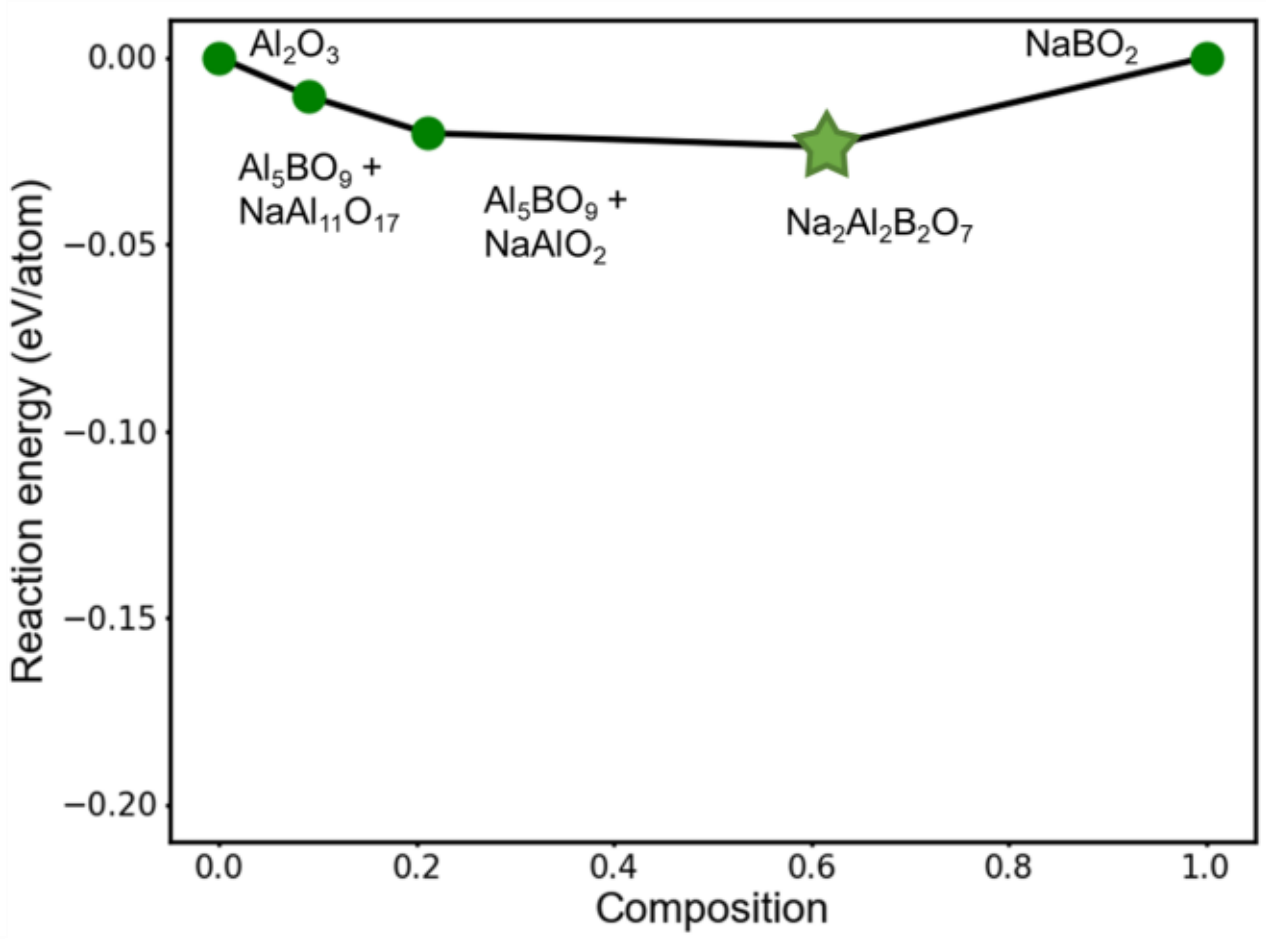

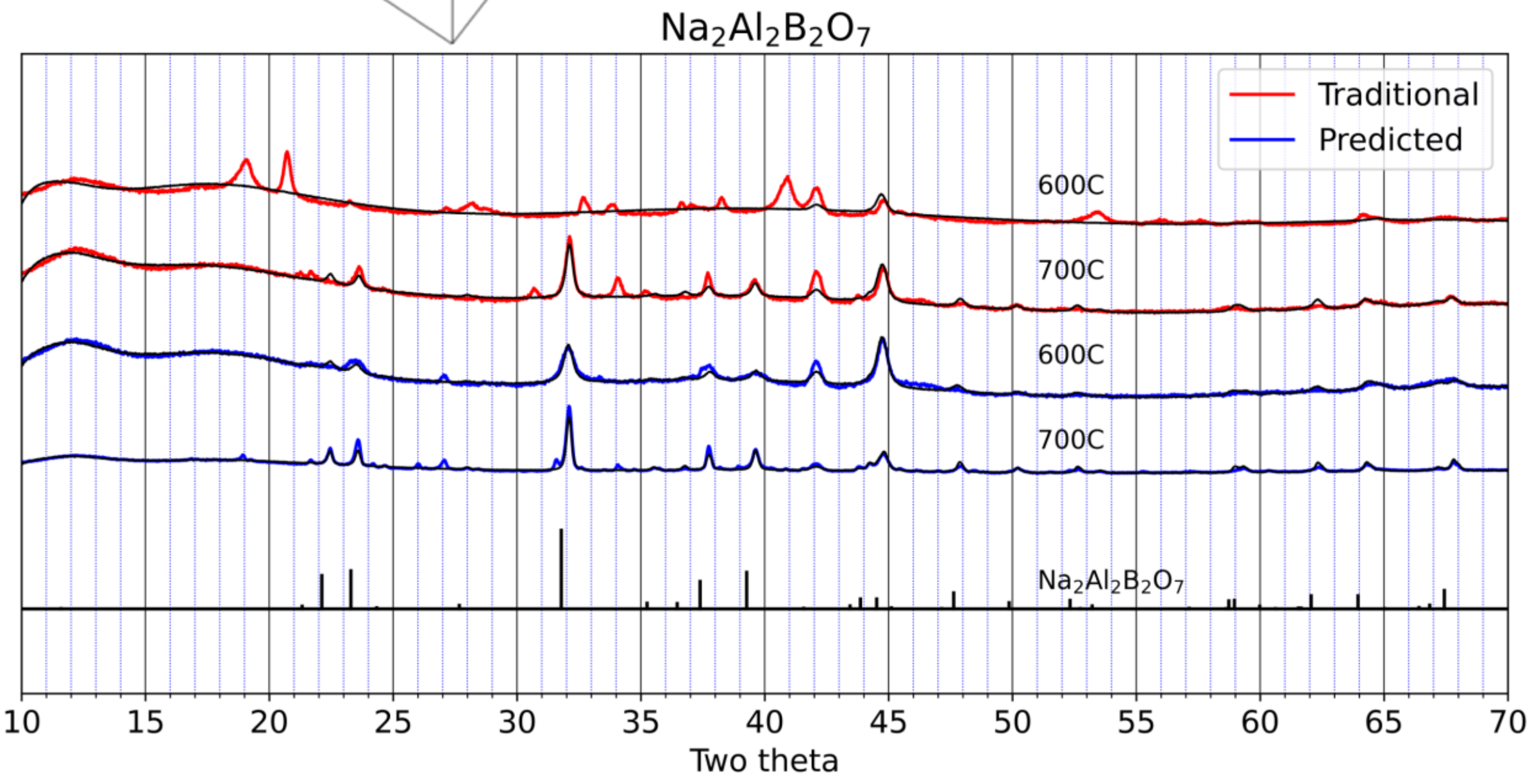

| Compound | Precursor type | Precursors | Temp (°C) | Target intensity (e6) | Residual intensity (e6) | Target phase fraction |
|---|---|---|---|---|---|---|
| $Na_2Al_2B_2O_7$ | Traditional | $Al_2O_3$, $B_2O_3$, $Na_2CO_3$ | 600 | 0.00 | 1.11 | 0.00 |
| | | | 700 | 0.46 | 0.54 | 0.46 |
| | Predicted | $Al_2O_3$, $NaBO_2$ | 600 | 0.48 | 0.43 | 0.52 |
| | | | 700 | 1.28 | 0.84 | 0.60 |

## $NaSiBO_4$

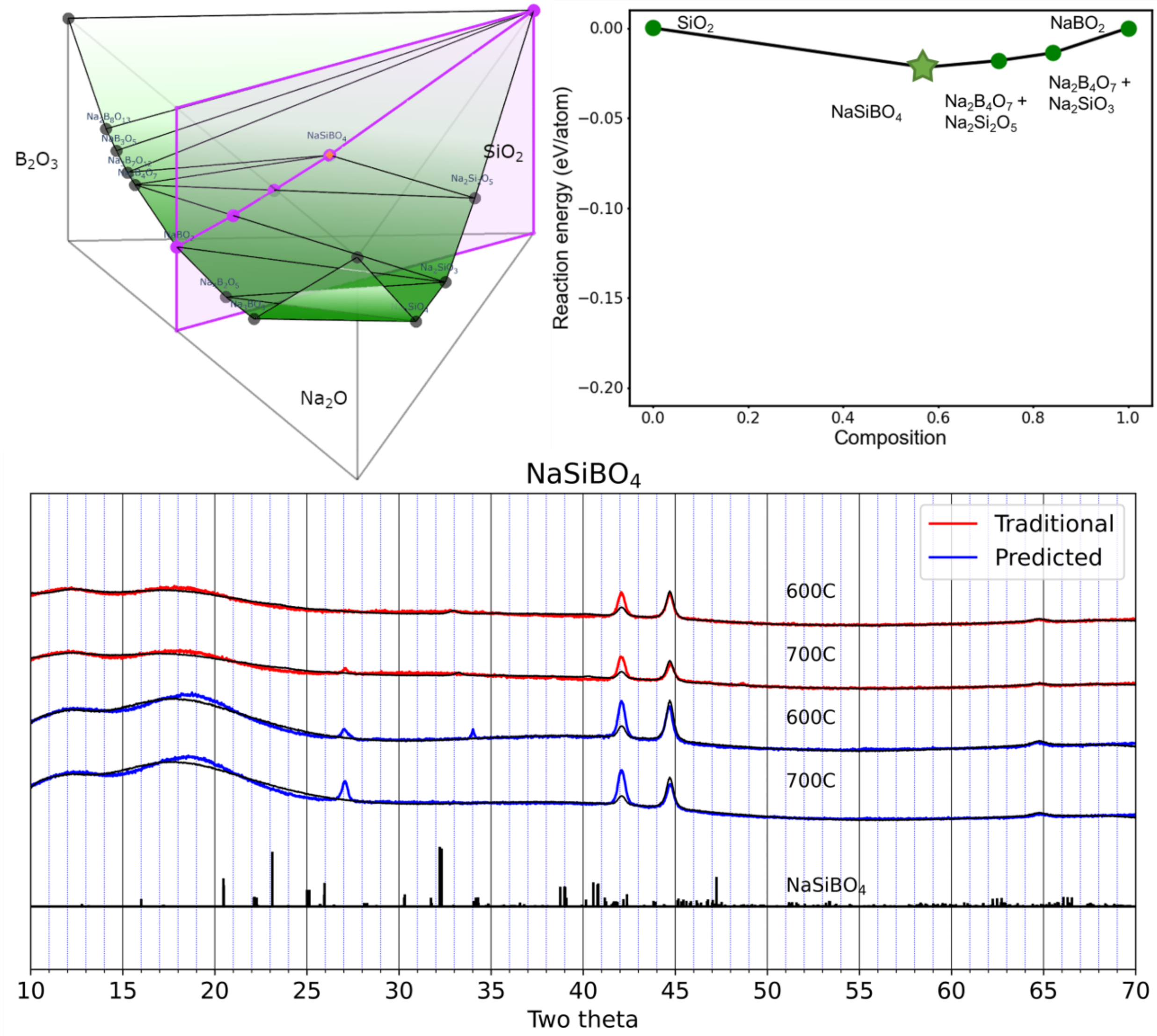

| Compound | Precursor type | Precursors | Temp (°C) | Target intensity (e6) | Residual intensity (e6) | Target phase fraction |
|---|---|---|---|---|---|---|
| $NaSiBO_4$ | Traditional | $B_2O_3$, $Na_2CO_3$, $SiO_2$ | 600 | 0.09 | 0.40 | 0.18 |
| | | | 700 | 0.08 | 0.45 | 0.15 |
| | Predicted | $NaBO_2$, $SiO_2$ | 600 | 0.00 | 0.70 | 0.00 |
| | | | 700 | 0.00 | 0.79 | 0.00 |

# $KTiNbO_5$

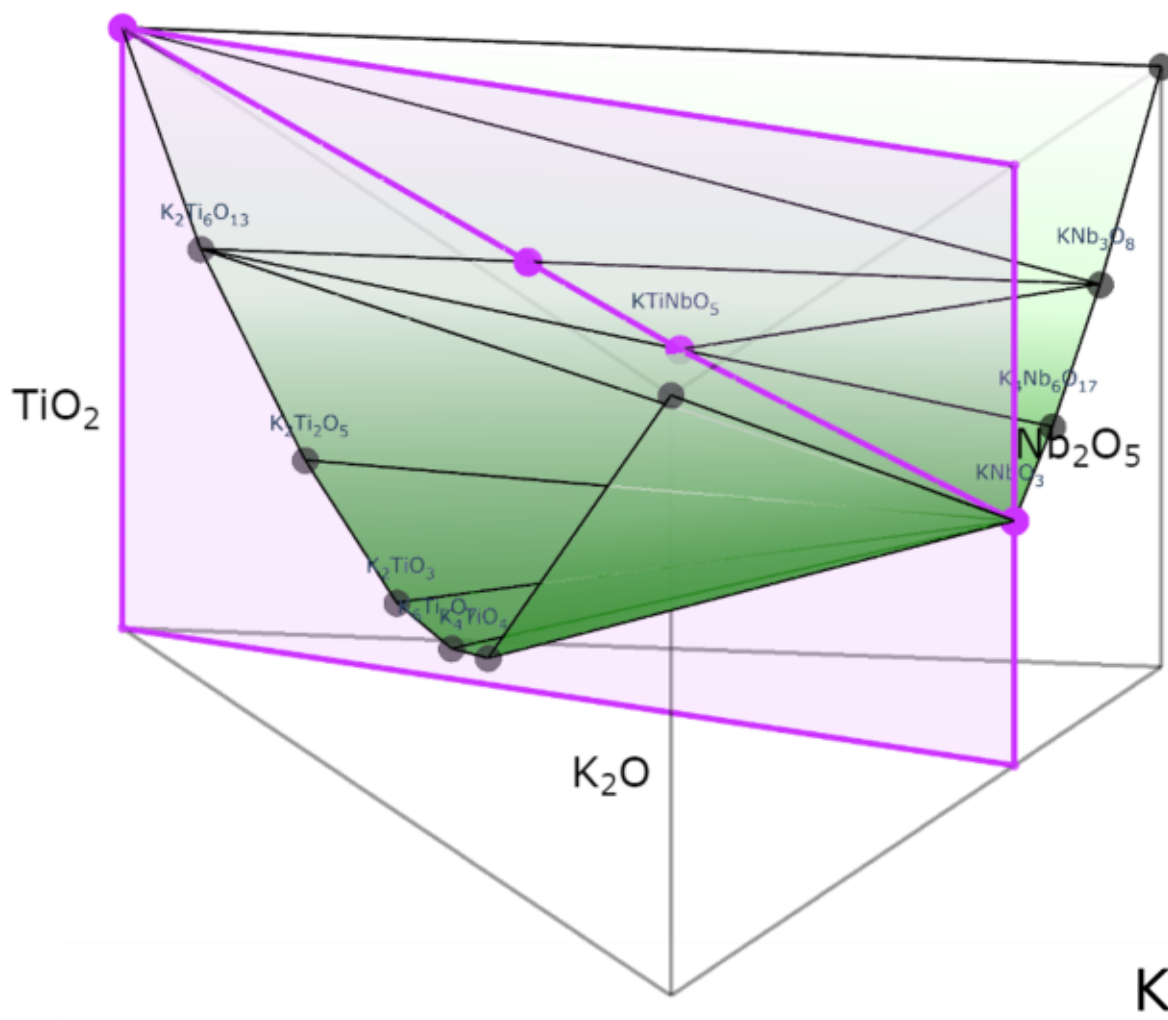

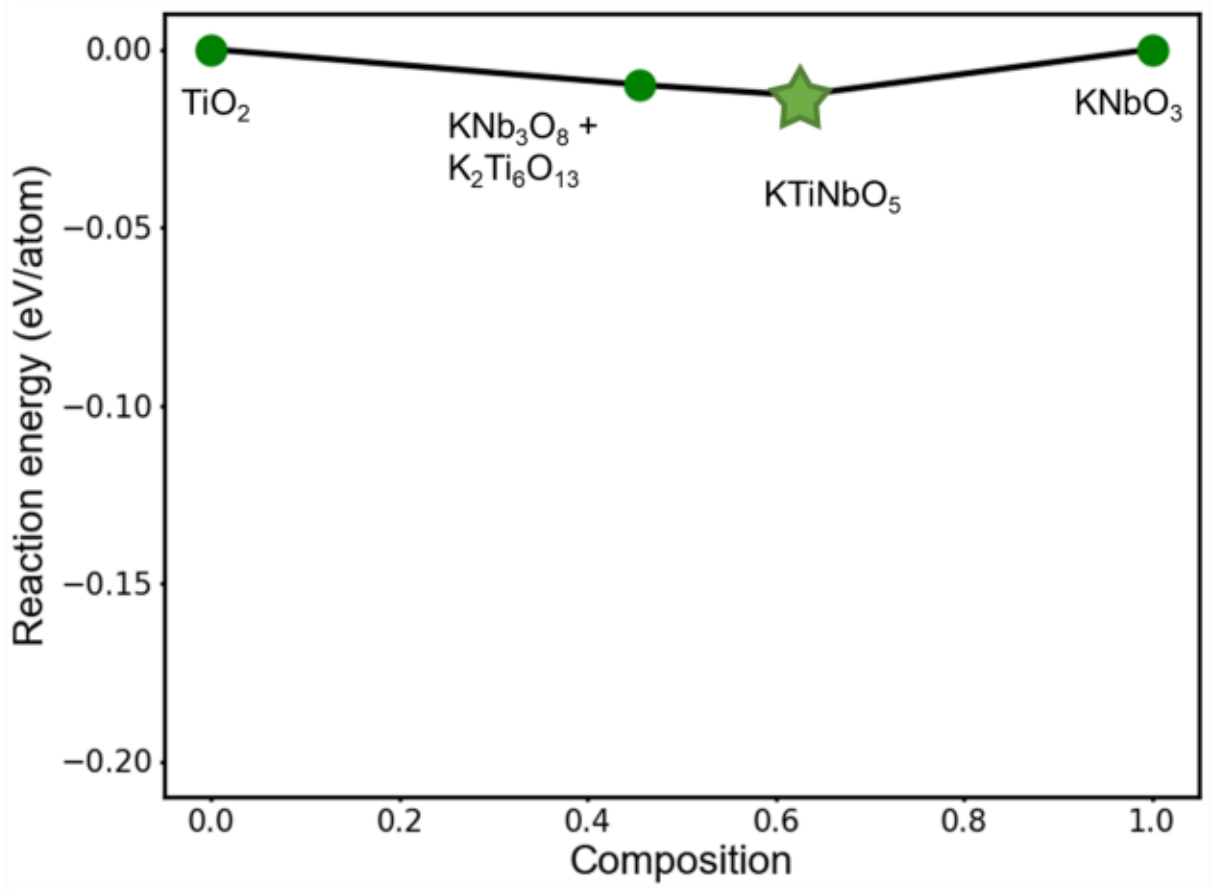

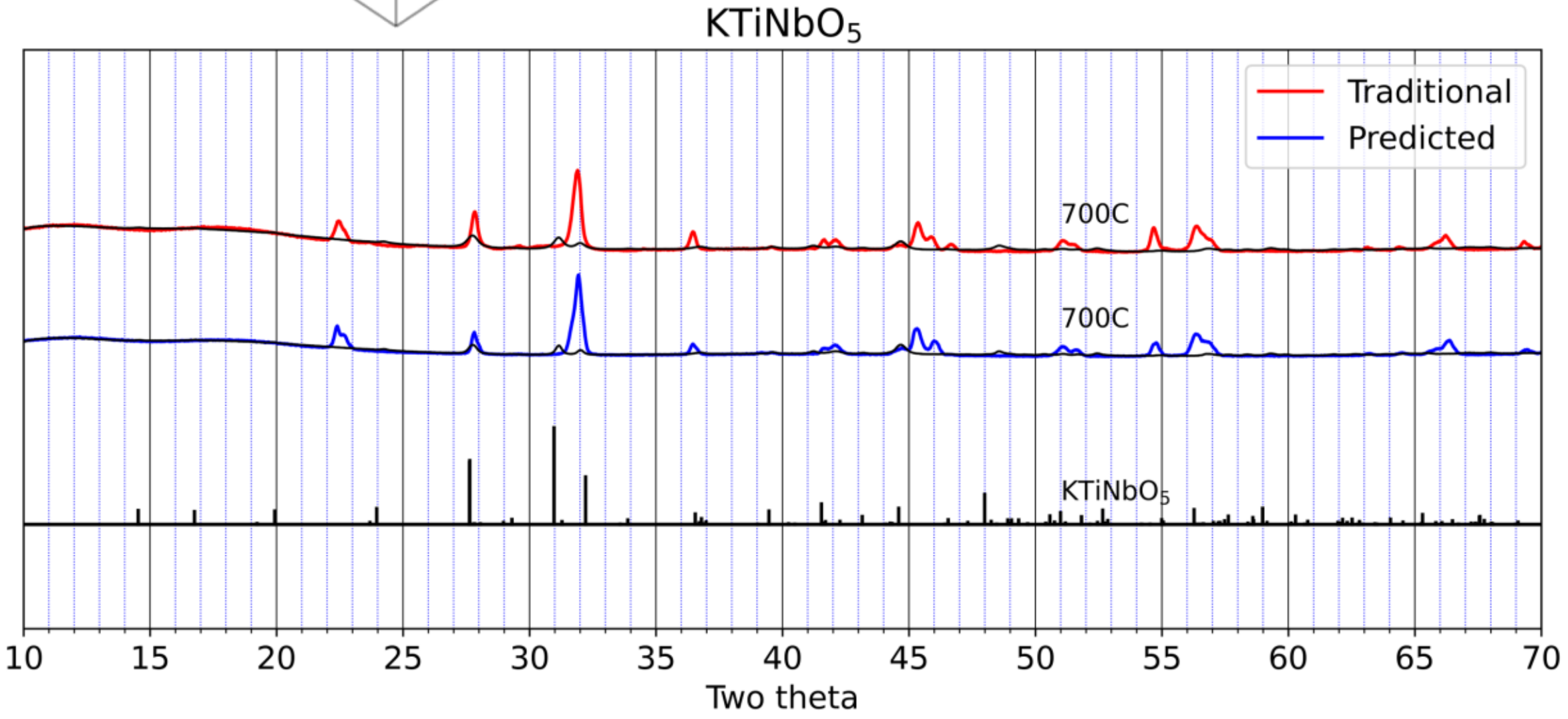

| Compound | Precursor type | Precursors | Temp (°C) | Target intensity (e6) | Residual intensity (e6) | Target phase fraction |
|---|---|---|---|---|---|---|
| $KTiNbO_5$ | Traditional | $K_2CO_3$, $Nb_2O_5$, $TiO_2$ | 700 | 0.78 | 2.06 | 0.27 |
| | Predicted | $KNbO_3$, $TiO_2$ | 700 | 0.50 | 2.21 | 0.18 |

## $LiSi_2BO_6$

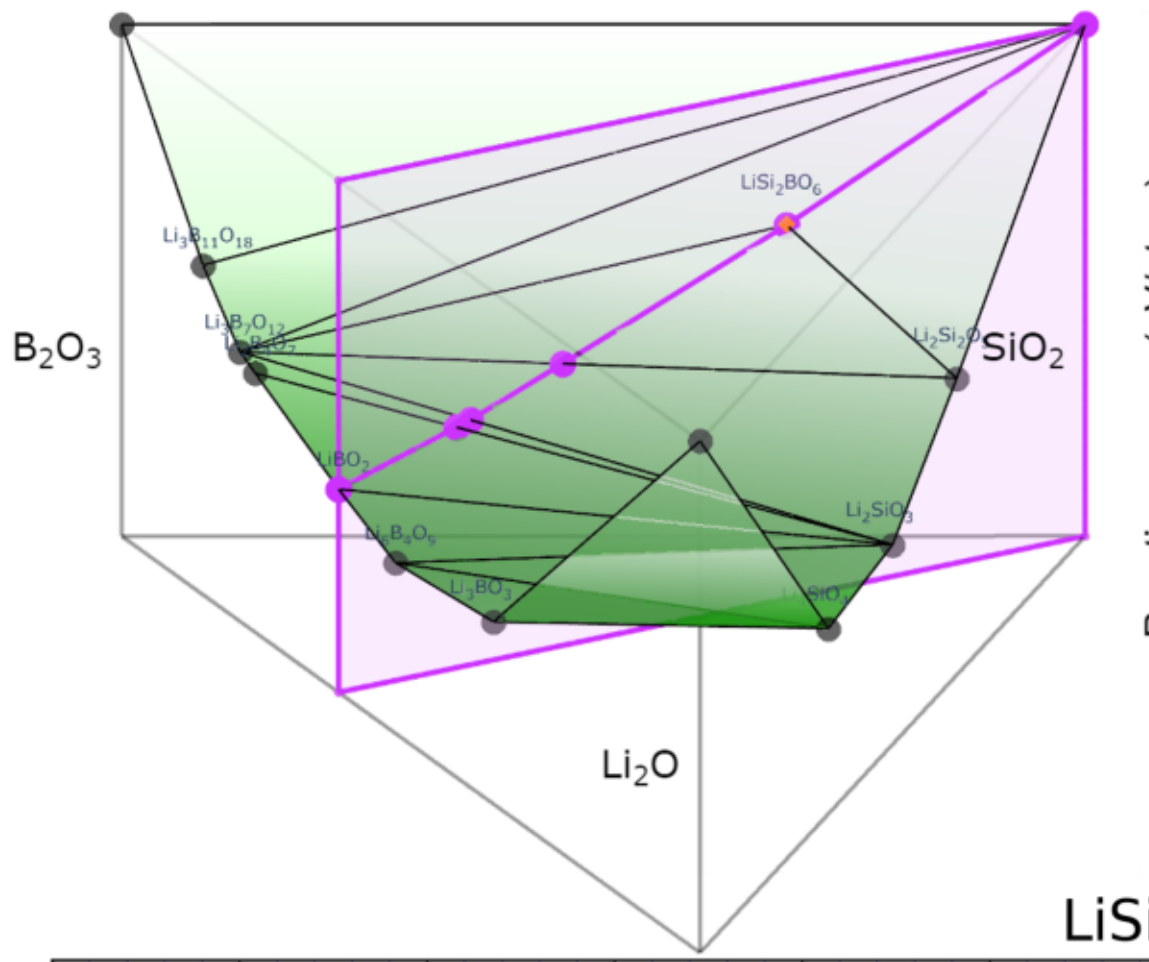

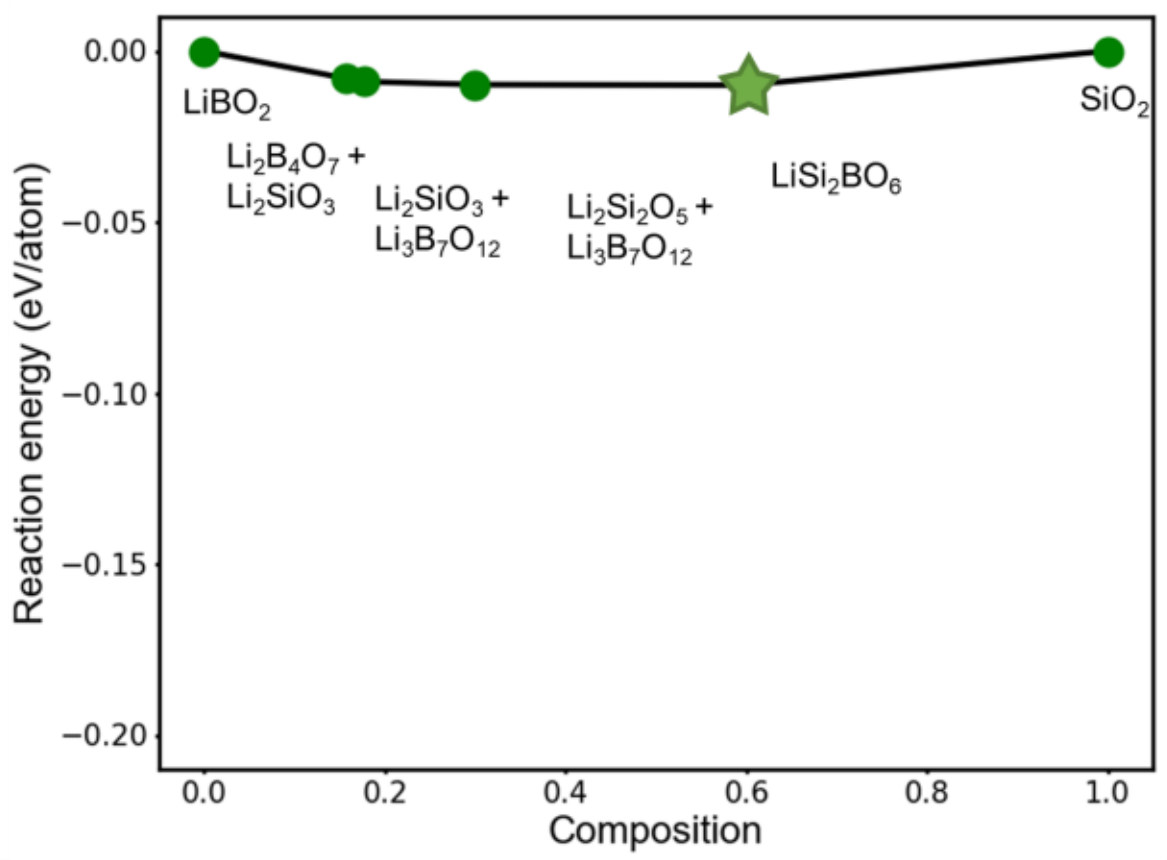

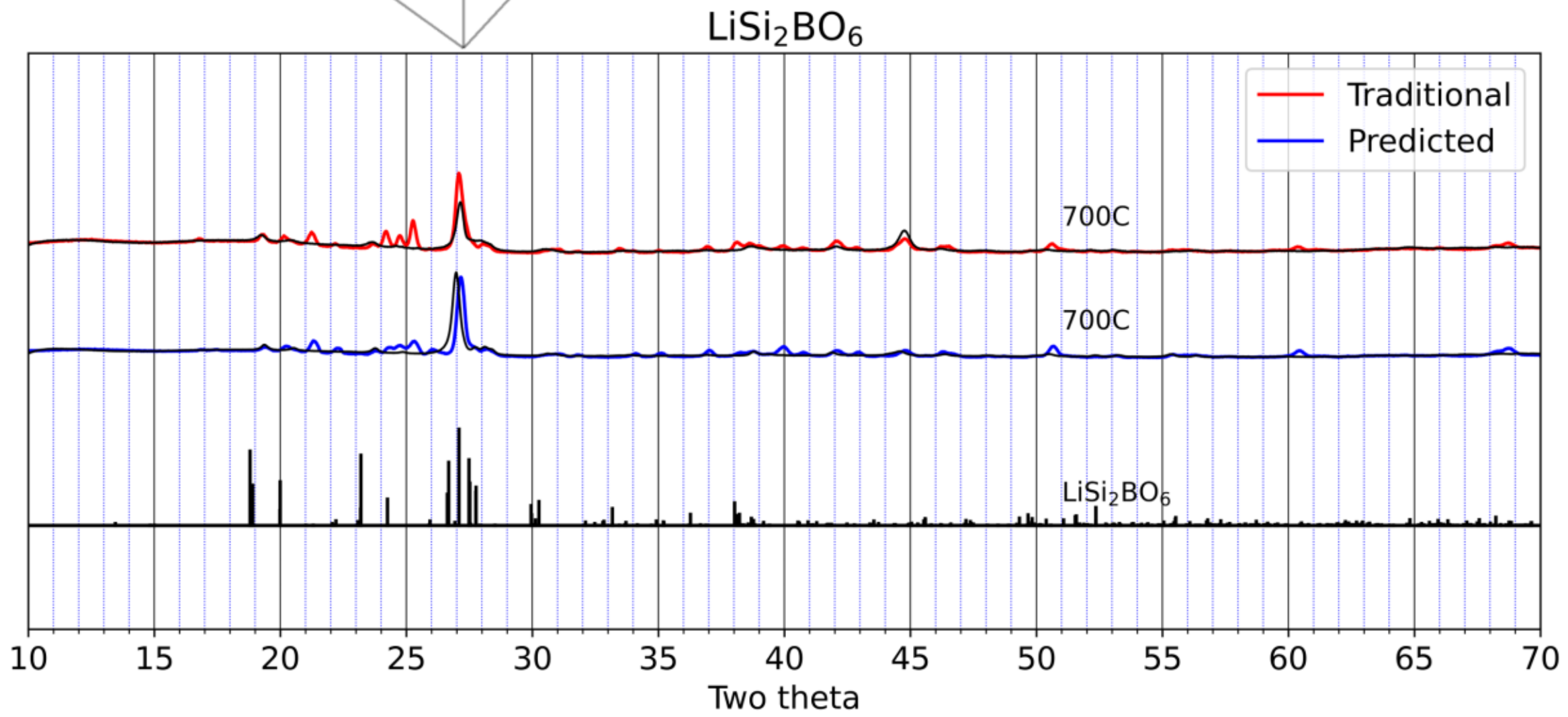

| Compound | Precursor type | Precursors | Temp (°C) | Target intensity (e6) | Residual intensity (e6) | Target phase fraction |
|---|---|---|---|---|---|---|
| $LiSi_2BO_6$ | Traditional | $B_2O_3$, $Li_2CO_3$, $SiO_2$ | 700 | 1.34 | 1.59 | 0.46 |
| | Predicted | $LiBO_2$, $SiO_2$ | 700 | 0.94 | 2.56 | 0.27 |

## $LiNbWO_6$

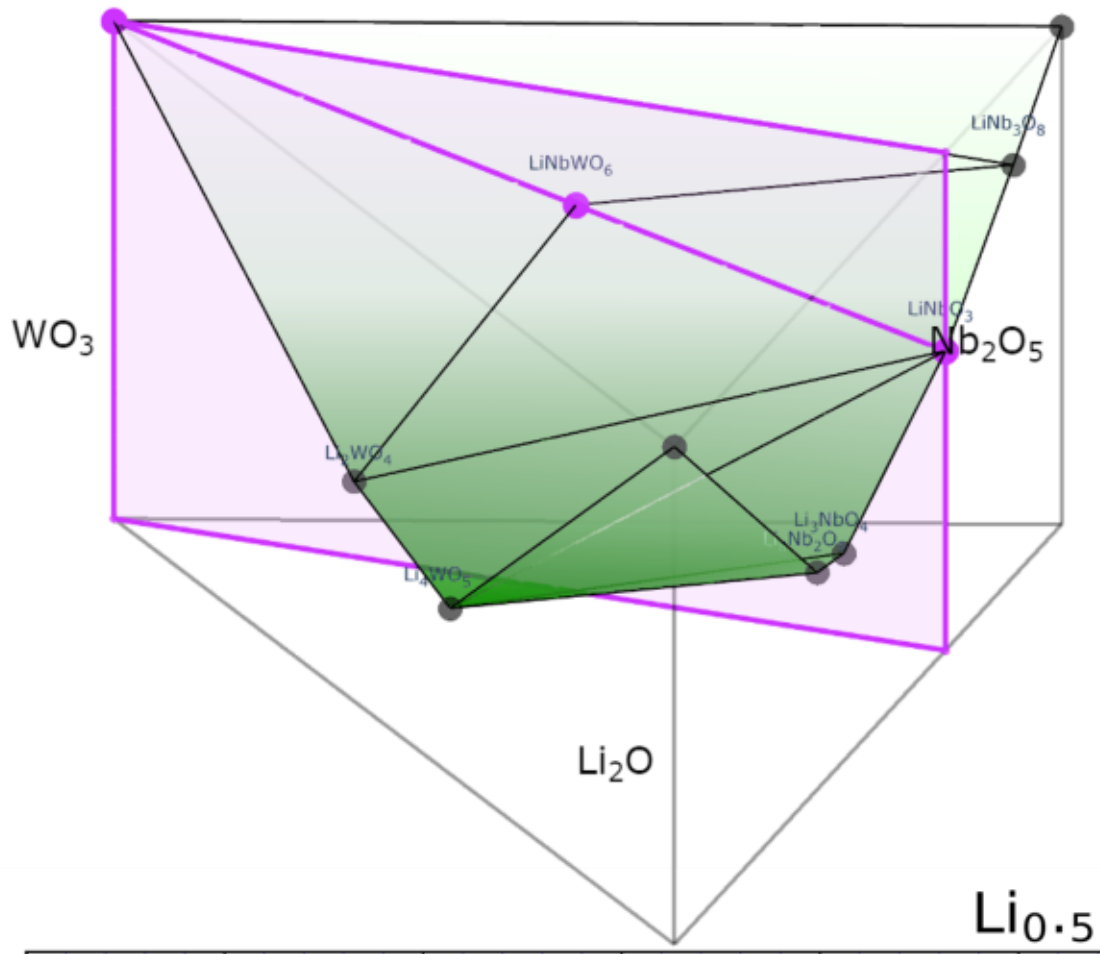

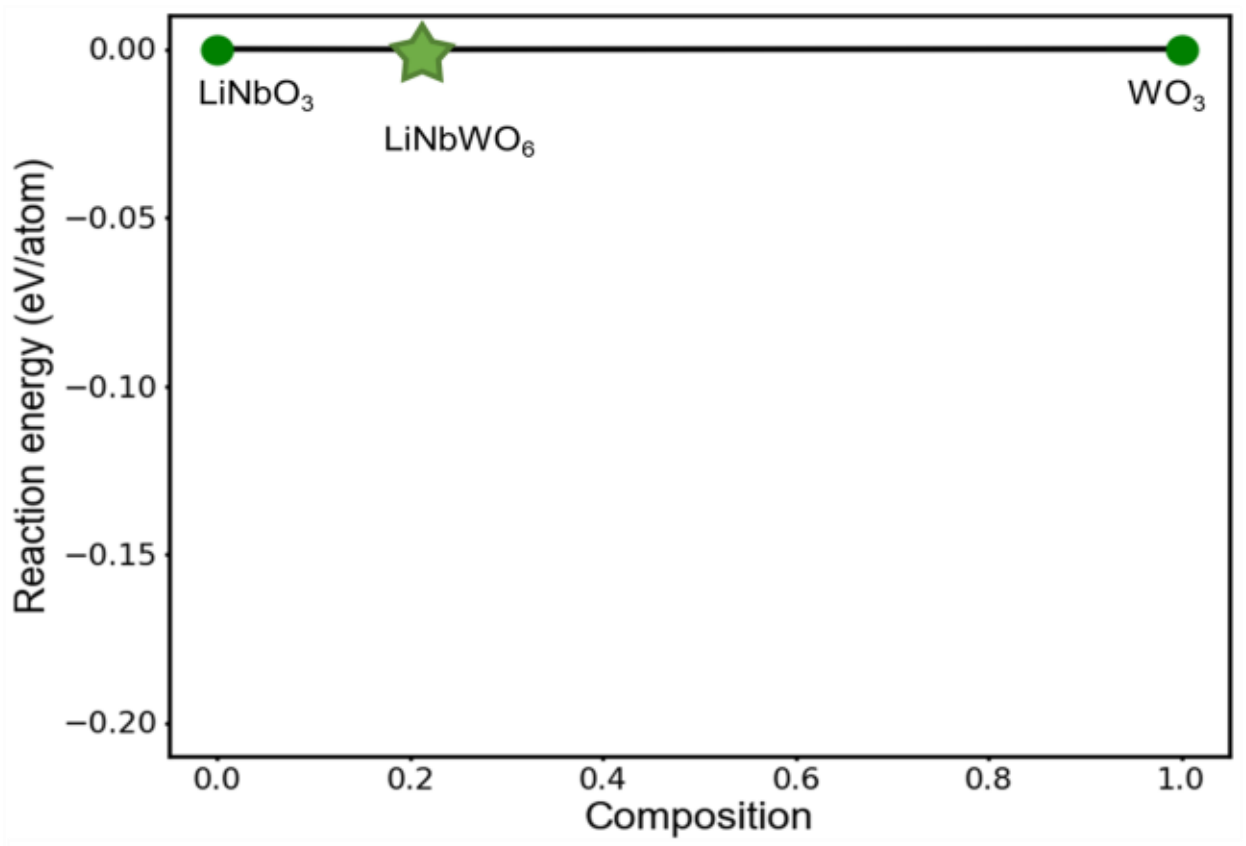

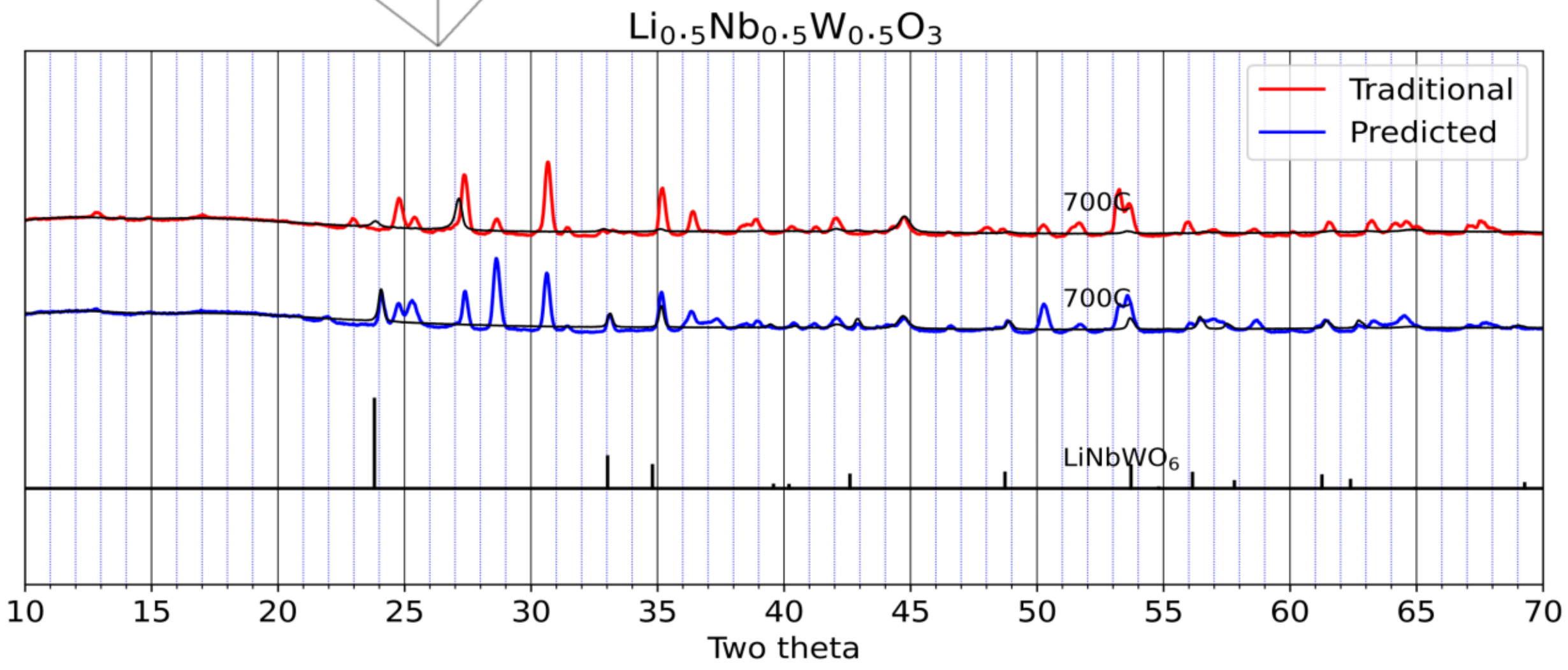

| Compound | Precursor type | Precursors | Temp (°C) | Target intensity (e6) | Residual intensity (e6) | Target phase fraction |
|---|---|---|---|---|---|---|
| $LiNbWO_6$ | Traditional | $Li_2CO_3$, $Nb_2O_5$, $WO_3$ | 700 | 0.14 | 2.52 | 0.05 |
| | Predicted | $LiNbO_3$, $WO_3$ | 700 | 0.41 | 2.05 | 0.17 |

## $LiZnBO_3$

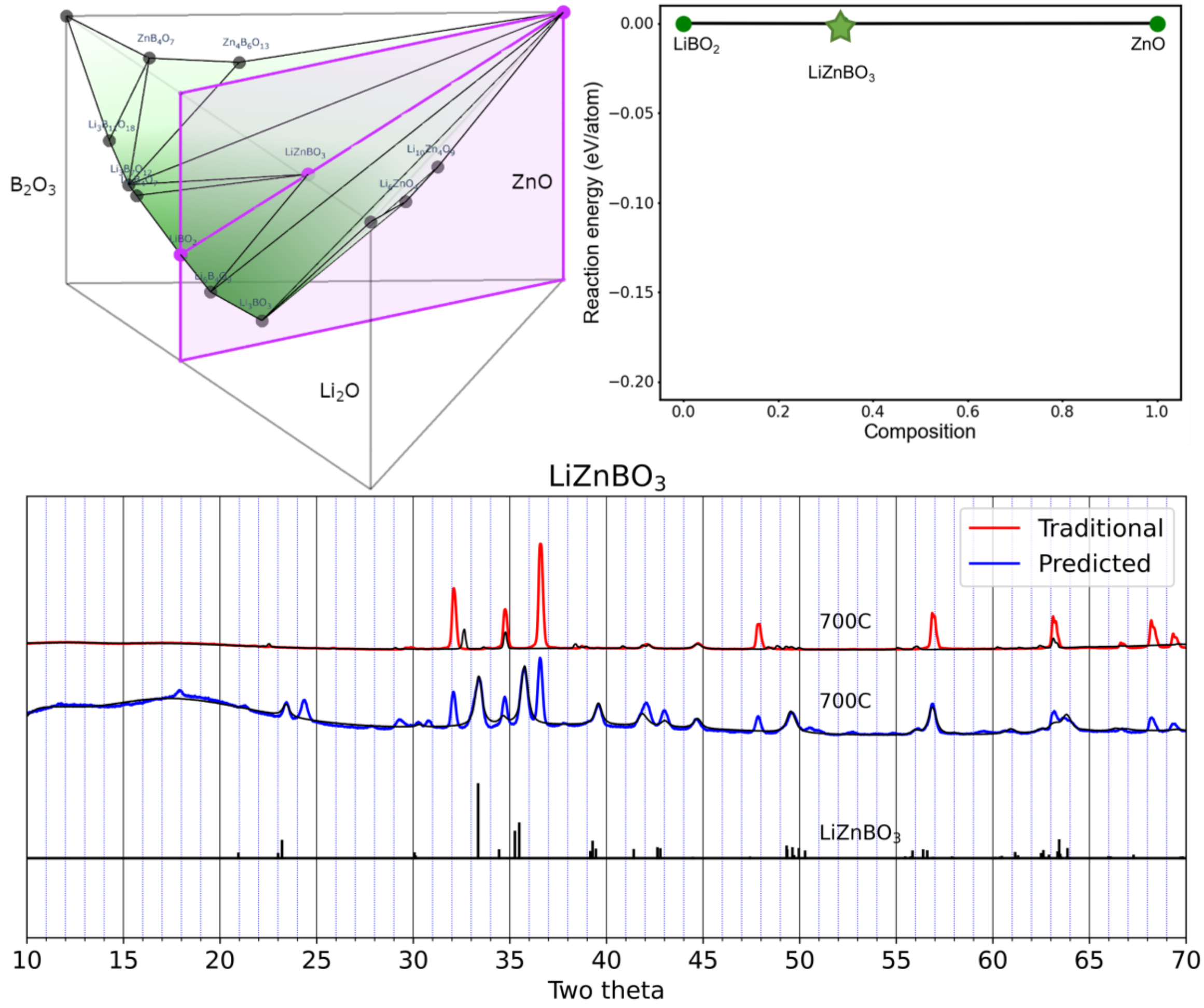

| Compound | Precursor type | Precursors | Temp (°C) | Target intensity (e6) | Residual intensity (e6) | Target phase fraction |
|---|---|---|---|---|---|---|
| $LiZnBO_3$ | Traditional | $B_2O_3$, $Li_2CO_3$, ZnO | 700 | 0.74 | 4.07 | 0.15 |
| | Predicted | $LiBO_2$, ZnO | 700 | 2.11 | 1.93 | 0.52 |